\documentclass[preprint,notoc]{JHEP3}
\usepackage{axodraw}

\newif\ifUglyJHEPoperators
\UglyJHEPoperatorstrue

\ifUglyJHEPoperators
\def\PlusBreak#1{+ \nonumber \\          
          &&  \hphantom{#1} \! \null   +}
\def\MinusBreak#1{- \nonumber \\         
          &&  \hphantom{#1} \!  \null  -}
\def\TimesBreak#1{\times \nonumber \\    
          &&  \hphantom{#1} \!  \null  \times}

\def\PlusBreakLab#1#2{+ \label{#1} \\      
          &&  \hphantom{#2} \! \null   +}

\def\TimesBreakLab#1#2{\times \label{#1} \\      
          &&  \hphantom{#2} \! \null   \times}

\else
\def\PlusBreak#1{\nonumber \\          
          &&  \hphantom{#1} \! \null   +}
\def\MinusBreak#1{\nonumber \\         
          &&  \hphantom{#1} \!  \null  -}
\def\TimesBreak#1{\nonumber \\    
          &&  \hphantom{#1} \!  \null  \times}

\def\PlusBreakLab#1#2{\label{#1} \\      
          &&  \hphantom{#2} \! \null   +}

\def\TimesBreakLab#1#2{\label{#1} \\      
          &&  \hphantom{#2} \! \null   \times}
\fi

\newbox\charbox
\newbox\slabox
\newbox\ourfigbox
\def\s#1{{      
        \setbox\charbox=\hbox{$#1$}
        \setbox\slabox=\hbox{$/$}
        \dimen\charbox=\ht\slabox
        \advance\dimen\charbox by -\dp\slabox
        \advance\dimen\charbox by -\ht\charbox
        \advance\dimen\charbox by \dp\charbox
        \divide\dimen\charbox by 2
        \raise-\dimen\charbox\hbox to \wd\charbox{\hss/\hss}
        \llap{$#1$}
}}

\def\SizedFigureWithCaption#1#2#3{%
\setbox\ourfigbox
  \hbox{\epsfysize #1 \epsfbox{#2}}
\hbox to \wd\ourfigbox{\vbox{\noindent\copy\ourfigbox\break
\vskip -6mm      \hbox to \wd\ourfigbox{\hss#3\hss}}}
}

\def\Tr{\mathop{\rm Tr}\nolimits}
\def\si{\sigma}
\def\L{{\Bigl(}}
\def\R{{\Bigr)}}

\def\BigglBl{\Biggl[}
\def\BiggrBr{\Biggr]}
\def\bigglB{\biggl[}

\def\bigglP{\biggl(}
\def\biggrP{\biggr)}
\def\BiglB{\Bigl[}

\def\ksl{\not{\hbox{\kern-2.3pt $k$}}}

\def\cg{c_\Gamma}
\def\cgh{\hat c_\Gamma}
\def\e{\epsilon}
\def\eps{\epsilon}

\def\deltaplus{\delta^{+}}
\def\Prop{D}
\def\lcproj{d}
\def\tree{{(0)}}
\def\oneloop{{(1)}}
\def\twoloop{{(2)}}

\def\Lloop{{(L)}}
\def\lloop{{(l)}}
\def\Llloop{{(L-l)}}
\def\Ord{{\cal O}}
\def\rsn{r_S}
\def\rsnren{{\overline r}_S}
\def\Neqfour{{{\cal N}=4}}
\def\Neqone{{{\cal N}=1}}
\def\ca{{\cal A}}

\def\Nc{N_c}

\def\Nf{{N_{\! f}}}
\def\Nfsq{{N_{\! f}^2}}
\def\la{\langle}
\def\ra{\rangle}
\def\RS{{\scriptscriptstyle\rm R\!.S\!.}}

\def\bom#1{{\mbox{\boldmath $#1$}}}
\def\MSbar{\overline{\rm MS}}

\def\FDH{{\rm FDH}}
\def\Absp{\mathop{\rm Absp}}
\def\lr{\leftrightarrow}

\def\Li{\mathop{\rm Li}\nolimits}

\def\alphas{\alpha_s}

\def\Eqn#1{Equation~(\ref{#1})}
\def\Eqns#1#2{Equations~(\ref{#1}) and~(\ref{#2})}
\def\eqn#1{eq.~(\ref{#1})}
\def\eqns#1#2{eqs.~(\ref{#1}) and~(\ref{#2})}

\def\spa#1.#2{\left\langle#1\,#2\right\rangle}
\def\spb#1.#2{\left[#1\,#2\right]}
\def\lor#1.#2{\left(#1\,#2\right)}
\def\sand#1.#2.#3{%
\left\langle\smash{#1}{\vphantom1}^{-}\right|{#2}%
\left|\smash{#3}{\vphantom1}^{-}\right\rangle}

\def\pol{\varepsilon}
\def\fig#1{figure~{\ref{#1}}}
\def\Fig#1{Figure~{\ref{#1}}}

\def\sect#1{section~{\ref{#1}}}
\def\Sect#1{Section~\ref{#1}}

\def\Sect#1{Section~\ref{#1}}

\def\app#1{appendix~{\ref{#1}}}

\def\iscol#1#2{{\buildrel#1\parallel#2\over\longrightarrow}}

\def\Split{\mathop{\rm Split}\nolimits}
\def\SplitRen{\mathop{\rm Split}_R\nolimits}
\def\SplitOp{\mathop{\bf Split}\nolimits}
\def\Gr{{\rm Gr}}
\def\n{n}    
\def\F{ {\bom f}_{\! P}}

\def\del{\partial}
\def\Spec{{\rm Spec}}
\def\Btie{{\rm Btie}}
\def\Sset{{\rm Sset}}
\def\Btri{{\rm Btri}}
\def\Fish{{\rm Fish}}
\def\Wedge{{\rm Wedge}}
\def\WedgeF{{\rm WedgeF}}
\def\LBtri{{\rm LBtri}}
\def\Zig{{\rm Zig}}
\def\Ptri{{\rm Ptri}}    
\def\LPBWedge{{\rm LPBWedge}}  
\def\LPBDtri{{\rm LPBDtri}}  
\def\FDtri{{\rm FDtri}}
\def\JDtri{{\rm JDtri}}
\def\MDtri{{\rm MDtri}}
\def\ubar{{\overline u}} 

\preprint{
  SLAC--PUB--10414\\
  UCLA/04/TEP/12\\
  Saclay SPhT--T04/051\\
  NSF-KITP-04-43\\
  April, 2004}

\title{ Two-Loop $g \rightarrow gg$ Splitting Amplitudes in QCD }

\author{Zvi Bern\thanks{Research supported by the US Department of 
        Energy under grant DE-FG03-91ER40662.} \\
	Department of Physics and Astronomy, UCLA \\
	Los Angeles, CA 90095-1547, USA\\
	E-mail: \email{bern@physics.ucla.edu}}

\author{Lance J. Dixon\thanks{Research supported by the US Department of 
        Energy under grant DE-AC03-76SF00515.}\\
	Stanford Linear Accelerator Center, Stanford University\\
	Stanford, CA 94309, USA\\
	E-mail: \email{lance@slac.stanford.edu}}

\author{David A. Kosower \\
	Service de Physique Th\'eorique, CEA--Saclay\\
	F-91191 Gif-sur-Yvette cedex, France \\
	E-mail: \email{kosower@spht.saclay.cea.fr}}

\abstract{ Splitting amplitudes are universal functions governing the
collinear behavior of scattering amplitudes for massless particles.
We compute the two-loop $g \rightarrow gg$ splitting amplitudes in
QCD, $\Neqone$, and $\Neqfour$ super-Yang-Mills theories, which describe the
limits of two-loop $n$-point amplitudes where two gluon momenta become
parallel.  They also represent an ingredient
in a direct $x$-space computation of DGLAP evolution kernels at
next-to-next-to-leading order.  To obtain the splitting
amplitudes, we use the unitarity sewing method.  In contrast to the
usual light-cone gauge treatment, our calculation does not rely on the
principal-value or Mandelstam-Leibbrandt prescriptions, even though
the loop integrals contain some of the denominators typically
encountered in light-cone gauge.  We reduce the integrals to a set of
13 master integrals using integration-by-parts and Lorentz invariance
identities.  The master integrals are computed with the aid of
differential equations in the splitting momentum fraction $z$.  The
$\e$-poles of the splitting amplitudes are consistent with a
formula due to Catani for the infrared singularities of two-loop
scattering amplitudes.  This consistency essentially provides an
inductive proof of Catani's formula, as well as an ansatz for
previously-unknown $1/\e$ pole terms having non-trivial color structure.
Finite terms in the splitting amplitudes determine the collinear behavior 
of finite remainders in this formula. }


\newif\ifWithFigs
\WithFigstrue

\begin{document}

\section{Introduction}
\label{IntroSection}

Gauge theories form the backbone of the standard 
${\rm SU}(3)\times {\rm SU}(2)\times {\rm U}(1)$ model of particle
interactions.  The computation of perturbative corrections in gauge
theories is thus central to testing the standard model at high-energy
colliders.  Such computations are technically complicated, so a general 
understanding of properties of the results is very useful.

In the past decades, a number of new approaches have been developed to
cope with this complexity, including helicity
methods~\cite{SpinorHelicity}, color decompositions~\cite{%
Color,TreeSixPoint,ManganoParkeXu,OneloopColor,DDDM}, 
recursion relations~\cite{Recursive}, ideas based on string
theory~\cite{BKgggg,CSW}, and the unitarity-based
method~\cite{Neq4Oneloop,Neq1Oneloop,BernMorgan,LoopReview}.  The latter
technique has been applied to numerous calculations, most recently the
two-loop calculation of all helicity amplitudes for gluon--gluon
scattering~\cite{AllPlusTwo,BDDgggg}.

The subject of two-loop calculations has seen tremendous technical
progress in the last five years.  Much of the progress has been facilitated
by new techniques for performing loop integrals.
Smirnov~\cite{PBScalar} and Tausk~\cite{NPBScalar} gave closed-form
expressions for the all-massless planar and non-planar double box 
integrals, respectively.
Smirnov and Veretin~\cite{PBReduction}
and Anastasiou {\it et al.}~\cite{NPBReduction} provided algorithms
for reducing the corresponding tensor integrals.  
More general reduction and evaluation
techniques for integrals have followed as
well~\cite{Lorentz,GRMass,GRMassCont,Laporta,NewerIntegralTechniques}. 
Using these techniques, several groups have computed the basic 
two-loop QCD (and also QED) amplitudes for four external partons~\cite{%
BhabhaTwoloop,GOTYqqqq,GOTYqqgg,GOTYgggg,gggamgamPaper,PhotonPaper,%
AGTYphotons,BGMvdB,BDDgggg,BDDqqgg,GTYqqgg,Gloverqqqq}, 
and for three partons and one external vector 
boson~\cite{Twoloopee3Jets,Twoloopee3JetsHel,GRMassCont}.
These amplitudes and matrix elements constitute one of the building blocks 
for next-to-next-to-leading order (NNLO) computations in perturbative QCD, 
in particular of the cornerstone processes 
$e^+ e^- \rightarrow 3{\rm\ jets}$ and 
$p{\overline p} \rightarrow 2{\rm\ jets}$.

The two-loop amplitudes contain both ultraviolet and infrared singularities.
Both are conventionally regulated using dimensional regularization.
The infrared singularities correspond to the circulation of gluons
in the loops that are nearly on shell and either soft or collinear with
one of the external momenta (or to the circulation of other particles
in the loops that are nearly on shell and collinear with one
of the external momenta).  The form and functional dependence of
these singularities was predicted by Catani~\cite{CataniTwoloop}.  The
{\it a priori\/} knowledge of these infrared singularities has been
of great value in the explicit computations of two-loop amplitudes cited
above.  

The singularities in the virtual corrections have a counterpart in
the infrared singularities of phase-space integrals 
of real-emission amplitudes.  In the sum over all relevant 
(physically indistinguishable) amplitudes, unitarity in the form of
the Kinoshita--Lee--Nauenberg theorem dictates that the singularities
must cancel.  The universality of the singularities in the integrals
over real-emission amplitudes in turn reflects the universality of
factorization of real-emission amplitudes in soft or collinear limits.
For next-to-next-to-leading order (NNLO) calculations, the double-emission
limits of tree amplitudes, and the single-emission limits of
one-loop amplitudes, allow one to organize the singular phase-space
integrations in a process-independent way.
The universal functions governing these limits are all 
known~\cite{DoubleSoft,CampbellGlover,CataniGrazzini,%
BernKilgoreSplit,KosowerUwerSplit,WeinzierlSingleUnres}.

In this paper, we will compute the analogous functions,
{\it splitting amplitudes\/}, governing the
universal behavior of two-loop amplitudes as two gluon momenta
become collinear.  For example, in the time-like case where
momenta $k_a$ and $k_b$ are both outgoing, we let $k_a \to z k_P$, 
$k_b \to (1-z) k_P$, where $k_P = k_a+k_b$ is the momentum of 
the nearly on-shell intermediate gluon $P$, and $z$ is the 
longitudinal (or light-cone) momentum fraction carried by gluon $a$.
Splitting amplitudes will be useful
in checking two-loop computations beyond four external legs. 
A putative result for a higher-point amplitude must satisfy non-trivial
constraints as momenta become collinear.
They can also play a role
in an alternative method of computing the NNLO corrections to
the Altarelli--Parisi kernel governing the $Q^2$ evolution of
parton distributions and fragmentation functions~\cite{AP}.   
The computation
of this kernel is of great importance to a program of precision 
extraction of parton distribution functions from experimental data.  It
has been the object of an ongoing effort by Moch, Vermaseren and 
Vogt~\cite{NNLOPDFApprox,MVV}, just recently
completed~\cite{MVVNNLO}.
The Mellin moments of the Altarelli--Parisi kernel are
anomalous dimensions of leading-twist operators whose matrix elements
give rise to parton distributions.  One can compute them in a traditional
manner, by computing ultraviolet divergences of loop corrections.
Factorization implies, however,
that one could compute the kernels in an infrared
approach, directly in $x$-space~\cite{KosowerUwerKernel,deFG}.  
As in the computation of differential cross sections, there are both 
`virtual' and `real-emission' contributions to the kernel.  
The splitting amplitudes we compute in the present paper provide the 
doubly-virtual contributions to the NNLO kernel for evolution
of the gluon distribution, $P_{gg}^{(2)}(x)$ (at $x\neq 1$).
Because the gluons have definite helicity, evolution of polarized
distributions is equally accessible.

The computation of the Altarelli--Parisi kernel in $\Neqfour$
supersymmetric gauge theories is also of interest, since its Mellin
moments are the anomalous dimensions of classes of operators.  The
study of such anomalous dimensions is important to investigations of
the anti-de~Sitter/conformal field theory duality~\cite{BeisertEtc}.
The NNLO anomalous dimensions were very recently extracted~\cite{Lipatov}
from the corresponding QCD calculation~\cite{MVVNNLO}, by adjusting the
fermion content to match $\Neqfour$ super-Yang-Mills theory, and then
keeping only those terms that are leading in ``transcendentality.''
(In $x$-space, a logarithm or a factor of $\pi$ is assigned one unit of
transcendentality, and an $n^{\rm th}$ degree polylogarithm $\Li_n(x)$
is assigned $n$ units.  The leading-in-transcendentality terms at two loops 
carry 4 units of transcendentality.)
Our results confirm that the leading-transcendentality procedure of 
ref.~\cite{Lipatov} does correctly extract the $\Neqfour$ result
from that of QCD, for the fully virtual NNLO contributions to the 
Altarelli--Parisi kernel, which arise from the two-loop splitting amplitude 
(and from the square of the one-loop splitting amplitude).

We gave the result for the splitting amplitude in $\Neqfour$ supersymmetric 
gauge theory in an earlier Letter~\cite{TwoloopN4}.  
We will document that calculation in the present paper.
The computation revealed an unexpected relation between 
splitting amplitudes at different loop orders:
the two-loop splitting amplitude can be expressed algebraically
in terms of the one-loop and tree splitting amplitudes, 
through ${\cal O}(\e^0)$, where $\e$ is the parameter of dimensional 
regularization, $D=4-2\e$.  In the planar (large-$\Nc$) limit,
the four-point two-loop amplitude can be expressed in a similar
`iterative' form.  This is very surprising
because in a general massless field theory, the analytic structure of 
a two-loop amplitude can be considerably more complicated than that
of one-loop amplitudes.  Thus the two-loop amplitude in this theory 
is much simpler than expected.
The splitting amplitude relation also led to a conjecture of
a similar relation between one- and two-loop amplitudes 
with an arbitrary number of external legs.

To calculate the $g\to gg$ splitting amplitudes,
we have used the unitarity-based method.  The method is useful in 
general loop calculations in
gauge theories.  Its advantage over Feynman-diagram calculations is of
course most obvious in those calculations which simply cannot be done
by conventional techniques, such as that of infinite series of loop
amplitudes~\cite{Neq4Oneloop,Neq1Oneloop,AllNGravity}.  The present
calculation furnishes another example where the method has a clear
advantage over conventional diagrammatic techniques.
While the splitting amplitudes could be computed by conventional 
diagrammatic techniques, such a computation at two loops would 
probably require the use of light-cone gauge, because this gauge 
has simple collinear factorization properties, even in the presence of 
infrared singularities~\cite{CSS}.
(An analysis in a covariant gauge is likely to be very difficult; for example,
a generalization of the analysis in ref.~\cite{BernChalmers} would require
explicit knowledge of higher-point two-loop integrals.)
As is well-documented in the
literature~\cite{Leibbrandt,AndrasiTaylor}, use of light-cone gauge is
fraught with subtleties and technical complications.  Indeed, following
the standard methods for dealing with light-cone gauge Feynman diagrams, 
along with any of the popular
prescriptions needed to avoid ill-defined integrals, would lead to
an answer with a surviving dependence on the prescription parameters,
which cannot describe the collinear behavior of a gauge-invariant
and prescription-parameter-independent amplitude.
We will discuss these issues in more detail in section~\ref{FeynmanSection}.
The unitarity-based method avoids these complications, and ensures that
the calculation can be done in a straightforward way using ordinary
dimensional regularization.  The insights furnished by the unitarity
method also allow the systematic cancellation of ill-defined
integrals appearing in more traditional light-cone gauge
calculations~\cite{FermionSplit}.

To evaluate the loop momentum integrals we used
integration-by-parts~\cite{IBP} and Lorentz invariance~\cite{Lorentz}
identities implemented via the Laporta algorithm~\cite{Laporta,AIR} to
solve the system of equations.  With this technique
the integrals can be reduced
to a set of 13 master integrals.  The master integrals are computed by
constructing a set of differential equations in $z$, along the 
lines of refs.~\cite{PBReduction,Lorentz}.

We will organize the calculation in a color-stripped formalism, in
which the color factors are separated from the kinematic content of
amplitudes.  At loop level, this leads to a hierarchy of terms, from
terms leading in the number of colors, down through subleading
contributions.  The leading-color terms, which would dominate in the
$\Nc\rightarrow\infty$ limit, correspond to planar diagrams.  These
leading-color terms contain only a single color trace, with additional
explicit powers of $\Nc$.  The two-loop splitting amplitude enters
only into the collinear limit of single-trace terms (whether leading
or subleading in color).  The collinear limits of terms with multiple
traces depend only on the tree-level and one-loop collinear splitting
amplitudes.  The full color-dressed splitting amplitude also
factorizes into a color factor multiplied by a factor which is a
function solely of the external momenta and helicities.  The color
factor for the $g\to gg$ splitting amplitude, in the language of
Feynman diagrams, is always just a Lie algebra structure constant, 
$f^{abc}$.  At two loops, the pure-glue result is purely leading-color; 
there are no subleading-color corrections.  Adding quarks in the loops
does generate subleading-color terms.  However, non-planar diagrams 
do not give any contribution to the $g\to gg$ collinear behavior at 
two loops, because their color factors all vanish.  
(This feature will not hold for $g\to gg$ at three loops, and already
fails to hold at two loops for splittings with external quarks, 
$q\to qg$ and $g\to q\bar{q}$.)

As noted above, Catani~\cite{CataniTwoloop} gave a formula which
predicted the infrared singularities of renormalized two-loop amplitudes,
which appear as poles through fourth order in $\e$.
One can take the collinear limit of the $n$-point formula, and
compare it with the $(n-1)$-point formula, to obtain a prediction
for the infrared singularities of the two-loop splitting amplitude.
The Catani formula is expressed in terms of color-space operators.
One term arising at order $1/\e$ (denoted by $\hat{\bom H}_n^{(2)}$) 
contains a rather non-trivial color structure, and was known previously
only for the four-point case.  We have constructed a simple ansatz for its 
generalization to $n$-point amplitudes.
A comparison of the Catani formula with our result requires an 
untangling of color correlations.
We have performed this analysis, and find complete agreement.  
One can interpret this agreement as a proof of Catani's formula
(including our ansatz for $\hat{\bom H}_n^{(2)}$)
for the case of two-loop $n$-gluon amplitudes.  The proof is
inductive in the number of legs $n$, and requires certain reasonable
assumptions about the types of functions that can appear in 
singular terms. (The functions should not vanish in all collinear 
limits. Such vanishing is unlikely to happen for functions which are equal
to the tree amplitude times pure logarithms or polylogarithms,
for example.)
This proof complements the verification based on resummation given 
in ref.~\cite{StermanIR}.

Given the consistency of our results with Catani's formula, 
we can subtract the singular behavior in $\e$, to obtain
a set of relations which control the finite remainder terms in the
formula.  We have checked that these relations are satisfied,
up to overall normalization, by the finite remainder for 
the two-loop $H\to ggg$ amplitude~\cite{Koukoutsakis}
for the case of identical-helicity gluons.

In the next section, we review the structure of gauge-theory amplitudes
in their collinear limits.  In section~\ref{FeynmanSection}, we
consider a hypothetical Feynman-diagram calculation, both in a covariant
gauge and in light-cone gauge.  We discuss the difficulties that would be
encountered in these calculations, and how a unitarity-based method
can circumvent them.  In section~\ref{UnitaritySewingSection}, we 
review the unitarity-based sewing method, and present a detailed algorithm.
(Appendix~\ref{RelabelingAppendix} describes a simple relabeling
algorithm, used in the sewing process
to bring integrands into a canonical form.)
In section~\ref{SplittingIntegrandsSection}, we discuss the application
of the sewing algorithm to the calculation of the two-loop splitting amplitude.
In section~\ref{IntegralsSection}, we describe the calculation of the
required two-loop integrals, and the reduction of tensor integrals.  
We present our results in section~\ref{ResultSection}.
These contain the complete set of helicity-decomposed $g\to gg$ splitting
amplitudes in QCD with $\Nf$ fermions, as well as in $\Neqone$ and
$\Neqfour$ supersymmetric gauge theories.  We present separately
the collinear behavior of the finite terms in a two-loop amplitude.
In section~\ref{CataniComparison}, we present the comparison of 
the singular parts of the color-trivial terms
to those predicted by Catani's formula, as well as the collinear 
behavior of the finite remainder terms.  (The precise definition
of `color-trivial' is given in that section.  The full color dependence
of the singular terms, including the ansatz for $\hat{\bom H}_n^{(2)}$,
and their collinear behavior, are discussed in
appendix~\ref{CataniCollinearSection}.)
We discuss the dressing of the splitting amplitudes with color factors,
and the collinear behavior of the color-non-trivial parts of amplitudes,
in section~\ref{FullColorSection}.  We give some concluding remarks
in the final section.

\section{Splitting Amplitudes}
\label{SplittingSection}

Gauge-theory amplitudes are singular when external momenta become
soft, or when a number of momenta become collinear.  In these limits,
the amplitudes factorize in a universal way.  
In collinear limits, 
the factorization is governed by splitting amplitudes depending only
on the legs becoming collinear, and not on the remaining legs of
the hard process.  The surviving hard amplitude, in turn, depends only
on the merged leg, whose momentum is the sum of the collinear momenta.
In this paper, we will consider the splitting amplitude governing the behavior
of amplitudes as two momenta become collinear.  (The tree-level behavior
as three or four momenta become collinear has been derived by
Campbell and Glover~\cite{CampbellGlover}, 
Catani and Grazzini~\cite{CataniGrazzini},
and by Del~Duca et al.~\cite{DDFM}.  The one-loop behavior as three momenta
become collinear has recently been considered by Catani, De~Florian
and~Rodrigo~\cite{CDFR}.)

We find it most natural to discuss collinear factorization in the
context of a trace-based color decomposition of the $n$-gluon 
amplitudes~\cite{Color,TreeSixPoint,ManganoParkeXu,OneloopColor}. 
Although the external gluons are in the adjoint representation, 
this color decomposition is given in terms of traces of matrices
$T^a$ in the fundamental representation of ${\rm SU}(\Nc)$, 
which we normalize by $\Tr(T^a T^b) = \delta^{ab}$. 
We begin by discussing the behavior of terms leading in the 
number of colors, $\Nc$, where we scale the number of fermions, $\Nf$, 
with $\Nc$.  The full color behavior is a straightforward extension, 
which we defer to~\Sect{FullColorSection}.

The loop expansion of the $n$-gluon amplitude is
\begin{equation}
 {\cal A}^{a_1\ldots a_n}_n(k_1,\lambda_1;\ldots;k_n,\lambda_n)
 = g^{n-2} \sum_{L=0}^\infty 
   \biggl[ g^2 {2 e^{-\e \gamma} \over (4\pi)^{2-\e} } \biggr]^L
  {\cal A}^{\Lloop \, a_1\ldots a_n}_n(k_1,\lambda_1;\ldots;k_n,\lambda_n)\,,
\label{LoopExpansion}
\end{equation}
where $a_i$ is the color index of the $i$-th external gluon, and 
the factor of $[2 e^{-\e\gamma}/(4\pi)^{2-\e}]^L$, with 
$\gamma = -\psi(1) = 0.5772\ldots$, corresponds to the normalization 
convention of ref.~\cite{CataniTwoloop}.
The trace-based color decomposition of the $L$-loop amplitude is
\begin{eqnarray}
{\cal A}^{\Lloop \, a_1\ldots a_n}_n(k_1,\lambda_1;\ldots;k_n,\lambda_n)
&=& \Nc^L
 \sum_{\sigma\in S_n/Z_n} 
 \Tr(\si(1)\ldots\si(n)) A^\Lloop_n(\si(1),\ldots,\si(n))
\PlusBreak{~}
  \Ord(\Nc^{L-1}) \,,~~~~ 
\label{LeadingColorAmplitude}
\end{eqnarray}
where $A_n^\Lloop$ are $L$-loop color-ordered (sub-)amplitudes; where
$\Tr(1\ldots n) \equiv \Tr(T^{a_1}\ldots T^{a_n})$; 
and where $\sigma$ runs over
the non-cyclic permutations $S_n/Z_n$ of $\{1,2,\ldots,n\}$.  The
latter correspond to the set of inequivalent traces.  The permutation
$\si$ acts both on the gluon momenta $k_i$ and helicity labels
$\lambda_i$, implicit on the right-hand side of~\eqn{TreeAmplitude}.
At tree level ($L=0$), expression (\ref{LeadingColorAmplitude}) is
exact, and has no subleading-color corrections.  At loop-level
($L>0$), there are subleading-color terms containing products of two
or more traces (see~\eqns{OneLoopGenColor}{TwoLoopGenColor}).  Also,
the amplitudes must be evaluated in $D=4-2\e$ dimensions to regulate
the virtual singularities.  We extract in a prefactor the leading 
$\Nc^L$ behavior.  The normalization of $A_n$, and hence of the
splitting amplitudes we present below, differs from that in
refs.~\cite{Neq4Oneloop,Neq1Oneloop}.  Note that $A_n^{(L>0)}$ will in
general contain terms proportional to powers of $\Nf/\Nc$ and
$1/\Nc^2$.

%
\FIGURE[t]{
{\epsfysize 1.2 truein \epsfbox{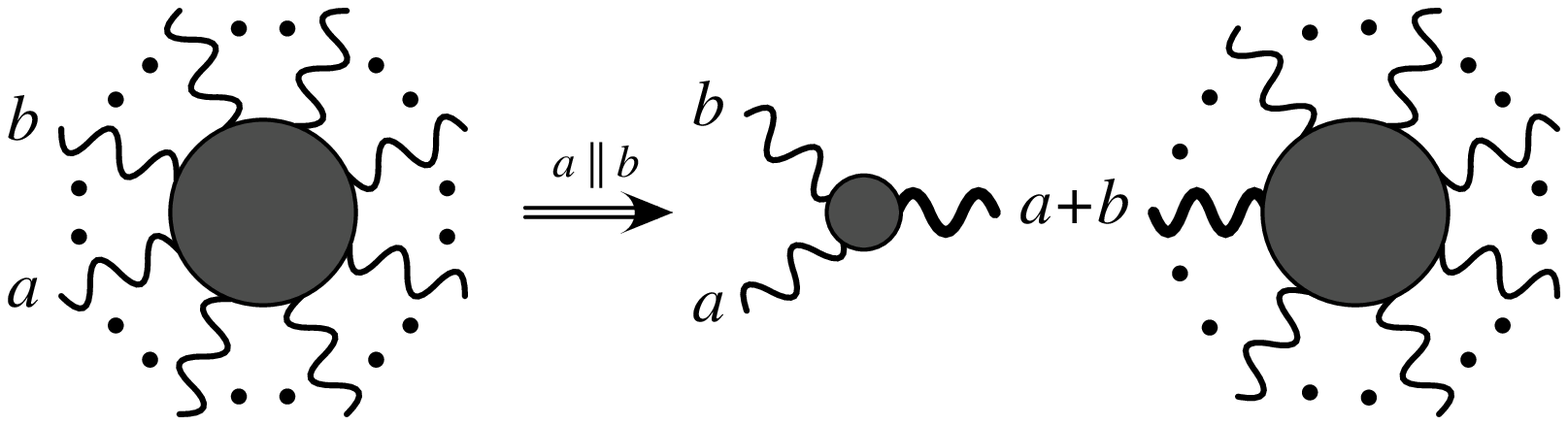}}
\caption{The collinear factorization of a tree-level amplitude.
The thick line represents a slightly off-shell gluon.}
\label{CollinearTreeFigure}
}

The color-ordered amplitudes in 
\eqn{LeadingColorAmplitude} have a universal behavior as legs 
become collinear~\cite{KosowerCollinearProof}. At tree level,
the amplitudes behave as~\cite{MPReview}
\begin{equation}
A_{n}^{\tree}(\ldots,a^{\lambda_a},b^{\lambda_b},\ldots)
 \mathop{\longrightarrow}^{a \parallel b}
\sum_{\lambda=\pm} 
  \Split^{\tree}_{-\lambda}(z; a^{\lambda_a},b^{\lambda_b})\,
      A_{n-1}^{\tree}(\ldots,P^\lambda,\ldots) \,,
\label{TreeSplit}
\end{equation}
in the limit where the momenta $k_a \rightarrow z k_P$ and $k_b
\rightarrow (1-z) k_P$ with $k_P = k_a + k_b$. Here
$\Split^{\tree}_{-\lambda}(z;a^{\lambda_a},b^{\lambda_b})$ is a
tree-level splitting amplitude.  Legs $a$ and $b$ carry helicities
$\lambda_a$ and $\lambda_b$, while the merged leg $P$ carries
helicity $\lambda$.  In the sum $\lambda$ runs over the two helicities
of the intermediate state.  The factorization of an
$n$-point tree amplitude into a splitting amplitude and an
$(n-1)$-point amplitude~(\ref{TreeSplit}) is depicted schematically in
\fig{CollinearTreeFigure}. 

The pure-glue tree-level splitting amplitudes
are~\cite{ParkeTaylor,ManganoParkeXu,MPReview}
\begin{eqnarray}
\Split^{\tree}_{-}(z; a^{-},b^{-}) &=&  0,
\label{TreeSplitAmplitudemm}\\
\Split^{\tree}_{-}(z; a^{+},b^{+})
            &=& {1\over \sqrt{z (1-z)}\spa{a}.b}, 
\label{TreeSplitAmplitudepp}\\
\Split^{\tree}_{-}(z; a^{+},b^{-})
            &=& -{z^2\over \sqrt{z (1-z)}\spb{a}.b},
\label{TreeSplitAmplitudepm}\\
\Split^{\tree}_{-}(z; a^{-},b^{+})
            &=& -{(1-z)^2\over \sqrt{z (1-z)}\spb{a}.b} \,.
\label{TreeSplitAmplitudemp}
\end{eqnarray}
These splitting amplitudes are expressed in terms of 
spinor inner products~\cite{SpinorHelicity,MPReview}, 
$\spa{i}.j = \langle i^- | j^+\rangle$ and 
$\spb{i}.j = \langle i^+| j^-\rangle$,
where $|i^{\pm}\rangle$ are massless Weyl spinors of momentum $k_i$,
labeled by the sign of the helicity.  The spinor products
are antisymmetric, with
norm $|\spa{i}.j| = |\spb{i}.j| = \sqrt{\mathstrut s_{ij}}$, where 
$s_{ij} = 2k_i\cdot k_j$.  A key advantage of the spinor formalism 
is that it makes the square-root behavior of the splitting amplitudes
manifest. The remaining tree-level splitting amplitudes, with $\lambda=-$,
may be obtained from the above ones by parity, which states 
(for general loop order $L$) that
\begin{equation}
 \Split^{(L)}_{-(-\lambda)}(z; a^{-\lambda_a},b^{-\lambda_b})
 = - \Split^{(L)}_{-\lambda}(z; a^{\lambda_a},b^{\lambda_b})
      \Big|_{\spa{a}.{b} \lr \spb{a}.{b}} \,.
\label{ParityRelation}
\end{equation}
In quoting explicit results in this paper, we use parity to 
assume that the intermediate state $P$ always has positive 
helicity, $\lambda=+$.  As we shall discuss further in
\sect{FullColorSection}, Bose symmetry implies that
the color-stripped splitting amplitude is antisymmetric
under exchange of its two arguments (including $z\leftrightarrow 1-z$),
\begin{equation}
 \Split^{(L)}_{-\lambda}(1-z; b^{\lambda_b},a^{\lambda_a})
 = -\Split^{(L)}_{-\lambda}(z; a^{\lambda_a},b^{\lambda_b}) \,.
\label{BoseRelation}
\end{equation}
This relation allows us to obtain the results for $P^+ \to a^+ b^-$
from those for $P^+ \to a^- b^+$.

%
\FIGURE[t]{
{\epsfysize 1.2 truein \epsfbox{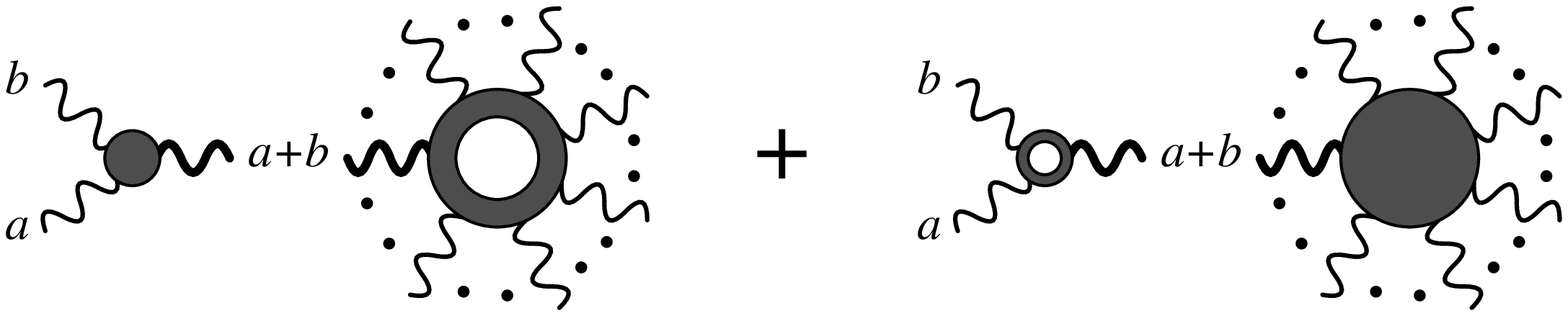}}
\caption{Two types of terms appear in the collinear factorization 
of a one-loop amplitude. The thick lines represent slightly 
off-shell gluons.}
\label{CollinearOneloopFigure}
}

At one loop the structure is similar.  In this case the 
the collinear limits are~\cite{Neq4Oneloop,BernChalmers,KosowerCollinearProof}
\begin{eqnarray}
A_{n}^{\oneloop}(\ldots,a^{\lambda_a},b^{\lambda_b},\ldots)
 \mathop{\longrightarrow}^{a \parallel b} &&
\sum_{\lambda=\pm} \biggl(
  \Split^{\tree}_{-\lambda}(z; a^{\lambda_a},b^{\lambda_b})\,
      A_{n-1}^{\oneloop}(\ldots,P^\lambda,\ldots)
\PlusBreak{ \sum_{\lambda=\pm} \bigglP }
  \Split^{\oneloop}_{-\lambda}(z; a^{\lambda_a},b^{\lambda_b})\,
      A_{n-1}^{\tree}(\ldots,P^\lambda,\ldots) \biggr) \;, \hskip 1 cm 
\label{OneloopSplit}
\end{eqnarray}
where $\Split^{\oneloop}_{-\lambda}$ is a one-loop splitting amplitude.
\Fig{CollinearOneloopFigure} displays~\eqn{OneloopSplit} schematically.

Except for $\Split_{-}(z;a^{-},b^{-})$, which vanishes at tree level
(along with $\Split_{+}(z; a^{+},b^{+})$), the ratio of the $g \to gg$
one-loop splitting amplitudes to the tree-level ones is well-defined.  
The ratio depends trivially on the Lorentz invariant $s_{ab}$, 
and non-trivially on $z$, but it does not involve spinor products. 
For the helicity configuration whose tree 
splitting amplitude vanishes we have,
\begin{eqnarray}
\Split^{\oneloop}_{-}(z; a^{-},b^{-}) & = &
   + \cgh \, \sqrt{z(1-z)} \, {\spa{a}.{b}\over{\spb{a}.{b}}^2} \, 
  {2 \over (1-2\eps)(2-2\eps)(3-2\eps)}
  \TimesBreak{ \sqrt{z(1-z)} }
 \left({\mu^2\over-s_{ab}}\right)^{\eps}
     \left(1-\eps\delta_R -{\Nf\over \Nc}\right)\,,
\label{OneloopHelFlip}
\end{eqnarray}
where
\begin{equation}
\cgh = {e^{\e \gamma} \over 2}\,  {\Gamma(1+\e) \Gamma^2(1-\e) \over 
           \Gamma(1-2\e) } \,.
\label{cghDef}
\end{equation}
Notice that in the $\e$-expansion of $(\mu^2/(-s_{ab}))^{\e}$,
$\ln(-s_{ab})$ appears.  For time-like kinematics, this expression 
has an imaginary part according to the prescription,
\begin{equation}
\ln(-s_{ab}) = \ln s_{ab} - i\pi, \qquad s_{ab} > 0.
\label{STimeLike}
\end{equation}

The remaining splitting amplitudes, for $\lambda=+$, are conveniently 
written in terms of their ratios to the tree-level ones,
\begin{equation}
\Split^\oneloop_{-}(z; a^{\lambda_a},b^{\lambda_b}) =  
 \rsn^{\oneloop \, \lambda_a\lambda_b}(z, s_{ab})
 \times {\Split}^{\tree}_{-}(z; a^{\lambda_a},b^{\lambda_b}),
\label{OnelooprS}
\end{equation}
where
\begin{eqnarray}
\rsn^{\oneloop \, ++, \, {\rm QCD}}(z,s)  &=& 
 \rsn^{\oneloop, \, \Neqfour}(z,s)
\PlusBreak{}
 \cgh \, \biggl({\mu^2\over-s}\biggr)^{\eps} \,
    {2 z (1-z)\over(1-2\eps)(2-2\eps)(3-2\eps)}
    \left(1-\eps\delta_R -{\Nf\over \Nc}\right) \,, ~~~~
\label{OneloopSplitExplicitpp} \\
\rsn^{\oneloop \, -+, \, {\rm QCD}}(z,s)&=& 
  \rsn^{\oneloop, \, \Neqfour}(z,s) \,,
\label{OneloopSplitExplicitmp}
\end{eqnarray}
and
\begin{eqnarray}
\rsn^{\oneloop,\, \Neqfour}(z,s) & = & 
   \cgh \biggl({\mu^2\over-s}\biggr)^{\eps}
    {1\over\eps^2}\left[-\left({1-z\over z}\right)^\eps 
    {\pi\eps\over \sin(\pi\eps)} + \sum_{m=1}^\infty 2\eps^{2m-1} 
    {\Li}_{2m-1}\left({z\over z-1}\right)\right] \nonumber \\
& = & \cgh  \biggl[ - {1\over\eps^2}\L{\mu^2\over z(1-z)(-s)}\R^{\eps}
 + 2 \ln z\,\ln(1-z) - \zeta_2 \biggr] + \Ord(\e)\,.
\label{OneloopSplitExplicitNeq4}
\end{eqnarray}
(Here, $\zeta_n$ is the Riemann zeta function.)
The label $\Neqfour$ on $\rsn^{\oneloop, \, \Neqfour}$ means that it is
the appropriate function for $\Neqfour$ super-Yang-Mills theory, for
both $++$ and $-+$ cases (see also \sect{Neq4MasterDecompExample}).
At one loop this expression also happens to serve as the splitting
amplitude for pure $\Neqone$ super-Yang-Mills theory, for both
non-vanishing cases,
\begin{equation}
\rsn^{\oneloop \, ++, \, \Neqone}(z,s)  
= \rsn^{\oneloop \, -+, \, \Neqone}(z,s) 
= \rsn^{\oneloop, \, \Neqfour}(z,s) \,.
\label{OneloopSplitExplicitNeq1}
\end{equation}
This relation will be violated at two loops.  The parameter $\delta_R$
selects the particular variant of dimensional regularization
(see \eqn{Dsdef}).  For
$\delta_R=1$ the scheme is the 't~Hooft-Veltman (HV)~\cite{HV} scheme,
while for $\delta_R = 0$ it is the four-dimensional
helicity~\cite{BKgggg,TwoloopSUSY} (FDH) scheme.  We always quote
results for supersymmetric theories in the FDH scheme, which
is related to, but distinct from, Siegel's dimensional reduction
scheme~\cite{DR}.  (As noted above, the normalization of the splitting
amplitudes differs from that in ref.~\cite{Neq4Oneloop}; 
to recover the earlier normalization for $\Split^\oneloop$, 
replace $\cgh$ by $\cg$, defined in eq.~(2.12) of that reference.
To recover the earlier normalization of $\rsn^\oneloop$, replace $\cgh$ by
unity.)

These one-loop splitting amplitudes were first obtained from the
collinear limits of five-point amplitudes~\cite{Neq4Oneloop}.
Subsequently they were obtained to all orders in $\e$, as required for
NNLO calculations~\cite{BernChalmers,BernKilgoreSplit,KosowerUwerSplit}.
We will express divergent parts of the two-loop splitting
amplitudes in terms of one-loop quantities. Hence we have retained all
the terms in the $\e$ expansion in these expressions.

At two loops, the subject of this paper, 
the amplitudes behave as~\cite{KosowerCollinearProof},
\begin{eqnarray}
A_{n}^{\twoloop}(\ldots,a^{\lambda_a},b^{\lambda_b},\ldots)
 \mathop{\longrightarrow}^{a \parallel b}&&
\sum_{\lambda=\pm} \biggl(
  \Split^{\tree}_{-\lambda}(z; a^{\lambda_a},b^{\lambda_b})\,
      A_{n-1}^{\twoloop}(\ldots,P^\lambda,\ldots)
\PlusBreak{ \sum_{\lambda=\pm} \bigglP }\vphantom{ \sum_{\lambda=\pm}\biggl() }
 \Split^{\oneloop}_{-\lambda}(z; a^{\lambda_a},b^{\lambda_b})\,
      A_{n-1}^{\oneloop}(\ldots,P^\lambda,\ldots)
\PlusBreak{ \sum_{\lambda=\pm} \bigglP }
 \Split^{\twoloop}_{-\lambda}(z; a^{\lambda_a},b^{\lambda_b})\,
      A_{n-1}^{\tree}(\ldots,P^\lambda,\ldots) \biggr) \;,
\label{TwoloopSplit}
\end{eqnarray}
where $\Split^{\twoloop}_{-\lambda}$ is the two-loop splitting
amplitude.  One of the goals of this paper is to calculate this
two-loop splitting amplitude in QCD.  To do so we will use the
unitarity sewing method~\cite{Neq4Oneloop,Neq1Oneloop,BernMorgan}, as
applied to splitting amplitudes~\cite{KosowerUwerSplit}.  In a
previous paper we presented the result of this calculation for the
special case of $\Neqfour$ super-Yang-Mills theory.  For this theory
the unique two-loop splitting amplitude has the remarkable property of
being an iteration of the one-loop
result~(\ref{OneloopSplitExplicitNeq4}), which led to a conjecture
that a similar iterative property holds for the planar contributions
to amplitudes~\cite{TwoloopN4}. This conjecture was shown to be
correct for the four-point amplitude using a previously-derived~\cite{BRY}
expression for the two-loop integrand.  In~\sect{ResultSection} we
shall present the explicit values of the two-loop splitting amplitudes
for $g\rightarrow gg$ in QCD, as well as in $\Neqfour$ and $\Neqone$
supersymmetric gauge theories.

As in the one-loop case, we write the two-loop splitting
amplitudes in terms of their ratios $\rsn$ to the corresponding tree-level
splitting amplitudes (when the latter do not vanish).  Taking $\lambda=+$ 
by parity, we define $\rsn^{\twoloop}$ via,
\begin{equation}
\Split^\twoloop_{-}(z; a^{\lambda_a}, b^{\lambda_b}) =  
 \rsn^{\twoloop \, \lambda_a\lambda_b}(z, s_{ab})
 \times \Split^{\tree}_{-}(z; a^{\lambda_a}, b^{\lambda_b}) \,.
\label{TwolooprS}
\end{equation}
We give the explicit expansions of the
functions $\rsn^{\twoloop \, \lambda_a\lambda_b}$, as Laurent 
expansions in $\e$, for the various theories in question, in
eqs.~(\ref{rSNeq4}), (\ref{rSNeq1ppm}), (\ref{rSNeq1mpm})
and (\ref{QCDppmmpm}), and the expression for
$\Split^\twoloop_{-}(z; a^{-}, b^{-})$ in \eqn{QCDmmm}.

Although we shall not discuss higher-loop splitting amplitudes here, 
we remark that the obvious $L$-loop generalization of the collinear 
behavior,
\begin{equation}
A_{n}^{\Lloop}(\ldots,a^{\lambda_a},b^{\lambda_b},\ldots) 
\mathop{\longrightarrow}^{a \parallel b}
\sum_{l=0}^L \sum_{\lambda=\pm}
  \Split^{\lloop}_{-\lambda}(z; a^{\lambda_a},b^{\lambda_b})\,
      A_{n-1}^{\Llloop}(\ldots,P^\lambda,\ldots) \,,
\end{equation}
can indeed be proven correct~\cite{KosowerCollinearProof}.  These
splitting amplitudes govern the collinear behavior of the entire
amplitude, including all multiple color trace terms.  We will 
discuss this more fully in \sect{FullColorSection}.  With our
normalizations the leading-color contributions to $\Split^\lloop$
are of order $\Nc^0$, but in general the splitting amplitudes
have contributions
of higher order in $1/\Nc^2$, as well as quark-loop contributions of 
order $(\Nf/\Nc)^p$, with $p\le L$, or higher in $1/\Nc^2$.


\section{Difficulties with Feynman Diagram Approach}
\label{FeynmanSection}

%
\FIGURE[t]{
{\epsfxsize 1.45 truein \epsfbox{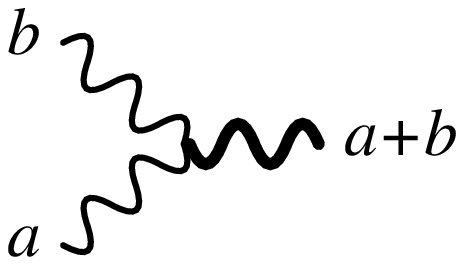}}
\caption{The three-point vertex diagram for obtaining the tree-level
splitting amplitude.  The thick line represents an off-shell gluon.}
\label{TreeThreeVertexFigure}
}

Before turning to our calculation of the two-loop $g\to gg$ splitting
amplitudes using the unitarity-based sewing method, it is instructive to
consider how one would proceed using a standard Feynman diagram approach.
Heuristically, one might try to `factorize' an $n$-point amplitude on
the collinear pole, {\it i.e.} to construct the $L$-loop splitting
amplitudes by summing up all $L$-loop Feynman diagrams with three external
legs, one of which is off-shell.  While this might seem to be the most
straightforward approach, a number of complications arise in practice, as
we shall see in this section.  In~\sect{SplittingIntegrandsSection} we
shall sidestep these complications using the unitarity-based sewing
procedure~\cite{Neq4Oneloop,Neq1Oneloop,BernMorgan,LoopReview},
outlined in~\sect{UnitaritySewingSection}.  This method has previously
been applied to splitting amplitudes at one
loop~\cite{KosowerCollinearProof,KosowerUwerSplit}.

\subsection{Tree-Level Splitting Amplitudes}

At tree level, factorization works without any subtleties.
That is, we can compute the splitting amplitude directly from the 
Feynman diagram three-point vertex depicted in~\fig{TreeThreeVertexFigure}.  
The only diagrams for an $n$-point amplitude with a pole in $s_{ab}$
are those containing this vertex as one factor. Such calculations 
have appeared elsewhere, for example in ref.~\cite{MPReview}.

As two color-adjacent momenta $k_a$ and $k_b$ become collinear, we factorize
an $n$-point tree amplitude on the $s_{ab}$ kinematic pole in terms
of a three-point vertex and an $(n-1)$-point amplitude,
\begin{eqnarray}
  A_n^{\tree} (1,2, \ldots,a,b,\ldots, n)\, & \iscol{a}{b}& \, 
\pol_\mu(a) \pol_\nu(b) \, V^{\mu\nu\rho}
    \left(i\sum_{\lambda}{\pol^{-\lambda}_\rho(P)\pol^{\lambda}_\sigma(P)
    \over s_{ab}}\right) 
\TimesBreak{ \pol_\mu(a) }
 {\partial\over\partial\pol^\lambda_\sigma(P)} 
  A^{\tree}_{n-1}(1,2, \ldots, P^\lambda, \ldots, n)\,,
\label{FormalTreeCollinear}
\end{eqnarray}
where the kinematics is the same as in~\eqn{TreeSplit}.  In setting up
the calculation the merged leg should be left slightly off-shell.
Otherwise the putative splitting amplitude would be ill-defined, since
the $1/s_{ab}$ pole diverges.  This basic structure is independent of
the particle type, although we have written \eqn{FormalTreeCollinear}
for the case of an intermediate gluon.  Here $\pol_\sigma^\lambda(P)$ is
the polarization vector for the gluon $P$ with helicity $\lambda$.
For the case of $g\to gg$, the ordinary Feynman gauge three-point vertex is
\begin{equation}
V^{\mu \nu \rho}(k_a,k_b)
= {i \over \sqrt{2}} \, \Bigl( \eta^{\mu\nu} \, (k_a-k_b)^\rho 
  + \eta^{\nu\rho} \, (2 k_b + k_a)^\mu 
  - \eta^{\mu\rho} \, (2 k_a +  k_b)^\nu \Bigr) \,. 
\label{FeynmanThreeVertex}
\end{equation}
After inserting an explicit
representation of the helicity states~\cite{SpinorHelicity}, we obtain
from~\eqn{FormalTreeCollinear} precisely the collinear
behavior~(\ref{TreeSplit}), together with the explicit values of the
splitting
amplitudes~(\ref{TreeSplitAmplitudemm})--(\ref{TreeSplitAmplitudemp}).  In
this limit the helicity algebra in the numerator of the vertex causes it
to vanish like $\sqrt{\mathstrut s_{ab}}$, partially canceling the pole 
in $s_{ab}$.
Physically, this cancellation is due to an angular momentum mismatch
between $\lambda$ and $\lambda^a+\lambda^b$.  Overall, we are left with
the $1/\sqrt{\mathstrut s_{ab}}$ behavior evident in the splitting 
amplitudes.  Other
gauge choices, such as light-cone gauge or the non-linear Gervais-Neveu
gauge~\cite{GN}, give the same final result as Feynman gauge.

\subsection{Difficulties at Loop Level}

%
\FIGURE[t]{
{\epsfxsize 1.2 truein \epsfbox{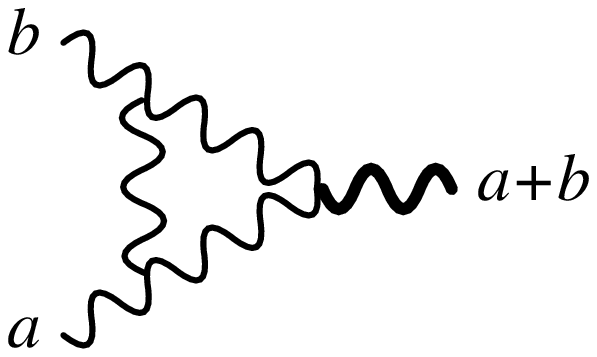}} \hskip 1 cm 
{\epsfxsize 1.7 truein \epsfbox{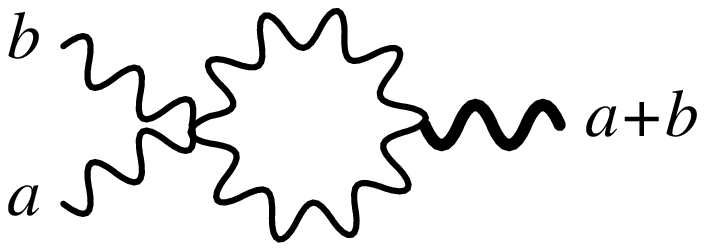}} \hskip 1 cm 
{\epsfxsize 2.0 truein \epsfbox{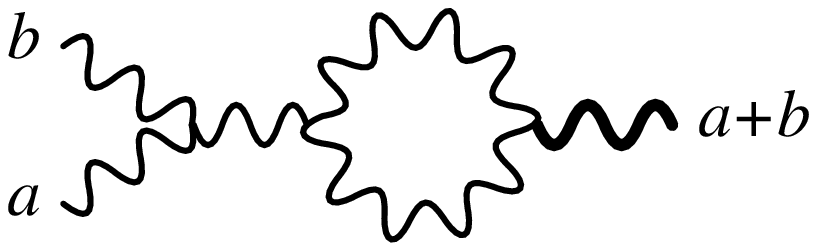}} 
\caption{Three non-vanishing Feynman diagrams contributing to the 
pure-glue splitting amplitude. The thick gluon line is slightly off shell.}
\label{LCGaugeOneloopDiagFigure}
}

The simplicity of a conventional diagrammatic calculation at tree
level might lead one to believe that a similar approach should work
at loop level.  Such a calculation would involve Feynman diagrams of 
the type depicted in~\fig{LCGaugeOneloopDiagFigure}.  This expectation,
however, turns out to be incorrect.  At loop level, the
situation is not quite this simple.

%
\FIGURE[t]{
{\epsfxsize 1.5 truein \epsfbox{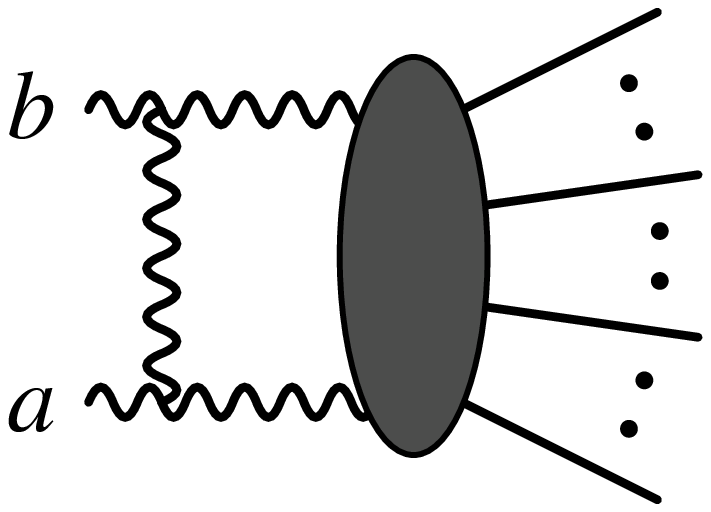}}
\caption{In covariant gauges, diagrams without single-particle 
factorization can contribute to the splitting amplitude.
}
\label{NonFactorizingFigure}
}

In covariant gauges, such as Feynman gauge, one immediately runs into
trouble~\cite{BernChalmers}, because there are contributions to
collinear behavior from Feynman diagrams of the form shown in
\fig{NonFactorizingFigure}, with no single-particle pole in $s_{ab}$.
The required pole emerges only after carrying out the loop integration.
Moreover, the $\e$-expansion of some of the integrals
has discontinuous behavior as $s_{ab} \to 0$, because of the
interchange of this limit with the limit $\e \to 0$.
The appearance of more complicated loop integrals
is reflected in the structure of the known results for the splitting
amplitudes~(\ref{OneloopSplitExplicitpp})--(\ref{OneloopSplitExplicitNeq4}),
which contain polylogarithms in the variable $z$.  Such functions
simply cannot occur in the integrals encountered in the triangle or 
bubble graphs in \fig{LCGaugeOneloopDiagFigure}, which in 
a covariant gauge produce only logarithms, in the variable $s_{ab}$.
Clearly, the polylogarithms must come from elsewhere.   
They do in fact arise from non-factorizing diagrams of the type shown 
in~\fig{NonFactorizingFigure}~\cite{BernChalmers}.  
This peculiar absence of factorization is tied to the presence of soft and
collinear virtual divergences in the theory, reflected as poles in $\e$
in the loop amplitudes.  At one loop, the non-factorizing pieces may 
be reconstructed using knowledge of all the possible integral functions 
that appear in amplitudes, along with the universal structure of 
the infrared divergences.
Indeed, such a reconstruction was used to prove universal
factorization of the one-loop amplitudes~\cite{BernChalmers} and to
compute their explicit values~\cite{BernKilgoreSplit}.  This
reconstruction, however, does not generalize straightforwardly to
higher loops.

As the next logical choice of a method for calculating a splitting 
amplitude directly from Feynman diagrams, 
one might turn to light cone-gauge.  This gauge is
known~\cite{LightConeFactorization,CSS} to have simple factorization
properties. A key advantage of light-cone gauge is that only physical
states propagate.  This can help clarify various formal properties.
Moreover, light-cone gauge vertex integrals do contain
polylogarithms of the type appearing in the one-loop splitting amplitude.
Light-cone gauge has a long history and has been used to address a wide
variety of problems.  For example, the first proofs of factorization
in QCD between the hard and soft parts of a process were 
performed in this gauge~\cite{CSS}. Another important example is
the next-to-leading order (NLO) calculation of the Altarelli--Parisi
evolution kernel in $x$-space~\cite{LightConeFactorization,CFP,FP,EV}.  
Light-cone gauge has also been used for more formal purposes, such as the 
proof of finiteness of maximally supersymmetric ($\Neqfour$)
Yang-Mills gauge theories~\cite{MandelstamN4}.

In describing the collinear limit of amplitudes, light-cone gauge is 
useful because, as a physical gauge, it can prevent the non-factorizing
graphs of~\fig{NonFactorizingFigure} from contributing.  In a general
covariant gauge, such graphs would inevitably mix under residual gauge 
transformations with the graphs in~\fig{LCGaugeOneloopDiagFigure}.

In light-cone gauge, the one-loop splitting amplitude is given by the
sum of the three Feynman diagrams shown
in~\fig{LCGaugeOneloopDiagFigure}. In this gauge there are no ghost
contributions. Furthermore, all cactus diagrams, as well as bubbles
attached to the massless external lines $a$ and $b$, vanish in
dimensional regularization, leaving only the three diagrams shown.
The light-cone gauge Feynman vertices are the same as those in Feynman
gauge. The propagator is now
\begin{equation}
\Prop_{\mu \nu} = -{i \lcproj_{\mu\nu}\over p^2 + i \e}\, 
\label{LightConePropagator}
\end{equation}
where the light-cone projector is,
\begin{equation}
\lcproj_{\mu\nu} = \eta_{\mu \nu} - 
{\n_\mu p_\nu + p_\mu \n_\nu \over p \cdot \n }.
\label{LightConeProjector}
\end{equation}
Here $p$ is the particle momentum,
$\eta$ is the Minkowski metric, and $n$ is a null vector
($n^2 = 0$) defining the light-cone direction. 

In carrying out a light-cone gauge calculation, one quickly 
runs into well-known technical difficulties~\cite{Leibbrandt} arising
from regions of loop integration where
the light-cone denominator $p \cdot n \equiv p^+$ vanishes.  The
light-cone denominators introduce a new set of singularities in the
Feynman integrals, some of which are not regulated by dimensional 
regularization.  In order for a generic light-cone gauge diagram to
be well-defined, a prescription for dealing with these singularities is
needed.  For example, the principal-value (PV) prescription replaces
\begin{equation}
{1\over p \cdot n} \rightarrow  \lim_{\delta \rightarrow 0} 
 {1\over 2} \biggl( {1\over  p \cdot n + i \delta }
                  + {1\over  p \cdot n - i \delta } \biggr) \,,
\label{PVdeltaeq}
\end{equation}
where $\delta$ is a regulator parameter.  Another choice, better
founded in field theory, is the Mandelstam-Leibbrandt
prescription~\cite{MandelstamN4,Leibbrandt}.  A comparison of these
two prescriptions may be found in ref.~\cite{KunsztLC}.

The introduction of an additional prescription in a splitting
amplitude calculation is problematic for a number of reasons:
\begin{itemize}

\item The additional prescription complicates the calculation, and
requires the computation of more difficult Feynman integrals.

\item At higher loops the validity of the
prescription is less clear~\cite{AndrasiTaylor}.

\item After expanding in small $\delta$, 
the results contain factors of $\ln\delta$, where $\delta$ is
the prescription parameter. In general these cancel only after
combining virtual and real emission contributions.
\end{itemize}

Further to the last point, a calculation of the splitting amplitudes
that retains dependence on a prescription parameter cannot match the
splitting amplitudes extracted from the collinear limits of scattering
amplitudes.  An on-shell $n$-point scattering amplitude is
gauge invariant, and depends only on the external momenta and on the
dimensional regulator parameter $\e$, not on any light-cone
prescription parameter $\delta$.  The same is true of its collinear
limits.  Thus loop splitting amplitudes defined via the
collinear limits of scattering amplitudes cannot depend on the
parameter $\delta$.  

If we had been computing the Altarelli--Parisi kernel, which receives
contributions from both virtual and real emission contributions, then
the dependence on $\delta$ would cancel between
them~\cite{CFP,FP,EV,KunsztLC}.  However, since we are interested in
computing the collinear behavior of the virtual contributions on their
own, in general the $\delta$ dependence would not cancel.
This difficulty makes it unclear how the desired
splitting amplitudes can emerge from a light-cone gauge calculation.

It is helpful to consider a few examples in order to illustrate 
the structure of integrals with light-cone denominators,
and to help explain how we will later sidestep these difficulties,
using the unitarity-based sewing method.

\subsection{One-Loop Light-Cone Integral Examples}
\label{OneLoopLightConeIntegralExamples}

%
\FIGURE[t]{
{\hbox to 2 cm{}\epsfxsize 2.7 truein \epsfbox{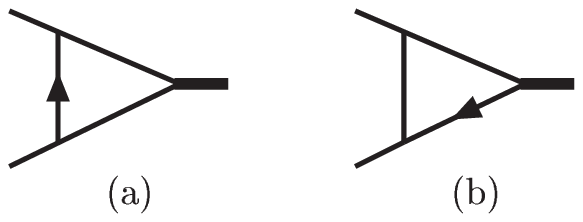}
\hbox to 2cm{}}
\caption{One-loop triangle diagrams containing
light-cone denominators.  The arrow indicates the line containing a
light-cone denominator.  Integral (a) contains an unregulated singularity,
but does not appear in the unitarity-based sewing method.}
\label{OneLoopExampleFigure}
}

Consider the scalar integrals shown in \fig{OneLoopExampleFigure}.
In the figure, an arrow on an internal line indicates the insertion of the 
light-cone factor $1/(p_i\cdot\n)$, with $p_i$ the momentum carried by 
the marked line.  Such integrals appear in a light-cone 
gauge calculation of the one-loop splitting amplitude, using the 
Feynman diagrams of~\fig{LCGaugeOneloopDiagFigure}.  We shall see
that there is a big difference between the two types of integrals 
in~\fig{OneLoopExampleFigure}; one is properly regulated by
dimensional regularization without any additional prescriptions,
and one is not~\cite{KunsztLC}.

The explicit expressions for the two integrals are,
\begin{eqnarray}
J^{(a)} (z, s) &=& - i \int {d^D p \over \pi^{D/2} } 
{1\over p^2 (p-k_1)^2 (p-k_1-k_2)^2 \, (p - k_1)\cdot \n } \,, 
\label{JaMomDef}\\
J^{(b)} (z, s) &=& - i \int {d^D p \over \pi^{D/2} }
{1\over p^2 (p-k_1)^2 (p-k_1-k_2)^2 \,  p\cdot \n }\,,
\label{JbMomDef}
\end{eqnarray}
where $s = (k_1+k_2)^2 = 2k_1\cdot k_2$ and, for consistency with 
the collinear kinematics, $z$ is defined by
\begin{equation}
k_1 \cdot \n = z \, (k_1 + k_2) \cdot \n \,,
\qquad 
k_2 \cdot \n = (1-z) \, (k_1 + k_2) \cdot \n \,.
\label{kidotn}
\end{equation}
For $J^{(a)}(z,s)$ the light-cone denominator involves the loop
momentum between the two massless legs, while for $J^{(b)}(z,s)$ the
light-cone denominator involves a loop momentum adjacent to the merged
off-shell external leg.

In order to evaluate these integrals it is helpful to make use of 
their properties as $n$ is rescaled~\cite{KosowerUwerSplit}, 
which imply that
\begin{eqnarray}
 J^{(a)} (z, s) & = & {1\over (-s)^{1+\e} \, 
                   (k_1 + k_2) \cdot \n} f^{(a)}(z)\,,  \nonumber\\
\hskip 2 cm
J^{(b)} (z, s) & = &{1\over (-s)^{1+\e} \, 
                   (k_1 + k_2) \cdot \n} f^{(b)}(z) \,,
\label{Scaling}
\end{eqnarray}
where $f^{(a)}$, $f^{(b)}$ are functions to be determined.  
Without loss of generality we may set 
\begin{equation}
 (k_1 + k_2)\cdot \n\ = -s \,.
\label{Replace_n}
\end{equation}
At the end of the evaluation we can use \eqn{Scaling} to replace 
one factor of $1/(-s)$ with $1/[(k_1 + k_2)\cdot\n]$.  
(The splitting amplitudes are independent of $n$, so the light-cone 
vector $n$ must cancel from final expressions.)

One might be tempted to switch to light-cone coordinates, as is commonly 
done when performing light-cone gauge calculations. In these coordinates 
we take $p^+ = p \cdot n$ and
$p^- = p \cdot n^*$ where $n^*$ is dual to $n$, {\it i.e.}
\begin{equation}
n_\mu = (n_0, \vec n)\,, 
\hskip 3 cm 
n_\mu^* = (n_0, -\vec n)\,. 
\end{equation}
In light-cone coordinates, the integral $J^{(b)}(z,s)$, for example, 
is given by
\begin{equation}
J^{(b)} (z, s) = - i \int {d^{D-2} p \over \pi^{D/2}} dp^+ dp^- 
{1\over p^2 \,(p-k_1)^2 \, (p-k_1-k_2)^2 \, p^+} \,,
\end{equation}
which is ill-defined because of the unregulated longitudinal integral
$\int dp^+/p^+$.  However, with this choice of coordinates we have
switched to a version of dimensional regularization where only the
transverse coordinates are regulated.  This does not correspond to
covariant dimensional regularization: $dp^+ dp^- d^{D-2}p \not = d^D
p$.  This difficulty is relatively minor, and may be dealt with by
reverting to covariant dimensional regularization.  Alternatively, we
could introduce the principal-value prescription, to justify
intermediate steps, but then remove it prior to performing the final
Feynman-parameter integration (see for example eq.~(34) of
ref.~\cite{KunsztLC}). In either case, care is required because of the
ill-defined nature of expressions.

Even covariant dimensional regularization, however, does not
suffice to properly regulate all light-cone integrals. 
Let us first compare the behavior of $J^{(a)}$ and $J^{(b)}$ 
in their momentum-space forms~(\ref{JaMomDef}) and (\ref{JbMomDef}).
In momentum space, singularities arise whenever two or
three denominator factors vanish.  The light-cone
denominator vanishes when $p$ becomes proportional to $n$, 
$p \rightarrow \zeta n$.  In this region $p^2$ also vanishes, 
but the other two denominators do not vanish.
The singularity that arises here thus looks very much like the
collinear singularity that arises when $p$ becomes collinear with
$k_1$, $p \rightarrow \zeta k_1$.  This singularity is regulated by 
covariant dimensional regularization, and so we may expect the same 
to be true for the new singularity that arises in the presence of
the light-cone denominator. This will indeed turn out to be the case. 
In contrast, for $J^{(a)}$, after shifting the momentum $p' = p - k_1$ 
we can see that in the soft region $p' \rightarrow 0$ we have {\it four} 
vanishing denominators, which is indicative of the difficulties that 
will be encountered in evaluating this integral.

We can see the difficulty with $J^{(a)}$ more explicitly 
using its Feynman-parametrized form.  We begin by Schwinger parametrizing,
\begin{eqnarray}
J^{(a)}(z, s) &=&  -i \int_0^\infty \prod_{i=1}^4 d t_i
\int {d^{4-2\e} p \over \pi^{2-\e} }
 \exp\Bigl[ (t_1 + t_2 + t_3) p^2 
          - 2 p \cdot (t_2 k_1 + t_3 k_1+ t_3 k_2)
\MinusBreak{ \exp\BiglB (t_1 + t_2 + t_3) p^2 - 2 p \cdot~~ }
  t_4 p\cdot n + t_3 s +  t_4 k_1 \cdot n \Bigr]  \,.
\end{eqnarray}
To integrate out the loop momentum we perform the shift
\begin{equation}
p = p' + {t_2 k_1 + t_3 k_1+ t_3 k_2 \over T} + {t_4 n  \over 2T} \,,
\end{equation}
where $T = t_1+ t_2 + t_3$. (Note that the Schwinger parameter 
$t_4$ associated with the light cone denominator is absent from $T$.)
Wick rotating and then integrating out the shifted loop momentum $p'$ gives
\begin{eqnarray}
J^{(a)}(z, s) & = & 
   \int_0^\infty \prod_{i=1}^4 d t_i \, T^{-D/2}
\exp \biggl[  - { (t_2+t_3) t_3 s \over T } 
             - { t_4 (t_2 k_1 + t_3 k_1+ t_3 k_2) \cdot n \over T }
\PlusBreak{ \exp \bigglB  - { (t_2+t_3) t_3 s \over T } -~~ }
 t_3 s + t_4 k_1 \cdot n \biggr]  \nonumber \\
&=&
 \int_0^\infty \prod_{i=1}^4 d t_i \, T^{-D/2} 
\exp \biggl[ { t_1 t_3 s \over T} 
            + { s t_4 (t_3 (1-z) - t_1  z) \over T } \biggr] \,, 
\end{eqnarray}
where we have used the replacements~(\ref{kidotn}) and (\ref{Replace_n})
to obtain the last line.
The integral over the Schwinger parameter $t_4$ associated with the 
light-cone denominator is now trivial, and yields
\begin{equation}
J^{(a)}(z,s) = \int_0^\infty \prod_{i=1}^3 d t_i \, T^{-D/2+1}
 \, {1\over s ( t_1 z - t_3 (1-z)) }
  \exp \biggl[ { t_1 t_3 s \over T } \biggr] \,.
\end{equation}
As usual we may convert the Schwinger parameters to Feynman parameters
by defining $a_i = t_i/T$ and integrating out the overall scale $T$,
yielding a compact Feynman parameter representation,
\begin{equation}
J^{(a)}(z, s) =   
- \Gamma(1+\e) (-s)^{-2-\e} 
\int_0^1 \prod_{i=1}^3 d a_i \, 
\delta\Bigl(1 - \sum_{j=1}^3 a_j \Bigr)
{\bigl( a_1 a_3 \bigr)^{-1-\e} \over a_1 z - a_3 (1-z) } \,.
\label{Integral_a_final}
\end{equation}
The reader will observe that for time-like kinematics,
the integrand blows up inside the region of integration.
For example, if $z = 1/2$ 
there is a singularity at $a_1 = a_3$.  This singularity could
be regulated by analytically continuing in $z$; in the space-like
region, where $z>1$, it is in fact absent.  This is not, however, the only
singularity in the integrand.  It is also singular in the corner
of the integration region where $a_2\rightarrow 1$, as can be
made manifest by changing variables 
$a_2 = 1-v$, $a_1 = v u$, $a_3 = v (1-u)$, with jacobian $v$,
\begin{equation}
J^{(a)}(z, s) =   
 \Gamma(1+\e) (-s)^{-2-\e} 
\int_0^1 dv du\; v^{-2-2\e} \, 
   { \bigl[ u(1-u)\bigr]^{-1-\e} \over 1-u-z}\,.
\label{Integral_a_finalII}
\end{equation}
The singularity as $v\rightarrow0$ is stronger than the $v^{-1-\e}$
that would correspond to a logarithmic divergence, and would give rise
to a pole in $\e$.  Formally, analytic continuation in $\e$ will
regulate this divergence (in fact the $v$ integral will not give rise
to a pole at all), but this requires a large analytic continuation,
and effectively happens through the subtraction of an infinite
constant.  (There is of course no associated
bare coupling here into which such an 
infinite constant could be absorbed.)
Note that this pathology is independent of $z$, and so
cannot be cured by analytically continuing in the latter variable.

The principal-value prescription~(\ref{PVdeltaeq})
is a widely-used method to deal with
this problem. For this integral it would
give rise to a `naked' $1/\delta$ singularity.  
Absent such a prescription, which as discussed above
we must avoid for other reasons, care would be
required in a complicated calculation to ensure that all integrals are
continued in a consistent manner, and that these continuations do not
violate any symmetries.  It would be much simpler if we do not have to
confront this issue at all.  As we shall see, our approach to the
calculation indeed allows us to avoid integrals like $J^{(a)}$
altogether.

In contrast, the integral $J^{(b)}$ in \fig{OneLoopExampleFigure}(b) 
is properly regulated by dimensional regularization.  
To see this explicitly, follow similar steps as in the computation
of $J^{(a)}$ to obtain,
\begin{equation}
J^{(b)}(z, s) =   
\Gamma(1+\e) (-s)^{-2-\e} 
\int_0^1 \prod_{i=1}^3 d a_i \, \delta\Bigl(1 - \sum_j a_j\Bigr)
{\bigl(a_1 a_3 \bigr)^{-1-\e} \over a_2 z + a_3 } \,.
\label{Integral_b_final}
\end{equation}
In this case, the integrand diverges no worse than $a_i^{-1-\e}$ near
any boundary, and so dimensional regularization renders the integral
finite for small negative $\e$ without infinite subtractions.
Indeed, this integral is well-defined and its value
is~\cite{KosowerUwerSplit},
\begin{eqnarray}
J^{(b)}(z,s) &=& 
 2 \, {\Gamma(1+\e) \Gamma^2(1-\e) \over \Gamma(1-2 \e)} 
 (-s)^{-2-\e} 
{1\over z}\biggl\{ {1\over \e^2}
- {1\over \e} \ln z 
+ {1 \over 2} \ln^2 z + \Li_2(1-z)
\biggr\}
\PlusBreak{ }
  \Ord(\e) \,,
\label{Jbresult}
\end{eqnarray}
so there is no need for an additional prescription here.
We will also encounter the integral $J^{(b)}(1-z)$.  
In the space-like case, with $1-z<0$,
the latter's integrand will be singular inside
the region of integration.  As explained above, this can be regulated
by analytic continuation in $z$ from $z<1$.

Although it is not necessary, it is still possible to use an
additional prescription for dealing with the light-cone denominator
singularity in $J^{(b)}$.  Had we, for example, used the principal-value
prescription we would have obtained instead
\begin{equation}
J_{\rm PV}^{(b)}(z,s) =  \Gamma(1+\e) (-s)^{-2-\e}
 \int_0^1 \prod_{i=1}^3 da_i \delta\Bigl(1-\sum_j a_j\Bigr)
      (a_1 a_3)^{-1-\e}
{ a_2 z + a_3 \over (a_2 z + a_3)^2 + (\delta/s)^2 } \,,
\end{equation}
following the discussion in {\it e.g.} ref.~\cite{KunsztLC}.
For $\delta \rightarrow 0$ this integral evaluates to 
\begin{eqnarray}
J_{\rm PV}^{(b)}(z,s) &=& 
 (-s)^{-2-\e}\, {\Gamma(1+\e) \over z}
 \biggl\{ {1 \over \e}
          \Bigl( {1\over \e} + \ln\Bigl|{\delta\over s} \Bigr| 
      - 2 \ln z \Bigr) 
 - \ln\Bigl|{\delta \over s} \Bigr| \, \ln z 
\PlusBreak{ (-s)^{-2-\e}\, {\Gamma(1+\e) \over z}
 \bigglB }
 \ln^2 z + \Li_2(1-z) 
\biggr\}
+  \Ord(\e) \,.
\label{JbresultPV}
\end{eqnarray}
Compared with the result in dimensional regularization
(\ref{Jbresult}), the $1/\e$ singularity arising from the light-cone
denominator has been traded for a $\ln\delta$ singularity.  The 
$\ln\delta$ would then appear in the result for the splitting amplitude.
As mentioned earlier,
such a result cannot match the splitting amplitude describing the collinear
limits of gauge-invariant, dimensionally-regulated one-loop amplitudes, 
because the latter depend only on $\e$, not $\delta$.  We therefore wish to 
avoid additional prescriptions, which are in any event
unnecessary for $J^{(b)}$.  They would be required only for integrals
like $J^{(a)}$.

What distinguishes the integrals (a) and (b) in
figure~\ref{OneLoopExampleFigure}?
Could it be that the appearance of ill-defined integrals like (a) is
purely an artifact of the gauge choice and would not appear in a more
physical construction of the splitting amplitudes? We will see that the
answer to the latter question is yes.

%
\FIGURE[t]{
{\epsfxsize 2.0 truein \epsfbox{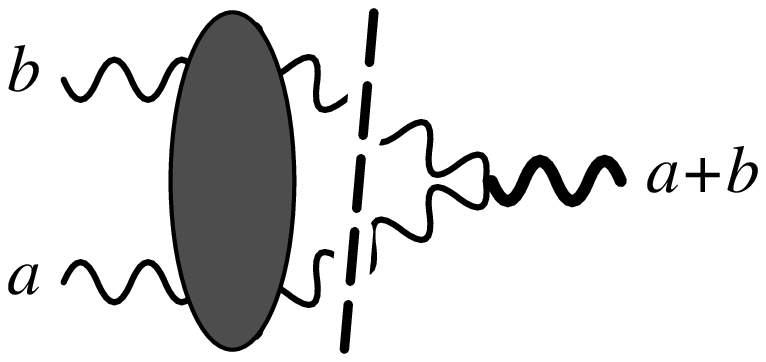}}
\caption{The two-particle cut of a one-loop splitting amplitude. The
cut is represented by the dashed line. On the left-hand side all legs,
including the cut ones, are on-shell.  On the right-hand side the
merged leg represented by a thick line is slightly off-shell.
\break
}
\label{OneLoopSplittingCutFigure}
}

A general difficulty with diagrammatic approaches is that
gauge-invariant results are chopped up and separated into gauge-dependent 
pieces by the decomposition into diagrams.  For example, for a fully
on-shell scattering amplitude calculation at tree level, each
light-cone gauge Feynman diagram contains light-cone denominators.
Yet by gauge invariance their sum must be free of such denominators.
Unitarity then implies that similar cancellations should happen
at loop level.  For splitting amplitudes, one leg is off-shell.
Hence the cancellation of light-cone denominators is not complete;
nevertheless, there will be a partial cancellation. 
In order to understand which light-cone denominators may
appear and which should not, we use unitarity.

On dimensional grounds, the result for the splitting amplitude
must have the form
\begin{equation}
f(\e,z) \, (-s)^{-\e},
\label{fezs}
\end{equation}
for some $f(\e,z)$, where the epsilonic power of $(-s)$ follows from
the integration measure $d^{4-2\e}p$.  In the time-like region,
$s>0$, we have 
$(-s)^{-\e} = 1 - \e \ln(-s) + \ldots 
            = 1 - \e (\ln s - i\pi) + \ldots$,
so any desired order in the Laurent expansion in $\e$ of $f(\e,z)$ 
can be computed by extracting the absorptive part of the function to
one higher order in $\e$.  If we can compute the absorptive part,
via unitarity cuts, to all orders in $\e$, we can completely 
determine $f(\e,z)$.
The two-particle cut of the one-loop splitting amplitude is
depicted in~\fig{OneLoopSplittingCutFigure}. Using the Cutkosky
rules~\cite{Cutting} we obtain,
\begin{eqnarray}
\Absp \Bigl[ \Split^\oneloop(z;a,b) \Bigr] &\sim&
\label{TwoParticleCut}
\\&&\hskip -25mm\int {d^D p \over (2\pi)^{D-2}} 
\deltaplus(\ell_1^2)\deltaplus(\ell_2^2)\,
 A_4^{\tree}(a,b, \ell_2^\nu, -\ell_1^\mu)
\times
\lcproj_{\mu \alpha}(\ell_1) \, \lcproj_{\nu\beta} (\ell_2)
\times V^{\alpha \beta \gamma}(\ell_1,-\ell_2)\,,
\nonumber
\end{eqnarray}
where $\ell_1 = p$, $\ell_2 = p - k_a - k_b$, the vertex
$V^{\alpha\beta\gamma}$ is defined in \eqn{FeynmanThreeVertex},
$\lcproj_{\mu\nu}$ is the physical state projector given in
\eqn{LightConeProjector}, and $\deltaplus(\ell^2) =
\Theta(\ell^0)\delta(\ell^2)$.  This is the only non-trivial cut of
the one-loop splitting amplitude and therefore yields the complete
absorptive part.  On the left-hand side of the cut, $A_4^{\tree}$ is a
gauge-invariant amplitude --- all legs including the cut ones are
on-shell.  Any light-cone denominator appearing to the left of the cut
must therefore be spurious: gauge invariance dictates that light-cone
denominators cannot appear after combining all Feynman diagrams,
because such denominators would not appear in a covariant gauge.
Because the full one-loop splitting amplitude can be reconstructed
from its absorptive part, the light-cone denominator appearing in the
integral in \fig{OneLoopExampleFigure}(a) must also be spurious
(absent from the sum over all diagrams) in the full calculation, not
just the absorptive part.  In contrast, light-cone denominators on the
cut lines are introduced by the Cutkosky rules via the sum over
polarizations across the cut line,
\begin{equation}
\sum_{\sigma} \pol_\mu^{(\sigma)}(p) \pol_\nu^{(\sigma)*}(p)
=  - \eta_{\mu\nu} + { p_\mu \n_\nu + \n_\mu p_\nu \over p\cdot \n}
= - d_{\mu\nu} \,.
\label{PolarizationSum}
\end{equation}
This physical state projector is the on-shell version of the one appearing
in the light-cone gauge propagator (\ref{LightConePropagator}).  
The light-cone denominators that appear here can survive.  This is the type of
light-cone denominator that appears in the integral of 
\fig{OneLoopExampleFigure}(b).  (Indeed, as noted earlier, {\it some\/}
light-cone denominators must survive in order to get an answer of
the sufficient polylogarithmic complexity.)

The unitarity argument indicates that only light-cone denominators
associated with a cut line need survive.  (In the multi-loop case,
denominators associated with lines to the right of all cuts can
also survive.) Thus it should be possible to perform calculations where
dangerous light-cone denominators of the sort depicted in
\fig{OneLoopExampleFigure}(a) do not appear. 
The unitarity-based sewing method, which we present in 
\sect{UnitaritySewingSection}, has exactly this property.

\subsection{Two-Loop Light-Cone Integrals}
\label{TwoLoopLightConeIntegrals}

At two loops we encounter a similar situation:
some integrals are properly regulated solely by covariant
dimensional regularization, and others are not.

\FIGURE[t]{
\hbox to \hsize{\hfil%
\SizedFigureWithCaption{1 truein}{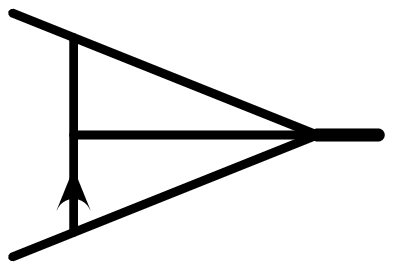}{(a)}
\SizedFigureWithCaption{1 truein}{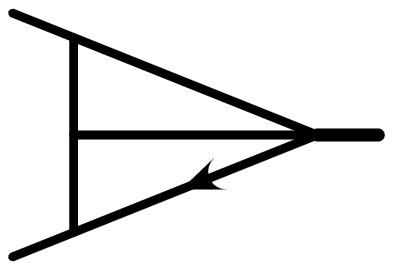}{(b)}
\hfil}
\label{TwoloopExampleFigure}
\caption{Sample three-point integrals at two loops containing
light-cone denominators. Integral (a) is ill-defined without additional
prescriptions, but integral (b) is rendered finite by dimensional
regularization alone.}
}

As concrete examples consider the two-loop integrals in
\fig{TwoloopExampleFigure}, given by
\begin{eqnarray}
L^{(a)}(z,s) &=& i \!
    \int {d^D p\over (2\pi)^{D}} {d^D q\over (2\pi)^{D}}\,
             {1\over p^2 (p-k_1)^2 (p+q)^2 (q+k_1+k_2)^2 (q+k_1)^2
                     (p-k_1)\cdot n} \,,
\hskip 1 cm 
\label{TwoloopExampleInta}\\
L^{(b)}(z,s) &=& i \!
    \int {d^D p\over (2\pi)^{D}} {d^D q\over (2\pi)^{D}}\,
             {1\over p^2 (p-k_1)^2 (p+q)^2 (q+k_1+k_2)^2 (q+k_1)^2
                     p\cdot n} \,.
\label{TwoloopExampleIntb}
\end{eqnarray}
In our calculation of the two-loop splitting amplitude from the unitarity
sewing method, we encounter only integrals similar to $L^{(b)}$.
Following similar steps as at one loop, we obtain the 
Feynman-parameterized form,
\begin{eqnarray}
L^{(a)}(z,s) &=&\Gamma(1+2\e)(-s)^{-2-2\e} 
\nonumber\\
&&\hphantom{\Gamma(1)}\times  
 \int_0^1 \prod_{i=2}^6 da_i\;\delta\Bigl(1-{\sum_j a_j}\Bigr)
{\Delta^{3\e} (a_2 a_3 a_6)^{-1-2\e}
\over z [a_6 (a_2+a_3) + a_3 (a_2+a_4)]-a_2 a_6} \,, \hskip 10mm
\end{eqnarray}
where
\begin{equation}
\Delta = (a_2+a_4)(a_3+a_5) + a_6 (a_2+a_3+a_4+a_5).
\end{equation}
This integral has an insufficiently regulated divergence in
the region $a_5\rightarrow 1$.  To see this, make the change
of variables $a_5 = 1-v$, $a_{j=2,3,4,6} = v b_j$, for which the
jacobian is $v^3$:
\begin{eqnarray}
L^{(a)}(z,s) &=&\Gamma(1+2\e)(-s)^{-2-2\e} 
\nonumber\\ &&\hskip 1mm\times
  \int_0^1 dv \hskip-0.7em \prod_{i=2,3,4,6} 
\hskip -0.8em db_i\;\delta\Bigl(1-{\sum_j b_j}\Bigr)
v^{-2-3\e} {\Delta_b^{3\e} (b_2 b_3 b_6)^{-1-2\e}
\over z [b_6 (b_2+b_3) + b_3 (b_2+b_4)]-b_2 b_6} \,, \hskip 10mm
\end{eqnarray}
where
\begin{equation}
\Delta_b = v b_3 (b_2+b_4) + v b_6 (b_2+b_3+b_4) + (1-v)(b_2+b_4+b_6).
\end{equation}

Like $J^{(a)}$ of the previous subsection, the integral has a power-law
divergence for $v\sim 0$ which would require a large analytic 
continuation, or equivalently the subtraction of an infinite constant.
This divergence is again independent of $z$, and hence cannot be
cured by analytic continuation in that variable.

On the other hand, for integral (b) in \fig{TwoloopExampleFigure} we
have the Feynman parametrized form
\begin{eqnarray}
L^{(b)}(z,s) &=& -\Gamma(1+2\e)(-s)^{-2-2\e}
\nonumber\\ &&\hphantom{\Gamma(1)}\times
  \int_0^1 \prod_{i=2}^6 da_i\;\delta\Bigl(1-{\sum_j a_j}\Bigr)
{\Delta^{3\e} (a_2 a_3 a_6)^{-1-2\e}
\over a_2 a_6 + z [a_4 a_6 + a_5 (a_2+a_4+a_6)]}.\hskip 10mm
\end{eqnarray}
In this case the integral is well-defined; at all boundaries,
the integrand goes like $v^{-1-m\e}$, $m=1,\ldots,4$.  
For example, as $a_3 = 1-v \to 1$, the presence of $a_3$ in the numerator
lessens the singularity to $v^{-1-\e}$.

%
\FIGURE[t]{
{\hbox to 2 cm{}\epsfxsize 2.2 truein \epsfbox{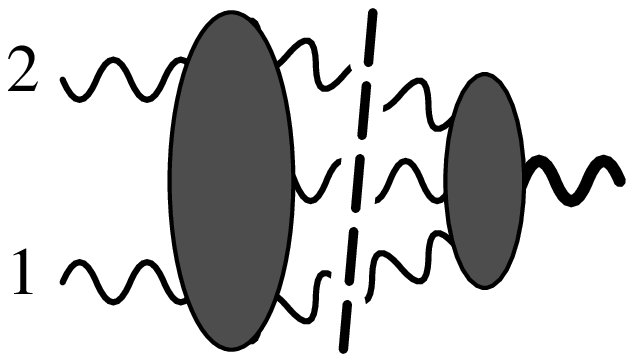}%
\hbox to 2 cm{}}  
\caption{The three-particle cut of a two-loop splitting amplitude.  On the
left-hand side of the cut, the amplitude is gauge invariant
since all legs including the cut ones are on-shell.}
\label{TwoLoopSplittingCutThreeFigure}}

But need we concern ourselves with the possible appearance of 
ill-defined integrals like $L^{(a)}$?
As at one loop, light-cone denominators at two loops can be separated
into two categories, depending on whether they appear in unitarity cuts
or not. For example, consider the three-particle cut of a two-loop splitting
amplitude shown in \fig{TwoLoopSplittingCutThreeFigure}.  The three-particle
cuts of the two integrals in \fig{TwoloopExampleFigure} are shown in
\fig{TwoloopCutExampleFigure}.  Since all legs of the five-point
amplitude on the left-hand side of the cut, including the cut ones,
are fully on-shell, then following the same logic as in the
one-loop case, the light-cone denominator appearing in the integral in
\fig{TwoloopExampleFigure}(a) is a light-cone gauge artifact which can
be eliminated. (This type of argument cannot be used on the right-hand side
of the cut, because the merged leg is off-shell.) The light-cone
denominator appearing in \fig{TwoloopExampleFigure}(b), on the other
hand, is allowed because it corresponds to a physical-state projector
on a cut line.

\FIGURE[t]{
\hbox to \hsize{\hfil%
\SizedFigureWithCaption{1 truein}{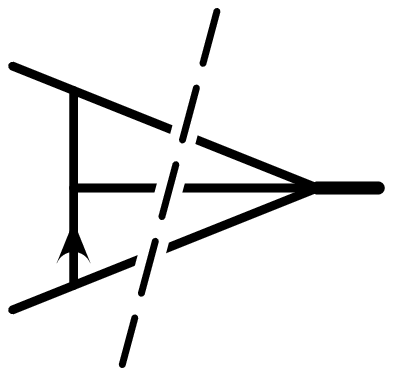}{(a)}
\SizedFigureWithCaption{1 truein}{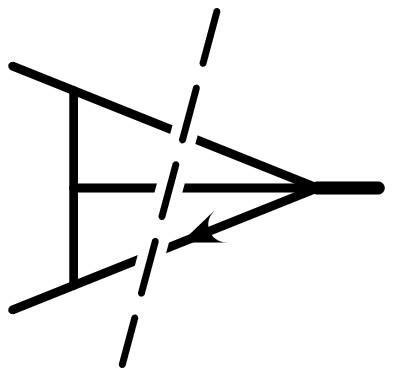}{(b)}
\hfil}
\label{TwoloopCutExampleFigure}
\caption{The three-particle cut of the two-loop three-point integrals
of \fig{TwoloopExampleFigure}.}
}

The use of the unitarity-based sewing method, which we describe
in detail in the following section, allows us to avoid the use
of any prescription for light-cone denominators in the calculations
in following sections.  But one can also imagine applying the
insights above to a more standard diagrammatic calculation in light-cone 
gauge.  One could proceed as follows: introduce one of the standard
prescriptions for dealing with light-cone denominator singularities.
Then, attempt to combine diagrams {\it algebraically\/} to remove
those light-cone denominators which by unitarity cannot appear
in the desired quantity.  Once all singularities leading to ill-defined
integrals have canceled, one can remove the additional prescription
(for example, by taking $\delta\rightarrow 0$ in the PV prescription),
and only then perform the loop integrals.

\FIGURE[t]{
{\epsfxsize 3.5 truein \epsfbox{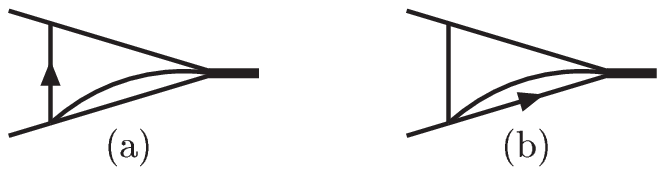}}
\label{TwoloopExampleBFigure}
\caption{Additional three-point integrals at two loops containing
light-cone denominators. Both integrals are well-defined using
dimensional regularization alone.}
}

The unitarity arguments above can be applied not only to ill-defined
integrals, but also to rule out certain well-defined integrals. 
As an example, consider the two integrals in \fig{TwoloopExampleBFigure}.
Both turn out to be well-defined (for integral~(a), this is clear
from momentum-space power-counting).  If we examine their cuts,
shown in \fig{TwoloopCutExampleBFigure}, however, we see that
integral~(a) has a light-cone denominator to the left of the cut,
and hence cannot appear.  Integral~(b) has a light-cone denominator
on the cut, and hence is not ruled out by the unitarity argument.
Indeed, it is the master integral $\Wedge(z,s)$ 
of~\fig{PlanarMasterFigure}, whose explicit expansion in $\e$ 
is given in~\eqn{WedgeLaurent}.

\FIGURE[t]{
{\epsfxsize 3.5 truein \epsfbox{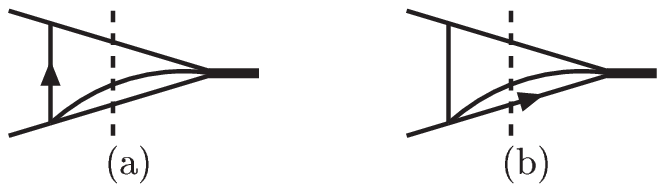}}
\label{TwoloopCutExampleBFigure}
\caption{The three-particle cuts of the two-loop three-point integrals
of \fig{TwoloopExampleBFigure}.}
}

In summary, the unitarity cuts point to a method that sidesteps the
prescription issues associated with light-cone denominators,
because only a restricted set of integrals appear.  In the next
section we explain in some detail how to construct loop-momentum integrands
using the unitarity method. In our calculation of two-loop
splitting amplitudes in sections~\ref{SplittingIntegrandsSection} and
~\ref{IntegralsSection}, such loop integrals are reduced to a linear
combination of master integrals.   There are several equivalent
bases that are convenient for different aspects of the calculation;
one of the equivalent forms contains only integrals well-defined
using dimensional regularization alone, and not requiring any
additional prescriptions.  All singular boundaries lead to
logarithmic singularities in the $\e\rightarrow 0$ limit, just
as for the integral~$J^{(b)}$ discussed in the previous subsection,
or the integral~$L^{(b)}$ above.


\section{Review of the Unitarity-Based Sewing Method}
\label{UnitaritySewingSection}

\subsection{Overview}
The unitarity of the scattering matrix
in a quantum field theory is the statement that probability
is conserved.  It is an essential property of any sensible and
consistent theory.  It relates the non-forward part $T$ of the 
scattering matrix $S$ to its square, $-i (T-T^\dagger) = T^\dagger T$,
where $T$ is defined via $S= 1 + i T$.
In Feynman diagrams, unitarity is expressed by the Cutkosky
rules~\cite{Cutting,PeskinSchroeder}, which express the `imaginary' or
absorptive part of a diagram\footnote{%
By `imaginary' we mean the discontinuities across branch cuts.}
in terms of phase-space integrals over
products of lower-loop diagrams.  The product is given by `cutting',
replacing specified sets of propagators by delta functions in the
propagator momentum.  Loop amplitudes are computed, of course, by
summing over appropriate collections of Feynman diagrams.  Their
absorptive parts are given by sums of products of lower-loop
diagrams. Collecting all diagrams on each side of a cut into
amplitudes, we see that the absorptive parts of loop {\it
amplitudes\/} are just sums over products of lower-loop {\it
amplitudes\/}.

This observation is particularly powerful in gauge theories
(and in gravity as well).  In gauge
theories, there are extensive cancellations between different diagrams
in the computation of scattering amplitudes for on-shell states.
These simplifications can be made manifest at early stages of a
tree-level calculation using the spinor helicity representation for
gauge-boson polarization vectors.  (These simplifications
may be understood using twistor space~\cite{CSW}.) 
The final answers in massless
theories are particularly simple, sometimes simpler than the
expression for a single Feynman diagram out of the hundreds or
thousands that contribute, and have a natural expression in terms of
spinor products.  We may then express the cut of an on-shell one-loop
amplitude, given by a product of on-shell tree amplitudes (or a sum of
such products), in simple form as well.  This simplicity carries
through order-by-order in perturbation theory.
The sewing technique aims
to exploit this simplicity, by turning the process around, and
building loop amplitudes out of their cuts, in turn given by
lower-loop amplitudes.

The full amplitude can in principle be reconstructed using
dispersion relations.  For general gauge theories in four dimensions,
the dispersive reconstruction of an amplitude suffers from an additive
ambiguity related to divergent ultraviolet behavior.  One can add a
rational function, free of cuts, to the amplitude.  This problem has
traditionally hampered the use of dispersion relations to obtain complete
amplitudes.  It is solved
in massless theories\footnote{In massive theories, there is an
additional source of ambiguities (from masses inside bubbles on
external legs).  When there is only one mass in a calculation, 
this problem can be resolved through simple
adjustments~\cite{BernMorgan}.} through the use of dimensional
regularization, which effectively tames the ultraviolet behavior of
the bare integrand~\cite{vanNeerven} and thereby removes the need for
explicit subtractions.  This represents a third role for the
dimensional regulator beyond its usual roles as a regulator for
ultraviolet and infrared divergences.  The sewing technique we review
is equivalent to the use of dispersion relations in dimensional
regularization, although for practical purposes it is preferable to
make use of ordinary Feynman-integral techniques for performing the
necessary integrations rather than doing explicit dispersion
integrals.  (At one loop, for example, knowledge of the complete
decomposition of $n$-point integrals in dimensional
regularization in terms of a basis of known integrals~\cite{Pentagons}
reduces the
problem to an algebraic one.)  Sewing back together cut amplitudes,
with the cut lines on shell but treated exactly in $D=4-2\e$ dimensions,
will reproduce the full gauge-theory answer.  (The different ways
of continuing the amplitude to $D$ dimensions correspond to the use
of different variants of dimensional regularization, such as 
CDR~\cite{CDR}, HV~\cite{HV}, or 
FDH~\cite{BKgggg,TwoloopSUSY}.)
One can then expand
in $\e$ to obtain the answer through ${\cal O}(\e^0)$, including 
the rational terms.  A more
pedestrian way to understand how the rational terms are included properly
is to observe that in dimensional regularization, these terms are not
purely rational, but rather are rational functions of the momentum
invariants, multiplied by $(-s)^{L\e}$ at $L$ loops, for some invariant
$s$ ({\it e.g.} as in \eqn{fezs}).  At ${\cal O}(\e)$, this factor contains
a logarithm, and hence an imaginary part for $s>0$.  
Only when the $\e\rightarrow0$
limit is taken at the end of the calculation, does the term become
purely a rational function, free of discontinuities.
If one calculates the cuts to all orders in $\e$, the full amplitude
can also be reconstructed to all orders in $\e$.

Yet another way to understand the uniqueness of cut reconstruction is
as follows:  In reconstructing a loop amplitude from a single cut, there is
always an ambiguity with respect to adding terms in the numerator of the
loop-momentum integrand which vanish on the cut in question,
{\it e.g.} those terms proportional to $\ell_1^2$ and $\ell_2^2$
where $\ell_1$ and $\ell_2$ are momenta crossing a two-particle cut.
However, after merging the information from all cuts, the only remaining 
ambiguities are those terms which are free of cuts in {\it every} channel.  
For the all-massless theories we consider, such integrals always take the form
of external leg corrections; that is, they are scale-free integrals which
vanish identically in dimensional regularization.

\def\hyph{\hbox{-}}
\def\scut{s\rm\hyph cut}

\subsection{Sewing at One Loop}
\label{OneLoopSewingSubsection}
Consider first the sewing method at one loop.
Before explaining it in generality, 
it will be useful to examine the procedure in a simple example.
In each example here and in later subsections, 
we will make contact with standard methods by starting
with an amplitude expressed in terms of conventional Feynman diagrams.
In the examples, we will work in a massless
$\Tr\phi^3$ field theory (with $\phi$ transforming under the
adjoint of ${\rm SU}(\Nc)$), but as discussed above
the method applies to general theories, and indeed is relatively more
powerful precisely in field theories with many redundant variables in
their covariant form, such as gauge theories and gravity.

\FIGURE[t]{
\hbox{\vbox to 2truein{\vskip -10pt%
\SizedFigureWithCaption{0.8 truein}{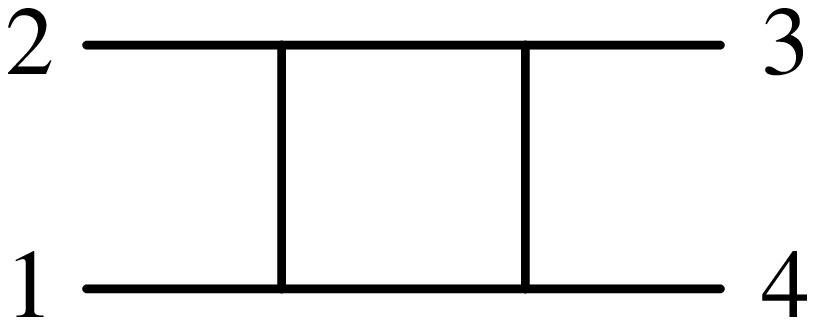}{(a)}
\SizedFigureWithCaption{0.8 truein}{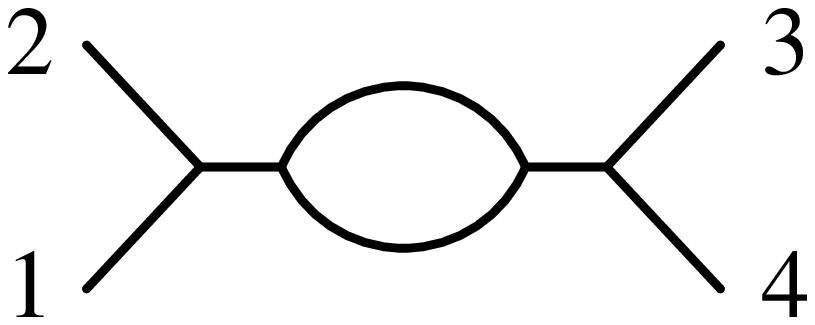}{(b)}
}}
\hbox{\vbox to 2truein{\vskip -10pt%
\SizedFigureWithCaption{1.8625 truein}{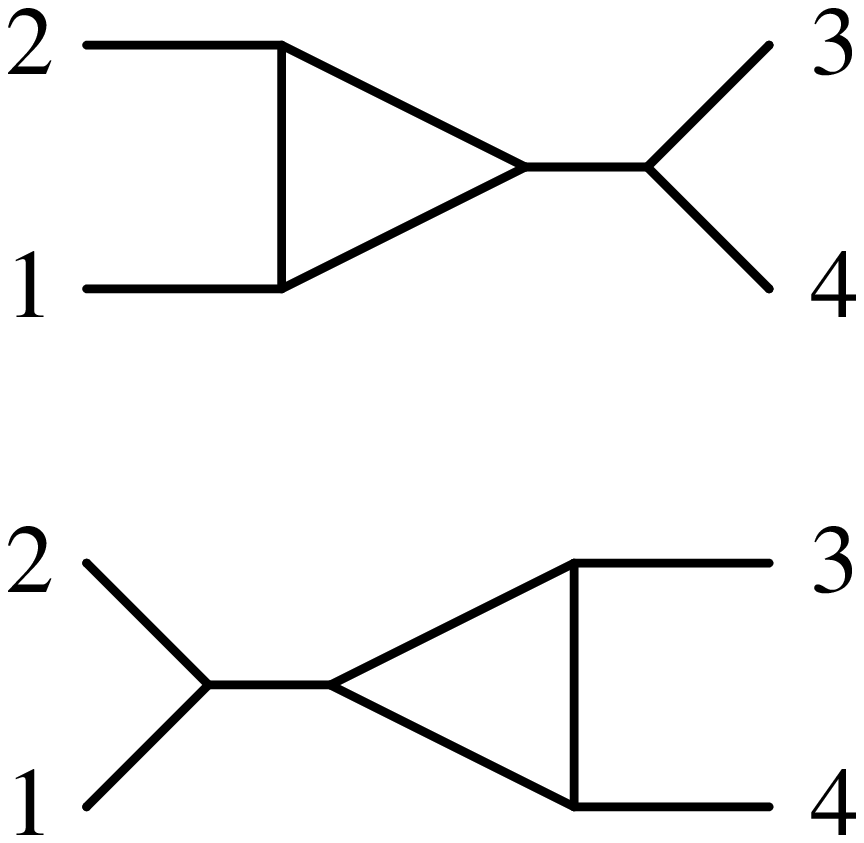}{(c)}
\vfil}}
\label{OneLoopPhi3Figure}
\caption{Color-ordered Feynman diagrams for the one-loop four-point 
amplitude in $\Tr\phi^3$
field theory: (a) the box diagram (b) the $s$-channel bubble diagram
(c) the $s$-channel triangle diagrams.  The two $t$-channel triangle 
diagrams and 
the $t$-channel bubble diagram are not shown explicitly.  Bubbles on
external legs vanish in dimensional regularization and are not shown here
either.}
}

Let us start by considering the four-point one-loop amplitude.  The
full amplitude has Bose symmetry, which just as in the gauge-theory
case we can exhibit most concisely by rewriting it 
as a sum over color permutations of a more basic quantity.  The leading-color
contributions (leading in a $1/\Nc$ expansion) involve only planar
diagrams.  We will focus on these contributions in the examples.
In particular, we focus on the coefficient of the color trace 
$\Tr(T^{a_1} T^{a_2} T^{a_3} T^{a_4})$ given by color-ordered
diagrams~\cite{MPReview,LoopReview} with the 1234 ordering of legs.

\FIGURE[t]{
\hbox{\vbox to 2truein{\vskip -10pt%
\SizedFigureWithCaption{0.8 truein}{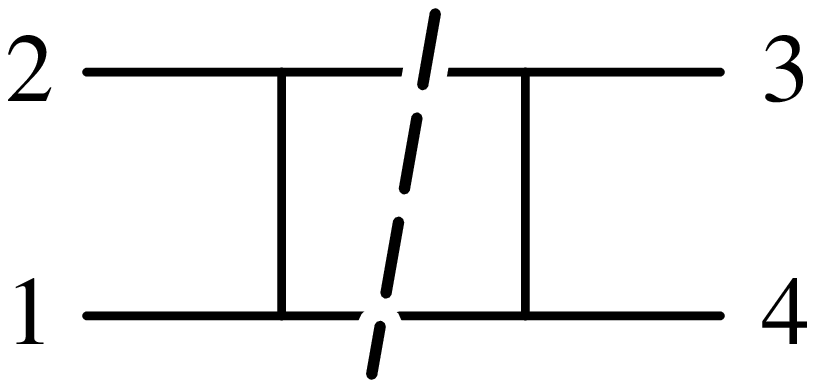}{(a)}
\SizedFigureWithCaption{0.8 truein}{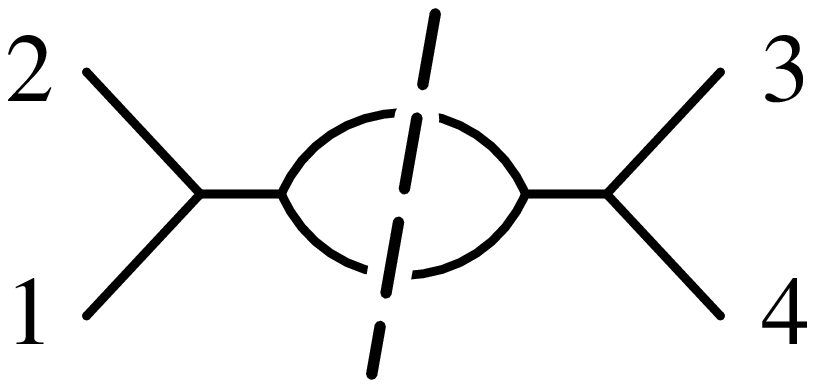}{(b)}
}}
\hbox{\vbox to 2truein{\vskip -10pt%
\SizedFigureWithCaption{1.8625 truein}{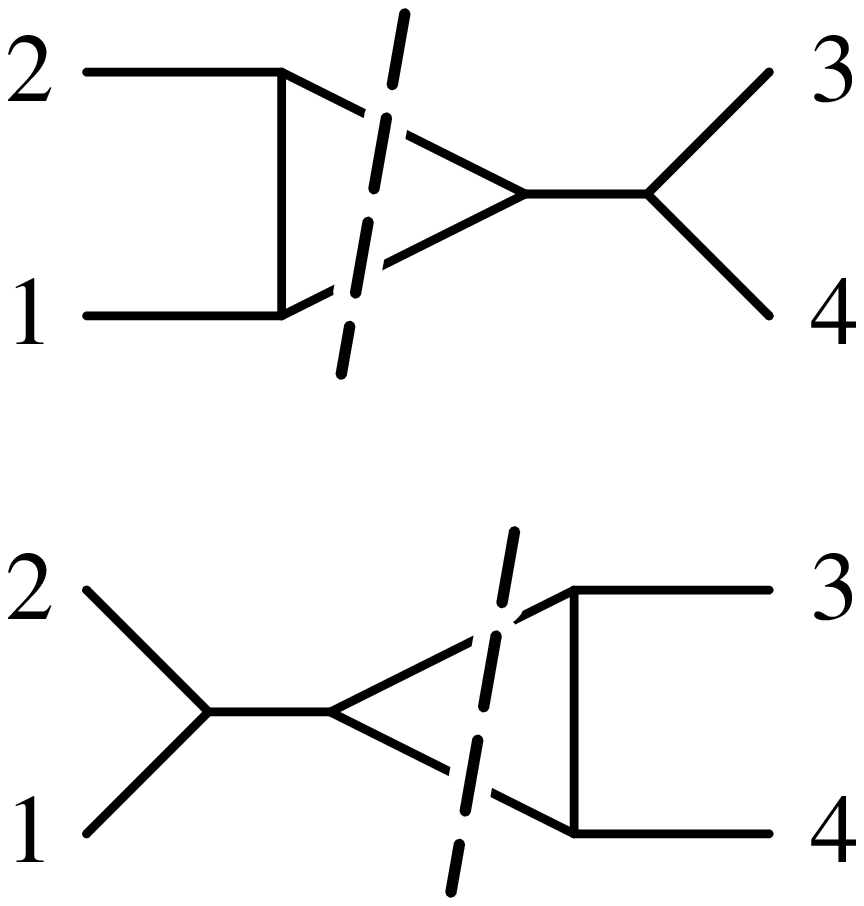}{(c)}
\vfil}}
\label{OneLoopPhi3CutFigure}
\caption{Diagrams depicting the $s$-channel cut of the one-loop 
four-point amplitude in $\Tr\phi^3$
field theory.}
}

There are seven color-ordered diagrams contributing to this one-loop partial
amplitude.
The ones with cuts in the $s$ channel
are depicted in \fig{OneLoopPhi3Figure}.  The ordered
amplitude has cuts in only two channels, $s$ and $t$.  If we examine
the $s$ channel, we see that only four of the diagrams contribute to
the cut: the box, two triangles, and one of the bubbles.  Similarly,
four diagrams contribute to the cut in the $t$ channel.  The
$s$-channel cut may be obtained by replacing the propagators cut in
\fig{OneLoopPhi3CutFigure} via
\begin{equation}
{1\over p^2 + i\epsilon} \rightarrow -2\pi i \deltaplus(p^2).
\end{equation}
This replacement converts the loop integral to one over the phase
space of the two cut legs, which are placed on shell.  We can
also see that the sum of terms factors, so that on each side
of the cut we obtain a tree amplitude as the sum of diagrams, 
as shown in \fig{OneLoopSewingExampleCutSumFigure}.  In each channel,
the cut is thus given by a phase space integral of the product
of two tree amplitudes,
\begin{eqnarray}
A^{(1)}(1,2,3,4) &=& 
\int {d^D\ell_1 d^D\ell_2\over (2\pi)^{D-2}}
    \deltaplus(\ell_1^2)\deltaplus(\ell_2^2)\delta^D(\ell_1+\ell_2+k_1+k_2)
\TimesBreak{\int {d^D\ell_1 d^D\ell_2\over (2\pi)^{2D-2}}}
   A^{(0)}(1,2,\ell_2,\ell_1) A^{(0)}(-\ell_1,-\ell_2,3,4) \,,
\nonumber
\end{eqnarray}
where 
\begin{equation}
A^{(0)}(1,2,3,4) = -i \Bigl( {1\over s_{12}} + {1\over s_{14}} \Bigr) \,,
\end{equation}
and we have suppressed powers of the three-scalar coupling.

\FIGURE[b]{
{\epsfysize 2.4 truein \epsfbox{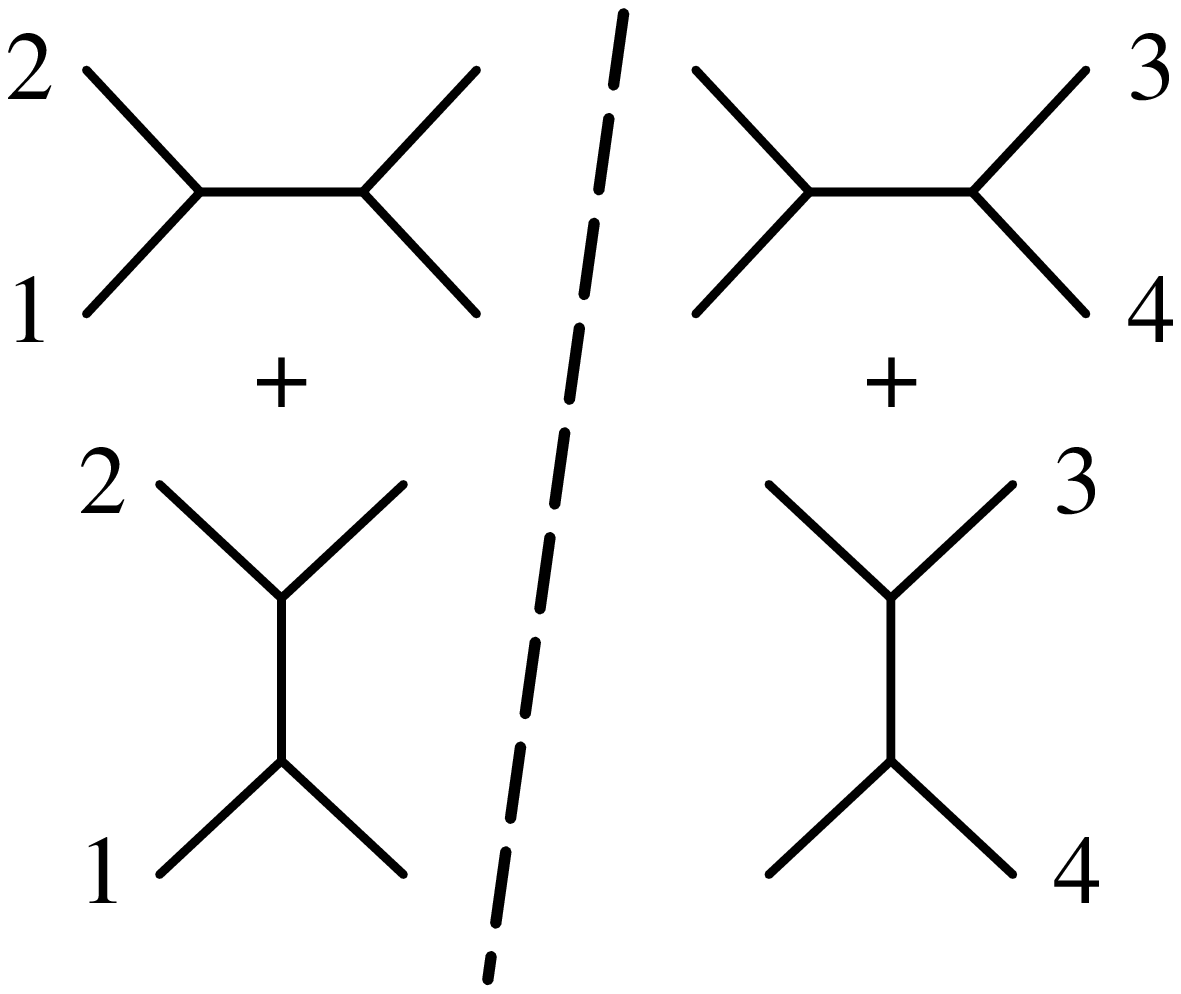} \hskip 1 cm }
\caption{The $s$-channel cut of the one-loop 
four-point amplitude in $\Tr\phi^3$
field theory, seen as a product of tree amplitudes.}
\label{OneLoopSewingExampleCutSumFigure}
}
The sewing procedure reverses this process.  We start, for example, in
the $s$ channel.  Multiply the tree amplitude on the left-hand side
of \fig{OneLoopSewingExampleCutSumFigure} by that on the right-hand side,
\begin{equation}
A^{(0)}(1,2,\ell_2,\ell_1) A^{(0)}(-\ell_1,-\ell_2,3,4)
= - \Bigl({1\over s_{12}}+{1\over s_{1\ell_1}}\Bigr)
\Bigl({1\over s_{34}}-{1\over s_{\ell_1 4}}\Bigr),
\end{equation}
where $\ell_1 = -\ell_2-k_1-k_2$.  Put in the two propagators crossing
the cut, and integrate over the loop momentum $\ell \equiv \ell_1$,
to yield,
\begin{eqnarray}
&&\int {d^D\ell\over (2\pi)^D}\;
 \Bigl({1\over s_{12}}+{1\over s_{1\ell}}\Bigr) {1\over \ell^2}
\Bigl({1\over s_{34}}-{1\over s_{\ell4}}\Bigr)
{1\over (\ell+k_1+k_2)^2}\nonumber\\
&&\hphantom{X} = \int {d^D\ell\over (2\pi)^D}\;
 \biggl[
{1\over s_{12}^2} {1\over \ell^2 (\ell+k_1+k_2)^2}
+{1\over s_{12}} {1\over \ell^2(\ell-k_4)^2 (\ell+k_1+k_2)^2}
\PlusBreakLab{OneLoopPhi3SChannel}{X  = \int {d^D\ell\over (2\pi)^D}\;\biggl[]}
{1\over s_{12}} {1\over (\ell+k_1)^2 \ell^2 (\ell+k_1+k_2)^2}
+{1\over (\ell+k_1)^2 \ell^2 (\ell-k_4)^2 (\ell+k_1+k_2)^2}
\biggr]\,.\hskip .5 cm \null \nonumber
\end{eqnarray}
Similarly, from the $t$-channel cut, we obtain,
\begin{eqnarray}
&&\hskip -10pt\int {d^D\ell\over (2\pi)^D}\;
 \biggl[
{1\over s_{23}^2} {1\over\ell^2 (\ell+k_2+k_3)^2}
+{1\over s_{23}} {1\over \ell^2(\ell-k_1)^2 (\ell+k_2+k_3)^2}
\PlusBreak{\hskip -10pt\int {d^D\ell\over (2\pi)^D}\;  []}
{1\over s_{23}} {1\over (\ell+k_2)^2 \ell^2 (\ell+k_2+k_3)^2}
+{1\over (\ell+k_2)^2 \ell^2 (\ell-k_1)^2 (\ell+k_2+k_3)^2}
\biggr]\,.  \hskip 1 cm \null 
\label{OneLoopPhi3TChannel}
\end{eqnarray}
The first three terms have no cut in the $s$ channel, but the last term
does: it is given by the residue of the poles as $\ell-k_1$ and $\ell+k_2$ 
simultaneously go on shell,
\begin{equation}
-{1\over 2 (\ell-k_1)\cdot k_1\, 2 (\ell-k_1)\cdot k_4}.
\end{equation}
Alternatively, we can shift $\ell\rightarrow \ell+k_1$, upon which
the last term becomes identical to the last term in 
\eqn{OneLoopPhi3SChannel}.  

We cannot simply add the contributions from the $s$ and $t$ channels,
because this would correspond to the sum of eight diagrams, double-counting
the box diagram~\ref{OneLoopPhi3Figure}(a), given by
the last terms in eqs.~(\ref{OneLoopPhi3SChannel}) 
and~(\ref{OneLoopPhi3TChannel}).  Accordingly, we must find
a function which has the correct cuts in all channels.  We can do this
either before or after integration, although in general it is easier to
do it before loop integration.  One way is simply to sum both contributions,
and then remove the overlap: terms in one cut channel 
which also have a `cut' ---
in the sense of having the propagators which give rise to a cut --- in the
other channel.  One can alternatively think of this as `merging' the two
expressions, taking a term if present in either cut or in both, but taking it
only once in the latter case.

The net effect is to drop one of the two equivalent terms;
we obtain the sum of the remaining
terms for the ordered one-loop amplitude,
\begin{eqnarray}
&&\hskip -10pt A^{(1)}(1,2,3,4) =\nonumber\\
&&\int {d^D\ell\over (2\pi)^D}\;
 \biggl[
{1\over s_{12}^2} {1\over \ell^2 (\ell+k_1+k_2)^2}
+{1\over s_{12}} {1\over \ell^2(\ell-k_4)^2(\ell+k_1+k_2)^2}
\PlusBreak{\int {d^D\ell\over (2\pi)^D}\; \biggl[]}
{1\over s_{12}} {1\over (\ell+k_1)^2 \ell^2 (\ell+k_1+k_2)^2}
+{1\over (\ell+k_1)^2 \ell^2 (\ell-k_4)^2 (\ell+k_1+k_2)^2}
\PlusBreak{\int {d^D\ell\over (2\pi)^D}\; \biggl[]}
{1\over s_{23}^2} {1\over \ell^2 (\ell+k_2+k_3)^2}
+{1\over s_{23}} {1\over \ell^2(\ell-k_1)^2(\ell+k_2+k_3)^2}
\PlusBreakLab{OneLoopPhi3Complete}{\int {d^D\ell\over (2\pi)^D}\; \biggl[]}
{1\over s_{23}} {1\over (\ell+k_2)^2 \ell^2 (\ell+k_2+k_3)^2}
\biggr],\nonumber
\end{eqnarray}
exactly as would have emerged from a Feynman-diagram computation.  Of
course, in a $\phi^3$ field theory, there are no cancellations between
different diagrams, so the sewing method is also equivalent in complexity
to the usual approach.  In gauge theories, the sewn on-shell tree amplitudes
are much simpler objects than the one-loop diagrams, and so the sewing
approach helps minimize the complexity of intermediate steps.

In the above examples, the procedures for sewing and removing
any overlaps or double-counting are completely mechanical.  Note
that none of them make any reference (to use an old-fashioned language)
to double dispersion relations. Indeed, only in an abstract sense
are dispersion relations used at all, since we do not perform the
dispersion integrals explicitly, but rather implicitly via construction
of appropriate Feynman integrals.
We will next explain how to formalize these procedures, and then
give an algorithm which can be used to implement them in practice.

\def\promote{%
\rlap{$\displaystyle\left\lceil\vphantom{\sum}\right.$}%
{\displaystyle\left\rfloor\vphantom{\sum}\right.}}
\def\promoteB{\promote}
\def\promoteC{\promote_{\kern -3pt C}}
\def\CutProj{{\cal P}}
To formalize the sewing procedure, introduce the basic {\it promotion}
operator $\promoteB A$.  It will be applied to products of
amplitudes (or to terms from a product).  It represents
the combined operations of summing
over helicity states (and over different particle states if appropriate),
re-expressing spinor products in terms of the cut momentum, multiplication
by the cut-crossing propagators, and completion of dot products in
denominators to standard propagator denominators.  
It does not introduce the phase-space
integral over the product of amplitudes.  The result of the promotion
operation is an integrand which depends on the external momenta and 
on the loop momentum.  Note that the sum over intermediate states must
in general be carried out in $D=4-2\e$ dimensions, and that it is
a sum only over physical states.  This implicitly introduces a physical
projection operator.  In calculations of full amplitudes, the 
resulting operators
leave little trace, but in calculations of splitting amplitudes such as
the one we carry out in the present paper, these operators will give
rise to light-cone-like denominators in integrals.

\def\Channels{{\cal C}}
\def\nChannels{n_\Channels}
\def\First{a}\def\Last{b}
We will also need to introduce the cut-projection
operator $\CutProj_s$, which yields the part of its argument that has
a cut in the $s$ channel, where $s$ denotes an arbitrary invariant
of $m$ consecutive external momenta.  At one loop, it extracts the joint pole
term in two propagator denominators ${1\over \ell_1^2\ell_2^2}$,
where $(\ell_1+\ell_2)^2 = s$.  It corresponds to requiring that
a pair (any pair) of propagators yielding a cut in the $s$ channel
be present in the diagram.

In a computation with $n$
external massless momenta, there are in general $n(n-3)/2$
independent invariants in $D$ dimensions.  Denote by $\Channels$
the ordered set of these invariants,
\begin{equation}
\Channels = \{s_{12},s_{23},s_{123},\ldots\}.
\label{CChannels}
\end{equation}
(We will denote the number of elements in $\Channels$ by $\nChannels$.)
We will sew the channels in the specified order, with the notation
$s_j\in \Channels$ denoting the $j$-th invariant in $\Channels$.
The optimal ordering (from the viewpoint of computational efficiency)
depends on the process and the particle content of the theory, 
but of course the final answer is independent
of this ordering.  
Let $K_j=k_{\First_j}+\cdots k_{\Last_j}$ be the momentum whose
square is the given invariant $s_j$.  The first momentum 
(within the cyclic order
of external momenta) we will denote $\First_j$,
and the last momentum by $\Last_j$.  The momentum before $\First_j$ will
be labeled $\First_j-1$, and the one after $\Last_j$, $\Last_j+1$.

The analytic behavior in different invariants is independent even if the
invariants are related by Gram determinant conditions arising from 
the restriction to four dimensions.
Thus even if we take all external momenta to be in four dimensions, we must
still take the full set of $D$-dimensional invariants.  
 
\def\Integrand{I}
The complete integrand of the one-loop amplitude 
$A^{(1)}(1,\ldots,n)$
is then given by the sum over all channels,
\begin{eqnarray}
\Integrand^{(1)} &=& \sum_{j=1}^{\nChannels} 
{\textstyle\prod_{l=1}^{j-1}}\Bigl(1- \CutProj_{s_l}\Bigr)
\CutProj_{s_j}
   \promoteB 
A^{(0)}(\ell,\First_j,\ldots,\Last_j,-\ell-K_j)
\TimesBreak{ \Pi_{l=1}^{j-1}\Bigl(1-\CutProj_{s_l}\Bigr)\biggl[]\promoteB
 \hskip 10 mm \null }
A^{(0)}(\ell+K_j,\Last_j+1,\ldots,\First_j-1,
        -\ell) \,.
\label{OneLoopSewingIngredient}
\end{eqnarray}
That is, we sum over all channels, each time removing all terms already
found in previous channels.  
The amplitudes in this equation must in general
have the sewn legs ($\ell$ or $\ell+K_j$) in
$D$ dimensions.  Whether the external legs are taken to be in four
dimensions or in $D$ dimensions depends on the variant of dimensional
regularization employed.  In practice, it is best to use a four-dimensional
scheme, and convert later if necessary.  

Indeed, there are several practical aspects not addressed 
by the formal expression
above.  These include questions of diagram labeling, classification, and
the use of a basis for organizing numerators of terms in the integrand.
As the formal expression hints,
 none of these tools are intrinsically required by the unitarity-based
sewing method.  The cut projection can be performed by extracting residues
of poles; and the promotion involves simple algebraic manipulations.
Furthermore, non-manifestly vanishing expressions still vanish, and do
not affect the final answer.
In a practical calculation, however, we would like the cut projection
to be simple, ideally just amounting to the identification of
the formal coefficient of a pole.  We would like to avoid the appearance
of complicated expressions which actually vanish.
Furthermore, the integrand produced by the sewing method will 
ultimately be fed to
an integration machinery which does require the corresponding diagrams
to be labeled and classified by topology.  We therefore might as well
incorporate these aspects into an algorithm.  

\FIGURE[b]{
\hbox{%
\SizedFigureWithCaption{1 truein}{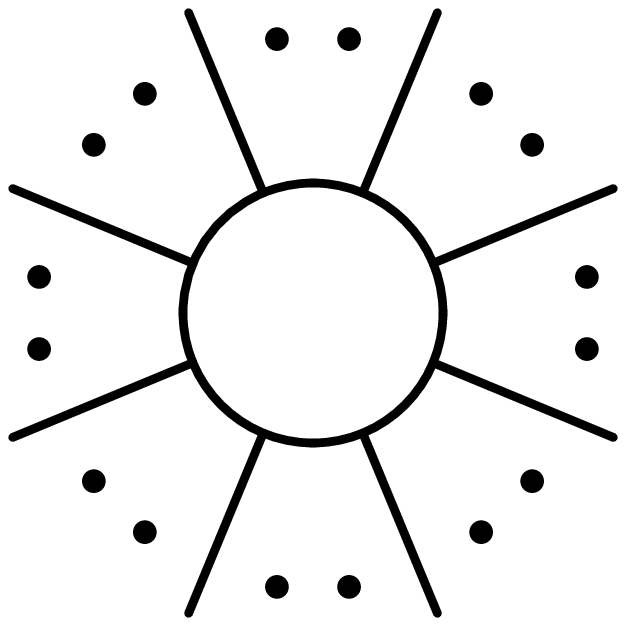}{(a)}
\SizedFigureWithCaption{1 truein}{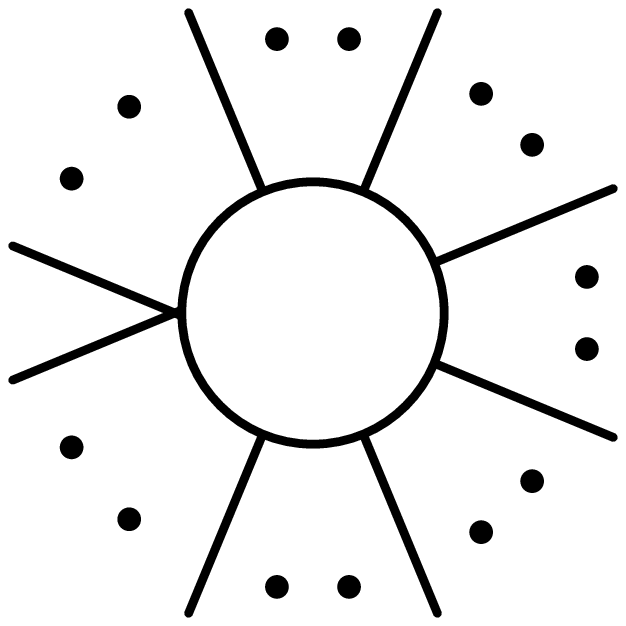}{(b)}
\SizedFigureWithCaption{1 truein}{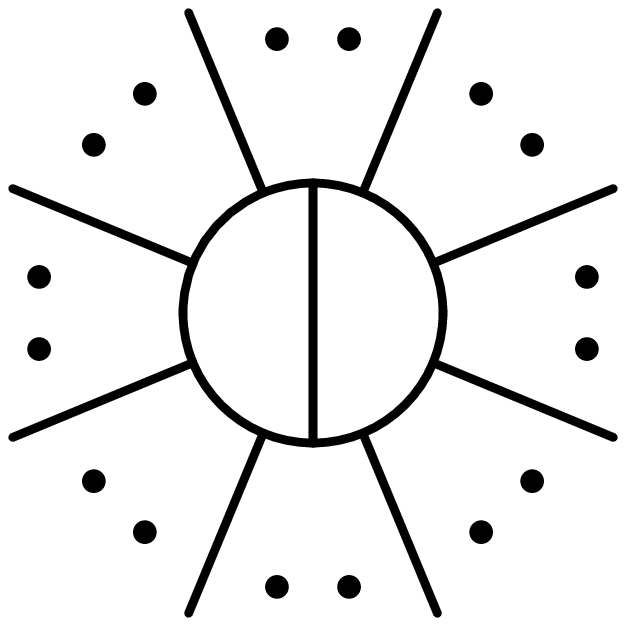}{(c)}
\SizedFigureWithCaption{1 truein}{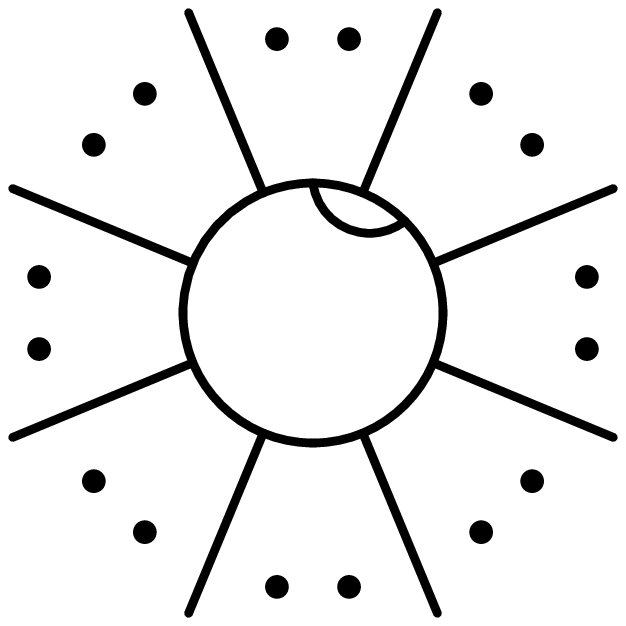}{(d)}
}
\caption{Examples of parent and daughter diagrams:
(a) a parent diagram at one loop; (b) a daughter diagram
at one loop, which is a daughter of the diagram in (a);
(c) and (d) parent diagrams at two loops.}
\label{ParentFigureExample}
}

To do so, start with all color-ordered graphs with a maximal number of
propagators containing the loop momenta.  We will call these `parent'
diagrams.  All other topologies can be obtained by canceling
propagators, that is by multiplying the numerator by an inverse
propagator.  These we will call `daughter' diagrams; below, we will
also include the parent in its set of all daughters.  
At one loop, if all external particles are massless, there is 
in fact only one parent diagram.  (Recall that we are restricting
attention to processes where all internal masses vanish.)
If we have two or more massive external particles (for
example, $W$ bosons), we will have
different parent diagrams corresponding to the different ways of
attaching the massive legs to the loop.  At two loops, there are
$n(n+1)/2$ planar parent diagrams when all external particles are
massless.  (Integrals that reduce to products of one-loop integrals,
such as bowtie integrals, have intrinsically two-loop integrals as
parent diagrams.)  Examples are shown in
\fig{ParentFigureExample}.

Each cut will
in general start with a different labeling of any given parent diagram, because
the loop momentum $\ell$ may denote a different propagator.  The
algorithm will make use of a simple relabeling operation, reviewed in
appendix~\ref{RelabelingAppendix}, to bring these into canonical form.
The labeling of propagators must incorporate a notation for the parent
diagram, because the algebraic relations of dot products
to inverse propagators differ from diagram to diagram.

\def\ubar{\bar u}
The external gluon legs we may choose to treat using formal polarization
vectors $\pol_i(k_i)$, or using the spinor-helicity method.  The external
fermion legs we may choose to treat using formal spinor wavefunctions
$u(k_i)$, $\ubar(k_i)$, or again using a helicity basis.  Either choice
(or a mixture) may be employed with the algorithm we will present below.
The basis one should use for expressing numerator polynomials
depends slightly on the external leg treatment, because there can
be different numbers of independent invariants from which the 
polynomials are built.  The algebraic processing of
expressions will also be somewhat different.  
In all cases, the basis at one loop will contain all inverse propagators
containing the loop momentum.
These are sufficient to express all dot products of the loop momentum with
external momenta.  Note that Levi-Civita tensors involving the loop momentum
can be converted to Gram determinants (and thence to dot products)
 by multiplying by another Levi-Civita tensor involving only external momenta.
(The latter object is just another constant as concerns the manipulations we
perform.)  We will also need the square of $(-2\e)$-dimensional components
of the loop momentum, $(\ell_{-2\e})^2$.
We must add dot products of the loop momentum with
formal external polarization vectors (if any).  These will not
give rise to expressions that can cancel propagator denominators.  (We
{\it could\/} have taken these to be in $D$ dimensions, after all, in which
case they would clearly be independent.)  In all algebraic
manipulations, one
should be sure to use momentum conservation, eliminating one external
momentum, and re-expressing invariants in terms of an independent set,
in order to avoid the appearance of zero in obscure forms.  (For formal
expressions $\pol_j\cdot k_i$, one should pick a momentum other than $k_j$
to eliminate, so as to impose the on-shell conditions too.)

For example, in computing the an $n$-point gluon amplitude, we can pick
the standard labeling to have the loop momentum between legs $n$ and
$1$.  If we treat all external legs in the spinor-helicity basis,
then the basis set will simply be,
\begin{equation}
\{\ell^2, (\ell-k_1)^2, (\ell-k_1-k_2)^2, \ldots,
(\ell-k_1-\cdots-k_{n-1})^2,
(\ell_{-2\e})^2\};
\end{equation}
if we choose to treat legs $1,\ldots,j$ using formal polarization vectors,
we should add
\begin{equation}
\{\pol_1\cdot\ell,\ldots,\pol_j\cdot\ell\}
\end{equation}
to this set.

Amplitudes with fermions in a loop arise from sewing amplitudes
with pairs of external fermions on either side of a cut. Sewing will include 
closing a fermion loop. (The usual minus sign must be included 
explicitly.)  This yields a spinor
trace, which can be expanded in terms of dot products and
Levi-Civita tensors.  The latter can be converted to dot products
as described above.  Internal fermions thus do not require any new
basis elements.

Amplitudes with external fermions will
contain in different terms, factors of a `spinor string' consisting of
an external spinor wavefunction (either formal $u$, $\ubar$ or in the
helicity basis), a product of gamma matrices dotted into various $D$-
or four-vectors (momenta, polarization vectors, the light-cone vector,
or other spinor strings), and an ending spinor wavefunction.  Roughly
speaking, we need to perform sufficient manipulations on these to
ensure that no difference of two such objects contains a factor of an
inverse propagator.  A basis for spinor strings involving spinor
wavefunctions can be obtained by commuting loop momenta to the left;
commuting $\s{\ell}_{-2\e}$, if present, to the next position; and
commuting formal polarization vectors (if present) to an ordered
sequence following them.  The spinor string can end with either a
spinor carrying an external momentum or the light-cone vector, or
another external fermion wavefunction.   Alternatively, one can convert
the spinor string to a trace by multiplying by appropriate spinorial
factors involving only external momenta and spins, and then expanding
the trace into dot products and Levi-Civita tensors as above.  (The
spinorial factors are just spinor products in a helicity basis.) In
this case, no spinor strings are needed in the basis.

The projection $\CutProj_{s_j}$ of the promoted integrand in
\eqn{OneLoopSewingIngredient} back onto the same channel which was
sewn, ensures that no (spurious) terms lacking a cut in the sewn
channel are generated.  This projection is not really needed in the
purely formal expression, but ensures that algebraic manipulations
when working in the basis required for a practical algorithm do not
create unwanted terms.

The basis will be used to identify terms that have cuts in different 
channels.  A term with uncanceled propagators corresponding to the cut
channel will have a cut in that channel.

For a practical algorithm, one may proceed as follows:
\begin{itemize}
\item[1.] Form the ordered set of all independent channels in $D$ dimensions.
(That is, in determining the independence of different invariants, one
should use only momentum conservation, and not 
integer-dimension-specific Gram determinant relations.)

\item[2.] Associate a labeling of internal lines to each distinct 
Feynman-integral parent which has a daughter appearing in the integrand.
(Because of cancellations in gauge theories, not every topology that
appears in the set of the usual Feynman diagrams for a given process will
necessarily appear in the sewn integrand.)  

\item[3.] For each parent integral,
form a basis set for expanding numerators, consisting of
the inverse propagators; the square of the $(-2\e)$-dimensional
components of the loop momentum, $(\ell_{-2\e})^2$; and dot products of
the loop momentum with any formal external polarization vectors.
For external fermions with formal wavefunctions, spinor strings
should be added as described above.

\item[4.] Initialize the integrand's value to zero.
\end{itemize}
We will form the integrand by iterating over channels.  For a 
color-ordered amplitude, each channel corresponds to a consecutive set of
external momenta, $k_{\First_j},\ldots,k_{\Last_j}$, the cut invariant
being $s_j = (k_{\First_j}+\ldots+k_{\Last_j})^2$.
 
\FIGURE[t]{
\centerline{\epsfysize 1.4 truein \epsfbox{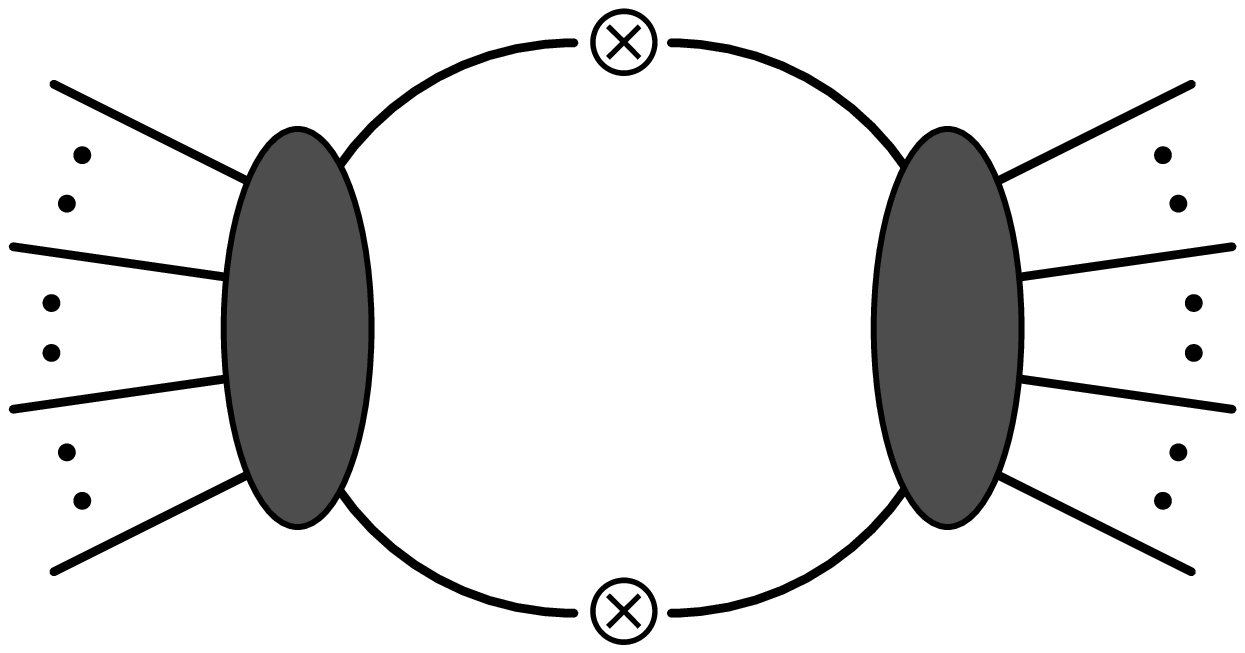}}
\label{OneLoopSewingGeneral1Figure}
\caption{Sewing together two tree-level amplitudes to produce a contribution
to the one-loop integrand.}
}

For each channel, 
\begin{enumerate}
\addtocounter{enumi}{4}
\item \label{OneLoopSewingStep} {\it Sew.}
Form the product of the two on-shell tree amplitudes 
$A^{(0)}(\ell_1,\First_j,\ldots,\Last_j,\ell_2)$ and
$A^{(0)}(-\ell_2,\Last_j+1,\ldots,n,1,\ldots,\First_j-1,-\ell_1)$,
where $\ell_1+\ell_2+K_j = 0$, summing over
the products of amplitudes for the different particle types and helicities
that can circulate in the loop.  The sum must in general be performed
in $D=4-2\e$ dimensions.  Use polarization-vector identities
(\eqn{PolarizationSum}, {\it etc.}),
\begin{eqnarray}
\sum_{\sigma} \pol_\mu^{(\sigma)}(\ell) \pol_\nu^{(\sigma)*}(\ell)
&\rightarrow& -\eta_{\mu\nu}
   +{\ell_\mu n_\nu + n_\mu \ell_\nu\over \ell\cdot n},
\nonumber\\ u(\ell)\ubar(\ell)  &\rightarrow& \s{\ell},
\qquad u(-\ell)\ubar(-\ell)  \rightarrow -\s{\ell},
\label{PolarizationIdentities}
\end{eqnarray}
to express everything in terms of the cut momentum $\ell$.
Here, $n$ is the light-cone reference vector.
(If all external legs are on shell, different $n_i$ may be used
for different cut legs if desired, and indeed the light-cone denominators 
can be removed algebraically.)
The dimensionality of $\eta$ --- the value of $\eta^\mu_\mu$ ---
depends on the variant of dimensional regularization (see \eqn{Dsdef}).
Multiply by $-1$ when a fermion loop is created by sewing.
Put in the propagators crossing the cut, $i\over\ell^2$ for each cut leg.
Complete dot products in the denominator
to form propagator denominators adjacent to the cut, 
$2\ell\cdot k_i \rightarrow \pm (\ell\pm k_i)^2$,
the $+$ corresponding to $k_i$ on the left side of the cut, the $-$ sign
to those on the right side of the cut.   
Rewrite the expression in terms of the basis set, and expand sums so 
that each term can be classified as the daughter of a single parent diagram.  
As mentioned above, the sewing procedure for a given cut is ambiguous with
respect to numerator terms which vanish on the cut in question,
{\it i.e.} those terms proportional to $\ell_1^2$ and $\ell_2^2$.  
After merging the different cuts (in step 9 below), the only remaining 
ambiguities will be those which have no cuts in any channel.  Such 
integrals have the topology of external leg corrections, and vanish for
dimensionally-regulated all-massless theories.
The sewing operation is depicted schematically in
\fig{OneLoopSewingGeneral1Figure}.  In \eqn{OneLoopSewingIngredient}, 
it corresponds to the promotion operator $\promoteB$.

\item {\it Put into canonical form.} Use momentum conservation 
to reduce the number of cut-crossing momenta (now loop momenta) 
appearing in spinor traces, and then expand spinor traces 
and all dot products in terms
of the basis.  (As explained above, this expansion may not always 
be necessary.)  When using the spinor-helicity method for external
polarization vectors, or explicit helicity states, we will obtain
spinor strings of the form 
$\langle j_1^-|\s{j}_2\cdots\s{\ell}\cdots\s{j}_3|j_4^\pm\rangle$.
As explained above, one can
 complete these to a trace, then convert the trace to
dot products and Levi-Civita tensors.  The latter should be converted
to Gram determinants (and thence to dot products) by multiplying (and dividing)
by a Levi-Civita tensor involving only external momenta.
(Spinor strings involving only external momenta need not be manipulated,
obviously.)

\item {\it Relabel.\/} For each term in the sewn expression for a given
channel, relabel the momenta to the standard labeling for its
parent integral.  Where required, insert factors of squared momenta in the
numerator and denominator to match a `parent' diagram.  In some cases,
it will be possible to obtain different parents by inserting different
factors; it doesn't matter which one is picked.

\item {\it Clean.}  Remove all terms which have no cut in the
current channel.  (Such terms might have been introduced by earlier algebraic
manipulations.)  That is, using a canonical basis as described above, remove
any terms which do not contain both cut propagators.  This step
corresponds to the operator $\CutProj_{s_j}$
in \eqn{OneLoopSewingIngredient}.

\item {\it Merge.} Remove all terms in the current-channel 
sewn expression that already
appear in the net integrand.  That is, remove any term which has cuts in
a previously-processed channel.  
Using a canonical basis as described above, it suffices to pick out
and remove terms that have a pair of propagators corresponding to
a cut in a previously-processed channel. This step
corresponds to the operator 
${\textstyle\prod_{l=1}^{j-1}}\Bigl(1- \CutProj_{s_l}\Bigr)$
in \eqn{OneLoopSewingIngredient}.

\item {\it Accumulate.} 
Add the remaining terms in the current-channel sewn
expression to the integrand.  This step corresponds to the sum 
in \eqn{OneLoopSewingIngredient}.

\item Continue with the next channel at step \#\ref{OneLoopSewingStep}.
\end{enumerate}

In special cases (for example, massless supersymmetric theories at
one loop), it may be possible to compute the cuts using four 
dimensional helicity states, and to make use of spinor-helicity
simplifications for the cut-crossing momenta~\cite{Neq4Oneloop,Neq1Oneloop}.  
(It is always possible to
use such simplifications for the external momenta at an early stage of
the calculation.)  In this case, one must re-express spinor products involving
$\ell_1$ and $\ell_2$ in terms of dot products of these momenta with
other vectors, 
$\spa{a}.{\ell_1}\spb{\ell_1}.b\rightarrow \sand{a}.{\s{\ell}_1}.{b}$.
(This is always possible because the cut-crossing momenta will appear
with opposite phase weight in the two tree amplitudes on either side
of the cut.)
It may also happen that some channels are redundant; this will
happen when all integrals appearing in the answer have cuts in multiple
channels.  
In the computation of splitting amplitudes, one side of the cut will
have an on-shell tree amplitude, while the other side contains a
tree-level splitting amplitude.  The sewing procedure will introduce
physical-projector denominators.
These resemble light-cone gauge denominators.
In the computation of on-shell amplitudes, these
physical-projector denominators will disappear
algebraically.  (This is a consequence of gauge invariance.)  In the
computation of splitting amplitudes, in contrast, these denominators
will survive into the integration, and in fact play a crucial role in 
obtaining the right sort of integral.  However, as discussed in 
section~\ref{FeynmanSection},
at one loop they will only arise in lines corresponding to cut momenta. 
As we have seen in the example of the $1\to2$ splitting amplitude, 
integrals with such projectors have no singularities beyond
those regulated by dimensional regularization.

In general, the integrand has cuts in many different channels.
Indeed, the resulting loop amplitudes are expressed in terms
of polylogarithmic functions that have discontinuities in several
different invariants.  Accordingly, any given term in the integrand
may have combinations of propagators leading to cuts in different channels,
and thus might emerge from sewing in any of those channels.  The
different ways of sewing must all yield the same answer.  The simplest
way of seeing this is to consider a gedanken calculation of a one-loop
amplitude from Feynman diagrams.  We can extract the cut of the
sum of all diagrams in any given channel.  Putting the cut legs on shell,
the sums of diagrams on either side will yield a
product of tree amplitudes. Because of their origin in a unique expression,
however, reconstructions of the analytic functions from different
channels will yield the same answer.

In terms of the cut projection operator introduced
earlier, this amounts to the statement that
\begin{equation}
\CutProj_{s_1} \promoteB A_1(s_2) A_2(s_2)
= \CutProj_{s_2} \promoteB A_3(s_1) A_4(s_1),
\label{CutConsistency}
\end{equation}
where $A(s)$ is an abbreviation for the amplitude with two adjacent
legs $\ell_{1,2}$ satisfying $2\ell_1\cdot\ell_2 = s$.
Checking the consistency of different cuts in intermediate steps, 
as expressed by this equation, is a good way of verifying the correctness 
of code implementing the algorithm.  

The consistency of different cuts also shows that the order
of evaluation of channels, {\it i.e.} the ordering in $\Channels$ 
in \eqn{CChannels}, does not affect the final result.  For
example, if we were to evaluate two channels in the order $\{s_1,s_2\}$,
we would obtain,
\begin{equation}
\promoteB A_3(s_1) A_4(s_1) 
+(1-\CutProj_{s_1})\promoteB A_1(s_2) A_2(s_2).
\end{equation}
(Recall that the projection back onto the sewn channel is not really
needed in the formal expression, and has been omitted here.)
In the other order, we would obtain
\begin{equation}
\promoteB A_1(s_2) A_2(s_2) 
+(1-\CutProj_{s_2})\promoteB A_3(s_1) A_4(s_1).
\end{equation}
The difference of the two evaluations is
\begin{equation}
\CutProj_{s_2}\promoteB A_3(s_1) A_4(s_1)
-\CutProj_{s_1}\promoteB A_1(s_2) A_2(s_2),
\end{equation}
which vanishes using \eqn{CutConsistency}.

Many terms (all, in an $\Neqfour$ supersymmetric theory) may
have cuts in more than one channel.  The cut consistency condition
can be used to cross-check these terms, and is often of great utility
in debugging computer code implementing a calculation.
Note that when using the spinor-helicity techniques, the use of non-trivial
(Gram determinant) identities may be required to show cut consistency.

Each sewn and integrated cut is gauge invariant independently when all
external legs are on shell.  Its absorptive part is, after all, equal
to the phase-space integral of a gauge-invariant non-forward matrix
element.  The dispersion integral of such a quantity is gauge
invariant as well.  At the integrand level, this is reflected in the
disappearance of any dependence on the light-cone vector $n$
introduced by the sum over gluon polarizations.  (One typically needs
to make use of momentum conservation to see this explicitly.)  The
same statement is not true if we are computing an object with
off-shell legs, as is the case for the splitting amplitude.  In that
case, some light-cone denominators will survive.  As we will discuss
in a later subsection, others will cancel algebraically, and it is
possible to predict in advance which must disappear.

The procedure we have described above is not the only way to merge
information from different cuts to yield a single function with
the correct cuts in all channels. 
One can also imagine performing the merging step after integration.
That is, one could integrate the sewn integrand in each channel separately,
and then search for a function whose branch-cut discontinuities match
those found in each of the separate invariants.  This requires the
use of a nonredundant basis of master integrals.   Operationally, one
would rewrite the result of integration in each channel in terms of this
basis, throwing away integrals with no discontinuity in the given channel.
One could verify cut consistency by checking that the coefficients of
a given master integral with discontinuities in multiple channels are
in fact the same in the different computations in these channels.  In
combining channels, one would then take the result from any of the channels
(that is, pick one, rather than adding together the different contributions).
This is in fact the procedure that was presented in ref.~\cite{Neq1Oneloop}.
It may be advantageous for some calculations, though in general it will
require the computation of superfluous integrals.  As with the basic
sewing procedure, there are adjustments that would
be required in a two-loop computation, which we will discuss in the next
subsection.

%
\FIGURE[t]{
\hbox{%
\SizedFigureWithCaption{1.46 truein}{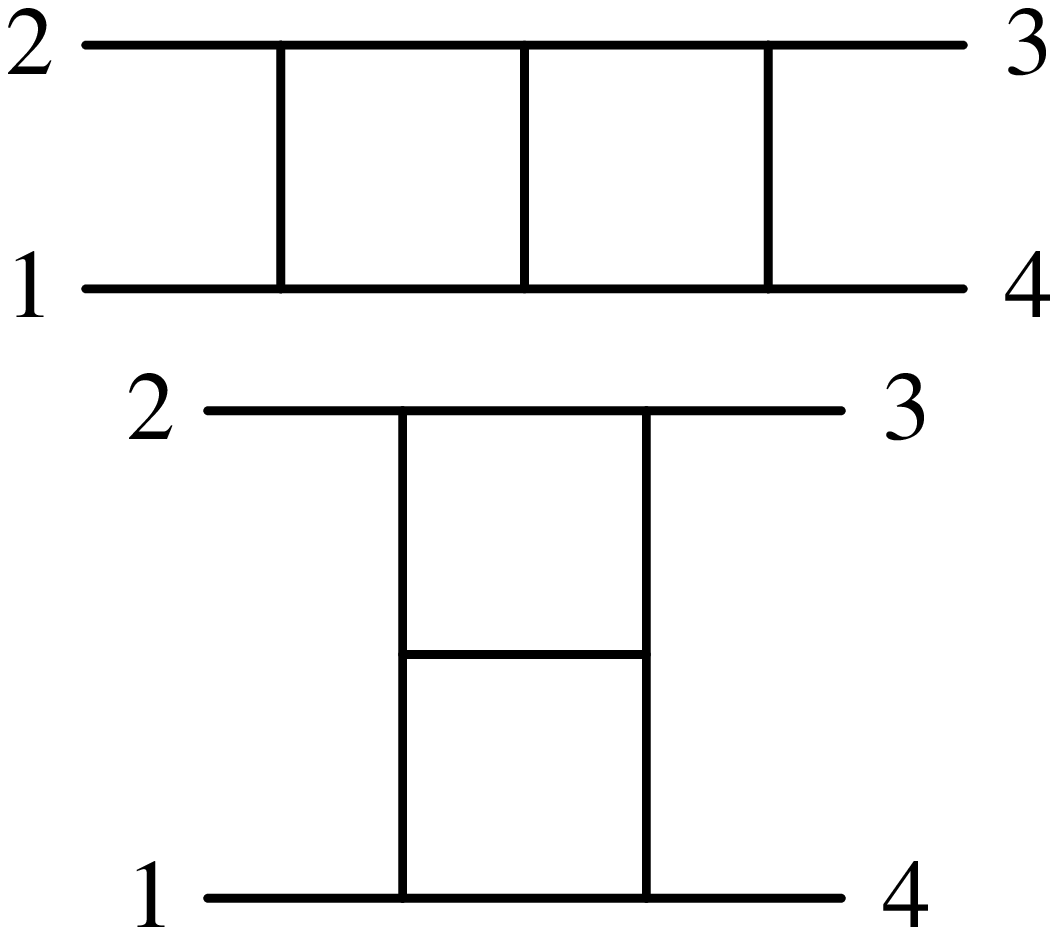}{(a)}
\SizedFigureWithCaption{1.22 truein}{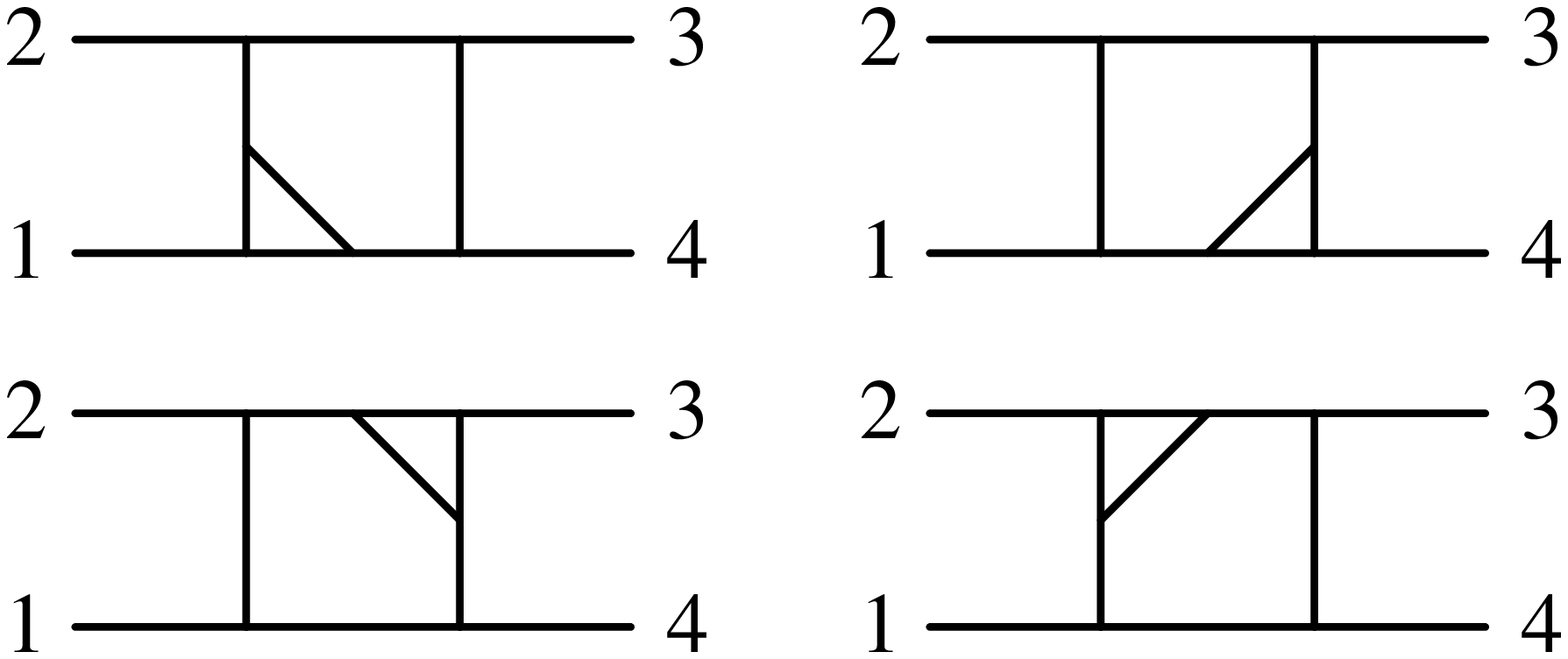}{(b)}
}\break%
\hbox{%
\SizedFigureWithCaption{1.22 truein}{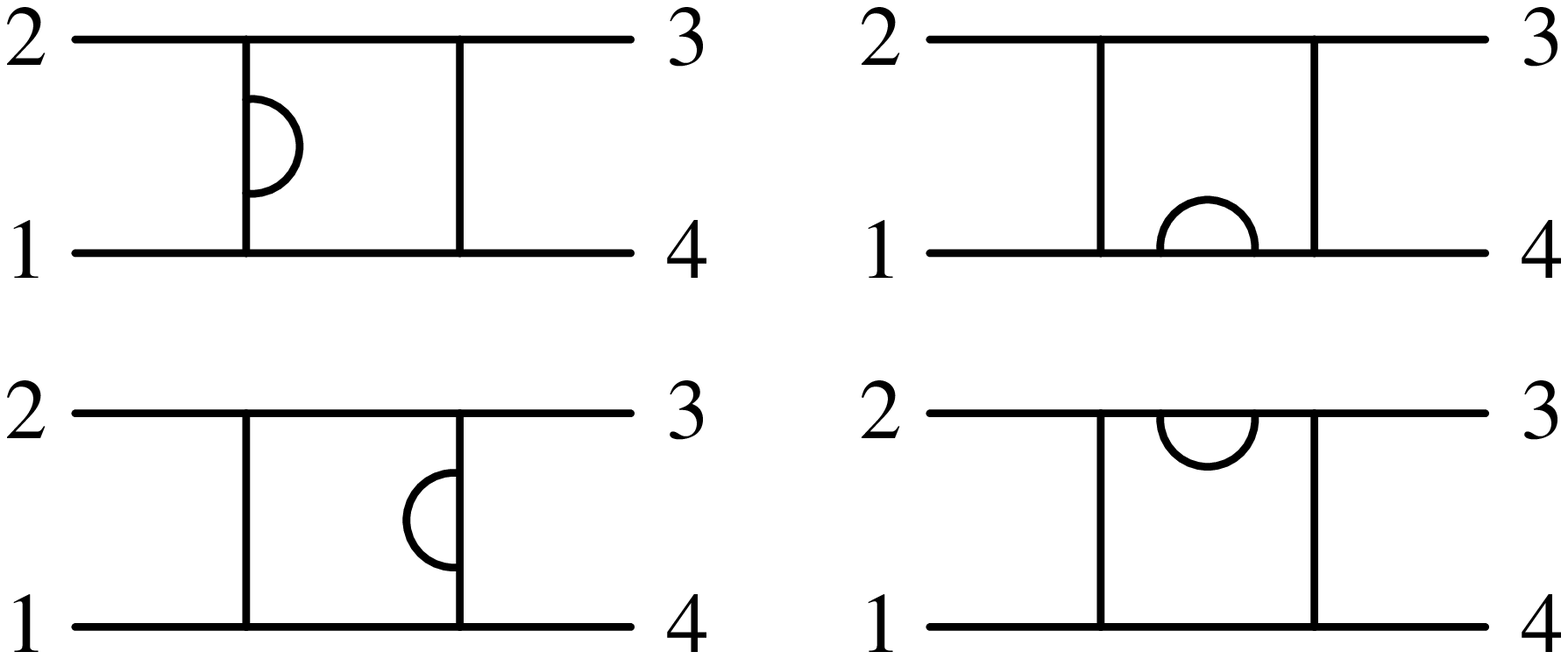}{(c)}
\SizedFigureWithCaption{1.22 truein}{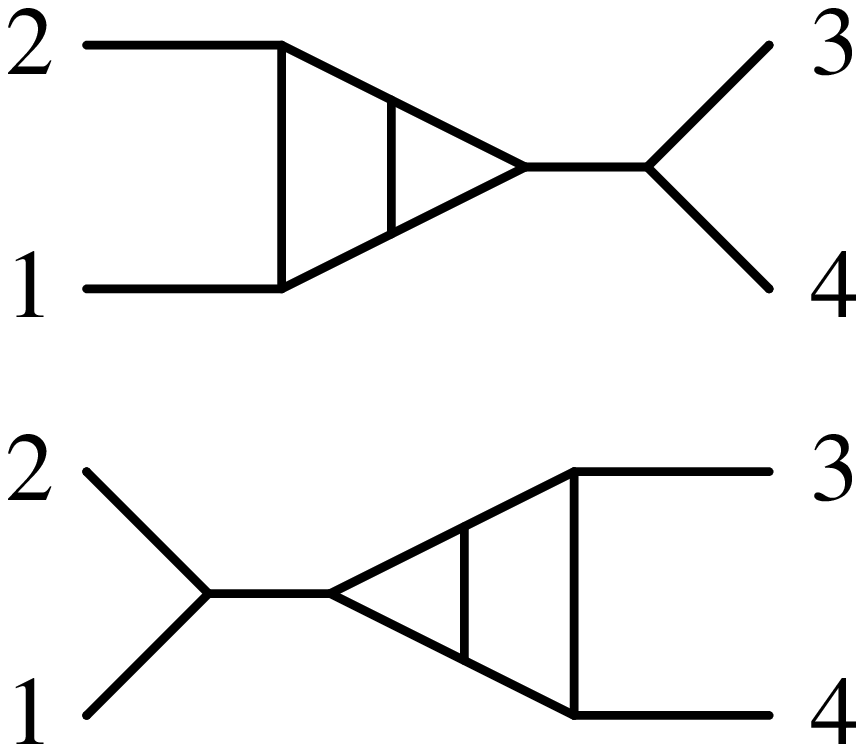}{(d)}
\SizedFigureWithCaption{1.22 truein}{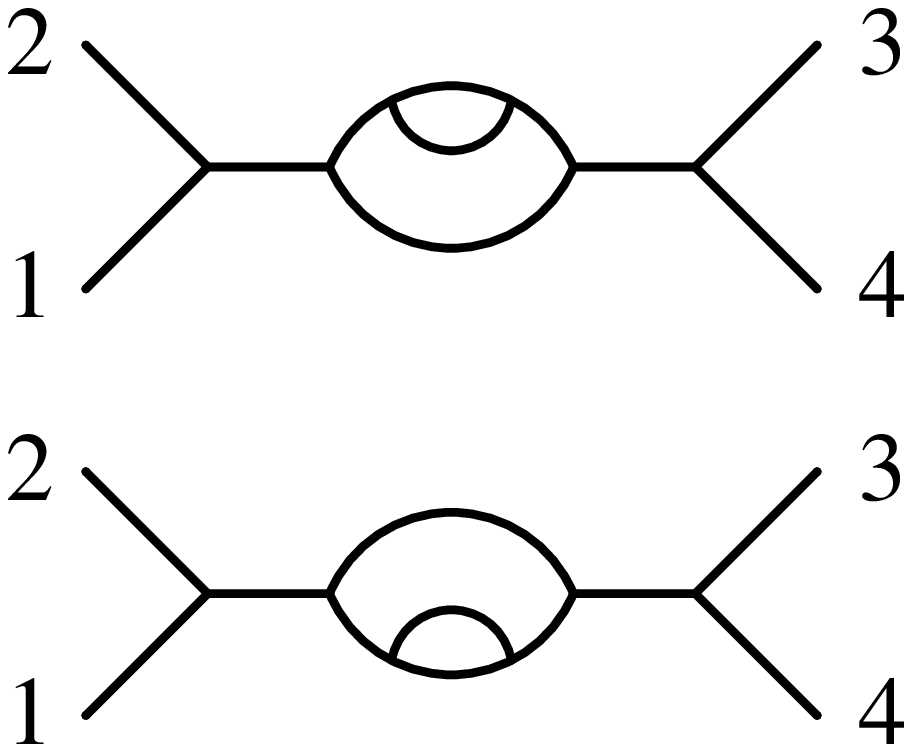}{(e)}
}\break%
\hbox{%
\SizedFigureWithCaption{1.86 truein}{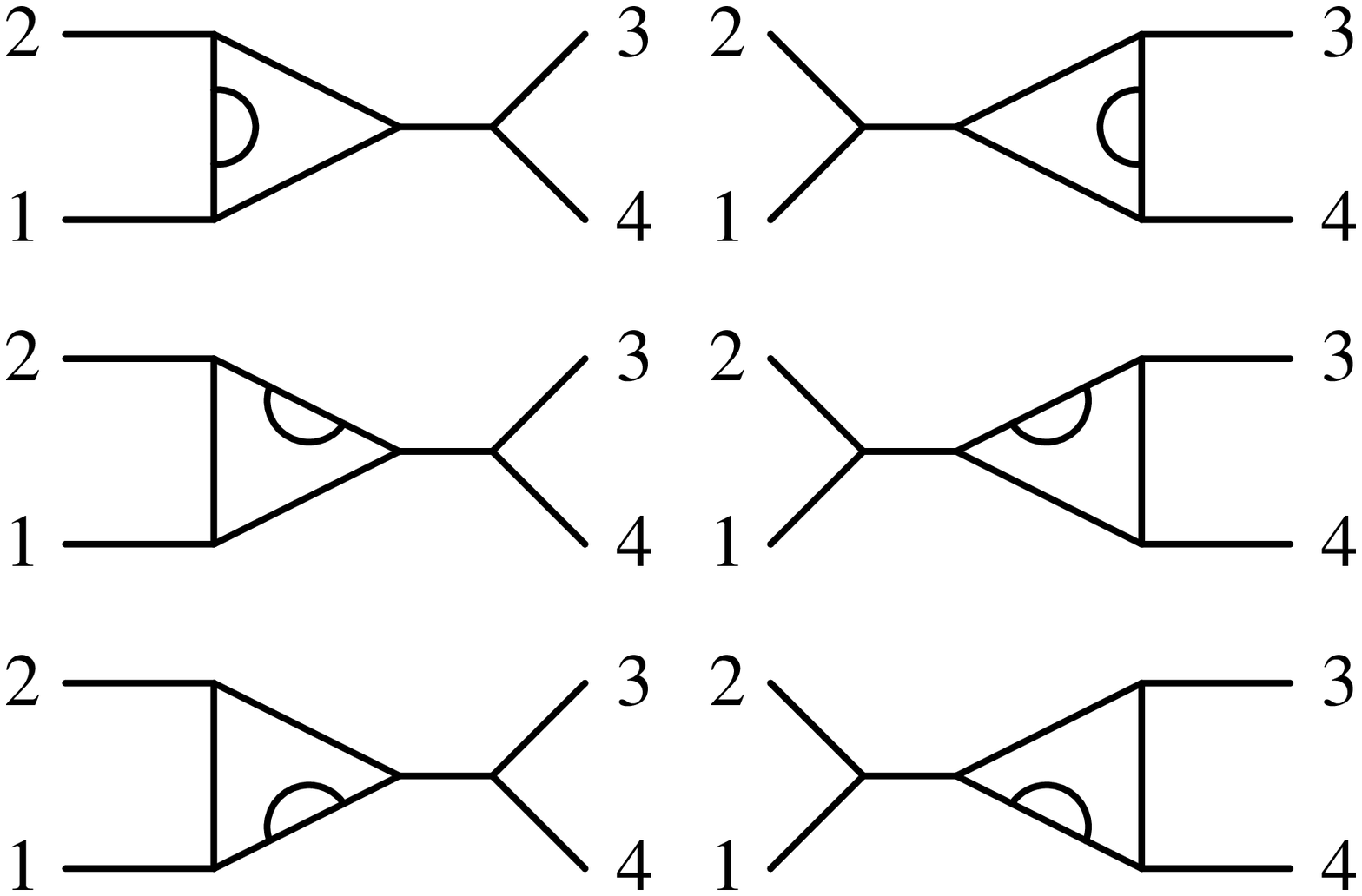}{(f)}
\vbox{
\halign{\hfil#\hfil\cr
\SizedFigureWithCaption{0.59 truein}{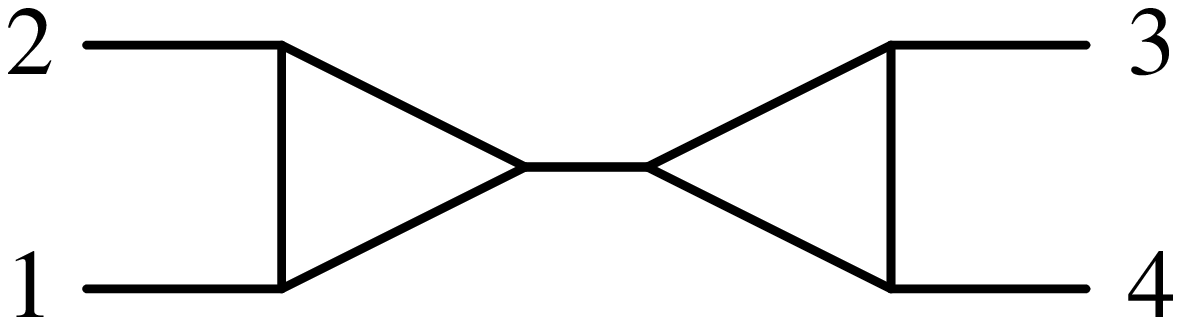}{(g)}
\cr
\noalign{\vskip -3mm}
\SizedFigureWithCaption{0.59 truein}{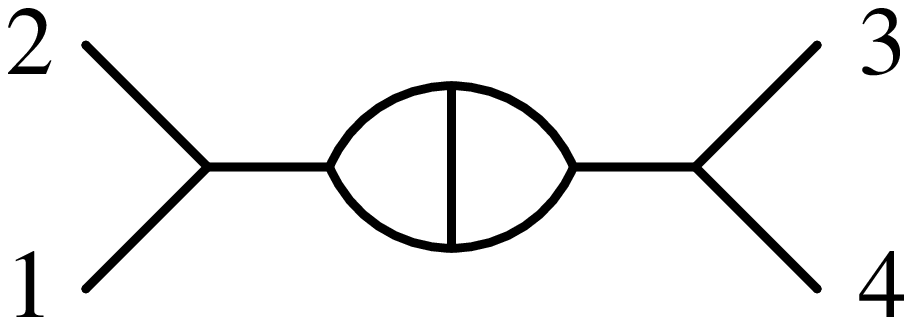}{(h)}
\cr
\noalign{\vskip -3mm}
\SizedFigureWithCaption{0.59 truein}{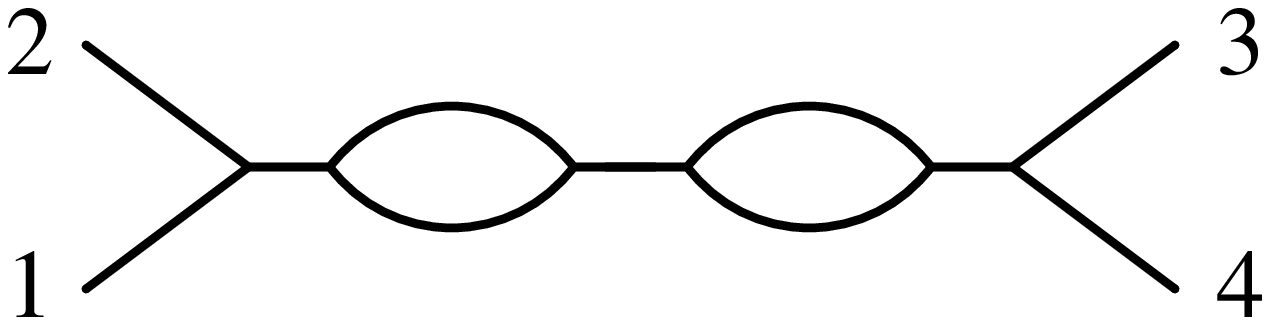}{(i)}
\cr}}}\break%
\hbox{%
\SizedFigureWithCaption{1.22 truein}{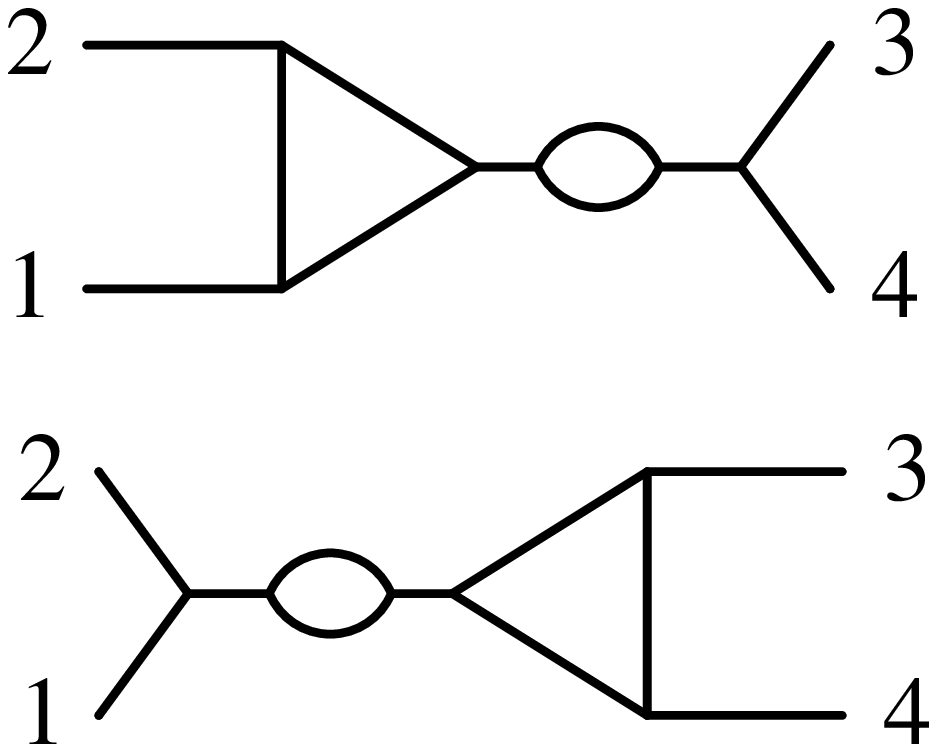}{(j)}
\SizedFigureWithCaption{1.22 truein}{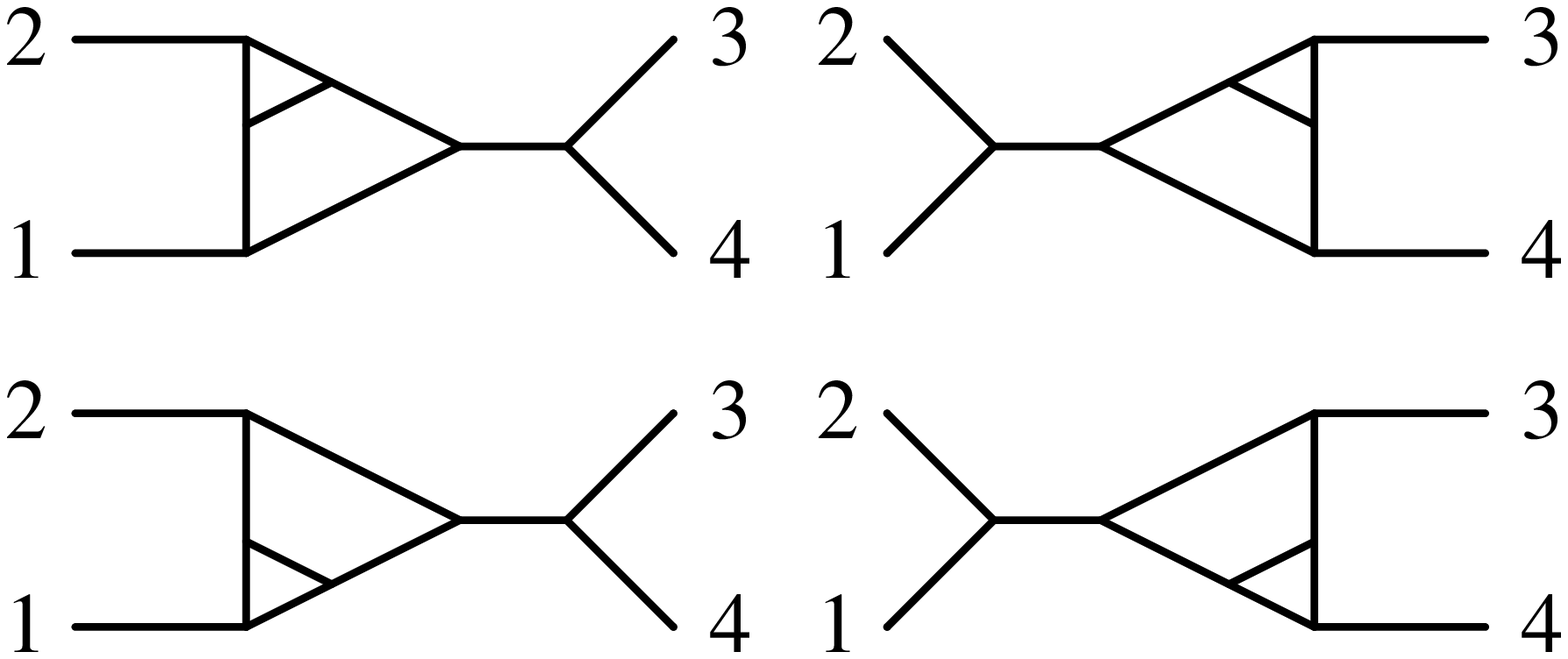}{(k)}
}
\label{TwoLoopPhi3Figure}
\caption{Color-ordered Feynman diagrams for the two-loop 
four-point amplitude in $\Tr\phi^3$
field theory: (a) planar box diagrams 
(b) triangle-in-box diagrams
(c) bubble-in-box diagrams
(d) $s$-channel ladder triangle diagrams 
(e) $s$-channel bubble-in-bubbles.  
(f) $s$-channel bubble-in-triangle diagrams 
(g) $s$-channel triangle-pair diagram
(h) $s$-channel lizard-eye bubble
(i) $s$-channel double bubble
(j) $s$-channel triangle-bubble diagram
(k) $s$-channel triangle-in-triangles.
The corresponding
 $t$-channel diagrams for (d--k) are not shown explicitly.}
}

\subsection{Sewing at Two Loops}

As for the one-loop case, we start our two-loop discussion 
with the example of a four-point amplitude.  We again
consider the contribution with a given ordering of the external legs, and
restrict attention to planar diagrams.
In the one-loop example, we saw that we must pay attention to potential
double-counting in assembling contributions from different channels.  At
two (and higher) loops, we must also confront potential double-counting
in contributions to a given channel.  To understand how this arises,
start once more with the Feynman diagrams for this planar ordered amplitude,
shown in \fig{TwoLoopPhi3Figure}.

Consider in particular the planar double box shown in
\fig{TwoLoopPhi3Figure}(a).  It has both two- and three-particle cuts.
The two-particle cuts, shown in 
\fig{DoubleBoxTwoParticleCutsFigure},
 contain a product of a one-loop amplitude and
a tree amplitude, with four external legs apiece.  There are two 
separate cuts.  

The three-particle cut in the planar double box is a sum of two terms,
shown in \fig{DoubleBoxThreeParticleCutsFigure}.
It corresponds to a product of two five-point tree amplitudes.
The complication here arises from the fact that a given term in
the two-loop amplitude may contribute to {\it both\/} terms in
the two-particle cut; {\it both\/} terms in the three-particle cut;
or to both two- and three-particle cuts.
  In such a case, when we reconstruct the
original integrand, we must count it only once.  
A simple analogy would be an integrand of the form $X^2$; since
 cutting is analogous to differentiation, we would have
\begin{equation}
\left.\bigl[ X^2 \bigr]\right|_{\rm cut} = 2 X\bigl|_{\rm cut} X,
\end{equation}
and promoting the latter back to an integrand requires a factor of 
$1/2$ just as it would for integration.
Note that if we denote the cut momenta by $\ell_{1,2,3}$, then the
terms which contribute to both two- and three-particle cuts 
necessarily contain the propagators
\begin{equation}
{1\over\ell_1^2\ell_2^2\ell_3^2 (\ell_1+\ell_2)^2 (\ell_2+\ell_3)^2} \,,
\end{equation}
using the labeling of legs in the three-particle cut.

\FIGURE[t]{
\centerline{
\epsfysize 2.0 truein \epsfbox{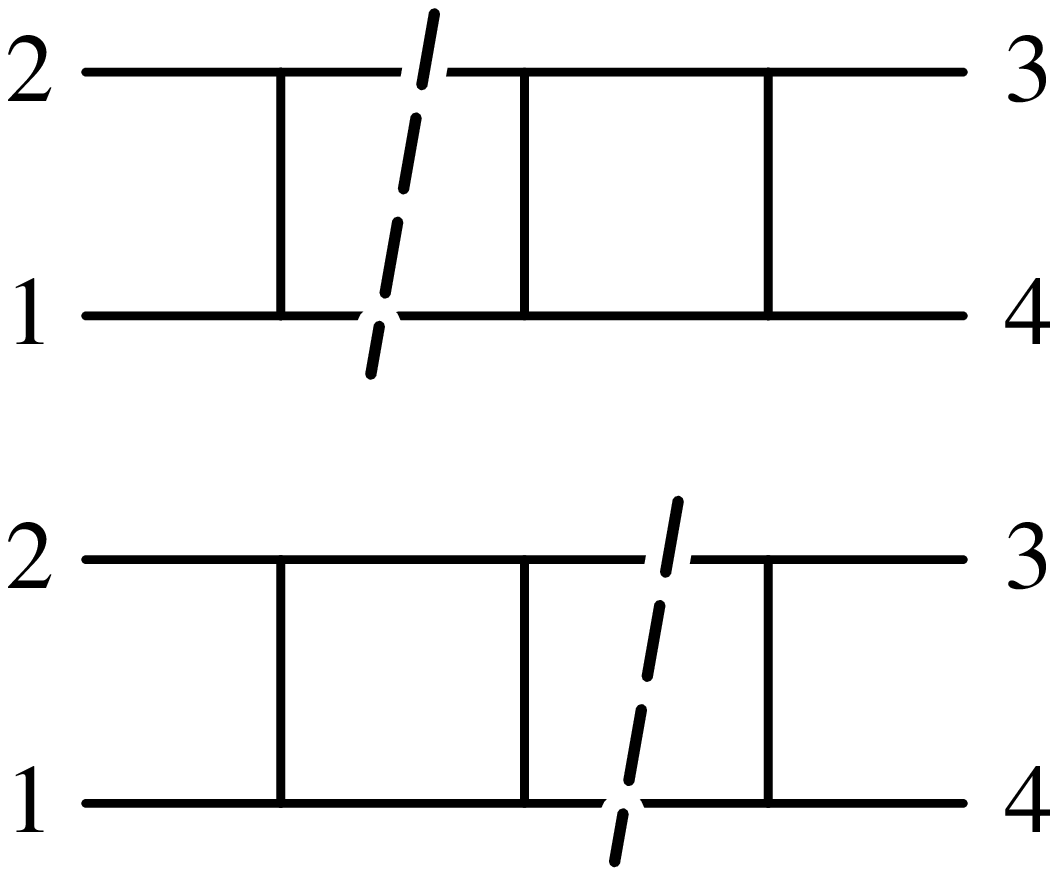}}
\label{DoubleBoxTwoParticleCutsFigure}  \hskip 10 cm 
\caption{The $s$-channel two-particle cuts of the
planar double box.}
}

We will focus here on the leading-color (planar) contributions to
the two-loop amplitude, though of course an analogous procedure applies
to the subleading contributions (which include non-planar integrals).
Let us begin the sewing procedure with the two-particle cuts in
the $s$ channel.  We must take the one-loop 
amplitude~(\ref{OneLoopPhi3Complete}), with 
legs $(3,4)$ replaced by legs $(\ell_2,\ell_1)$, and 
$\ell$ replaced by $p$ in the integral, and sew it
to the tree amplitude $A^{(0)}(-\ell_1,-\ell_2,3,4)$,
obtaining
\begin{eqnarray}
&& i \int {d^D p\over (2\pi)^D}{d^D \ell\over (2\pi)^D}\;
\TimesBreak{\int}
\biggl[
{1\over s_{12}^2} {1\over p^2 (p+k_1+k_2)^2}
+{1\over s_{12}} {1\over p^2(p-\ell)^2(p+k_1+k_2)^2}
\PlusBreak{\int \times \biggl[]}
{1\over s_{12}} {1\over (p+k_1)^2 p^2 (p+k_1+k_2)^2}
+{1\over (p+k_1)^2 p^2 (p-\ell)^2 (p+k_1+k_2)^2}
\PlusBreak{\int \times \biggl[]}
{1\over [(\ell+k_1)^2]^2} {1\over p^2 (p-\ell-k_1)^2}
+{1\over (\ell+k_1)^2} {1\over p^2(p-k_1)^2 (p-\ell-k_1)^2}
\PlusBreakLab{TwoLoopPhi3SChannelTwoCutA}{\int \times \biggl[]}
{1\over (\ell+k_1)^2} {1\over (p+k_2)^2 p^2 (p-\ell-k_1)^2}
\biggr]\,
{1\over(\ell+k_1+k_2)^2\ell^2} 
\Bigl({1\over s_{34}}+{1\over (\ell-k_4)^2}\Bigr)
\nonumber
\end{eqnarray}
where we have relabeled $\ell_1\rightarrow \ell$. 
Restricting attention to those terms with explicit powers of
$1/s_{12}^2$ or $1/s_{12}^3$, we have
\begin{eqnarray}
&& i \int {d^D p\over (2\pi)^D}{d^D \ell\over (2\pi)^D}\;
{1\over s_{12}^2 (\ell+k_1+k_2)^2\ell^2} 
\biggl[
{1\over s_{12}} {1\over p^2 (p+k_1+k_2)^2}
+{1\over p^2(p-\ell)^2(p+k_1+k_2)^2}
\PlusBreak{~~~~~~~~~~~~~~~~~~~~~~~}
{1\over (p+k_1)^2 p^2 (p+k_1+k_2)^2}
+{1\over p^2 (p+k_1+k_2)^2 (\ell - k_4)^2}
\biggr]+\cdots
\label{TwoCut:s12A}
\end{eqnarray}
These terms are sufficient to illustrate
the issues associated with potential double-counting.

\FIGURE[t]{
\centerline{\epsfysize 2.0 truein \epsfbox{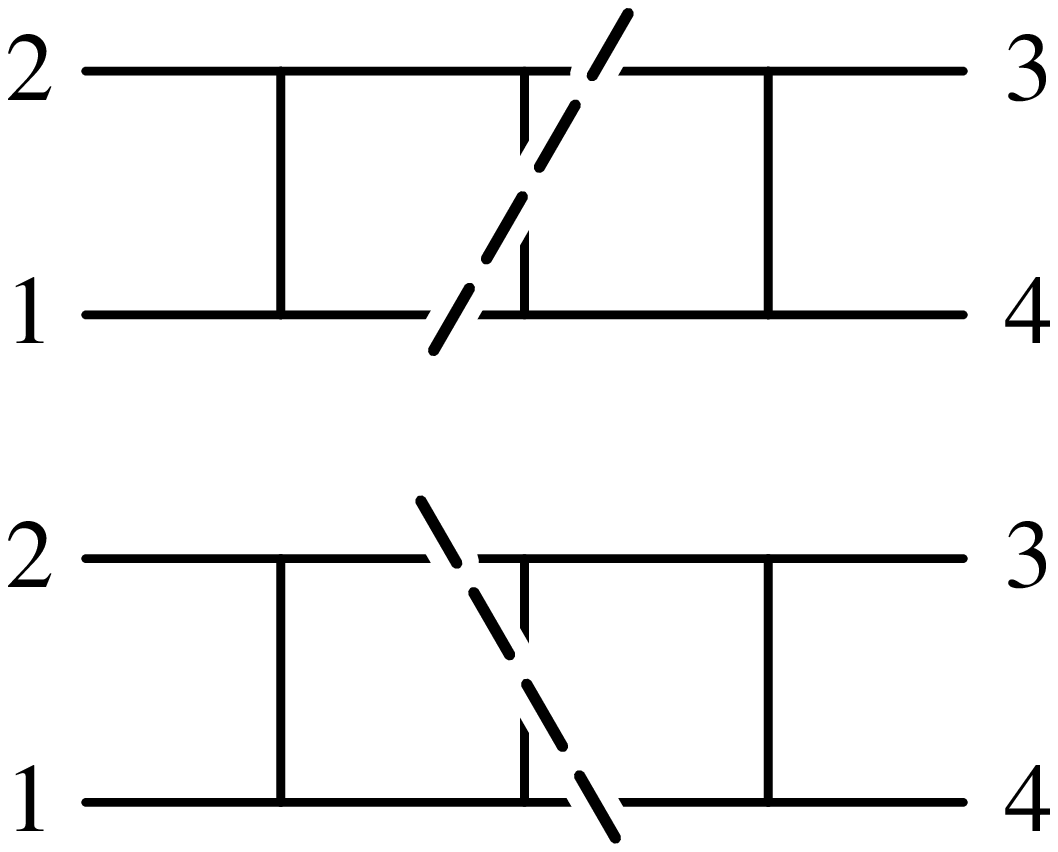}}
\label{DoubleBoxThreeParticleCutsFigure}
\caption{The $s$-channel three-particle cuts of the
planar double box.}
}

\FIGURE[t]{
\centerline{\epsfysize 0.88 truein \epsfbox{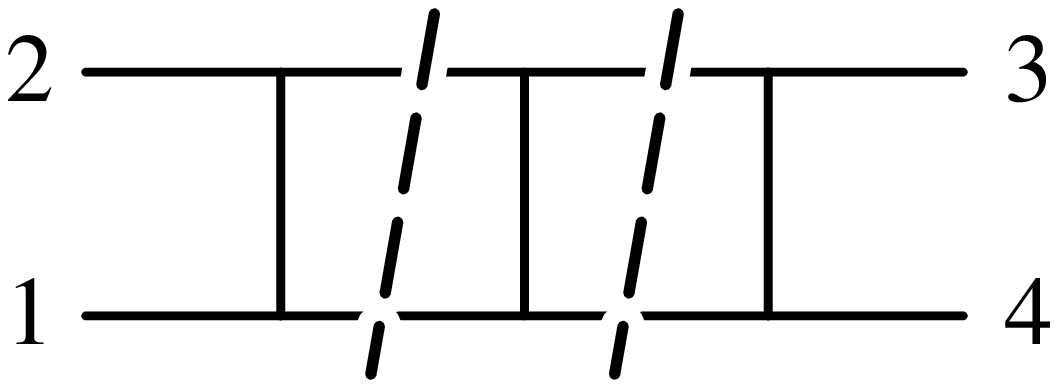}}
\label{DoubleBoxDoubleCutFigure}
\caption{The $s$-channel `double' two-particle cuts of the
planar double box.}
}

There is a similar contribution from sewing $A^{(0)}(1,2,\ell_2,\ell_1)$
to $A^{(1)}(-\ell_1,-\ell_2,3,4)$.
If we now take the cut of the expression~(\ref{TwoLoopPhi3SChannelTwoCutA}),
however, we discover that we can cut not only the $\ell$ loop, but also
the $p$ loop.  The corresponding terms would appear not only in this
sewing, but also in the other contribution (with the loop amplitude
on the right-hand side of the cut).
This would be a source of double-counting.  To correct for
it, we must subtract from the sum of the two contributions those terms
that can be cut both ways.  These are the terms extracted by a `double cut'
(shown in \fig{DoubleBoxDoubleCutFigure}),
again in the sense of requiring that the propagators giving rise to 
a two-particle cut be present in both the $\ell$ and $p$ loops,
\begin{equation}
\Bigl({i\over s_{12}}+{i\over s_{p1}}\Bigr)
\Bigl({i\over s_{12}}-{i\over s_{p\ell}}\Bigr)
\Bigl({i\over s_{34}}-{i\over s_{\ell4}}\Bigr),
\end{equation}
or promoted back to a two-loop integral,
\begin{eqnarray}
&& i \int {d^D p\over (2\pi)^D}{d^D \ell\over (2\pi)^D}\;
{1\over s_{12}^2 (\ell+k_1+k_2)^2\ell^2 p^2 (p+k_1+k_2)^2} 
\TimesBreak{\int}
\biggl[ {1\over s_{12}} +{1\over (p+k_1)^2} + {1\over (p-\ell)^2}
+ {1\over (\ell-k_4)^2}
\biggr]+\cdots \,,
\label{TwoCut:s12B}
\end{eqnarray}
which is identical to the terms in \eqn{TwoCut:s12A}, obtained from the
first of the two-particle cut contribution.  Counting the contribution
only once then gives \eqn{TwoCut:s12B}
as the result of combining the two two-particle cut contributions.

Next, we must consider the three-particle cuts.  We begin with the
product of two five-point amplitudes,
\begin{equation}
A^{(0)}(1,2,\ell_3,\ell_2,\ell_1) =
i \biggl[ {1\over s_{\ell_2\ell_3} s_{\ell_1 1}}
+{1\over s_{\ell_1\ell_2} s_{12}}
+{1\over s_{\ell_1\ell_2} s_{\ell_3 2}}
+{1\over s_{\ell_2\ell_3} s_{12}}
+{1\over s_{\ell_1 1} s_{\ell_3 2}} \biggr] \,,
\end{equation}
and $A^{(0)}(-\ell_1,-\ell_2,-\ell_3,3,4)$.

\FIGURE[t]{
\centerline{
\epsfysize 2.0 truein \epsfbox{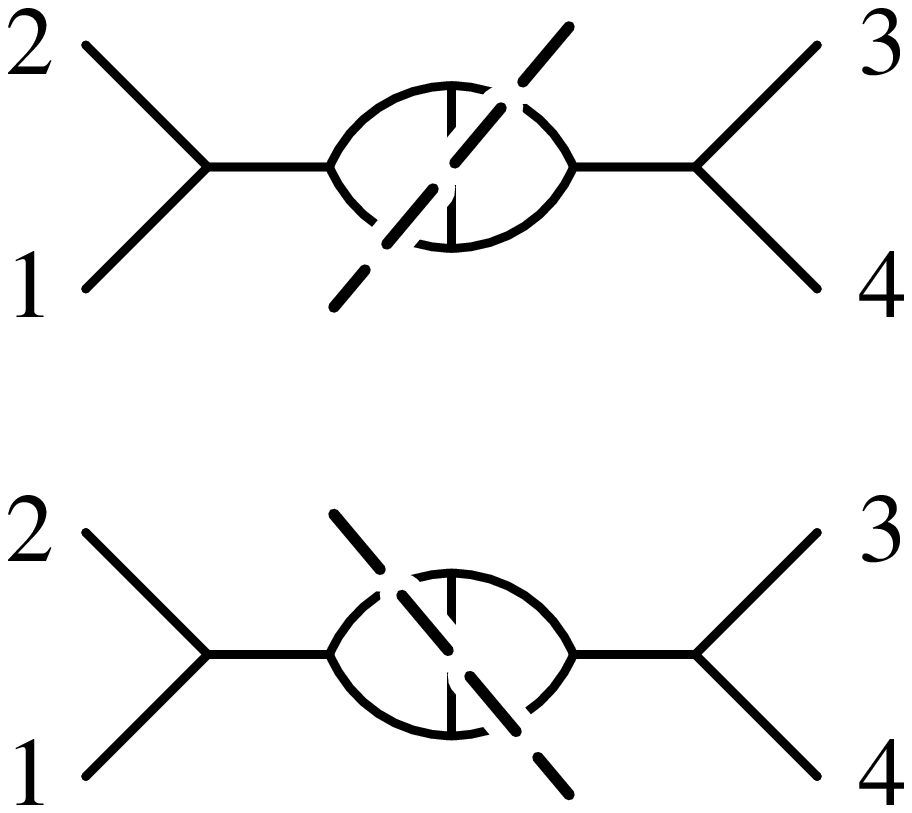}}
\label{LizardEyeBubbleThreeParticleCutsFigure}
\caption{The $s$-channel three-particle cuts of the lizard-eye bubble.}
}

Let us focus on the terms containing a factor of $1/s_{12}^2$.  These are
\begin{eqnarray}
&& i \int {d^D \ell_1\over (2\pi)^D} {d^D \ell_3\over (2\pi)^D}\;
{1\over s_{12}^2} {1\over\ell_1^2\ell_3^2 (\ell_1+\ell_3+k_1+k_2)^2}
\TimesBreakLab{s12SqrFactors}{\int\times }
    \biggl(
  {1\over [(\ell_3+k_1+k_2)^2]^2}
 +{2\over (\ell_3+k_1+k_2)^2 (\ell_1+k_1+k_2)^2}
 +{1\over [(\ell_1+k_1+k_2)^2]^2}\biggr).\hbox to 20pt{} \nonumber
\end{eqnarray}
The first and last of these terms correspond to the diagrams of 
\fig{TwoLoopPhi3Figure}(e).
The middle term corresponds to the diagram of \fig{TwoLoopPhi3Figure}(h); 
but it appears in
the product with a (superfluous) factor of 2.  This factor is due precisely
to the fact that this term can be cut in two different ways, corresponding
to the two terms depicted in \fig{LizardEyeBubbleThreeParticleCutsFigure}.
  We must remove this double counting,
by subtracting those terms which contribute twice; this leaves us with
\begin{eqnarray}
&&i \int {d^D \ell_1\over (2\pi)^D} {d^D \ell_3\over (2\pi)^D}\;
{1\over s_{12}^2} {1\over\ell_1^2\ell_3^2 (\ell_1+\ell_3+k_1+k_2)^2}
\TimesBreakLab{TwoLoopPhi3NetThreeParticleCut}{\int\times }
    \biggl(
  {1\over [(\ell_3+k_1+k_2)^2]^2}
 +{1\over (\ell_3+k_1+k_2)^2 (\ell_1+k_1+k_2)^2}
 +{1\over [(\ell_1+k_1+k_2)^2]^2}\biggr).\hbox to 20pt{}
\nonumber
\end{eqnarray}

Finally, we must combine the two- and three-particle cuts.  Again, we
can add the two contributions, and remove terms which appear in both,
for example by removing terms in the three-particle cuts which have 
(any) two-particle cut.  Of the terms listed explicitly 
in \eqn{TwoLoopPhi3NetThreeParticleCut},
only the middle term has a two-particle cut.  Removing it (thereby
performing the required merging), and relabeling $\ell_1\rightarrow \ell$,
$\ell_3\rightarrow -p-k_1-k_2$,  we obtain,
\hskip -20pt\begin{eqnarray}
&& i \int {d^D p\over (2\pi)^D}{d^D \ell\over (2\pi)^D}\;{1\over s_{12}^2}
{1\over \ell^2 (p+k_1+k_2)^2}
\TimesBreak{int}
\biggl[
{1\over s_{12}} {1\over p^2 (\ell+k_1+k_2)^2}
+ {1\over p^2(p-\ell)^2(\ell+k_1+k_2)^2}
\PlusBreak{\int\times \biggl[]}
{1\over (p+k_1)^2 p^2 (\ell+k_1+k_2)^2}
+{1\over p^2 (\ell-k_4)^2  (\ell+k_1+k_2)^2}
\PlusBreakLab{TwoLoopPhi3SChannelNet}{\int\times \biggl[]}
{1\over (p-\ell)^2 [p^2]^2}
+ {1\over (p-\ell)^2 [(\ell+k_1+k_2)^2]^2}
\biggr]+\cdots,
\nonumber
\end{eqnarray}
again in exact agreement with the terms that would emerge from a
Feynman-diagram calculation.  (The reader may wonder
why the first and last terms in 
\eqn{TwoLoopPhi3NetThreeParticleCut}, which correspond to the diagrams
in \fig{TwoLoopPhi3Figure}(e),
do not contain two-particle cuts.  Naively, these diagrams
do contain two-particle cuts; but a closer inspection shows that such a cut
would have a bubble on an external line; such bubbles are
scale-free and hence vanish in dimensional regularization.)

\FIGURE[t]{
\centerline{\epsfysize 0.8 truein \epsfbox{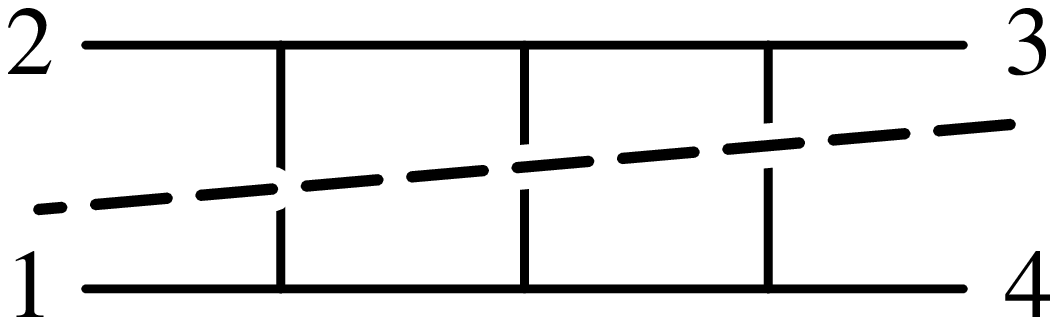}}
\label{DoubleBoxTChannelThreeParticleCutFigure}
\caption{The $t$-channel cut of the
planar double box.}
}

Note that not all contributions will require these subtractions to match
overlaps; for example, the $t$-channel cut of the planar double box shown in
\fig{DoubleBoxTChannelThreeParticleCutFigure} has only one contribution, 
a three-particle cut.  

\def\doublecut{||}
\def\crosscut{\times}
In practice, it appears to be more
efficient to start with the two-particle cuts, and then merge in additional
terms from the three-particle cuts.  In addition to the cut projection
operator $\CutProj_s$, which yields the part of its argument that has
any cut (two- or three-particle) in the specified channel $s$, we will also
make use of four additional projection operators.  These are
$\CutProj_s^{(2)}$, which extracts terms containing a two-particle cut;
$\CutProj_s^{(3)}$, those containing a three-particle cut;
$\CutProj_{s\doublecut}^{(2)}$, those containing a `double cut', that is whose
two-particle cut can be cut again 
(for example, \fig{DoubleBoxDoubleCutFigure});
and $\CutProj_{s\crosscut}^{(3)}$, those containing a contribution to both
terms in the three-particle cut (corresponding to terms which have 
all the propagators cut in \fig{DoubleBoxDoubleCutFigure} and the center 
propagator in addition).  Note that these projection operators do
not necessarily remove terms with other cuts; for example, 
$\CutProj_s^{(2)}$ may yield an expression containing three-particle cuts
in addition to the two-particle cut.  In cases where there is only a single
two-particle cut, or where there is only a single contribution to the 
three-particle cuts, $\CutProj_{s\doublecut}^{(2)}$ and 
$\CutProj_{s\crosscut}^{(3)}$ are understood to vanish.

Using these projection operators, we can define a complete promotion operator
$\promoteC$
for a given channel,
\begin{eqnarray}
&&\promoteC A^{(\cdot)}(\{\ell_i\},\First_j,\ldots,\Last_j)
         A^{(\cdot)}(\Last_j+1,\ldots,\First_j-1,\{-\ell_i\}) =
\nonumber\\&&\hphantom{\promote AA}
\Bigl(1-{1\over2}\CutProj_{s\doublecut}^{(2)}\Bigr)
\promoteB \Bigl[
A^{(0)}(\ell,\First_j,\ldots,\Last_j,-\ell-K_j)
\TimesBreak{\promote AA\Bigl(1-{1\over2}\CutProj_{s\doublecut}^{(2)}\Bigr)
            \promoteB\Bigl[][]} 
A^{(1)}(\ell+K_j,\Last_j+1,\ldots,\First_j-1,
        -\ell)
\PlusBreakLab{CompletePromotion}
    {\promote AA\Bigl(1-{1\over2}\CutProj_{s\doublecut}^{(2)}\Bigr)\promoteB} 
A^{(1)}(\ell,\First_j,\ldots,\Last_j,-\ell-K_j)
\TimesBreak{\promote AA\Bigl(1-{1\over2}\CutProj_{s\doublecut}^{(2)}\Bigr)
            \promoteB\Bigl[][]} 
A^{(0)}(\ell+K_j,\Last_j+1,\ldots,\First_j-1,
        -\ell)
          \Bigr]
\PlusBreak{\promote AA}
\Bigl(1-\CutProj_s^{(2)}\Bigr)
\Bigl(1-{1\over2}\CutProj_{s\crosscut}^{(3)}\Bigr)
\nonumber\\&&
 \hphantom{\promote AA \Bigl(1-{1\over2}\CutProj_{s\doublecut}^{(2)}\Bigr)}
\promoteB
A^{(0)}(\ell_1,\ell_2,\First_j,\ldots,\Last_j,
        -\ell_1-\ell_2-K_j)
\TimesBreak{\promote AA\Bigl(1-{1\over2}\CutProj_{s\doublecut}^{(2)}\Bigr)
            \promoteB \Bigl[]} 
A^{(0)}(\ell_1+\ell_2+K_j,
        \Last_j+1,\ldots,\First_j-1,
        -\ell_2,-\ell_1) \,,
\nonumber
\end{eqnarray}
where $A^{(1)}$ denotes the {\it integrand\/} for the one-loop amplitude.
Here we consider only the planar case, so that all cut-crossing momenta
are color-adjacent; but the construction generalizes in a straightforward
way to non-planar amplitudes.  As in the example discussed earlier, the
role of the factors of $1/2$ in front of $\CutProj_{s\doublecut}^{(2)}$
and $\CutProj_{s\crosscut}^{(3)}$ is to remove double-counting that
occurs in cutting, ensuring that each term contributes to the integrand
with the correct coefficient.

Using this complete promotion operator, the complete integrand for
the two-loop amplitude $A^{(2)}(1,\ldots,n)$ is given by an expression
very similar in form to that for the one-loop 
amplitude~(\ref{OneLoopSewingIngredient}),
\begin{eqnarray}
\Integrand^{(2)} &=& \sum_{j=1}^{\nChannels} 
{\textstyle\prod_{l=1}^{j-1}}\Bigl(1- \CutProj_{s_l}\Bigr) \CutProj_{s_j}
   \promoteC
A^{(\cdot)}(\{\ell_i\},\First(s_j),\ldots,\Last(s_j))
\TimesBreakLab{TwoLoopSewingFormula}{\sum_{s_j\in \Channels} 
             \Pi_{l=1}^{j-1}\Bigl(1-\CutProj_{s_l}\Bigr)\promoteB }
A^{(\cdot)}(\Last(s_j)+1,\ldots,\First(s_j)-1,\{-\ell_i\})\,,
\nonumber\end{eqnarray}
but with the complete promoted cut being used where the basic one was
in the one-loop case.

A practical algorithm is similar in structure to that at one loop,
but requires additional merging over two- and three-particle cuts.
The first three steps (construct ordered set of all channels,
determine labelings, initialize integrand to zero) are the same.
As in the one-loop case,
from a formal point of view, the relabelings below are not required
by the method, but since we will ultimately be feeding the resulting
integrands to an integration machinery, we might as well incorporate
the standardization at an early stage in the calculation.

\FIGURE[t]{
\SizedFigureWithCaption{1.4 truein}{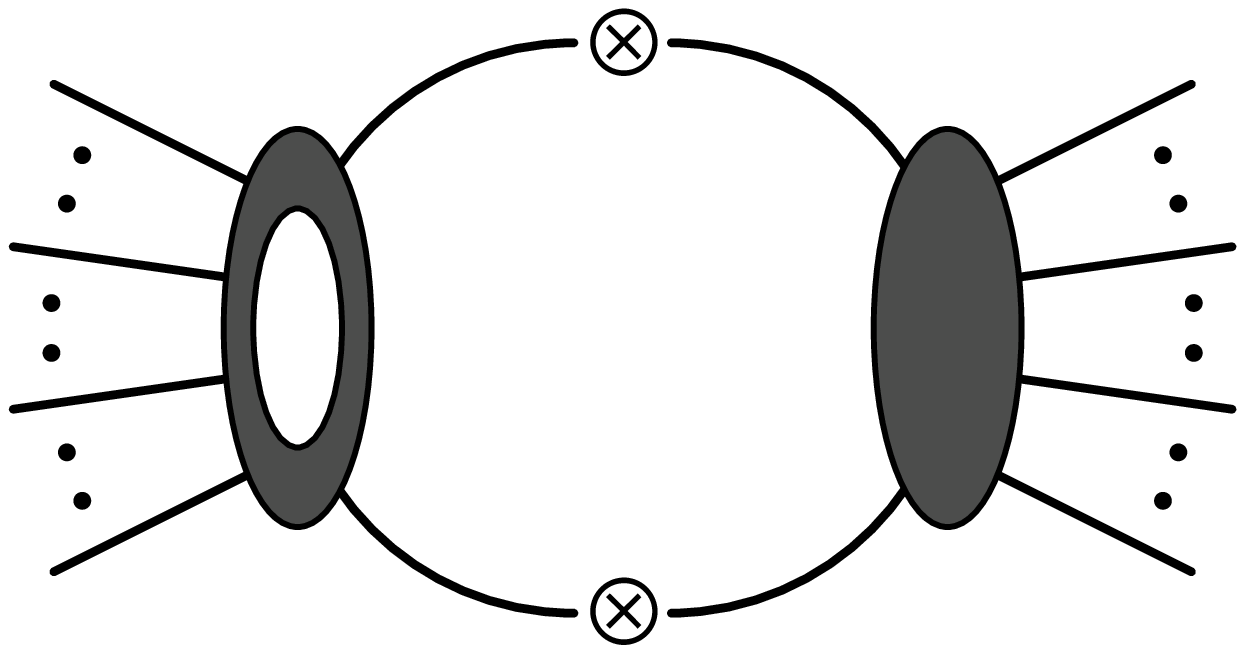}{(a)}
\hskip 5truemm
\SizedFigureWithCaption{1.4 truein}{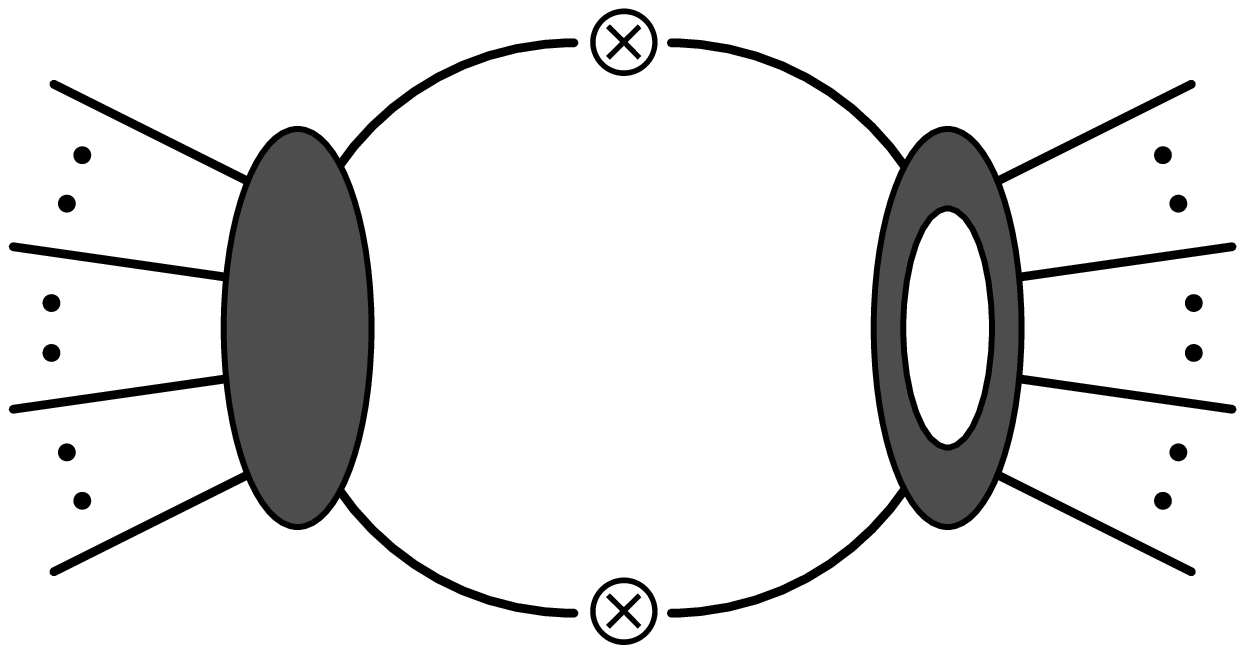}{(b)}
\label{TwoLoopSewingGeneral1Figure}
\caption{The two contributions wherein one sews a one-loop amplitude to
a tree-level amplitude to produce a contribution
to the two-loop integrand.}
}

In the two-loop case, `irreducible numerators' appear.
These are dot products of loop momenta and external momenta that 
cannot be written as linear combinations of inverse propagators.
Such terms must be added to the
basis of expressions described just before the one-loop algorithm.
With this modification,
the algorithm again continues after the first four set-up steps
 by iterating over all channels.
For each channel $(k_{\First_j},\ldots,k_{\Last_j})$,
\begin{enumerate}
\addtocounter{enumi}{4}
\item \label{FirstSewingStep}{\it Sew first two-particle cuts.}
Form the product of the integrand for the on-shell 
one-loop amplitude
$A^{(1)}(\ell_1,\First_j,\ldots,\Last_j,\ell_2)$ and the tree amplitude
$A^{(0)}(-\ell_2,\Last_j+1,\ldots,n,1,\ldots,\allowbreak\First_j-1,-\ell_1)$ 
(where $\ell_1+\ell_2+K_j = 0$), summing over
the different particle types and helicities
that can circulate in the loop.  The sum must in general be performed
in $D=4-2\e$ dimensions.  Use the polarization-vector
identities~(\ref{PolarizationIdentities})
to express everything in terms of the cut momenta $\ell_1, \ell_2$ and the
light-cone vector $n$.  (As in
the one-loop case, this will introduce physical projectors into
the integrand.)   
Multiply by $-1$ for each fermion loop is created by sewing.
Put in the propagators crossing the cut, $i\over\ell^2$ for each cut
leg.
Note that in general four-dimensional Fierz identities may {\it not\/}
be used.  Fermion traces
over internal fermion lines should be performed first, re-expressing
the result in dot products, with summation
over gluon polarizations performed afterwards.
Rewrite terms using the basis expressions, and expand sums so that each term
can be classified as the daughter of a single parent diagram.  
Complete dot products in the denominator
to form propagator denominators adjacent to the cut, 
$2\ell\cdot k_i \rightarrow \pm (\ell\pm k_i)^2$,
the $+$ corresponding to $k_i$ on the left side of the cut, the $-$ sign
to those on the right side of the cut.   The sewing operation is
depicted schematically in \fig{TwoLoopSewingGeneral1Figure}(a), 
and corresponds to the promotion of the second term in brackets
in \eqn{CompletePromotion}.

\item \label{FirstCanonicalStep} {\it Put into canonical form.} Use 
momentum conservation to reduce the number of cut-crossing momenta
(now loop momenta) appearing in spinor traces, and then expand spinor 
traces and all dot products in terms
of the basis.  (As explained above, this expansion may not always 
be necessary.)  When using the spinor-helicity method for external
polarization vectors, or explicit helicity states, we will obtain
spinor strings of the form 
$\langle j_1^-|\s{j}_2\cdots\s{\ell}\cdots\s{j}_3|j_4^\pm\rangle$.
As explained earlier, one can
complete these to a trace, then convert the trace to
dot products and Levi-Civita tensors.  The latter can be converted
to Gram determinants (and thence to dot products) by multiplying (and dividing)
by a Levi-Civita tensor involving only external momenta.
(Spinor strings involving only external momenta need not be manipulated,
obviously.)

\item \label{FirstCutResult} {\it Relabel.\/} For each term obtained in 
step~\ref{FirstCanonicalStep},
 relabel the momenta to the standard labeling for its
parent integral.  Where required, insert factors of squared momenta in the
numerator and denominator to match a `parent' diagram.  In some cases,
it will be possible to obtain different parents by inserting different
factors; it doesn't matter which one is picked.

\item  \label{SecondSewingStep} {\it Sew second two-particle cuts.}
Form the product of the on-shell tree amplitude
$A^{(0)}(\ell_1,\First_j,\ldots,\Last_j,\ell_2)$ and the 
integrand for the on-shell one-loop amplitude
$A^{(1)}(-\ell_2,\allowbreak\Last_j+1,\ldots,n,1,\ldots,\First_j-1,-\ell_1)$ 
(where $\ell_1+\ell_2+K_j = 0$), again summing over
the different particle types and helicities
that can circulate in the loop.  As in step~\ref{FirstSewingStep},
put in fermi minus signs and the cut propagators, 
complete dot products to propagators,
and rewrite terms using basis expressions.
The sewing operation is
depicted schematically in \fig{TwoLoopSewingGeneral1Figure}(b),
and corresponds to the promotion of the first term in brackets
in \eqn{CompletePromotion}.

\item \label{SecondCanonicalStep} {\it Put into canonical form.} 
Use momentum conservation,
the cut condition, and expansion of spinor traces, along the lines
of step~\ref{FirstCanonicalStep}.

\item \label{SecondCutResult} {\it Relabel.\/} For each term obtained in 
step~\ref{SecondCanonicalStep},
 relabel the momenta to the standard labeling for its
parent integral, inserting factors of squared momenta in the
numerator and denominator where required.

\item \label{MergeTwoParticleCuts} %
{\it Merge two-particle cuts, removing double-counting.}
To the result obtained in step~\ref{FirstCutResult}, add those terms obtained
in step~\ref{SecondCutResult} not present in 
the result from step~\ref{FirstCutResult}.
(Equivalently, add the results of steps~\ref{FirstCutResult} 
and~\ref{SecondCutResult},
and then subtract terms present in both
expressions. These latter terms are those with two pairs of propagators
present, {\it i.e.} those cut through in the `double' two-particle cut.) 
This step corresponds to the operator 
$(1-{1\over2}\CutProj_{s\doublecut}^{(2)})$ in \eqn{CompletePromotion},
and yields the sewn two-particle cut in the given channel.

\FIGURE[t]{
\centerline{\epsfysize 1.4 truein \epsfbox{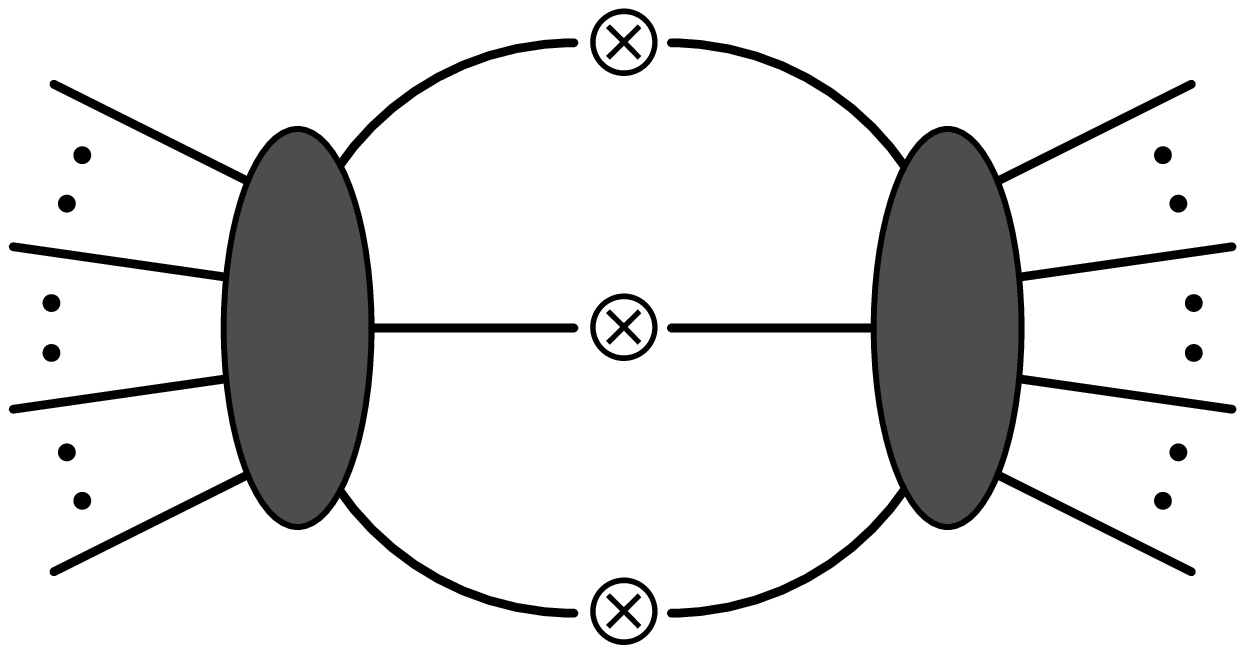}}
\label{TwoLoopSewingGeneral3Figure}
\caption{Sewing together two tree-level amplitudes to produce 
the three-particle cut contribution
to the two-loop integrand.}
}

\item \label{ThirdSewingStep} {\it Sew three-particle cuts.}
Form the product of on-shell tree amplitudes 
$A^{(0)}(\ell_2,\ell_1,\First_j,\allowbreak\ldots,\Last_j,\ell_3)$ and
$A^{(0)}(-\ell_3,\Last_j+1,\ldots,n,1,\ldots,\First_j-1,-\ell_1,-\ell_2)$ 
(where $\ell_1+\ell_2+\ell_3+K_j = 0$), summing over
the different particle types and helicities for the $\ell_i$.
  As in step~\ref{FirstSewingStep},
put in fermi minus signs and the cut propagators, 
complete dot products to propagators,
and rewrite terms using basis expressions.
This operation is depicted schematically in 
\fig{TwoLoopSewingGeneral3Figure},
and corresponds to the promotion operation in the last term in 
\eqn{CompletePromotion}.

\item \label{ThirdCanonicalStep} {\it Put into canonical form.} 
Use momentum conservation,
the cut condition, and expansion of spinor traces, along the lines
of step~\ref{FirstCanonicalStep}.

\item {\it Relabel.\/} For each term obtained in 
step~\ref{ThirdCanonicalStep},
 relabel the momenta to the standard labeling for its
parent integral, inserting factors of squared momenta in the
numerator and denominator where required.

\item \label{MergeThreeParticleCuts} {\it Remove double-counting.}
Multiply by one-half those terms (if any) which contribute to both
terms in the three-particle
cuts.  These are the terms that contain all five propagators
$1/\ell_1^2$, $1/\ell_2^2$, $1/\ell_3^2$,
$1/(\ell_1+\ell_2)^2$, and $1/(\ell_2+\ell_3)^2$.  
As before, in general, identification of these terms may require
expansions of numerators, and use of momentum-conservation identities.
This step corresponds to the operator 
$(1-{1\over2}\CutProj_{s\crosscut}^{(3)})$ in \eqn{CompletePromotion},
and yields the sewn three-particle cut in the given channel.

\item {\it Merge two- and three-particle cuts.}
To the two-particle cut obtained in step~\ref{MergeTwoParticleCuts}, 
add those terms
in the three-particle cut not present in the two-particle cut.  (Equivalently,
add the two expressions, then subtract those terms in the three-particle
cut also present in the two-particle cut, corresponding to the action of
the operator $\CutProj_{s}^{(2)}$ in \eqn{CompletePromotion}.)  
This yields the complete sewn expression~(\ref{CompletePromotion})
in the current channel.

\item {\it Clean.}  Remove all terms which have no cut in the
current channel.  (Such terms might have been introduced by earlier algebraic
manipulations.)  Using a canonical basis as described above, remove
any terms which do not contain both cut propagators.  
This step
corresponds to the operator $\CutProj_{s_j}$
in \eqn{TwoLoopSewingFormula}. 

\item {\it Merge.} Remove all terms in the current-channel 
sewn expression that already
appear in the net integrand.  That is, remove any term which has cuts in
a previously-processed channel.  
Using a canonical basis as described above, it suffices to pick out
and remove terms that have a pair or triplet of propagators corresponding to
a cut in a previously-processed channel.
This step
corresponds to the operator 
${\textstyle\prod_{l=1}^{j-1}}\Bigl(1- \CutProj_{s_l}\Bigr)$
in \eqn{TwoLoopSewingFormula}.

\item {\it Accumulate.} 
Add the remaining terms in the current-channel sewn
expression to the integrand.  This step corresponds to the sum 
in \eqn{TwoLoopSewingFormula}.

\item Continue with the next channel at step \#\ref{FirstSewingStep}.
\end{enumerate}

As in the one-loop case, one could alternatively do the merging after
integration rather than before.  One again needs a nonredundant basis
of master integrals; but here, one needs to adjust the coefficients of
some integrals to account for the double-counting issues discussed in the
earlier example, and handled in steps~\ref{MergeTwoParticleCuts}
and~\ref{MergeThreeParticleCuts} of the two-loop algorithm above.  The
master integrals must be chosen so that each is associated with a
definite overall correction for double-counting; and one must keep
track of the original set of cut lines in each term, alongside the
integral result.  In performing integral reductions, one must
eliminate integrals in which cut propagators are cancelled.
In the computation of a given cut in an amplitude, one first needs to
merge the different two-particle cuts, as there is now more than one
contribution.  Those master integrals with two contributions to the
three-particle cuts in the given channel would have their coefficients
decreased by a factor of two.  The three-particle cuts must then be
merged with the two-particle cuts to obtain the full set of terms for
the given channel.  One can check cut consistency between different
channels just as at one loop.  Master integrals with discontinuities in
two or more channels must appear with the same coefficient in each channel. 
The merging of contributions from different channels also proceeds in the
same manner as at one loop: for those master integrals with
discontinuities in multiple channels, one would take the result from
any of the channels (that is, pick one, rather than adding together
the different contributions).

\subsection{Merging with Legs Off Shell}
\label{MergingWithOffShellSubsection}

When computing splitting amplitudes, we have an off-shell leg in the
problem, and not all light-cone denominators cancel from the final
integrand as they do for fully on-shell scattering amplitudes.
For two-loop splitting amplitudes, one side of the cut will have an
on-shell tree or one-loop amplitude integrand, while the other side
contains a tree-level splitting amplitude or one-loop splitting
amplitude integrand.  The three-particle cuts require the use of
$1\rightarrow 3$ tree-level splitting amplitudes, which should be 
calculated in light-cone gauge.  This introduces additional light-cone gauge
denominators into certain integrals.  As at one loop, sewing gluons
across the cut using \eqn{PolarizationIdentities} will introduce
similar denominators.  These denominator factors will survive into the
integration, and in fact play a crucial role in obtaining the right
`complexity' of integrals.  However, as we shall discuss in
section~\ref{SplittingIntegrandsSection}, they will only arise in
lines corresponding to cut momenta or connected directly to the
off-shell vertex.

The presence of non-canceling light-cone denominators does not alter
the merging procedure described above: these
denominators simply go along for the ride.  
Following the discussion in
section~\ref{FeynmanSection}, we may expect all surviving light-cone
denominators to produce integrals properly regulated by covariant
dimensional regularization, and this is indeed the case in
the calculation we have performed.
In contrast, a light-cone gauge Feynman diagram approach would
contain ill-defined diagrams requiring an additional prescription such as
the PV or ML prescriptions.  The ill-defined contributions
may cancel in the sum over all diagrams before integration,
if a great deal of care were taken to align momenta properly across
different diagrams, as guided by the unitarity cuts.

As discussed in~\sect{OneLoopSewingSubsection}, 
if all legs are on shell then all light-cone
denominators cancel from the integrands of each cut.  Even if some
legs are off shell, some light-cone denominators (introduced
by the physical-state projector a cut) can be canceled
prior to integration by combining information from different cuts.
The possibility of these cancellations is dictated by cut consistency.
If a given term in an integrand has a cut in more than one channel
then it must appear with the same coefficient in each such cut. A 
light-cone denominator absent in a term in any one of the cuts 
must also cancel in all other cuts.  
This cancellation may, however, not be manifest.  To cancel
the denominators, in general, momentum conservation rearrangements are
required.  We note that in performing the calculation it is
generally helpful to explicitly cancel as many light cone-denominators
as possible, to reduce the number and complexity
of integral types that need to be evaluated.  (This goes beyond the
automatic absence of dangerous light-cone denominators discussed
at the end of \sect{FeynmanSection}.)

\FIGURE[t]{
\hbox{%
\SizedFigureWithCaption{0.8 truein}{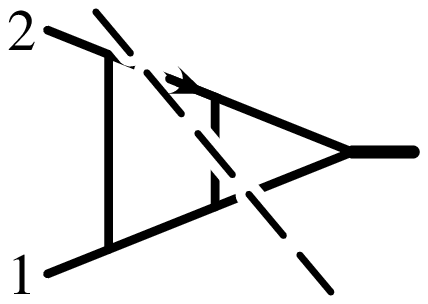}{(a)}
\hfil\hskip 2 truemm
\SizedFigureWithCaption{0.8 truein}{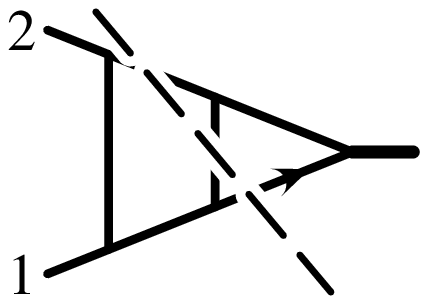}{(b)}
\hfil\hskip 2 truemm
\SizedFigureWithCaption{0.8 truein}{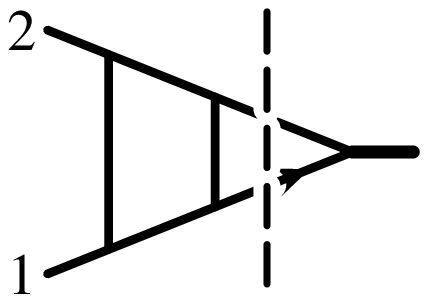}{(c)}%
}
\label{TwoLoopLCExampleFigure}
\caption{The three-particle cut (a) contains a light-cone denominator
on the leg indicated by an arrow, but the two-particle cut (c) will 
not contain one on that leg.  Cut consistency demands that the light-cone
denominator indicated in (a) must cancel from all terms where the
two-particle cut (c) does not vanish. The 
light-cone denominator arising in the three-particle cut indicated
in (b) is compatible with the ones in the two-particle cut (c) and therefore
does not need to cancel.}
}

As an example, consider the contributions to the two-loop splitting
amplitude depicted in \fig{TwoLoopLCExampleFigure}, with both two- and
three-particle cuts.  In the cuts a light-cone denominator will appear
on each sewn gluon line.  The three-particle cut contains
contributions with the light-cone denominators indicated in
\fig{TwoLoopLCExampleFigure}(a) and (b).  However, if these
contributions also have the two-particle cut indicated in
\fig{TwoLoopLCExampleFigure}(c), then the light-cone denominator
indicated in (a) must cancel algebraically in the three-particle cut
after suitably combining terms using momentum conservation.  
The reason is cut consistency: in terms which have both two- and
three-particle cuts, we must obtain identical results.  In the two-particle
cut (c) everything to the left of the cut is gauge invariant, and
cannot contain light cone-denominators.  (If we consider, instead
of \fig{TwoLoopLCExampleFigure},
a contribution for which either of the two cut propagators in (c) are 
absent, then we cannot determine from that two-particle cut
whether the denominator (a) should cancel or not.
However, by considering the other contribution to the
three-particle cut we find that the absence of the top cut 
propagator in (c) allows the denominator in (a) to be present;
but in the absence of the bottom cut propagator alone, the 
denominator in (a) has to cancel.) 
More examples of arguments of this type can be found in the 
discussion of the types of integral topologies encountered in 
\fig{ProjFigure} in \sect{IntegralsSection}.


\section{Generation of Splitting Amplitude Integrands}
\label{SplittingIntegrandsSection}

\FIGURE[t]{
\hbox{%
\SizedFigureWithCaption{0.8 truein}{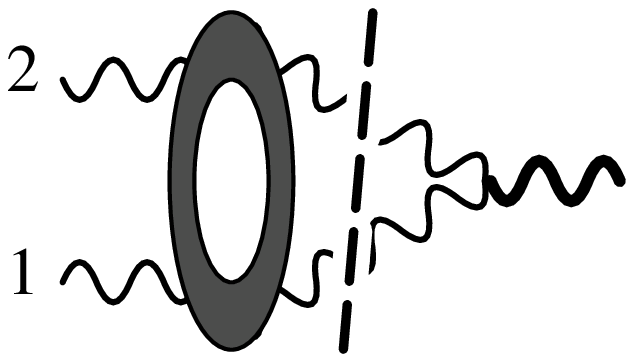}{(a)}
\hfil\hskip 2 truemm
\SizedFigureWithCaption{0.8 truein}{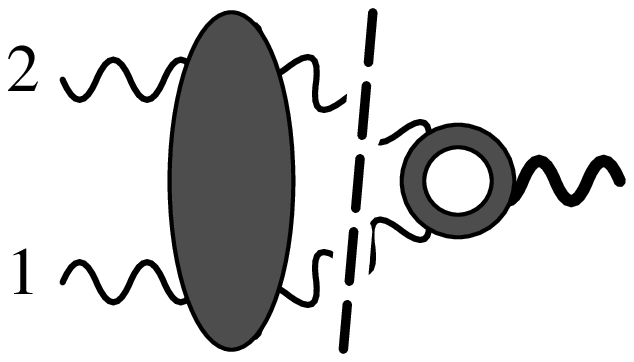}{(b)}
\hfil\hskip 2 truemm
\SizedFigureWithCaption{0.8 truein}{TwoLoopSplittingCut3.eps}{(c)}%
}
\label{TwoLoopSplittingCutFigure}
\caption{The three contributing cuts.}
}

Using the unitarity-based method described in the previous section we have
generated the integrand for the two-loop splitting amplitude.  Here,
instead of sewing together two on-shell amplitudes, we sew an on-shell
amplitude to a splitting amplitude.  The tree splitting amplitudes are
given by a current with one leg off shell~\cite{Recursive}, or
equivalently, an amplitude with one leg off shell divided by the
squared momentum of that leg.  In general, the current should be evaluated
in light-cone gauge.  It can be computed either
recursively~\cite{Recursive}, or via color-ordered Feynman rules.  
In a 
computation of the one-loop splitting amplitude, 
only the $1\rightarrow2$ tree splitting amplitude enters;
in the computation of the two-loop one, both the $1\rightarrow2$ and
$1\rightarrow3$ splitting amplitudes enter.  The $1\rightarrow2$
tree-level splitting amplitude is basically just a vertex, and so has
the same expression in both covariant and light-cone gauges; but the
expression for the $1\rightarrow3$ amplitude is {\it not\/} the same
in different gauges~\cite{CataniGrazzini,NNLOPS}.  

The $1\rightarrow3$ splitting amplitude governs the universal behavior
of tree-level gauge-theory amplitudes in triply-collinear limits,
where three color-adjacent momenta $k_{1,2,3}$ become collinear, and
all invariants $t_{123}$, $s_{12}$, $s_{23}$ are comparably small.
These limits have been described previously in
refs.~\cite{CampbellGlover,CataniGrazzini}.

There are three basic types of contributions we need to consider:
\begin{itemize}
\item[(a)] two-particle cuts with an on-shell one-loop four-point amplitude
integrand sewn onto the tree-level $1\rightarrow2$ splitting amplitude, 
depicted in \fig{TwoLoopSplittingCutFigure}(a);

\item[(b)] two-particle cuts with an on-shell tree-level four-point
amplitude sewn onto the one-loop $1\rightarrow2$ splitting amplitude
integrand, depicted in \fig{TwoLoopSplittingCutFigure}(b);

\item[(c)] three-particle cuts with an on-shell tree-level five-point
amplitude sewn onto the tree-level $1\rightarrow 3$ splitting amplitude,
depicted in \fig{TwoLoopSplittingCutFigure}(c).
\end{itemize}

In our calculation, there is a cut in only one channel 
(the invariant $s_{12}$ in \fig{TwoLoopSplittingCutFigure}), 
so cross-channel projection and consistency issues
do not arise.  However, the three contributing cuts do need to 
be constructed and merged
as described in the previous section, as there are terms that
are common to different contributions.  As in the one-loop case, the
light-cone denominators inserted by the physical projection operators on
the cut are crucial to getting the correct answer.  This
is also true
for the light-cone denominator on a line connected to the off-shell
vertex, contained in the $1\rightarrow 3$ splitting 
amplitude.  Its absence would result in different and
inconsistent integrands emerging from the two- and three-particle cuts.

In principle, non-planar topologies could enter into the $g\rightarrow
gg$ splitting amplitude, for example the crossed triangle graphs shown
in \fig{NonPlanarColor}.  However, it turns out that all such
non-planar graphs, for both pure-glue and fermion-loop contributions,
have vanishing color factors: one simply dresses the diagrams with
their color factors and performs the color algebra to demonstrate
this.  Accordingly, we only need to sew amplitudes into planar
configurations for our calculation.  As we shall show in
\sect{FullColorSection}, the only color factor that arises in the 
pure-glue contributions is the leading-color one, namely $C_A^2 = \Nc^2$,
for the simple reason that there are no other color Casimirs at this
order. The fermion-loop contributions can be divided into
leading-color ($C_A \Nf = \Nc \Nf$) and subleading-color ($(C_A - 2
C_F) \Nf = \Nf/\Nc$) terms.  Both types arise from planar diagrams.
Each planar diagram contributing to the subleading-color terms 
can be drawn with the virtual gluon on the inside of the fermion loop,
whereas each for the leading-color terms can be drawn with the 
virtual gluon on the outside.  (Some diagrams with bubble insertions 
can be drawn both ways, and are proportional to  $C_F = (\Nc^2-1)/(2\Nc)$.)
We discuss the full color dressing of the splitting amplitudes in
\sect{FullColorSection}.

For splitting amplitudes with external quarks,
$g \to q \bar{q}$ and $q\to qg$, the non-planar color factors
no longer vanish. Non-planar two-loop three-point integrals 
with light-cone denominators will be required.  However,
these integrals should be amenable to the same methods used
in the present computation.

%
\FIGURE[t]{
{\epsfxsize 4.0 truein \epsfbox{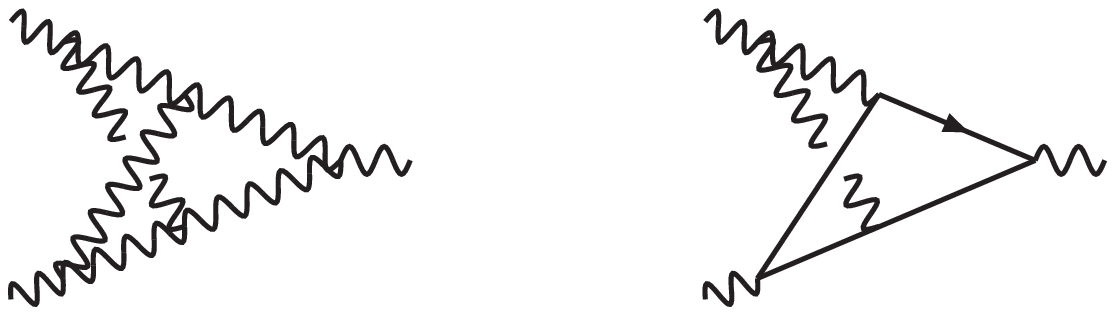}}
\caption{Examples of non-planar contributions to the two-loop
$g \to gg$ splitting amplitude, all of which have vanishing color
factors.}
\label{NonPlanarColor}}

In carrying out the calculation, we group terms in each sewn cut
into different integral topologies,
according to their propagators and light-cone denominators.
To reduce the number of independent topologies, we
perform a partial-fraction decomposition when certain multiple
light-cone-denominators appear.  For example, if $p_1+p_2 = k_1+k_2$
for two internal momenta $p_1$, $p_2$, we substitute
\begin{equation}
{1 \over p_1\cdot\n\ p_2\cdot\n}
 \to {1\over (k_1+k_2)\cdot\n} \biggl[ 
   {1 \over p_1\cdot\n} + {1 \over p_2\cdot\n} \biggr] \,.
\label{ParFrac}
\end{equation}
Although the physical-state projectors for a two-particle cut
generate two different light-cone denominators, 
the use of \eqn{ParFrac} allows us to consider 
only one at a time.  For triplets of light-cone
denominators associated with a three-particle cut, only two have to 
be considered at a time.  We also use a symmetry of the integrals 
under $k_1 \lr k_2$ ($z \lr 1-z$) to restrict the number of integral 
topologies to those described in \fig{ProjFigure} in the next section.  

After relabeling momenta circulating in the sewn cut
(see \app{RelabelingAppendix}) to match the labeling used by the
integration routine for a given topology (see the next section), 
we obtain an integrand of the form
\begin{equation}
{f(p^2, p\cdot q, q^2, p \cdot k_i, \pol_i \cdot p, \pol_j \cdot q) \over
 \prod_i p_i^2}\,, 
\label{GeneralTensorIntegral}
\end{equation}
ignoring factors that come out of the integral such as $\pol_i \cdot \pol_j$.
Here $p$ and $q$ are the loop momenta; $\prod_i p_i^2$ is shorthand for
the set of both Feynman propagators and light-cone denominators for 
the topology.  The polarization vectors for particles $i=1,2,P$ are
denoted by $\pol_i$, and are taken to satisfy the light-cone-gauge 
condition $\pol_i\cdot\n=0$, as well as the transverse condition 
$\pol_i\cdot k_i = 0$.  (Note that although $k_P$ is slightly off shell,
any terms arising from $\pol_P\cdot k_P\neq 0$ will not be sufficiently
singular in the $k_P^2\rightarrow 0$ limit to contribute to the splitting
amplitude, and hence we may as well set $\pol_P\cdot k_P$ to zero.)

In the dot products of polarization vectors with loop momenta, 
$\pol_i \cdot p$ and $\pol_i \cdot q$, the 
$D$-dimensional loop momenta are effectively projected into 
four dimensions, because the external physical
polarizations are four dimensional. 
Following the discussion of ref.~\cite{BKgggg}, we write
the loop momenta in these dot products as linear combinations of four
independent momenta: $k_1, k_2, n$, and the dual vector
\begin{equation}
v^\mu \equiv 
\pol^{\mu}_{~\nu_1\nu_2\nu_3} k_1^{\nu_1} k_2^{\nu_2} n^{\nu_3} \,.
\label{vdef}
\end{equation}
Then we have, for example,
\begin{eqnarray}
\pol_i \cdot p = c_1^p \, \pol_i \cdot k_1 + c_2^p \, \pol_i \cdot k_2 
               + c_n^p \, \pol_i \cdot \n + c_v^p \, \pol_i \cdot v \,,
\label{epExpansion}
\end{eqnarray}
where
\begin{eqnarray}
c_1^p &=& {1\over2 k_1\cdot k_2 \, k_1\cdot \n} \Bigl[ 
-  k_2 \cdot \n \, p\cdot k_1 +  k_1 \cdot \n \, p\cdot k_2 
       + k_1 \cdot k_2 \, p\cdot \n \Bigr] \,, 
\nonumber \\
c_2^p &=& {1\over2  k_1\cdot k_2 \,  k_2 \cdot \n} \Bigl[ 
  k_2 \cdot \n \, p\cdot k_1 - k_1\cdot \n \, p\cdot k_2 
      + k_1 \cdot k_2 \, p\cdot \n  \Bigr] \,, 
\nonumber \\
c_n^p &=& {1\over2 k_2\cdot \n \, k_1 \cdot \n} \Bigl[ 
   k_2 \cdot \n \, p\cdot k_1 + k_1\cdot \n \, p\cdot k_2 
        - k_1\cdot k_2 \, p\cdot \n \Bigr] \,, 
\nonumber \\
c_v^p &=& -{1\over 2  k_1\cdot k_2 \, k_1\cdot \n \, k_2 \cdot \n} 
            \varepsilon_{\mu\nu_1\nu_2\nu_3} 
   p^\mu k_1^{\nu_1} k_2^{\nu_2} \n^{\nu_3}
     \ =\  -{1\over 2  k_1\cdot k_2 \, k_1\cdot \n \, k_2 \cdot \n} p\cdot v\,.
\label{cpEqns}
\end{eqnarray}
After this substitution we use momentum conservation 
to express each integrand in terms of a basis consisting of:
\begin{enumerate}
\item inverse propagators; 
\item light-cone denominators and numerators; 
\item irreducible numerators (dot products of loop momenta with $k_1$ and
$k_2$ which cannot be written in terms of inverse propagators); and
\item dot products of loop momenta with the dual vector (\ref{vdef}),
coming from $c_v^p$ and $c_v^q$.
\end{enumerate}
Following the merging procedure described in the previous section, 
at the integrand level, we obtained a single expression with correct 
cuts in all channels.  
Using the formul\ae{} in section 4 of ref.~\cite{BKgggg}, we then replaced 
dot products of loop momenta with the dual vector, $p\cdot v$ and 
$q\cdot v$,  by other elements of the basis.  (These formul\ae{} 
are valid at the level of integrals, not integrands, so they should
only be applied after checking cut consistency.)
Typical propagator momenta, light-cone dot products,
and irreducible numerators that appear in the calculation are given
in \eqns{Lpidef}{LPBpidef}.  This produces an expression 
ready to be integrated.  We describe the integration method
in the next section.

\section{Integrals}
\label{IntegralsSection}


\subsection{Introduction}
\label{IntegralsIntroductionSection}

All the integrals encountered in our computation of the two-loop
splitting amplitudes are 3-point integrals with one external
massive leg, and two massless legs.  The massless legs carry
momenta $k_1$ and $k_2$.  The massive leg carries momentum $k_P = k_1+k_2$ 
with $k_P^2 = (k_1+k_2)^2 = 2k_1\cdot k_2 \equiv s$.  We consider
the time-like case, $s > 0$.  For the space-like case,
we take the splitting to be $(-k_1) \to (-k_P) + k_2$, with
$k_P^2 = s < 0$.  Here $(-k_P)$ and $k_2$ carry longitudinal 
momentum fractions $x$ and $1-x$ respectively, with $x=1/z$.  
The space-like case may be obtained from the time-like case 
using analyticity.  As there is no other dimensionful
parameter in the problem, the dependence on $s$ is determined 
by dimensional analysis to be $\propto (-s)^{-2\e}$.  To reach the
physical range $0 < x < 1$, it is also necessary to continue the 
momentum fraction $z$ to values larger than 1, as we shall discuss
in \sect{Spacelike}.

In addition to standard propagator factors of the form
$1/p^2$ in the denominator, and tensors in the numerator, there can also 
be denominator factors from light-cone gauge (or physical state
projection, of the form $1/(p\cdot n)$ where $n^\mu$ is the light-cone
gauge vector, $n^2=0$.  The vector $n$ is also used to define the
collinear momentum fraction $z$, according to the kinematic
relations~(\ref{kidotn}).  We also rescale $n$ so that it 
obeys~\eqn{Replace_n}, $k_P\cdot n = -s$.

Two-loop 3-point integrals containing light-cone denominator factors
have not been encountered previously, and contain non-trivial
dependence on $z$.  For the case of a gluon splitting to two gluons,
we shall see that the non-planar topologies all have vanishing color
factor, and hence the corresponding integrals do not need to be
calculated.  The integrals from the planar topologies can be reduced
to a set of 13 master integrals, using identities based on integration
by parts (IBP)~\cite{IBP} and Lorentz invariance~\cite{Lorentz}.  The
master integrals can be evaluated as Laurent expansions in $\e$ with
the aid of differential equations in $z$, along the lines of
refs.~\cite{PBReduction,Lorentz}.

The Laurent expansions of each master integral, through the order
in $\e$ required to obtain the splitting amplitudes to order $\e^0$, 
can be expressed in terms of logarithms, plus the polylogarithmic 
functions defined~\cite{Lewin} by,
\begin{eqnarray}
\Li_n(x) &=& \sum_{i=1}^\infty { x^i \over i^n }
          = \int_0^x {dt \over t} \Li_{n-1}(t)\,,  \\
\Li_2(x) &=& -\int_0^x {dt \over t} \ln(1-t) \,.
\label{PolyLogDef}
\end{eqnarray}
Here we need $\Li_n(x)$, for $n=2,3,4$, and the argument $x$ can be
$z$, $1-z$ or $-(1-z)/z$.  (For $n<4$, identities relate some of
these polylogarithms to each other.)  For the order $\e^1$ terms in 
the splitting amplitudes, the corresponding set of functions 
(with $n$ extending up to 5) is not sufficient; 
instead harmonic polylogarithms~\cite{RV,Lorentz} are required.

In section~\ref{ReductionProcedureSection} we describe the procedure
for reducing the two-loop three-point integrals with light-cone
denominators to master integrals, and give a list of master
integrals required for the $g\to gg$ splitting amplitude.
In section~\ref{DifferentialEquationSection} we illustrate how to derive
differential equations for the master integrals, and present the
differential equations.  We also give the Laurent expansions
of the master integrals.


\subsection{Reduction Procedure}
\label{ReductionProcedureSection}

A typical integral encountered is the two-loop nested double triangle 
integral with two light-cone denominators inserted,
\begin{equation}
L(\nu_1,\nu_2,\nu_3,\ldots,\nu_9) \equiv 
\int {d^{D}p\over \pi^{D/2}} \, {d^{D}q\over \pi^{D/2}} \, 
 \prod_{i=1}^9 { 1 \over \bigl( p_i^2 \bigr)^{\nu_i} } \,,
\label{LtriInt}
\end{equation}
where $D=4-2\e$,
\begin{eqnarray}
&&p_1 = q, \qquad p_2 = q+k_1+k_2, \qquad p_3 = p, \qquad p_4 = p-k_1-k_2, 
\nonumber \\
&& p_5 = p-k_1, \qquad p_6 = p+q, \nonumber \\
&&  p_7^2 = q\cdot n, \qquad p_8^2 = 2q\cdot k_1, 
\qquad p_9^2 = (p-k_1-k_2)\cdot n, 
\label{Lpidef}
\end{eqnarray}
and the $\nu_i$ are integers.
The momentum routings for this integral are depicted in 
\fig{PlanarTriFigure}(a).  The off-shell external momentum $k_1+k_2$
flows in from the right of the diagram, and splits into on-shell 
momenta $k_1$ and $k_2$ flowing out to the left.  The uncircled numbers
adorning the internal lines label the internal momenta $p_i$,
whose squares are the `ordinary' denominators $p_i^2$, $i=1,2,\ldots,6$,
appearing in~\eqn{LtriInt}.  The circled numbers correspond to the
light-cone denominators, in this case $p_7^2$ and $p_9^2$, which are
linear in $n$.  Associated with each circled number is an arrow on
an internal line.  The arrow is a reminder of the direction of
the internal momentum which is Lorentz-contracted with $n$ to form
the light-cone denominator.
One of the $p_i^2$ in~\eqn{LtriInt} is not shown 
in~\fig{PlanarTriFigure}(a):  $p_8^2 = 2q\cdot k_1$.  In this topology,
it only appears in the numerator, {\it i.e.} $\nu_8 \leq 0$.
It will appear as an irreducible numerator 
in the tensor integrals arising from the numerator
algebra generated in evaluating the cuts.  Its presence is also
required to close the IBP equations.

%
\FIGURE[t]{
{\epsfxsize 5 truein \epsfbox{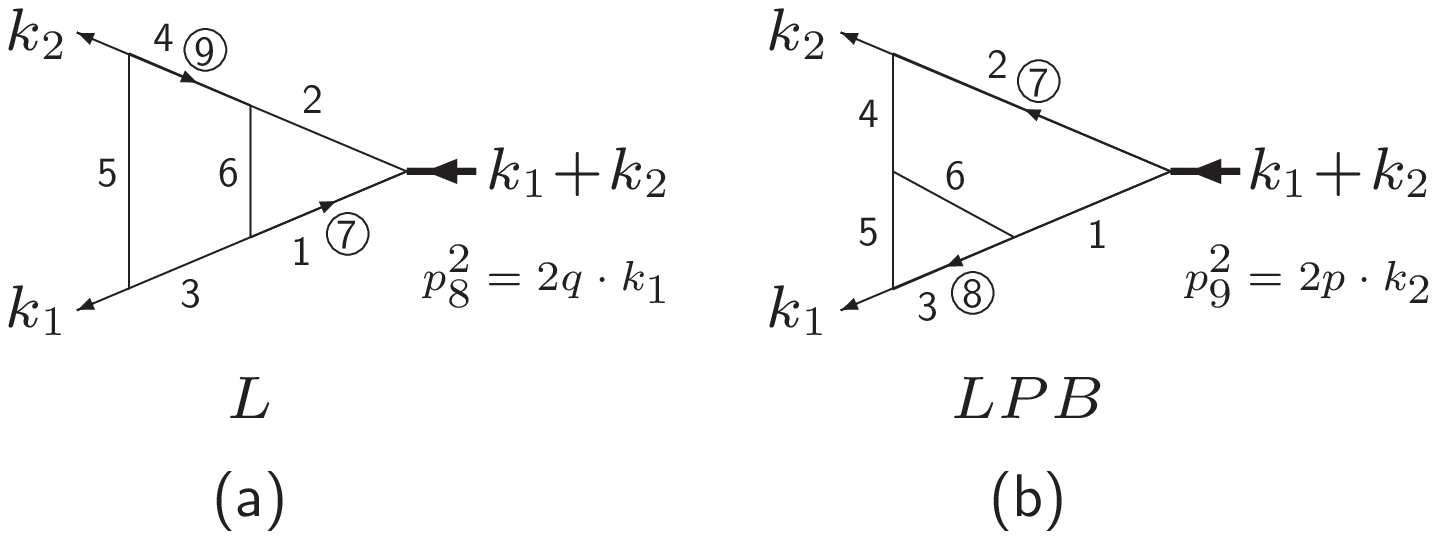}}
\caption{(a) The planar double triangle integral $L(\nu_i)$ defined
in~\eqn{LtriInt}. (b) The planar double triangle integral $LPB(\nu_i)$
defined in \eqn{LPBtriInt}.  In each case $k_1$ and $k_2$ are the two
outgoing massless external lines.  The internal lines, or `ordinary'
propagators, labelled with an uncircled integer $i$,
$i\in\{1,2,\ldots,6\}$, carry momenta $p_i^\mu$.
The special light-cone denominator factors $p_i\cdot n$ which can be present 
are marked with an arrow on an internal line, and the corresponding integer
label is circled. The arrow serves to remind one of the direction of the
loop-momentum used in their definition.  Propagators 8 (in case
(a)) and 9 (in case (b)) do not appear in denominators in these integrals. 
The expressions for $p_8^2$ and $p_9^2$ are given explicitly.}
\label{PlanarTriFigure}}

We also encounter the `pentabox' integral topology~\cite{PentaboxAGO} 
(but with one ordinary propagator cancelled), again with two 
light-cone denominator insertions,
\begin{equation}
LPB(\nu_1,\nu_2,\nu_3,\ldots,\nu_9) \equiv 
\int {d^{D}p\over \pi^{D/2}} \, {d^{D}q\over \pi^{D/2}} \, 
 \prod_{i=1}^9 { 1 \over \bigl( p_i^2 \bigr)^{\nu_i} } \,,
\label{LPBtriInt}
\end{equation}
where
\begin{eqnarray}
&&p_1 = q, \qquad p_2 = q+k_1+k_2, \qquad p_3 = p, \qquad p_4 = q+k_1, 
\nonumber \\
&& p_5 = p-k_1, \qquad p_6 = p+q, \nonumber \\
&&  p_7^2 = (q+k_1+k_2)\cdot n, \qquad p_8^2 = p\cdot n, 
\qquad p_9^2 = 2p\cdot k_2. 
\label{LPBpidef}
\end{eqnarray}
The momentum routings for this integral are depicted in 
\fig{PlanarTriFigure}(b). In this case $p_7^2$ and $p_8^2$
are the circled, light-cone denominators, and $p_9^2$ only
appears in the numerator, $\nu_9 \leq0$.

Besides the two types of integrals depicted in~\fig{PlanarTriFigure},
there are several more types, which differ from either $L$
or $LPB$ only in the location of the light-cone denominators.
All the light-cone configurations are shown in~\fig{ProjFigure}.
The leftmost two cases, $L$ and $LPB$, are equivalent to 
\fig{PlanarTriFigure}(a) and \fig{PlanarTriFigure}(b).  
The momentum routings and the labels for the external legs and for 
propagators 1 to 6 are exactly the same as in \fig{PlanarTriFigure}, 
so we have suppressed them in~\fig{ProjFigure}.
The remaining cases define the integrals $D(\nu_i)$,
$F(\nu_i)$, $G(\nu_i)$, $H(\nu_i)$, $J(\nu_i)$ and $M(\nu_i)$,
by analogy to \eqns{LtriInt}{LPBtriInt}.
As in the $L$ and $LPB$ cases, one extra propagator, linear in the 
loop momentum but not a light-cone denominator, is required for 
tensor integral reduction and for the IBP equations to close; 
the expression for this propagator is shown explicitly in the figure.

The symbol `$\times$' on a line in~\fig{ProjFigure} indicates that
that ordinary propagator never appears in the denominator, when the
light-cone denominators indicated by the arrows are present.  As
discussed in~\sect{MergingWithOffShellSubsection}, the unitarity cuts
guarantee these facts.  For example, suppose the propagator marked
with a `$\times$' in the $J$ topology were present.  Then such an
integral has a 2-particle cut to the right of the circled propagator
$8$.  (If propagator $2$, which also must be present in the 2-particle
cut, did not appear in the denominator, then the integral would become
a massless external leg integral, which vanishes trivially in
dimensional regularization.)  But we know that no light-cone
projectors should appear to the left of such a cut, so either the
circled propagator $8$ or the $\times$-marked propagator must be
absent in $J$.  If the circled propagator $8$ is absent from the
denominator, we consider the integral to belong to the $F$ topology
instead.

Note that we have not explicitly shown integrals that are related to 
the ones in~\fig{ProjFigure} by the external leg permutation 
$k_1 \lr k_2$, which also takes $z \lr 1-z$.  We always use
this permutation to map such integrals into those shown 
in~\fig{ProjFigure}.

%
\FIGURE[t]{
{\epsfxsize 5.8 truein \epsfbox{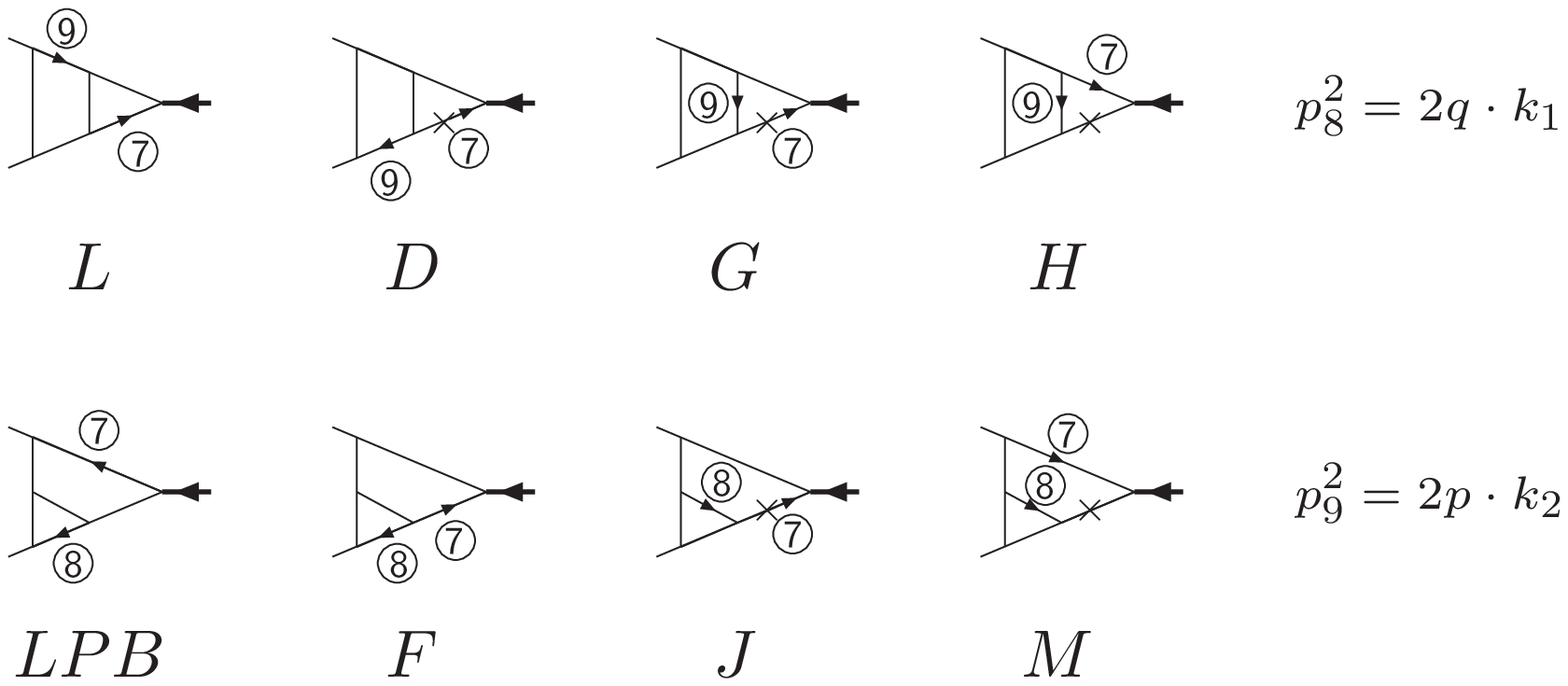}}
\caption{List of required light-cone denominator insertions for
the $g\to gg$ splitting amplitude.
The leftmost two cases are equivalent to \fig{PlanarTriFigure}(a) and 
\fig{PlanarTriFigure}(b).  The labels for the external legs and for 
propagators 1 to 6 are suppressed.  The expression for the extra 
propagator required for tensor reductions is the same for $L$, $D$, $G$
and $H$ topologies:  $p_8^2 = 2q\cdot k_1$.  The extra propagator for 
$LPB$, $F$, $J$ and $M$ topologies is $p_9^2 = 2p\cdot k_2$.
For the $D$, $G$, $H$, $J$ and $M$ topologies, the `$\times$' signifies that 
the marked ordinary propagator is not required in the denominator, 
only in the numerator.}
\label{ProjFigure}}

There are three independent external four-vectors in the light-cone integrals, 
$k_1$, $k_2$ and $n$.  Due to momentum conservation,
the same number of vectors appear in the two-loop four-point integrals 
needed for $2\to2$ scattering, $k_1 + k_2 \to k_3 + k_4$.
In fact, $n$ may be thought of as a fictitious momentum vector $k_3$,
if desired.  Then the derivation of IBP and Lorentz invariance
identities follows straightforwardly from previous work on 
four-point integrals~\cite{IBP,PBReduction,Lorentz,NPBReduction}.
The IBP equations for the $L$ topology are derived by considering
\begin{equation}
0 =  
\int {d^{D}p\over \pi^{D/2}} \, {d^{D}q\over \pi^{D/2}} \, 
{\del \over \del a^\mu} \biggl[
 b^\mu \prod_{i=1}^9 { 1 \over \bigl( p_i^2 \bigr)^{\nu_i} } \biggr] \,,
\label{LIBP}
\end{equation}
where the $p_i^2$ are given in \eqn{Lpidef}, 
$a^\mu \in \{ p^\mu, q^\mu \}$ are the two independent loop momenta,
and $b^\mu$ can be any of the five available vectors,
$b^\mu \in \{ p^\mu, q^\mu, k_1^\mu, k_2^\mu, n^\mu \}$.
Thus for each value of $\{ \nu_i \}$, there are $2 \times 5 = 10$ 
IBP equations, which are linear in the $L(\nu_i)$.  
They can be written as
\begin{eqnarray}
0 &=&
 (D - 2 \nu_3 - \nu_4 - \nu_5 - \nu_6 - \nu_9)
 + \nu_4 ( s - {\bf 3^-} ) {\bf 4^+}
 - \nu_5 {\bf 3^- 5^+}
\PlusBreak{ }
   \nu_6 ( {\bf 1^-} - {\bf 3^-} ) {\bf 6^+} 
 + \nu_9 \, s \, {\bf 9^+}
\,,
\label{LIBP1} \\
0 &=&
 \nu_3 - \nu_6
 + \nu_3 ( {\bf 1^-} - {\bf 6^-} ) {\bf 3^+}
 - \nu_4 ( s - {\bf 2^-} - {\bf 3^-} + {\bf 6^-} ) {\bf 4^+}
 + \nu_5 ( {\bf 1^-} + {\bf 3^-} - {\bf 6^-} + {\bf 8^-} ) {\bf 5^+}
\MinusBreak{ }
   \nu_6 ( {\bf 1^-} - {\bf 3^-} ) {\bf 6^+} 
 - \nu_9 \, {\bf 7^-} {\bf 9^+}
\,,
\label{LIBP2} \\
0 &=&
 - \nu_3 + \nu_5
 + \nu_3 {\bf 5^-}  {\bf 3^+}
 + \nu_4 ( s - {\bf 3^-} + {\bf 5^-} ) {\bf 4^+}
 - \nu_5 {\bf 3^-} {\bf 5^+}
\MinusBreak{ }
   \nu_6 ( {\bf 3^-} - {\bf 5^-} + {\bf 8^-} ) {\bf 6^+} 
 + \nu_9 \, z s \, {\bf 9^+}
\,,
\label{LIBP3} \\
0 &=&
 \nu_4 - \nu_5
 - \nu_3 ( s - {\bf 4^-} + {\bf 5^-} ) {\bf 3^+}
 - \nu_4 {\bf 5^-} {\bf 4^+}
 + \nu_5 {\bf 4^-} {\bf 5^+}
\PlusBreak{ }
   \nu_6 ( {\bf 1^-} - {\bf 2^-} + {\bf 4^-} - {\bf 5^-} + {\bf 8^-} ) 
         {\bf 6^+} 
 + \nu_9 \, (1-z) s \, {\bf 9^+}
\,,
\label{LIBP4} \\
0 &=&
 \nu_3 ( s - {\bf 9^-} ) {\bf 3^+}
 - \nu_4 {\bf 9^-} {\bf 4^+}
 + \nu_5 ( (1-z) s - {\bf 9^-} ) {\bf 5^+}
 + \nu_6 ( s - {\bf 7^-} - {\bf 9^-} )  {\bf 6^+} 
\,,
\label{LIBP5} \\
0 &=&
 \nu_1 - \nu_6
 + \nu_1 ({\bf 3^-} - {\bf 6^-} ) {\bf 1^+}
 - \nu_2 ( s - {\bf 1^-} - {\bf 4^-} + {\bf 6^-} ) {\bf 2^+}
 + \nu_6 ( {\bf 1^-} - {\bf 3^-} ) {\bf 6^+}
\PlusBreak{ }
   \nu_7 ( s - {\bf 9^-} ) {\bf 7^+}
 - \nu_8 ( {\bf 3^-} - {\bf 5^-} )  {\bf 8^+} 
\,,
\label{LIBP6} \\
0 &=&
 D - 2 \nu_1 - \nu_2 - \nu_6 - \nu_7 - \nu_8
 + \nu_2 ( s - {\bf 1^-} ) {\bf 2^+}
 - \nu_6 ( {\bf 1^-} - {\bf 3^-} ) {\bf 6^+}
\,,
\label{LIBP7} \\
0 &=& 
 - \nu_1 {\bf 8^-} {\bf 1^+}
 - \nu_2 ( s + {\bf 8^-} ) {\bf 2^+}
 - \nu_6 ( {\bf 3^-} - {\bf 5^-} + {\bf 8^-} ) {\bf 6^+}
 + \nu_7 \, z s \, {\bf 7^+}
\,,
\label{LIBP8} \\
0 &=&
 \nu_1 - \nu_2
 + \nu_1 ( s - {\bf 2^-} + {\bf 8^-} ) {\bf 1^+}
 + \nu_2 ( {\bf 1^-} + {\bf 8^-} ) {\bf 2^+}
 + \nu_6 ( {\bf 1^-} - {\bf 2^-} + {\bf 4^-} - {\bf 5^-} + {\bf 8^-} ) 
         {\bf 6^+} 
\PlusBreak{ }
   \nu_7 \, (1-z) s \, {\bf 7^+}
 - \nu_8 \, s \, {\bf 8^+}
\,,
\label{LIBP9} \\
0 &=&
 - \nu_1 {\bf 7^-} {\bf 1^+}
 + \nu_2 ( s - {\bf 7^-} ) {\bf 2^+}
 + \nu_6 ( s - {\bf 7^-} - {\bf 9^-} ) {\bf 6^+} 
 + \nu_8 \, z s \, {\bf 8^+}
\,.
\label{LIBP10}
\end{eqnarray}
Here ${\bf i^\pm}$ are operators taking $\nu_i \to \nu_i \pm 1$;
for instance, in \eqn{LIBP1} the expression $-\nu_5 {\bf 3^- 5^+}$ 
is shorthand for the term 
$-\nu_5 L(\nu_1,\nu_2,\nu_3-1,\nu_4,\nu_5+1,\nu_6,\nu_7,\nu_8,\nu_9)$.

Three Lorentz invariance identities for each value of $\{ \nu_i \}$
can also be derived, by requiring
\begin{equation}
0 =  
\int {d^{D}p\over \pi^{D/2}} \, {d^{D}q\over \pi^{D/2}} \, 
\eps^\mu_{j \,\nu} \biggl( k_1^\nu {\del \over \del k_1^\mu }
                   + k_2^\nu {\del \over \del k_2^\mu }
                   + n^\nu {\del \over \del n^\mu } \biggr)
  \prod_{i=1}^9 { 1 \over \bigl( p_i^2 \bigr)^{\nu_i} } \,,
\label{LLor}
\end{equation}
where
\begin{eqnarray}
\eps^\mu_{1 \, \nu} &=& k_1^\mu k_2^\nu -  k_2^\mu k_1^\nu,
\nonumber\\
\eps^\mu_{2 \, \nu} &=& k_1^\mu n^\nu -  n^\mu k_1^\nu,
\nonumber\\
\eps^\mu_{3 \, \nu} &=& k_2^\mu n^\nu -  n^\mu k_2^\nu.
\label{LLoreps}
\end{eqnarray}
The three identities are
\begin{eqnarray}
0 &=&
 ( \nu_2 + \nu_4 - \nu_5 - \nu_8 )
- \nu_2 ( s + {\bf 1^-} + 2 \, {\bf 8^-} ) {\bf 2^+}
- \nu_4 ( s  - {\bf 3^-} + 2 \, {\bf 5^-} ) {\bf 4^+} 
+ \nu_5 \, {\bf 3^- 5^+}
\PlusBreak{ }
  \nu_7 ( z s + z \, {\bf 1^-} - z \, {\bf 2^-} + {\bf 8^-} ) {\bf 7^+}
- \nu_9 ( z s - (1-z){\bf 3^-} - z \, {\bf 4^-} + {\bf 5^-} ) {\bf 9^+}
 \,,
\label{LLor1} \\
0 &=&
 z ( \nu_5 - \nu_7 + \nu_8 - \nu_9 )
+ \nu_2 ( {\bf 7^-} + {\bf 8^-} ) {\bf 2^+}
+ \nu_4 ( s  - {\bf 3^-} + {\bf 5^-} - {\bf 9^-}) {\bf 4^+} 
\MinusBreak{ }
  \nu_5 \, z \, {\bf 3^- 5^+}
+ \nu_9 \, z s \, {\bf 9^+}
 \,,
\label{LLor2} \\
0 &=&
 \nu_2 + \nu_4 - ( \nu_5 + \nu_8 ) z - ( \nu_7 + \nu_9 ) (1-z)
- \nu_2 ( s + {\bf 1^-} - {\bf 7^-} + {\bf 8^-} ) {\bf 2^+}
\MinusBreak{ }
  \nu_4 ( {\bf 5^-} + {\bf 9^-}) {\bf 4^+} 
+ \nu_5 ( (1-z) s + z \, {\bf 4^-} - {\bf 9^-} ) {\bf 5^+}
\MinusBreak{ }
  \nu_8 ( z s + z \, {\bf 1^-} - z \, {\bf 2^-} - {\bf 7^-} ) {\bf 8^+}
+ \nu_9 \, (1-z) s \, {\bf 9^+}
 \,.
\label{LLor3}
\end{eqnarray}
There are analogous sets of equation for the other integral topologies:
$LPB$, $D$, $G$, $H$, $F$, $J$ and $M$.

The next step is to solve the linear system of IBP and Lorentz equations
for each topology. We use a Gauss elimination algorithm first introduced
by Laporta~\cite{Laporta}.  We have used a customized
version of this algorithm~\cite{AIR}, written in MAPLE~\cite{maple} and 
FORM~\cite{form}.
After performing the reductions, we obtain 13 master integrals for the
$g\to gg$ problem:
\begin{eqnarray}
 L(1,1,1,1,0,0,0,0,0) &=& 
 \Spec(s) = { \e^2 \, s^2 \over (1-2\e)^2 } \, \Btie(s) \,,
\nonumber \\
 L(1,0,0,1,0,1,0,0,0) &=& LPB(0,1,1,0,0,1,0,0,0) = \Sset(s) \,,
\nonumber \\
 L(1,1,0,0,1,1,0,0,0) &=& \Btri(s) \,,
\nonumber \\
 L(0,1,1,1,0,1,1,0,0) &=& \Fish(s) \,,
\nonumber \\
 L(1,0,0,1,1,1,1,0,0) &=& - LPB(0,1,1,1,0,1,0,1,0) = \Wedge(z,s) \,,
\nonumber \\
 L(1,1,0,0,1,1,1,0,0) &=& \LBtri(z,s) \,,
\nonumber \\
 L(0,1,1,0,1,1,1,0,0) &=& \WedgeF(z,s) \,,
\nonumber \\
 L(1,1,1,0,1,1,1,0,0) &=& \Zig(z,s) \,,
\nonumber \\
 L(1,1,1,1,1,1,1,0,0) &=& \Ptri(z,s) \,,
\nonumber \\
 L(1,1,1,1,1,2,1,0,0) &=& \Ptri_2(z,s) \,,
\nonumber \\
 L(1,0,0,1,1,1,0,0,1) &=& - LPB(0,1,1,0,1,1,0,1,0) \vert_{z\to 1-z}
   = - \LPBWedge(z,s) \,,
\nonumber \\
 LPB(0,1,1,1,1,2,1,1,0) &=& \LPBDtri_2(z,s) \,,
\nonumber \\
 F(0,1,1,1,1,1,1,1,0) &=& \FDtri(z,s) \,.
\label{AllMasters}
\end{eqnarray}
($\Btie(s)$ is an integral with two independent one-loop triangles,
which is trivially related to $\Spec(s)$.)

The master integrals are depicted in \fig{PlanarMasterFigure}.  The first
three of the master integrals have no light-cone denominators present, so
they do not depend on $z$ and were encountered long ago in the computation
of the quark form factor~\cite{Gonsalves}.  The next master
integral, $\Fish(s)$, has a light-cone denominator, but nevertheless does
not depend on $z$.  The remaining master integrals depend on $z$.  The
next six come from the $L$ topology, although they can also come from
other topologies.  In the $L$ topology labeling the have light-cone
denominator 7, but not 9, present.  The next master integral after that,
$\LPBWedge(z,s)$ can come from either the $L$ or $LPB$ topology. but in
the $L$ labeling it has light-cone denominator 9, but not 7, present.
Finally, the last two master integrals come from the $LPB$ and $F$
topologies and have two light-cone denominators present, 7 and 8.

%
\FIGURE[t]{
{\epsfxsize 5.8 truein \epsfbox{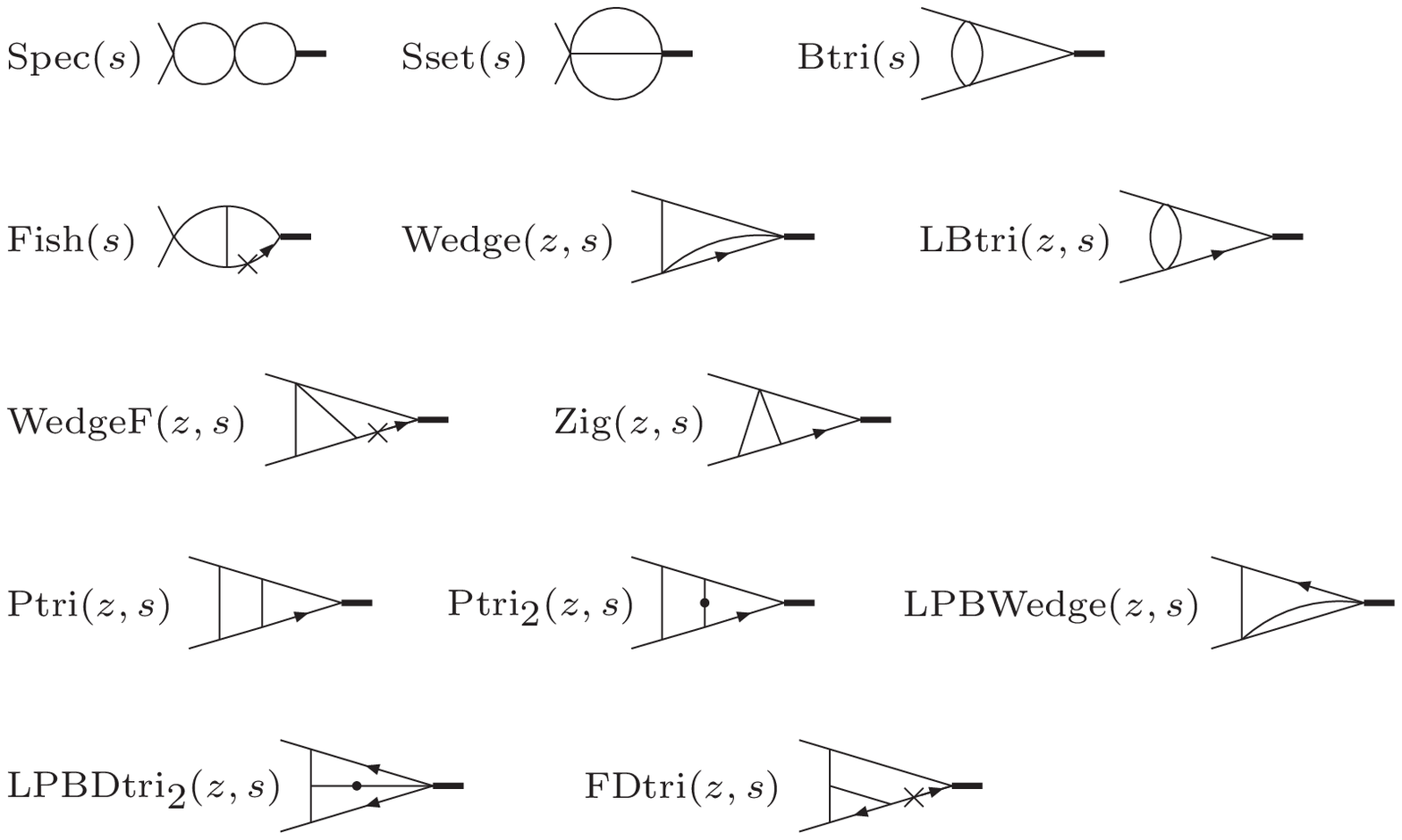}}
\caption{The planar master integrals required for computing two-loop
leading-color splitting amplitudes.  An arrow indicates a light-cone
propagator, a dot indicates a doubled ordinary propagator, and
a `$\times$' indicates an omitted ordinary propagator.}
\label{PlanarMasterFigure}}

In intermediate steps, two other master integrals appear,
$J(0,1,1,1,1,1,1,1,0) \equiv \JDtri(z,s)$
and $M(0,1,1,1,1,1,1,1,0) \equiv \MDtri(z,s)$.  However, there
is a partial fraction equation relating $\FDtri$, $\JDtri$ and $\MDtri$.  
Consider the identity
\begin{equation}
{ 1 \over (p+q)\cdot n } \, { 1 \over q\cdot n }
= { 1 \over p\cdot n }  \, { 1 \over q\cdot n }
+ { 1 \over (p+q)\cdot n }  \, { -1 \over p\cdot n } \,.
\label{FJMparfrac}
\end{equation}
It implies, after inspecting~\fig{ProjFigure}, that
\begin{equation}
{\rm JDtri}(z,s) = \FDtri(z,s) + {\rm MDtri}(1-z,s).
\label{FJMparfrac2}
\end{equation}
After using this relation to eliminate $\JDtri$, we find that the 
coefficient of $\MDtri$ cancels out of all $g\to gg$ splitting 
amplitude expressions.


\subsection{Differential Equations for Master Integrals}
\label{DifferentialEquationSection}

Since $s$ is the only dimensionful scale entering the 
master integrals, their dependence on $s$ is fixed simply
by dimensional analysis.
The $z$-dependence of the master integrals can be determined
by differentiating their Schwinger-parametrized form with respect
to $z$.  This produces an integral where some of the $\nu_i$ and the
dimension $D$ have been shifted by integers ($D$ is always shifted by
an even integer).  The effect of shifting
$D$ can also be converted into a shift of $\nu_i$.  Integrals
with shifted $\nu_i$ are again reducible to master integrals.
Thus we first derive `dimension-shifting' relations, and then
differential equations in $z$, for the master integrals.

Consider a case where all the $\nu_i$ are non-negative.
Schwinger parametrization is achieved by inserting (for $\nu_i >0$)
\begin{equation}
 {1\over \bigl( p_i^2 \bigr)^{\nu_i} } 
= {(-1)^{\nu_i}\over \Gamma(\nu_i)}
  \int_0^\infty dt_i \, t_i^{\nu_i-1} \exp(t_i p_i^2),
\label{Schwinger2}
\end{equation}
into the loop-momentum integrals, Wick rotating, and then performing 
the integrals over $p$ and $q$, to arrive at the generic result 
\begin{equation}
X(\nu_i) = 
 - \int_0^\infty \prod_{i=1}^{n_\nu} \biggl( dt_i 
 (-1)^{\nu_i} { t_i^{\nu_i-1} \over \Gamma(\nu_i) } \biggr)
  \times \Bigl[ \Delta(t_i) \Bigr]^{-D/2}
   \exp\Biggl[- {(-s)Q_X(z,t_i) \over \Delta(t_i)} \Biggr] \,.
\label{SchwingerTensor}
\end{equation}
Here $X$ stands for one of the integral topologies, $n_\nu$ is the
number of positive $\nu_i$, and $Q_X(z,t_i)$ is a cubic polynomial in the
$t_i$.  Also,
\begin{equation}
 \Delta(t_i) \equiv T_p T_q + T_p T_{pq} + T_q T_{pq} \,,
\label{DeltaDef}
\end{equation}
where $T_p$, $T_q$, $T_{pq}$ are the sums of Schwinger
parameters along the lines carrying loop momenta $p$, $q$, $p+q$,
respectively.  For the $L$, $D$, $G$ and $H$ integrals,
\begin{equation} 
T_p = t_3 + t_4 + t_5, \qquad T_q = t_1 + t_2, \qquad T_{pq} = t_6.
\label{TiLDef}
\end{equation}
For the $LPB$, $F$, $J$ and $M$ integrals,
\begin{equation} 
T_p = t_3 + t_5, \qquad T_q = t_1 + t_2 + t_4, \qquad T_{pq} = t_6.
\label{TiLPBDef}
\end{equation}
From \eqn{SchwingerTensor} we see that shifting $D \to D-2$ is equivalent
to inserting a factor of $\Delta$ into the Schwinger-parametrized result.
Breaking up $\Delta$ into monomials in the Schwinger parameters $t_i$,
these integrals can be rewritten using shifted indices $\nu_i$. 
For example,
\begin{eqnarray} 
&&\Ptri(z,s)|_{D\to D-2} = L(1,1,1,1,1,1,1,0,0)|_{D\to D-2} 
\nonumber\\ && \hskip0.2cm
= L(2,1,1,1,1,2,1,0,0) + L(1,2,1,1,1,2,1,0,0) + L(1,1,2,1,1,2,1,0,0)
\PlusBreak{ = L }
  L(1,1,1,2,1,2,1,0,0) + L(1,1,1,1,2,2,1,0,0) + L(2,1,2,1,1,1,1,0,0) 
\PlusBreak{ = L }
 L(1,2,2,1,1,1,1,0,0) + L(2,1,1,2,1,1,1,0,0) + L(1,2,1,2,1,1,1,0,0) 
\PlusBreak{ = L }
 L(2,1,1,1,2,1,1,0,0) + L(1,2,1,1,2,1,1,0,0).
\label{PTriShift}
\end{eqnarray}
The latter combination of integrals can be reduced to a linear combination
of $\Ptri(z,s)$, $\Ptri_2(z,s)$, and the eight master integrals
preceding them in \eqn{AllMasters} or \fig{PlanarMasterFigure}.
We have worked out such dimension-shifting relations for all of the 
master integrals, but we refrain from presenting them all here.
We shall give two examples in~\eqns{Ptri26shift}{LPBDtri26shift} below.

Once the $z$-dependence of $Q_X(z,t_i)$ in \eqn{SchwingerTensor} is 
known, the derivative of $X(\nu_i)$ with respect to $z$ may be computed
from the parametric representation~(\ref{SchwingerTensor}).
The polynomial $Q_X(z,t_i)$ is always linear in $z$,
\begin{equation}
Q_X(z,t_i) = Q_X^0(t_i) + z \, Q_X^1(t_i), 
\label{QLinear}
\end{equation}
where $Q_X^0$ and $Q_X^1$ are independent of $z$. 
For the integrals encountered in the calculation,
the $Q_X(z,t_i)$ have a definite sign for $0 < z < 1$.
Hence all the master integrals are real in the time-like region,
apart from the overall factor of $(-s)^{-2\e}$.
For the first ten master integrals in \eqn{AllMasters},
$Q_X^1(t_i) = t_5 t_6 t_7$, unless any of $\nu_5$, $\nu_6$, $\nu_7$
vanishes.  In these cases, $Q_X^1(t_i) = 0$ and the master integral is 
independent of $z$.  Thus $\Fish(s)$ is independent of $z$, despite 
containing a light-cone denominator.  For the non-trivial cases where
$\nu_5 \nu_6 \nu_7 > 0$, 
using~\eqn{SchwingerTensor}, the derivatives of the master integrals are 
simply obtained by inserting $(-s) Q_X^1/\Delta$ 
into the Schwinger-parametrized forms.  (There
is an extra minus sign from Wick rotation.) For example,
\begin{equation}
{\del \over \del z} \Ptri(z,s)
 = {\del \over \del z} L(1,1,1,1,1,1,1,0,0)
 = (-s) L(1,1,1,1,2,2,2,0,0)|_{D\to D+2} \,.
\label{diffPtriRAW}
\end{equation}
The shift $D\to D+2$ comes from the factor of $1/\Delta$.  The effect
of this shift is found by inverting the $D\to D-2$ shift computed
by inserting one factor of $\Delta$. 

For the integral $\LPBWedge(z,s)$, we have $Q_X^1(t_i) = t_4 (t_3+t_6) t_7$.
For $\LPBDtri_2(z,s)$ and $\FDtri(z,s)$, we find
$Q_X^1(t_i) = 
  [ t_5 (t_4+t_6) + t_4 (t_3+t_6) ] t_7
- [ t_4 (t_5+t_6) + t_5 (t_2+t_6) ] t_8$.

Reducing the right-hand side of equations like~(\ref{diffPtriRAW})
to linear combinations of master integrals, we obtain the following
set of differential equations:
\begin{eqnarray}
{\del\over\del z} \Wedge(z,s) &=& 
{ z-2\e \over z(1-z)} \Wedge(z,s)
 + {(1-2\e)(2-3\e)(1-3\e) \over \e^2 s^2\ z(1-z)} \Sset(s)
 \,,
\hskip0.4cm
\label{DiffWedge} \\
{\del\over\del z} \LBtri(z,s) &=&
 - \e {2-z \over z(1-z)} \LBtri(z,s)
                - {1-3\e \over s\ z(1-z)} \Btri(s)
 \,,
\label{DiffLBtri} \\
{\del\over\del z} \WedgeF(z,s) &=&
  - { 2 (1-2\e) - z (2-3\e) \over 2 \, z(1-z) } \WedgeF(z,s)
  + { \e \, s \over 2 (1-z) } \Zig(z,s)
\PlusBreak{ }
    {3\over2} { 1-2\e \over z(1-z) } \LBtri(z,s)
  + {3\over4} { (1-2\e)(1-3\e) \over \e s\ z(1-z) } \Btri(s)
\MinusBreak{ }
    { (1-2\e)(1-3\e)(2-3\e) \over 2 \, \e^2 s^2\ z(1-z) } \Sset(s)
 \,,
\label{DiffWedgeF} \\
{\del\over\del z} \Zig(z,s) &=&
- {3\over2} { \e \over s\ (1-z) } \WedgeF(z,s)
- { 2 + 4\e - z (2+\e) \over 2 \, z(1-z) } \Zig(z,s)
\MinusBreak{ }
  {3\over2} { 1-2\e \over s\ z(1-z) } \LBtri(z,s)
+ {3\over4} { (1-2\e)(1-3\e) \over \e s^2\ z(1-z) } \Btri(s)
\PlusBreak{ }
  {3\over2} { (1-2\e)(1-3\e)(2-3\e) \over \e^2 s^3\ z(1-z) } \Sset(s)
 \,,
\label{DiffZig} \\
{\del\over\del z} \Ptri(z,s) &=&
 { - 1 \over 3 - 2 z } \biggl[ 
{ 3 + 6 \e - 2 z \over z } \Ptri(z,s)
 + { 8 \e (1-2\e) \over s } \Ptri_2^{(6)}(z,s)
\PlusBreak{ }
  { 2 \e \over s } \Zig(z,s)
- { 2 \e \, z \over s^2\ (1-z) } \WedgeF(z,s)
\PlusBreak{ }
  { 2 (1-2\e) \, (9 - 7 z - 3 z^2) \over s^2\ z^2(1-z) } \LBtri(z,s)
\PlusBreak{ }
  { 2 \e \, (1-z) (9 - 4 z) \over s^2\ z^2 } \Wedge(z,s)
+ {2 \e \, (3-4 z) \over s^2\ z(1-z) } \Fish(s)
\PlusBreak{ }
  { (1-2\e) (1-3\e) \, (9 - 16 z + 5 z^2) \over \e \, s^3\ z^2 (1-z) }
         \Btri(s)
\MinusBreak{ }
  { (1-2\e) (1-3\e) (2-3\e) \, (9 + 26 z) \over \e^2 \, s^4\ z^2 }
         \Sset(s)
- { 12 \, \e \over s\; z } \Btie(s) \biggr]
 \,,~~~~~
\label{DiffPtriNEW} \\
{\del\over\del z} \Ptri_2^{(6)}(z,s) &=&
 { - 1 \over 3 - 2 z } \biggl[
- { s\over 2\ (1-z) } \Ptri(z,s)
\PlusBreak{ }
  { 3 (1-\e) - 2 (3-2\e) z + 2 z^2 \over z(1-z) } \Ptri_2^{(6)}(z,s)
\MinusBreak{ }
  { 1 \over 4\ (1-z)} \Zig(z,s)
- { 3 \over 4 \, s\ (1-z) } \WedgeF(z,s)
\MinusBreak{ }
  { 11 \, (1-2\e) \over 4 \, \e \, s\ z(1-z) } \LBtri(z,s)
- { 1 \over 2 \, s\ z } \Wedge(z,s)
\MinusBreak{ }
  { 3 \over 2 \, s\ z(1-z) } \Fish(s)
- { 5 \, (1-2\e) (1-3\e) \over 8 \, \e^2 \, s^2\ z(1-z) } \Btri(s)
\PlusBreak{ }
   { 2 \, (1-3\e) (2-3\e) \, (2 - \e - 2\e z) 
     \over \e^3 \, s^3\ z(1-z) } \Sset(s)
\PlusBreak{ }
  { 3 - 4\e z \over 2 \, (1-2\e)\ z(1-z) } \Btie(s) \biggr]
 \,,
\label{DiffPtri2NEW} \\
{\del\over\del z} \LPBWedge(z,s) &=& 
  { z+\e \over z(1-z)} \LPBWedge(z,s)
 - { (1-3\e)(2-3\e) \over \e s^2\ z(1-z)} \Sset(s)
 \,,
\label{DiffLPBWedge} \\
{\del\over\del z} \LPBDtri_2^{(6)}(z,s) &=&
- { 1 - 2 z \over z(1-z) } \LPBDtri_2^{(6)}(z,s)
\MinusBreak{ }
  { 1-2\e \over (1-3\e)\ s }
  \biggl( { \LPBWedge(z,s) \over z^2 } 
        - { \LPBWedge(1-z,s) \over (1-z)^2 } \biggr)
\PlusBreak{ }
  { \e \over (1-3\e) s\ z(1-z) }
    \Bigl( \Wedge(z,s) - \Wedge(1-z,s) \Bigr)
\PlusBreak{ }
  { (1-2\e)(2-3\e) \, (1 - 2 z) \over \e^2 \, s^3\ z^2 (1-z)^2 }
     \, \Sset(s)
 \,, 
\label{DiffLPBDtri2NEW} \\
{\del\over\del z} \FDtri(z,s) &=& 
- 2 { 1 - z + \e \over z (1-z) } \FDtri(z,s) 
- 2 { \e \over s\ z (1-z) } \Zig(z,s)
\PlusBreak{ }
  {  6 \, (1 - 2 \e) \over s^2\ z^2 (1-z) } \biggl[
      z \, \LPBWedge(1-z,s) - \LBtri(z,s) 
\MinusBreak{ {  6 \, (1 - 2 \e) \over s^2\ z^2 (1-z) } \bigglB z }
    { 1 - 3 \e \over 2 \e \, s } \Btri(s) \biggr]
 \,.
\label{DiffFDtri}
\end{eqnarray}
%
%
In eqs.~(\ref{DiffPtriNEW}), (\ref{DiffPtri2NEW}) and (\ref{DiffLPBDtri2NEW}), 
we have performed a change of basis: we exchange the $(4-2\e)$-dimensional
integral $\Ptri_2(z,s) \equiv \Ptri_2(z,s)|_{D=4-2\e}$ 
for its $(6-2\e)$-dimensional version, 
$\Ptri_2^{(6)}(z,s) \equiv \Ptri_2(z,s)|_{D=6-2\e}$,
and similarly for $\LPBDtri_2(z,s)$.  
All of the integrals in this new basis are well-defined using ordinary 
dimensional regularization, following the same analysis as 
in~\sect{TwoLoopLightConeIntegrals}.  The relations required to 
change the basis are examples of the dimension-shifting relations
discussed previously,
\begin{eqnarray}
&&\Ptri_2^{(6)}(z,s) = 
  { 3(1-z) + \e  (9-8z) \over 2 \, (1-2\e) (1+2\e) \, z (1-z) }
      \Btie(s)
\MinusBreak{ }
  { (1-3\e)(2-3\e)  \over 8 \, \e^3 (1+\e) (1+2\e)^2 \, s^3 \, z^2 (1-z) }
   \Bigl[ 9-47z+38z^2 + \e  (72-373z+270z^2) 
\PlusBreak{ (1-3\e) }
   \e^2 (171-964z+652z^2)
    + 2 \e^3 (63-379z+246z^2) \Bigr]
      \Sset(s)
\MinusBreak{ }
 { 1-3\e \over 16 \, \e^2 (1+\e) (1+2\e)^2 \, s^2 \, z^2 (1-z) }
  \Bigl[ 54-92z+34z^2 + \e  (7-8z)(54-31z) 
\PlusBreak{ (1-3\e) }
      \e^2 (828-1459z+578z^2)
    + 8 \e^3 (72-130z+53z^2) \Bigr]
      \Btri(s)
\MinusBreak{ }
 { 3-4z + \e (9-10z) \over 4 \, (1-2\e) (1+2\e) \, s \, z (1-z) }
      \Fish(s)
\PlusBreak{ }
  { 1-z \over 4 \, (1-2\e) (1+\e) (1+2\e) \, s \, z^2 }
   \Bigl[ 9-8z + 6 \e (9-7z) + \e^2 (63-46z) \Bigr]
      \Wedge(z,s)
\MinusBreak{ }
  { 1  \over 8 \, \e (1+\e) (1+2\e) \, s \, z^2 (1-z) }
  \Bigl[ 2 (27-37z+9z^2) + 3 \e (90-125z+36z^2) 
\PlusBreak{ (1-3\e) }
  \e^2 (288-403z+126z^2) \Bigr]
      \LBtri(z,s)
\PlusBreak{ }
  { 1  \over 8 \, (1-2\e) (1+\e) (1+2\e) \, s \, (1-z)  }
   \Bigl[ 2z - 3 \e (9-8z) - \e^2 (45-34z) \Bigr]
      \WedgeF(z,s)
\MinusBreak{ }
  { 1 \over 8 \, (1-2\e) (1+\e) (1+2\e) \, z (1-z) }
  \Bigl[ 2(9-11z+3z^2) + \e (102-125z+36z^2)
\PlusBreak{ (1-3\e) }
  \e^2 (120-145z+42z^2) \Bigr]
      \Zig(z,s)
\MinusBreak{ }
  { s \, ( 3-2z + \e (9-8z) ) \over 8 \, \e (1-2\e) \, z (1-z) }
      \Ptri(z,s)
- { s^2 \, (1+\e) \, (3-2z) \over 8 \, \e (1-2\e) (1+2\e) \, (1-z) }
      \Ptri_2(z,s)
 \,,
\label{Ptri26shift}
\end{eqnarray}
and
\begin{eqnarray}
&&\LPBDtri_2^{(6)}(z,s) = 
{ (1-2\e)(2-3\e) \over 4 \, \e^3 (1+2\e)^2 \, s^3 \,  z^2 (1-z)^2 } 
\TimesBreak{ (1-2\e)(1-2\e) }
 \Bigl[ 5z(1-z) - 2 \e (1 - 12z(1-z)) - 4 \e^2 (1 - 8z(1-z)) \Bigr]
     \Sset(s)
\MinusBreak{ }
 { z - 2\e(1-2z) \over 2\, (1+2\e) (1-3\e) \, s \, z^2 }
     \Wedge(z,s)
- { 1-z + 2\e(1-2z) \over 2 \, (1+2\e) (1-3\e) \, s \, (1-z)^2 }
     \Wedge(1-z,s)
\MinusBreak{ }
 { 1-2\e \over \e (1-3\e) \, s \, z } 
     \LPBWedge(z,s)
- { 1-2\e \over \e (1-3\e) \, s \, (1-z) } 
     \LPBWedge(1-z,s)
\MinusBreak{ }
 {s^2\over6}  { 1+\e \over (1-3\e)(1+2\e)(1+3\e) }
     \LPBDtri_2(z,s)
 \,.
\label{LPBDtri26shift}
\end{eqnarray}

We solve the differential equations~(\ref{DiffWedge})--(\ref{DiffFDtri}) 
as Laurent expansions in $\e$, beginning with the simplest cases, 
such as \eqn{DiffWedge}, which only depends on $\Wedge(z,s)$ 
and the previously known integral $\Sset(s)$.  
In a given differential equation (or a coupled pair
of equations, in the case of \eqns{DiffPtriNEW}{DiffPtri2NEW}),
we start with the most singular terms in the Laurent expansions,
and proceed until we reach the order required for the expansion of
the splitting amplitudes through $\Ord(\e^0)$.  This order corresponds
to transcendental weight 4, where each $\ln(x)$ or $\pi$ has weight 1,
and $\Li_n(x)$ and $\zeta_n$ ($\zeta_n \equiv \Li_n(1)$) have weight $n$.  
We insert into the differential equation an ansatz which is a linear
combination of such functions, where $x$ can be $z$, $1-z$, or $-(1-z)/z$.
We adjust the coefficients in the linear combination until the
differential equation is satisfied.  The constants of integration were
determined by different techniques.  In some cases the integrals could be
performed analytically, at least for one value of $z$.  It is also
possible to evaluate the Schwinger parameter integrals numerically by
Monte Carlo integration.  To fix the constant, it is often sufficient to
check that for a certain limiting value of $z$, $z=0$ or $z=1$, an
integral remains finite.

We first give the Laurent expansions for the previously known master
integrals, which constitute the inhomogeneous terms for the first set
of non-trivial differential equations:
\begin{eqnarray}
\Btie(s) &=&
 - \biggl[ (-s)^{-1-\e}
          { \Gamma(1+\e) \Gamma^2(-\e) \over \Gamma(1-2\e) } \biggr]^2
\nonumber\\
&=&
 { (-s)^{-2\e} \over s^2 } \biggl[ 
  - { 1\over \e^4 } + { \zeta_2\over \e^2 }
  + { 14 \over 3 } { \zeta_3\over \e } + { 21 \over 4 } \zeta_4 \biggr]
   + \Ord(\e)  \,,
\label{BtieLaurent}\\
\Sset(s) &=&
 (-s)^{1-2\e}
  { \Gamma(-1+2\e) \Gamma^3(1-\e) \over \Gamma(3-3\e) }
\nonumber\\
&=&
 { (-s)^{-2\e} \, s \over (1-2\e) (1-3\e) (2-3\e) } 
\biggl[ { 1 \over 2 \e } - {\zeta_2\over2} \e 
       - { 16 \over 3 } \zeta_3 \e^2 - { 57 \over 8 } \zeta_4 \e^3 
\biggr]
+ \Ord(\e^4)  \,,
\label{SsetLaurent}\\
\Btri(s) &=&
 - (-s)^{-2\e}
  { \Gamma(\e) \Gamma(2\e) \Gamma^2(1-\e) \Gamma^2(1-2\e) 
     \over \Gamma(2-2\e) \Gamma(2-3\e) }
\nonumber\\
&=&
 { (-s)^{-2\e} \over (1-2\e) (1-3\e) } 
\biggl[ - { 1 \over 2 \e^2 } - {\zeta_2\over2} 
       + { 13 \over 3 } \zeta_3 \e + { 41 \over 8 } \zeta_4 \e^2 
\biggr]
+ \Ord(\e^3) \,.
\label{BtriLaurent}
\end{eqnarray}
For the remaining integrals we just give the Laurent expansions, 
omitting an overall factor of $\exp(-2\gamma\e)$.  
The $z$-independent integral $\Fish(s)$ can be performed analytically, 
and the result is
\begin{equation}
\Fish(s) =
 - { (-s)^{-2\e} \over s } \biggl[ 
  { \zeta_2\over 2 \, \e^2 } + { 7 \over 2 }{ \zeta_3\over \e }
  + { 55 \over 4 } \zeta_4 \biggr]
 \,.
\label{FishLaurent}
\end{equation}
The remaining, $z$-dependent Laurent expansions are
\begin{eqnarray}
\Wedge(z,s) &=&
 { (-s)^{-2\e} \over s \, (1-z) } \biggl[ 
  { 1\over 2 \, \e^3 } - { \zeta_2\over 2 \, \e }
  - { 16 \over 3 } \zeta_3 \biggr] 
 \biggl[ \ln(z) 
 + 2 \, \e   \, \Li_2\Bigl(-{1-z \over z}\Bigr)
\PlusBreak{ ~~~~~~~~~~~~~~~~~ }
4 \, \e^2 \, \Li_3\Bigl(-{1-z \over z}\Bigr)
 + 8 \, \e^3 \, \Li_4\Bigl(-{1-z \over z}\Bigr) \biggr]
\,,
\label{WedgeLaurent}\\
\LBtri(z,s) &=&
 { (-s)^{-2\e} \over s \, (1-2\e) } 
\Biggl\{ { 1 \over 2 \e^3 } - { \ln z \over 2 \, \e^2 } 
       + { 1 \over 2 \, \e } \Bigl[ \Li_2(1-z) + \ln^2 z + \zeta_2 \Bigr]
\PlusBreak{ }
        \Li_3(z) + {1\over2} \Li_3(1-z)
\PlusBreak{ }
        {1\over2} \ln^2 z \ln(1-z)
      - {1\over3} \ln^3 z - {3\over2} \zeta_2 \ln z - {16\over3} \zeta_3
\PlusBreak{ }
    \e \biggl[ - \Li_4(z) - {1\over2} \Li_4(1-z) 
               - \Li_4\Bigl(-{1-z \over z}\Bigr)
\MinusBreak{ + \e \bigglB }
             { \zeta_2 \over 2 } 
             \Bigl(  \Li_2(z) + \ln z \ln(1-z) - 2 \ln^2 z \Bigr)
\PlusBreak{ + \e \bigglB }
             {1\over8} \ln^4 z - {1\over6} \ln^3 z \ln(1-z) 
           + {16\over3} \zeta_3 \ln z - {23\over8} \zeta_4 \biggr]
\Biggr\}
 \,,
\label{LBtriLaurent}\\
\WedgeF(z,s) &=&
 - { (-s)^{-2\e} \over s \, z } \Biggl\{ 
   {1 \over 2 \, \e^2} \Bigl( \Li_2(1-z) - \zeta_2 \Bigr)
\PlusBreak{ }
   {1 \over \e} \biggl[ - 3 \Li_3(z) 
                     + {1\over2} \Bigl( \Li_3(1-z) - \zeta_3 \Bigr)
           - 2 \Bigl( \Li_2(1-z) - \zeta_2 \Bigr) \ln z
\MinusBreak{ + {1 \over \e} \bigglB }
           {3\over2} \ln^2 z \ln(1-z) \biggr]
\MinusBreak{ }
    11 \Li_4(z) - {5\over2} \Li_4(1-z) - 3 \Li_4\Bigl(-{1-z \over z}\Bigr)
\PlusBreak{ }
  4 \, \ln z \Bigl( \Li_3(1-z) - \zeta_3 + 2 \, \Li_3(z) \Bigr) 
\PlusBreak{ }
  \Li_2(z) \biggl[ {3\over2} \Bigl( \Li_2(z) - \zeta_2 \Bigr) - 2 \ln^2 z 
                   + 3 \ln z \ln(1-z) \biggr]
\MinusBreak{ }
    {1\over8} \ln^4 z + {1\over6} \ln^3 z \ln(1-z) 
  + {3\over2} \ln^2 z \ln^2(1-z)
\MinusBreak{ }
    {3\over2} \zeta_2 \Bigl( \ln^2 z + \ln z \ln(1-z) \Bigr)
  - {11\over4} \zeta_4 \Biggr\}
\,,
\label{WedgeFLaurent}\\
\Zig(z,s) &=&
 { (-s)^{-2\e} \over s^2 \, z } \Biggl\{ 
 - { 1 \over 4 \, \e^4 }
 + { \ln z \over 2 \, \e^3 }
 - { 1 \over 2 \, \e^2 } \Bigl( \ln^2 z + {7\over2} \zeta_2 \Bigr)
\PlusBreak{ }
  {1 \over \e} \biggl[ {1\over3} \ln^3 z 
                     + {7\over2} \zeta_2 \ln z 
                     - {7\over3} \zeta_3 \biggr]
\PlusBreak{ }
    6 \Li_4(z) - 6 \Li_4(1-z) - 6 \Li_4\Bigl(-{1-z \over z}\Bigr)
  + 6 \Li_3(1-z) \ln z
\PlusBreak{ }
    3 \Li_2(z) \biggl[ {1\over2} \Li_2(z) + \ln z \ln(1-z) + \zeta_2 \biggr]
  - {5\over12} \ln^4 z + \ln^3 z \ln(1-z) 
\PlusBreak{ }
    {3\over2} \ln^2 z \ln^2(1-z)
  - \zeta_2 \Bigl( {13\over2} \ln^2 z - 3 \ln z \ln(1-z) \Bigr)
\MinusBreak{ }
    {4\over3} \zeta_3 \ln z
  - {487\over16} \zeta_4 \Biggr\}
\,,
\label{ZigLaurent}\\
\Ptri(z,s) &=&
 - { (-s)^{-2\e} \over s^3 \, z } \Biggl\{ 
   { 1 \over \e^4 }
 - 2 { \ln z \over \e^3 }
 + { 1 \over \e^2 } \Bigl( \Li_2(1-z) + 2 \ln^2 z + {3\over2} \zeta_2 \Bigr)
\PlusBreak{ }
   {1 \over \e} \biggl[ 8 \Li_3(z) - 3 \ln z \Li_2(z) 
\MinusBreak{ + {1 \over \e} \bigglB }
             {4\over3} \ln^3 z + \ln^2 z \ln(1-z) 
           - 5 \zeta_2 \ln z - {25\over6} \zeta_3 \biggr]
\MinusBreak{ }
    14 \Li_4(z) - 8 \Li_4(1-z) - 8 \Li_4\Bigl(-{1-z \over z}\Bigr)
  - \Li_3(1-z) \ln z
\PlusBreak{ }
   \Li_2(z) \biggl[ - {5\over2} \Li_2(z) - 5 \ln z \ln(1-z)
                   + 3 \ln^2 z + 6 \zeta_2 \biggr]
\PlusBreak{ }
   {1\over3} \ln^4 z + {2\over3} \ln^3 z \ln(1-z) 
  - {5\over2} \ln^2 z \ln^2(1-z)
\PlusBreak{ }
   \zeta_2 \Bigl( \ln^2 z + 6 \ln z \ln(1-z) \Bigr)
  + {28\over3} \zeta_3 \ln z
  + {5\over2} \zeta_4 \Biggr\}
\,,
\label{PtriLaurent}\\
\Ptri_2^{(6)}(z,s) &=&
 { 1 \over s^2 } \Biggl\{
   { 1 \over z }  \biggl[
   4 \Li_4(z) - 3 \Li_4(1-z) - 4 \Li_4\Bigl(-{1-z \over z}\Bigr) 
\PlusBreak{ { 1 \over s^2 } \BigglBl { 1 \over z } \bigglB }
    4 \Bigl( \Li_3(1-z) - \zeta_3 \Bigr) \ln z 
 + \Li_2(z)  \Bigl( \Li_2(z) + 2 \ln z \ln(1-z) \Bigr)
\MinusBreak{ { 1 \over s^2 } \BigglBl { 1 \over z } \bigglB }
   \ln^2 z  \Bigl( {1\over6} \ln^2 z - {2\over3} \ln z \ln(1-z) 
                 - \ln^2(1-z) + 2 \zeta_2 \Bigr)
 - 4 \zeta_4 \biggr]
\PlusBreak{ { 1 \over s^2 } \BigglBl }
  { 1 \over 1-z } \biggl[
   6 \Li_4(z) - 2 \ln z \Bigl( \Li_3(z) - \zeta_3 \Bigr)
 + \Li_3(1-z) \ln z 
\PlusBreak{ { 1 \over s^2 } \BigglBl + { 1 \over 1-z } \bigglB }
   {1\over2} \Li_2(z) 
     \Bigl( \Li_2(z) + 2 \ln z \ln(1-z) - 2 \zeta_2 \Bigr)
\PlusBreak{ { 1 \over s^2 } \BigglBl + { 1 \over 1-z } \bigglB }
   {1\over2} \ln^2 z \ln^2(1-z) - \zeta_2 \ln z \ln(1-z) 
 - {19 \over 4} \zeta_4 \biggr]
\Biggr\}
 \,,
\label{Ptri26Laurent}\\
\LPBWedge(z,s) &=&
 { (-s)^{-2\e} \over 2 \, s \, (1-z) (1-2\e) } \Biggl\{ 
   { 1 \over \e^3 }
 - { \ln(1-z) \over \e^2 }
 + { 1 \over \e } \Bigl( \Li_2(z) + {1\over2} \ln^2(1-z) - \zeta_2 \Bigr)
\PlusBreak{ }
     \Li_3(z) + \Li_3(1-z) + {1\over2} \ln z \ln^2(1-z) 
   - {1\over6} \ln^3(1-z) - {35\over3} \zeta_3
\PlusBreak{ }
    \e \biggl[ 
      \Li_4\Bigl(-{1-z \over z}\Bigr)
    - \zeta_2 \Bigl( \Li_2(z) + \ln z \ln(1-z) - {1\over2} \ln^2 z \Bigr)
\PlusBreak{ + \e \bigglB }
     {1\over24} ( \ln z - \ln(1-z) )^4 + {32\over3} \zeta_3 \ln(1-z) 
    - {25\over2} \zeta_4 \biggr] \Biggr\}
\,,
\label{LPBWedgeLaurent}\\
\LPBDtri_2^{(6)}(z,s) &=&
 { (-s)^{-2\e} \over 2 \, s^2 \, z(1-z) (1-3\e) } 
\Biggl\{ 
   { 1 \over \e } \biggl[ \ln z \Li_2(z) + \ln(1-z) \Li_2(1-z) 
\PlusBreak{ {(-s)^{-2\e}\over2\,s^2\,z(1-z)(1-3\e)}{1\over\e} \bigglB }
    {1\over2} \ln z \ln(1-z) \Bigl( \ln z + \ln(1-z) \Bigr) \biggr]
\PlusBreak{ }
    3 \biggl[ (\ln z + \ln(1-z)) 
         \Bigl( \Li_3(z) + \Li_3(1-z) - \zeta_3 \Bigr)
\MinusBreak{ + 3 \bigglB }
        \Li_2(z) \Li_2(1-z)
     - {1\over2} \Bigl( \ln^2 z \Li_2(z) 
                      + \ln^2(1-z) \Li_2(1-z) \Bigr)
\MinusBreak{ + 3 \bigglB }
      {1\over6} \ln z \ln(1-z) 
                 \Bigl( \ln^2 z + \ln^2(1-z)
                     - {9\over2} \ln z \ln(1-z) + 12 \zeta_2 \Bigr)
    \biggr] \Biggr\}
\,,
\nonumber\\&&\hskip1.0cm
\label{LPBDtri26Laurent}\\
\FDtri(z,s) &=&
 { (-s)^{-2\e} \over s^3 \, z^2 } 
\Biggl\{ 
   { 1 \over \e^4 } 
 - 2 { \ln z \over \e^3 } 
 + { 1 \over \e^2 } \Bigl( 3 \Li_2(1-z) + 2 \ln^2 z - {\zeta_2 \over 2} \Bigr)
\PlusBreak{ }
    { 1 \over \e } \biggl[ 
   3 \Li_3(z) + 6 \Li_3(1-z) + 3 \ln z \Li_2(z) 
\MinusBreak{ + { 1 \over \e } \bigglB }
    {4\over3} \ln^3 z + {9\over2} \ln^2 z \ln(1-z) 
  - 5 \zeta_2 \ln z - {61\over6} \zeta_3 \biggr]
\PlusBreak{ }
     6 \Li_4(z) + 3 \Li_4(1-z) - 9 \Li_4\Bigl(-{1-z \over z}\Bigr)
\MinusBreak{ }
    3 \ln z \Bigl( \Li_3(z) + \Li_3(1-z) \Bigr)
\PlusBreak{ }
   {3\over2} \Li_2(z) \Bigl( \Li_2(z) - \ln^2 z + 2 \ln z \ln(1-z) \Bigr) 
 + {7\over24} \ln^4 z 
\MinusBreak{ }
   2 \ln^3 z \ln(1-z) 
 + {3\over2} \ln^2 z \ln^2(1-z)
 + {\zeta_2\over2} \ln z^2 + {34\over3} \zeta_3 \ln z - {41\over4} \zeta_4
    \Biggr\}
\,.
\nonumber\\&&\hskip1.0cm
\label{FDtriLaurent}
\end{eqnarray}

The differential equations~(\ref{DiffWedge})--(\ref{DiffFDtri})
are valid to all orders in $\e$.   Whereas
eqs.~(\ref{WedgeLaurent})--(\ref{FDtriLaurent}) only give the solutions
through transcendental weight 4, we have obtained the solutions through 
weight 5, up to integration constants.  Beyond weight 4, the 
ordinary polylogarithms $\Li_n$ are insufficient to describe the 
solution space.  Instead, one can use the harmonic polylogarithms 
(HPLs)~\cite{RV,Lorentz}, denoted by ${\rm H}(\vec{m}_w;z)$.
For transcendental weight $w$ the vector $\vec{m}_w$ is a string of 
$w$ entries, which (for our application) can only take on the values
$0$ or $+1$.  The HPLs obey the differential equations,
\begin{equation}
{d \over dz} {\rm H}(\vec{m}_w;z) = f(a;z) {\rm H}(\vec{m}_{w-1};z) \,,
\label{HPLdiffeq}
\end{equation}
where $a = m_w$ is the leftmost component of $\vec{m}_w$,
$\vec{m}_{w-1}$ is obtained from $\vec{m}_w$ by omitting that component,
and 
\begin{equation}
f(0;z) = {1\over z} \,, \qquad f(+1;z) = { 1 \over 1 - z } \,.
\label{fadef}
\end{equation}
These $2^w$ HPLs suffice because the only true singularities in $z$
on the right-hand side of the differential equations are those 
given in \eqn{fadef}.  Actually, the differential 
equations~(\ref{DiffPtriNEW}) and (\ref{DiffPtri2NEW}) for $\Ptri(z,s)$
and $\Ptri_2^{(6)}(z,s)$ contain factors of $1/(3-2z)$ on the right-hand side.
However, these factors are artifacts of the change of 
basis~(\ref{Ptri26shift}) from $\Ptri_2(z,s)$ to $\Ptri_2^{(6)}(z,s)$.

The 32 HPLs at $w=5$ can be written in terms of ordinary logarithms and
polylogarithms, plus three more functions.  Although a few of the
$z$-dependent master integrals ($\Wedge$, $\LPBWedge$, and
$\LPBDtri_2^{(6)}$) do not require the three additional functions
at the $w=5$ level of their expansions, the generic master integral
requires all of them.


\section{Splitting Amplitude Results}
\label{ResultSection}

With the integrands obtained as described in
\sect{SplittingIntegrandsSection}, and using the integrals obtained in
the previous section, we can express the results for the splitting
amplitudes in a variety of gauge theories.  We will present results
for QCD as well for $\Neqfour$ and $\Neqone$ super-Yang-Mills theory. 
For QCD, both the integrands and the combination of master integrals 
are too lengthy to present here.  We will present only the results
expanded in $\e$ through $\Ord(\e^0)$. 

The corresponding expressions for the $\Neqfour$ maximally supersymmetric
Yang-Mills (MSYM) theory are relatively simple, and we shall present
them in detail in \sect{Neq4MasterDecompExample}.  We also give
the results expanded through $\Ord(\e^0)$, which we presented
previously~\cite{TwoloopN4}.
The MSYM results also serve as useful building blocks for the 
splitting amplitudes in $\Neqone$ super-Yang-Mills theory.  
In \sect{Neq1Results}
we give the $\Neqone$ results in terms of master integrals, and 
expanded in $\e$.  The $\Neqone$ results in turn form useful 
building blocks for representing the QCD results in \sect{QCDResults}.

In the numerator algebra for the loop-momentum polynomials, the 
dimensionality of the metric $\eta_{\mu\nu}$ appears, 
\begin{equation}
\eta^\mu_\mu \equiv D_s \equiv 4 - 2 \e \delta_R .
\label{Dsdef}
\end{equation}
As mentioned in \sect{SplittingSection},
setting $\delta_R=1$ defines the HV scheme,
which has $2$ physical states for external gluons, but 
$2-2\e$ physical states for internal gluons.
Setting $\delta_R=0$ defines the FDH scheme, which has $2$ physical states 
for both internal and external gluons, matching the number
of fermionic states, and which preserves 
supersymmetry~\cite{BKgggg,TwoloopSUSY}. 
We therefore present the splitting amplitude results for
supersymmetric theories in the FDH scheme.

The QCD results are given in both the HV and FDH schemes.  The
CDR scheme is often used in computations of unpolarized cross sections,
or amplitude interferences.  It has $2-2\e$ physical states for external 
as well as internal gluons. The HV-scheme results could be converted to 
the CDR scheme by including splitting amplitudes for gluons carrying 
epsilonic helicities.
In practice, the CDR results for the scattering amplitude 
interferences required for unpolarized cross sections agree, through
two loops, with 
the sum over helicities of the HV results, after the infrared 
singularities are removed using the Catani formula\cite{BDDgggg,BDDqqgg}. 
The tree and one-loop amplitudes in the Catani formula
are different in the two schemes, but the finite remainders agree.
Thus the collinear limits of the finite remainders in the CDR scheme
should be obtainable from the HV-scheme results in \sect{CataniComparison}.

The results we will present in this section are for bare
(`unrenormalized') splitting amplitudes. We will discuss their
renormalization in \sect{CataniComparison}, where we also present a
discussion of the collinear behavior of remainders after subtraction
of infrared divergences.

\subsection{$\Neqfour$ Super-Yang-Mills Theory Results}
\label{Neq4MasterDecompExample}

In super-Yang-Mills theory, scattering amplitudes
are heavily constrained by supersymmetry Ward 
identities (SWI)~\cite{SWI,Neq4SWI}, and this has implications
for the splitting amplitudes.
First of all, amplitudes with only 0 or 1 negative-helicity gluon 
vanish at any loop order $L$ in any supersymmetric theory~\cite{SWI},
\begin{equation}
  A_n^{(L), \, {\rm SUSY}}(1^\pm,2^+,,\ldots,n^+) = 0 \,.
\label{Neq1vanishn}
\end{equation}
Next consider the color-ordered amplitudes with two negative-helicity
gluons, 
\begin{equation}
  A_{n}^{(L), \, {\rm SUSY}}(1^+,2^+,\ldots,i^-,\ldots,j^-,\ldots,n^+)\,,
\label{Neq4MHVn}
\end{equation}
known as maximally helicity-violating (MHV) amplitudes.
In MSYM, the $\Neqfour$ SWI imply that \eqn{Neq4MHVn} is completely 
independent of the cyclic position of $i$ and $j$, up to a trivial
overall spinor-product factor of ${\spa{i}.{j}}^4$~\cite{Neq4SWI}.
In $\Neqone$ super-Yang-Mills theory, this relation holds at tree level,
but is violated at one loop.

What are the implications for supersymmetric splitting amplitudes?
First consider the vanishing amplitude with $1^-$ in \eqn{Neq1vanishn}.
Let two cyclicly-adjacent legs $a$ and $b$ become collinear,
so that the amplitude factorizes on the MHV amplitudes,
\begin{equation}
 \sum_{l=0}^L \Split_{+}^{(l), \, {\rm SUSY}}(z;a^+,b^+) 
      \times A_{n-1}^{(L-l)}(1^-,2^+,\ldots,P^-,\ldots,n^+) \,.
\label{SUSYhelflipvanish0}
\end{equation}
Since the amplitudes $A_{n-1}^{(L-l)}$ in are non-vanishing,
the helicity-flip splitting amplitudes for $P^- \to a^+ b^+$
must vanish to all orders in a supersymmetric theory.
Using parity, we have
\begin{equation}
\Split_{-}^{(L), \, {\rm SUSY}}(z;a^-,b^-) = 0 \,,
\label{SUSYhelflipvanish}
\end{equation}
for all $L$.  This result includes the tree-level 
vanishing~(\ref{TreeSplitAmplitudemm}), since tree-level $n$-gluon 
amplitudes are effectively ($\Neqfour$) supersymmetric.

Due to parity, we need to present results below only for the cases
where $P$ has positive helicity,
$P^+ \to a^{\lambda_a} b^{\lambda_b}$, which we shall denote by
`$\lambda_a \lambda_b$'.  Two of these four cases, $-+$ and $+-$,
are related to each other by exchanging legs $a$ and $b$, which
also exchanges $z \lr 1-z$.  As just noted, the $--$ case vanishes 
in any supersymmetric theory.
The two independent non-vanishing supersymmetric splitting amplitudes 
are for $P^+ \to a^+ b^+$ ($++$) and $P^+ \to a^- b^+$ ($-+$).
Since these two tree-level splitting amplitudes are nonzero, 
we define loop ratios $\rsn^{(L)}$ as
in \eqns{OnelooprS}{TwolooprS}, or for $L$ loops,
\begin{equation}
 \Split_{-\lambda}^{(L)}(a^{\lambda_a},b^{\lambda_b})
 = \rsn^{(L) \, \lambda_a,\lambda_b}(z,s_{ab}) \times 
      \Split_{-\lambda}^\tree(a^{\lambda_a},b^{\lambda_b}) \,.
\label{LLooprDef}
\end{equation}

In $\Neqfour$ supersymmetry, the $++$ and $-+$ cases are related to each 
other by the fact that the expression~(\ref{Neq4MHVn}) 
divided by ${\spa{i}.{j}}^4$ is independent of $i$ and $j$.  
Thus the collinear limit $P^- \to i^- (i+1)^+$ is essentially the same 
as that of $P^+ \to 1^+ 2^+$, up to overall spinor-product factors which are 
the same as at tree level.  In other words, a universal $\rsn$
function describes both cases:
\begin{equation}
 \rsn^{(L)\,++, \, \Neqfour}(z,s) = \rsn^{(L)\, -+, \, \Neqfour}(z,s) 
 \equiv \rsn^{(L),\, \Neqfour}(z,s) \,.
\label{rSNeq4universal}
\end{equation}
Using the Bose symmetry relation~(\ref{BoseRelation}) for $P^+ \to a^+ b^+$,
the universal function $\rsn^{(L),\, \Neqfour}(z,s)$ must be symmetric,
\begin{equation}
 \rsn^{(L),\, \Neqfour}(z,s) = \rsn^{(L),\, \Neqfour}(1-z,s) \,.
\label{rSNeq4symmetric}
\end{equation}
This symmetry is not manifest in the one-loop
expression~(\ref{OneloopSplitExplicitNeq4}), but it is easily 
demonstrated using polylogarithm identities.

After carrying out the unitarity-based sewing procedure for $\Neqfour$
super-Yang-Mills theory, the integral required for the two-loop 
$g\to gg$ splitting amplitude
can be written as the sum of four terms, corresponding
to the $L$, $D$, $LPB$ and $F$ integral topologies,
plus the same four terms with $z \to 1-z$:
\begin{equation}
\rsn^{(2),\, \Neqfour}(z,s) = I^{\Neqfour}_A(z) + I^{\Neqfour}_A(1-z) \,, 
\label{Neq4flipsplit}
\end{equation}
where
\begin{equation}
I^{\Neqfour}_A(z) = I_L + I_D + I_{LPB} + I_F,
\label{Neq4LLPBFsplit}
\end{equation}
and
\begin{eqnarray}
I_L &=& - { s^2 \over 8 }
\int {d^{D}p\over \pi^{D/2}} \, {d^{D}q\over \pi^{D/2}} \, 
\biggl[ s z + (1-z) ( p_1^2 - p_2^2 ) + p_8^2 
      + { p_2^2 ( s (1-z) + p_5^2 ) \over p_9^2 }  \biggr]
 \prod_{i=1}^7 { 1 \over p_i^2 } \,,~~~~~~~~
\label{ILdef} \\
I_D &=& { s^2 \over 8 }
\int {d^{D}p\over \pi^{D/2}} \, {d^{D}q\over \pi^{D/2}} \, 
{ s z + p_5^2 \over p_9^2 }
 \prod_{i=2}^7 { 1 \over p_i^2 } \,,
\label{IDdef} \\
I_{LPB} &=& - { s^2 \over 8 } z(1-z) 
\int {d^{D}p\over \pi^{D/2}} \, {d^{D}q\over \pi^{D/2}} \, 
 p_1^2
 \, \prod_{i=2}^8 { 1 \over p_i^2 } \,,
\label{ILPBdef} \\
I_F &=& { s^2 \over 8 } z
\int {d^{D}p\over \pi^{D/2}} \, {d^{D}q\over \pi^{D/2}} \, 
 ( s z + (1-z) p_1^2 )
 \, \prod_{i=2}^8 { 1 \over p_i^2 } \,.
\label{IFdef}
\end{eqnarray}
The labeling of the propagators $p_i^2$ is given in
\fig{PlanarTriFigure} and, in the case of the $D$ and $F$ topologies, 
\fig{ProjFigure}.  Note that the $z \lr 1-z$ 
symmetry~(\ref{rSNeq4symmetric}) is manifest in \eqn{Neq4flipsplit}.
The result~(\ref{Neq4flipsplit}) is given in the FDH scheme, $\delta_R=0$.

We can see explicitly from these expressions how the appearance of 
the light-cone propagators is restricted by unitarity.  In \eqn{ILdef},
whenever $p_9^2$ appears in the denominator, $p_2^2$ appears in the
numerator.  The appearance of $p_2^2$ eliminates any 
two-particle cut (through lines 1 and 2) or any three-particle cut
(through lines 2, 3 and 6) entirely to the right of propagator 9,
in the terms containing that light-cone denominator.
Exactly the same argument (after flipping the $L$ integral over by
letting $k_1 \lr k_2$) explains why $p_1^2$ cannot appear 
in the denominator of the $D$ integrand in \eqn{IDdef} along
with $p_9^2$. Finally, in \eqns{ILPBdef}{IFdef}, $p_1^2$ never 
appears in the denominator. If it had, there would have been a two
particle cut (through lines 1 and 2) entirely to the right of 
propagator 8, which is again forbidden by unitarity.

The result of reducing \eqn{Neq4LLPBFsplit} to master integrals is
\begin{eqnarray}
I^{\Neqfour}_A(z)  &=&
- { 1 \over 16 } \Biggl\{ 
   { s \over 3 - 2 z } \biggl[  
   4 s \, z (1-z) \, \Bigl( s\, \Ptri(z,s)
                          + 2 (1-2\e) \, \Ptri_2^{(6)}(z,s) \Bigr)
\MinusBreak{ -{1\over16}\BigglBl {s\over3-2z} \bigglB }  
   {2\over3} z (3-z) \Bigl( s \, \Zig(z,s) + 3 \, \WedgeF(z,s) \Bigr)
\MinusBreak{ -{1\over16}\BigglBl {s\over3-2z} \bigglB }  
   2 { 1-2\e \over \e }  \, (7 - z) \, \LBtri(z,s)
\MinusBreak{ -{1\over16}\BigglBl {s\over3-2z} \bigglB }  
   4 (1-z) (5 - 3 z) \, \Wedge(z,s)
 - 4 z \, \Fish(s)
\MinusBreak{ -{1\over16}\BigglBl {s\over3-2z} \bigglB }  
    { (1-2\e) (1-3\e) (2-3\e) (5 - 14 z) \over \e^3 \, s^2 } \, \Sset(s)
\MinusBreak{ -{1\over16}\BigglBl {s\over3-2z} \bigglB }  
    { (1-2\e) (1-3\e) \, (16 - 9 z) \over \e^2 \, s }  \, \Btri(s)
- 2 s (3 - 4 z)  \, \Btie(s) \biggr]
\PlusBreak{ -{1\over16}\BigglBl }  
    2 s { 1-2\e \over \e } \Bigl[ (1-z) \LPBWedge(z,s)
                      - z \LPBWedge(1-z,s) \Bigr]
\MinusBreak{ -{1\over16}\BigglBl }  
    4 s^2 z
    \biggl[ {1\over3} s z \FDtri(z,s) 
         + (1-3\e) (1-z) \LPBDtri_2^{(6)}(z,s) \biggr] \Biggl\}
\,.
\label{ZDSplitIntegrals}
\end{eqnarray}

Inserting the Laurent expansions~(\ref{BtieLaurent})--(\ref{FDtriLaurent})
into \eqn{ZDSplitIntegrals}, and adding the terms with $z \to 1-z$, we obtain,
\begin{eqnarray}
&&\rsn^{(2),\, \Neqfour}(z,s) \nonumber \\
&&\hskip-0.2cm =
  { 1 \over 4 } \biggl( { \mu^2 \over -s } \biggr)^{2\e} \Biggl\{
  { 1 \over 2 \, \e^4 } 
- { 1 \over \e^3 } \Bigl( \ln(z) + \ln(1-z) \Bigr)
+ { 1 \over \e^2 } \biggl( \ln^2(z) + \ln^2(1-z) + \zeta_2 \biggr)
\PlusBreak{{1\over4}\bigglP{\mu^2\over-s}\biggrP^{2\e}\BigglBl}
  { 1 \over \e } \biggl[ 2 \Bigl( \Li_3(z) + \Li_3(1-z) \Bigr)
                       - { 2 \over 3 } \Bigl( \ln^3(z) + \ln^3(1-z) \Bigr)
\PlusBreak{{1\over4}\bigglP{\mu^2\over-s}\biggrP^{2\e}~~~~~~~~}
   \biggl( \ln(z) \ln(1-z) - 2 \, \zeta_2 \biggr)
           \Bigl( \ln(z) + \ln(1-z) \Bigr)
     - {23 \over 6} \zeta_3 \biggr]
\MinusBreak{{1\over4}\bigglP{\mu^2\over-s}\biggrP^{2\e}\BigglBl}
  2 \Bigl( \Li_3(z) + \Li_3(1-z) - {17 \over 6} \zeta_3 \Bigr)
         \Bigl( \ln(z) + \ln(1-z) \Bigr) 
\PlusBreak{{1\over4}\bigglP{\mu^2\over-s}\biggrP^{2\e}\BigglBl}
   { 1 \over 3 } \Bigl( \ln^4(z) + \ln^4(1-z) \Bigr)
  - \biggl( \ln(z) \ln(1-z) - 2 \, \zeta_2  \biggr)
       \Bigl( \ln^2(z) + \ln^2(1-z) \Bigr)    
\MinusBreak{{1\over4}\bigglP{\mu^2\over-s}\biggrP^{2\e}\BigglBl}
    2 \, \zeta_2   \ln(z) \ln(1-z) + {7 \over 8} \zeta_4
 \Biggr\}
 \,. 
 \label{rSNeq4}
\end{eqnarray}
This result can be cast into the suggestive form~\cite{TwoloopN4},
\begin{equation}
r^{(2),\, \Neqfour}_S(\e;z, s) =
   {1 \over 2} \Bigl( r^{(1),\, \Neqfour}_S(\e;z, s) \Bigr)^2
     + f(\e) r^{(1),\, \Neqfour}_S(2\e;z, s)
     + \Ord(\e) \,,
\label{OneloopTwoloopSplit}
\end{equation}
where 
\begin{equation}
f(\e) \equiv (\psi(1-\e)-\psi(1))/\e =
  - (\zeta_2 + \zeta_3 \e + \zeta_4 \e^2 + \cdots)
\label{fDef}
\end{equation}
with $\psi(x) = (d/dx) \ln\Gamma(x)$, $\psi(1) = -\gamma$.
The iterative relation~(\ref{OneloopTwoloopSplit}), along
with a similar type of relation for the two-loop, leading-color, 
four-gluon amplitude in $\Neqfour$ super-Yang-Mills theory, 
has led to an ansatz based on collinear limits for 
the two-loop, leading-color, $n$-gluon amplitude 
in $\Neqfour$ super-Yang-Mills theory~\cite{TwoloopN4}.
At least for the case of maximal helicity violation,
we expect the following relation to hold,
\begin{equation}
M_n^{\twoloop}(\e) = 
 {1 \over 2} \Bigl(M_n^{\oneloop}(\e) \Bigr)^2
             + f(\e) \, M_n^{\oneloop}(2\e) - {5\over 4} \zeta_4 
             + \Ord(\e)\,.
\label{TwoloopOneloop}
\end{equation}
where $M_n^\Lloop(\e) = A_n^\Lloop / A_n^\tree$.

For $n=4$, we found that \eqn{TwoloopOneloop} was violated at order
$\e^1$~\cite{TwoloopN4}.  To be more specific,
$[M_4^{\oneloop}(\e)]^2$ contains at $\Ord(\e^1)$ only two-types of
$\Li_5$ polylogarithms, $\Li_5(-s/u)$ and $\Li_5(-t/u)$, whereas
$M_4^\twoloop(\e)$ contains in addition the independent function
$\Li_5(-s/t)$.  We have now examined the $z$-dependence of the order
$\e^1$ terms in $\rsn^{(2),\, \Neqfour}(z,s)$.  Again we find
functions not present in the square of the corresponding one-loop
quantity.  At order $\e^1$, $[\rsn^{(1),\, \Neqfour}(z,s)]^2$ contains
$\Li_5(-(1-z)/z)$, but no other $\Li_5$ functions.  On the other hand,
$\rsn^{(2),\, \Neqfour}(z,s)$ contains all three $\Li_5(x)$ functions
($x = z$, $1-z$, $-(1-z)/z$) as well as all three additional
non-$\Li_n$ functions required (in the basis we used).  The violation
of \eqn{TwoloopOneloop} at $\Ord(\e)$ is consistent with the intuition
that conformal symmetry underlies this result, and so it should
hold only for $D \rightarrow 4$.

\subsection{$\Neqone$ Super-Yang-Mills Theory Results}
\label{Neq1Results}

For pure $\Neqone$ super-Yang-Mills theory, the ratio $\rsn^{(2)\,
\lambda_a\lambda_b, \, \Neqone}(z,s)$ now depends on the helicity
configuration.  The two independent non-vanishing cases are $P^+ \to
a^+ b^+$ ($++$) and $P^+ \to a^- b^+$ ($-+$).  The linear combinations
of master integrals obtained are again relatively simple, and may be
expressed as,
\begin{equation}
\rsn^{(2)\,++, \, \Neqone}(z,s) = \rsn^{(2),\, \Neqfour}(z,s) 
+ I^{++, \, \Neqone}_A(z) + I^{++, \, \Neqone}_A(1-z),
\label{Neq1ppmA}
\end{equation}
where
\begin{eqnarray}
I^{++, \, \Neqone}_A(z) &=&
 - { 3 \over 8 } s \biggl[ 
 { \e \over 1-2\e } z \Bigl( s \, \Zig(z,s) + \WedgeF(z,s)
                  - 2 \Wedge(1-z,s) \Bigr)
\PlusBreak{ - { 3 \over 8 } s \bigglB }
  \LBtri(z,s) \biggr]
 \,,
\label{Neq1ppmAans}
\end{eqnarray}
and
\begin{eqnarray}
\rsn^{(2)\, -+, \, \Neqone}(z,s) &=& \rsn^{(2)\, ++, \, \Neqone}(z,s) 
\PlusBreak{ }
  { 3 \over 8 } s \Biggl\{ 
  2 \, s \, z \, \e \Bigl( \Ptri_2^{(6)}(z,s) - \Ptri_2^{(6)}(1-z,s) \Bigr)
\MinusBreak{ + { 3 \over 8 } s \BigglBl }
   { \e \over 1-2\e } \biggl[ 
    { z^2 \over 1-z } \Bigl( s \, \Zig(z,s) + \WedgeF(z,s) \Bigr)
\MinusBreak{ + { 3 \over 8 } s \BigglBl~~~~~~~~~~~~ }
     (1-z) \Bigl( s \, \Zig(1-z,s) + \WedgeF(1-z,s) \Bigr) \biggr]
\MinusBreak{ + { 3 \over 8 } s \BigglBl }
   { z\over 1-z } \LBtri(z,s) + \LBtri(1-z,s) 
\MinusBreak{ + { 3 \over 8 } s \BigglBl }
   { 1-3\e \over 2 \, \e^2 \, s^2 } { 1 - 2 z \over 1 - z } 
    \Bigl( 2 (2-3\e) \Sset(s) + s \, \e \, \Btri(s) \Bigr)   
   \Biggr\}
 \,.
\label{Neq1mpmA}
\end{eqnarray}
These expressions, like the $\Neqfour$ results, are evaluated in the 
FDH scheme with $\eta_\mu{}^\mu = D_s = 4$, or $\delta_R=0$.

At one loop, it happens that the splitting amplitude in $\Neqone$ 
super-Yang-Mills theory is the same for $-+$ as for $++$,  
\begin{equation}
\rsn^{(1)\, -+, \, \Neqone}(z,s) = \rsn^{(1)\, ++, \, \Neqone}(z,s).
\label{Neq1oneloopmpm}
\end{equation}
This relation is spoiled at two loops, but we expect the difference
to be finite as $\e \to 0$, due to the one-loop relation.
A form of \eqn{Neq1mpmA} which makes this property more manifest
can be obtained by switching basis from $\Zig(z,s)$ to
\begin{equation}
\Zig_2^{(6)}(z,s) \equiv L(1,1,1,0,1,2,1,0,0)\Bigr|_{D=6-2\e},
\label{Zig26def}
\end{equation}
and defining
\begin{equation}
I_B^{\Neqone}(z,s) \equiv
 (1-3\e) \Zig_2^{(6)}(z,s) 
   + { \e \over 1-2\e } \WedgeF(z,s)
   + \e \, s \, \Ptri_2^{(6)}(z,s).
\label{IBNeq1def}
\end{equation}
Then
\begin{equation}
\rsn^{(2)\, -+, \, \Neqone}(z,s) = \rsn^{(2)\, ++, \, \Neqone}(z,s) 
+ { 3 \over 4 } \, s \, z 
   \Bigl( I_B^{\Neqone}(z,s) - I_B^{\Neqone}(1-z,s) \Bigr).
\label{Neq1mpmB}
\end{equation}
The function $I_B^{\Neqone}(z,s)$ is finite as $\e \to 0$:
$\Ptri_2^{(6)}$ is finite, so it gives a vanishing contribution.
There are $1/\e$ poles which cancel between $\Zig_2^{(6)}$ and
$\WedgeF$, yielding
\begin{equation}
I_B^{\Neqone}(z,s) =
 - { 1 \over s (1-z) } \Bigl[
  2 \Bigl( \Li_3(z) - \zeta_3 \Bigr) 
 - \ln z \Bigl( \Li_2(z) - \zeta_2 \Bigr) \Bigr] 
 + \Ord(\e).
\label{IBNeq1limit}
\end{equation}

The result of performing the expansions in $\e$ of 
\eqns{Neq1ppmA}{Neq1mpmB} is,
\begin{eqnarray}
&&\rsn^{(2)\, ++, \, \Neqone}(z,s) = \rsn^{(2),\, \Neqfour}(z,s)
\PlusBreak{ }
  {3 \over 8} \biggl( { \mu^2 \over -s } \biggr)^{2\e} \Biggl\{
- { 1 \over 2 \, \e^3 } 
+ { 1 \over \e^2 } \Bigl( \ln(z) + \ln(1-z) - 1 \Bigr)
\PlusBreak{ ~~~~~~~~~~ }
  { 1 \over \e } \biggl[ - \Bigl( \ln(z) - \ln(1-z) \Bigr)^2
    + 2 \Bigl( \ln(z) + \ln(1-z) - 1 \Bigr) - { \zeta_2 \over 2 } \biggr]
\MinusBreak{ ~~~~~~~~~~ }
  12 \Bigl( \Li_3(z) + \Li_3(1-z) \Bigr)
+ 2 \Bigl( \ln(z) \Li_2(z) + \ln(1-z) \Li_2(1-z) \Bigr)
\PlusBreak{ ~~~~~~~~~~ }
  { 2 \over 3 } \Bigl( \ln^2(z) + \ln^2(1-z) 
    - 4 \ln(z) \ln(1-z) + {3 \over 2} \zeta_2 \Bigr) 
       \Bigl( \ln(z) + \ln(1-z) \Bigr)
\MinusBreak{ ~~~~~~~~~~ }
  2 \Bigl( \ln(z) - \ln(1-z) \Bigr)^2 
+ 4 \Bigl( \ln(z) + \ln(1-z) - 1 \Bigr)
+ {67 \over 3} \zeta_3 - \zeta_2 \Biggr\}
 \,,
\label{rSNeq1ppm}
\end{eqnarray}
\begin{eqnarray}
\rsn^{(2)\, -+, \, \Neqone}(z,s) &=& \rsn^{(2)\, ++, \, \Neqone}(z,s)
\PlusBreak{ }
  {3 \over 4} \Biggl\{
    2 \Bigl( \Li_3(1-z) - \zeta_3 \Bigr) 
        - \ln(1-z) \Bigl( \Li_2(1-z) - \zeta_2 \Bigr) 
\MinusBreak{ ~~~~~~ }
        { z \over 1-z } \biggl[ 2 \Bigl( \Li_3(z) - \zeta_3 \Bigr) 
                    - \ln(z) \Bigl( \Li_2(z) - \zeta_2 \Bigr) \biggr] 
  \Biggl\}
 \,.
\label{rSNeq1mpm}
\end{eqnarray}
%

\subsection{QCD Results}
\label{QCDResults}

We now present the $g\to gg$ splitting amplitudes in QCD,
for a general value of the regularization-scheme parameter $\delta_R$.
First we introduce some auxiliary functions describing the fermion-loop
contributions and residual dependence on $\delta_R$.

The functions $f_L$ describing the leading-color fermion-loop contributions
to the QCD splitting amplitudes are similar to the differences
between the $\Neqone$ and $\Neqfour$ splitting amplitudes, so we write them as
\begin{eqnarray}
&&f_L^{++}(z,s) =
 - {2 \over 9} \Bigl[ 
   \rsn^{(2)\, ++, \, \Neqone}(z,s) 
 - \rsn^{(2),\, \Neqfour}(z,s) \Bigr]
\PlusBreak{ }
  {1 \over 12} \biggl( { \mu^2 \over -s } \biggr)^{2\e} \Biggl\{
   {1 \over \e^2 } \Bigl( z (1-z) - {1 \over 6} \Bigr)
\PlusBreak{ ~~~~~~~~~~~~~~~~~ }
   {1 \over \e } \biggl[ 
     - z (1-z) \Bigl( \ln(z) + \ln(1-z) + {2 \over 3} \Bigr)
     + {1 \over 3} \Bigl( \ln(z) + \ln(1-z) \Bigr) - {17 \over 18 } \biggr]
\PlusBreak{ ~~~~~~~~~~~~~~~~~ }
   z (1-z) \biggl[ 
      { 1 \over 2 } \Bigl( \ln(z) - \ln(1-z) \Bigr)^2 
    - { 11 \over 3 } \Bigl( \ln(z) + \ln(1-z) \Bigr)
    - { 187 \over 6 } + { \delta_R \over 3 } \biggr]
\MinusBreak{ ~~~~~~~~~~~~~~~~~ }
    { 1 \over 3 } \Bigl( \ln(z) - \ln(1-z) \Bigr)^2 
  + \Bigl( { 17 \over 9 } - z \Bigr) \ln(z) 
  + \Bigl( { 8 \over 9 } + z \Bigr) \ln(1-z)  
\MinusBreak{ ~~~~~~~~~~~~~~~~~ }
    { \zeta_2 \over 6 } - { 151 \over 54 } \Biggr\}
 \,,
\label{fLppm}
\end{eqnarray}
and
\begin{eqnarray}
&&f_L^{-+}(z,s) =
 - {2 \over 9} \Bigl[ 
   \rsn^{(2)\, -+, \, \Neqone}(z,s) 
 - \rsn^{(2),\, \Neqfour}(z,s) \Bigr]
\PlusBreak{ }
  {1 \over 12} \biggl( { \mu^2 \over -s } \biggr)^{2\e} \Biggl\{
- {1 \over 6 \, \e^2 }
+ {1 \over \e } \biggl[ 
    {1 \over 3} \Bigl( \ln(z) + \ln(1-z) \Bigr) - {17 \over 18 } \biggr]
\PlusBreak{ ~~~~~~~~~~~~~~~~~ }
   { z (1+z) \over (1-z)^3 } \biggl[ 2 \Bigl( \Li_3(z) - \zeta_3 \Bigr) 
                        - \ln(z) \Bigl( \Li_2(z) - \zeta_2 \Bigr) \biggr]
\MinusBreak{ ~~~~~~~~~~~~~~~~~ }
    2 { z \over (1-z)^2 } \Bigl( \Li_2(1-z) - 2 \, \zeta_2 \Bigr) 
  + { \ln(z) \over 1 - z }
\MinusBreak{ ~~~~~~~~~~~~~~~~~ }
    { 1 \over 3 } \Bigl( \ln(z) - \ln(1-z) \Bigr)^2 
  + { 8 \over 9 } \Bigl( \ln(z) + \ln(1-z) \Bigr)
  - { \zeta_2 \over 6 } - { 151 \over 54 } \Biggr\}
 \,.
\label{fLmpm}
\end{eqnarray}

The functions $f_{SL}$ describing the subleading-color fermion-loop
and $f_{2}$ describing the double fermion-loop contributions are 
very simple:
\begin{eqnarray}
f_{SL}^{++}(z,s) &=&
 {1 \over 8 } z (1-z)
 \,, 
\label{fSLpp} \\
f_{SL}^{-+}(z,s) &=&
 0
 \,, 
\label{fSLmp} \\
f_{2}^{++}(z,s) &=&
 { 1 \over 18 } z (1-z) 
   \biggl( { \mu^2 \over -s } \biggr)^{2\e}
   \biggl[ { 1 \over \e } + { 16 \over 3} \biggr]
 \,, 
\label{f2pp} \\
f_{2}^{-+}(z,s) &=&
 0
 \,.
\label{f2mp}
\end{eqnarray}
There are also some functions describing the residual $\delta_R$ 
dependence of the QCD results:
\begin{eqnarray}
\Delta^{++}(z,s) &=& \Delta^{-+}(z,s)
+ {1 \over 12 } z(1-z) \biggl( { \mu^2 \over -s } \biggr)^{2\e}
  \biggl[ { 1 \over \e } - \ln(z) - \ln(1-z) + { 2 \over 3 } \biggr]
 \,,
\label{Deltapp} \\
\Delta^{-+}(z,s) &=& 
{1 \over 24 } \biggl( { \mu^2 \over -s } \biggr)^{2\e} \Biggl\{
   - { 1 \over 2 \, \e^2 } 
   + { 1 \over \e } \Bigl( \ln(z) + \ln(1-z) - { 4 \over 3 } \Bigr)
 -  \Bigl( \ln(z) - \ln(1-z) \Bigr)^2
\PlusBreak{ ~~~~~~~~~~~~~~ }
  { 8\over 3 } \Bigl( \ln(z) + \ln(1-z) \Bigr)
 - { \zeta_2 \over 2 } - { 35 \over 9 } \Biggr\}
 \,.
\label{Deltamp}
\end{eqnarray}
Then the unrenormalized two-loop QCD splitting amplitude $\rsn$ factors
are given for both $++$ and $-+$ by
\begin{eqnarray}
\rsn^{(2)\, \alpha, \, {\rm QCD}} &=& \rsn^{(2)\, \alpha, \, \Neqone}
 - \Bigl( 1 - {\Nf \over \Nc} \Bigr) \, f_L^\alpha
 - \Bigl( 1 + {\Nf \over \Nc^3} \Bigr) f_{SL}^\alpha
 - \Bigl( 1 - {\Nf^2 \over \Nc^2} \Bigr) f_2^\alpha
 + \delta_R \Delta^\alpha,
\nonumber \\ 
\alpha &=& \hbox{$++$ or $-+$}. 
\label{QCDppmmpm}
\end{eqnarray}

For the helicity-flip case, $P^+ \to a^- b^-$ ($--$),
we similarly define
\begin{eqnarray}
f_L^{--}(z,s) &=&
{1 \over 12 } \biggl( { \mu^2 \over -s } \biggr)^{2\e} \Biggl\{
   { 1 \over \e^2 } 
 - { 1 \over \e } \Bigl( \ln(z) + \ln(1-z) + { 2 \over 3 } \Bigr)
 + {1 \over 2} \Bigl( \ln(z) - \ln(1-z) \Bigr)^2
\MinusBreak{ ~~~~~~~~~~~~}
   {11 \over 3} \Bigl( \ln(z) + \ln(1-z) \Bigr)
 - 2 \Bigl( { \ln(z) \over 1-z } + { \ln(1-z) \over z } \Bigr)
 - { 187 \over 6 } + { \delta_R \over 3 } \Biggr\}
 \,,
\nonumber\\
\label{fLmm} \\
f_{SL}^{--}(z,s) &=&
 { 1 \over 8 }
 \,,
\label{fSLmm} \\
f_{2}^{--}(z,s) &=&
 { 1 \over 18 }
  \biggl( { \mu^2 \over -s } \biggr)^{2\e} 
  \biggl[ { 1 \over \e } + { 16 \over 3} \biggr]
 \,,
\label{f2mm} \\
\Delta^{--}(z,s) &=&
 {1 \over 12 } \biggl( { \mu^2 \over -s } \biggr)^{2\e}
  \biggl[ { 1 \over \e } - \ln(z) - \ln(1-z) + { 2 \over 3 } \biggr]
 \,.
\label{Deltamm}
\end{eqnarray}
Then the helicity-flip splitting amplitude in QCD is
\begin{eqnarray}
\Split_{-}^{(2),\, {\rm QCD}}(z;a^-,b^-) &=&
  \sqrt{z(1-z)} { \spa{a}.{b} \over {\spb{a}.{b}}^2 } \biggl[
 - \Bigl( 1 - {\Nf \over \Nc} \Bigr) \, f_L^{--}
 - \Bigl( 1 + {\Nf \over \Nc^3} \Bigr) f_{SL}^{--}
\MinusBreak{ ~~~~~~~~~~~~~~~~~~~~~~ }
   \Bigl( 1 - {\Nf^2 \over \Nc^2} \Bigr) f_2^{--}
 + \delta_R \Delta^{--} \biggr]
 \,. 
\label{QCDmmm}
\end{eqnarray}

We remark that the functions $f_{SL}$ and $f_2$ are so simple because
no light-cone projectors on internal propagators are required to
compute them.  Clearly $f_2$, representing the double fermion-loop
($\Nf^2$) contribution, has no internal gluon lines, and only the
simple master integral $\Btie(s)$ can appear.  There is an internal
gluon line in the graphs contributing to the subleading-color
single-fermion loop ($\Nf^1$) contribution $f_{SL}$.  However, this
gluon always appears {\it inside} the fermion loop.  All such graphs
contribute with equal weight to $f_{SL}$, exactly as if the gluon were
a photon.  If we think of this gluon as a photon, it becomes clear
that its gauge transformations can be separated from those of the
off-shell external gluon with momentum $k_P$, and indeed, a covariant
gauge could have been used for it.  Therefore, in this special case we
could have used a covariant propagator for the internal gluon, instead
of a light-cone gauge propagator, which means that only the
$z$-independent master integrals given in
eqs.~(\ref{BtieLaurent})--(\ref{BtriLaurent}) can appear in $f_{SL}$.
We have discussed in detail in sections~\ref{FeynmanSection}
and~\ref{UnitaritySewingSection} how unitarity can prevent certain
light-cone denominators from ever appearing.  The $f_{SL}$ terms are
examples of how color can occasionally do the same sort of thing.

\Eqns{QCDppmmpm}{QCDmmm} give the dependence of the QCD results
on $\Nf$ and $\Nc$.  The results can also be written in terms
of general group Casimir constants (after multiplying by the extracted
factor of $\Nc^2 = C_A^2$), using the substitution rules:
\begin{eqnarray}
 1 &\to& C_A^2 \,,
\nonumber\\
 {\Nf\over \Nc} &\to& 2 C_A T_R \Nf \,,
\nonumber\\
 {\Nf\over \Nc^3} &\to& 2 (C_A - 2 C_F) T_R \Nf \,,
\nonumber\\
 {\Nf^2\over \Nc^2} &\to& (2 C_A T_R \Nf)^2 \,.
\label{subsCasimirs}
\end{eqnarray}
To recover pure $\Neqone$ super-Yang-Mills theory, which contains
one Majorana fermion in the adjoint representation, we set $C_F=C_A$,
$T_R=C_A$, and $\Nf=1/2$.   Thus $\Nf/\Nc \to 1$ and $\Nf/\Nc^3 \to -1$,
so that for $\delta_R=0$ each term in \eqn{QCDmmm} vanishes,
and all but the first term in \eqn{QCDppmmpm} vanish, as required.

Motivated by ref.~\cite{Lipatov}, let us examine the degree 
of transcendentality of the above results, where $\ln(x)$ or $\pi$ 
has weight 1, and $\Li_n(x)$ and $\zeta_n$ ($\zeta_n \equiv \Li_n(1)$) 
have weight $n$.   Leading transcendentality is 4 units
for the $\Ord(\e^0)$ terms in the expansion, or more generally $4-m$ units
for the $\Ord(\e^{-m})$ terms.  The result in $\Neqfour$ super-Yang-Mills 
theory, \eqn{rSNeq4}, carries this maximum weight uniformly.
The $\Neqone$ super-Yang-Mills results~(\ref{rSNeq1ppm}) and (\ref{rSNeq1mpm})
are equal to the $\Neqfour$ result, plus terms of lower weight.
The additional functions $f_L^\alpha$, $f_{SL}^\alpha$, $f_2^\alpha$ 
and $\Delta^\alpha$ entering the QCD results~(\ref{QCDppmmpm}) 
are also of lower weight.  Hence the maximum-weight terms in the QCD,
$\Neqone$ and $\Neqfour$ two-loop splitting amplitudes all coincide.
It is easy to see that the same result (with $4-m$ replaced by $2-m$)
is true for the one-loop splitting amplitudes to $\Ord(\e^2)$ 
(in fact to all orders in $\e$).   So both types of two-particle final-state 
contributions to the NNLO Altarelli-Parisi kernel
--- two-loop times tree, and one-loop squared --- respect the 
transcendentality relation of ref.~\cite{Lipatov}.

\subsection{Continuation to Space-Like Region}
\label{Spacelike}

The preceding splitting amplitudes are for time-like 
kinematics, $k_P^2 = s > 0$.  In the remainder of this section
we briefly discuss the space-like case, $s < 0$.
In the application of splitting amplitudes to NNLO evolution
kernels, the time-like region is relevant for the $Q^2$ evolution 
of fragmentation functions.   The space-like region is relevant
for the evolution of parton distribution functions, which play a key role
in the prediction of collider cross sections. 
(As mentioned in the introduction, the NNLO kernels for unpolarized
space-like evolution have been computed very recently~\cite{MVVNNLO}.)

The time-like results can be continued analytically
into the space-like region.  We take the space-like splitting
process to be 
\begin{equation}
  (-k_1)^{-\lambda_1} \to (-k_P)^{-\lambda_P} + k_2^{\lambda_2} \,,
\label{SpaceConfig}
\end{equation}
with $k_P^2 = s < 0$.
It is obtained from the time-like process by crossing leg 1 into
the initial state, and leg $P$ into the final state. 
The physical helicities of legs 1 and $P$ flip under this crossing,
but we retain the uncrossed labeling.
Relative to $(-k_1)$, the vectors $(-k_P)$ and $k_2$ now carry longitudinal 
momentum fractions of $x$ and $1-x$ respectively.
Comparing the ratio $(k_1\cdot n)/(k_P\cdot n)$ between the time-like and
space-like cases, we identify $z=1/x$. 
In principle, all we need to do to obtain the space-like splitting
amplitude results from the above time-like formul\ae{}
is to let $s$ be negative, and substitute $z \to 1/x$, where 
$0 < x < 1$.   For example, the tree-level splitting amplitude
$\Split^{\tree}_{-}(z; a^{+},b^{+})$ given in \eqn{TreeSplitAmplitudepp}
becomes the space-like splitting amplitude for
$(-k_1)^- \to (-k_P)^- + k_2^+$, which we denote by giving $\Split^{\tree}$
the argument $x$,
\begin{equation}
\Split^{\tree}_{-}(x; a^{+},b^{+})
            = {x\over \sqrt{1-x} \spa{a}.b} \,. 
\label{SpaceTreeSplitAmplitudepp}
\end{equation}
(Note that in the construction of the space-like Altarelli-Parisi kernel,
after squaring the $x$-dependent part of the splitting amplitude there 
is an additional factor of $x$; see {\it e.g.} eq.~(6.23) of 
ref.~\cite{CFP}.)
There are overall phases stemming from the factors $\sqrt{1-z}$
and $\spa{a}.b$, but they can be associated with the external states,
and we neglect them in \eqn{SpaceTreeSplitAmplitudepp}.

At the loop-level, the preceding formul\ae{} for $\rsn^\Lloop(z,s)$ 
have been been written so that the logarithms and polylogarithms 
are manifestly real for $0 < z < 1$.  To continue them to $z > 1$, 
one can apply polylogarithm identities so that the only function 
which is not manifestly real is $\ln(1-z) = \ln(1-x) - \ln x \pm i \pi$.
For example, consider the one-loop splitting amplitudes 
$\rsn^{\oneloop, \, \Neqfour}(z,s)$ in \eqn{OneloopSplitExplicitNeq4},
which contain ${\Li}_{2m-1}({z\over z-1})$.
These functions develop branch cuts for $z>1$.  In this case,
though, we can use the (non-manifest) $z \lr 1-z$ symmetry of
$\rsn^{\oneloop, \, \Neqfour}(z,s)$, \eqn{rSNeq4symmetric},
to let $z \to 1-z$ in \eqn{OneloopSplitExplicitNeq4} before
substituting $z \to 1/x$.  (This amounts to using
an infinite sequence of polylogarithm identities.) 
We obtain,
\begin{eqnarray}
\rsn^{\oneloop,\, \Neqfour}(x,s) & = & 
   \cgh \biggl({\mu^2\over-s}\biggr)^{\eps}
    {1\over\eps^2}\left[ - [-(1-x)]^{-\eps} 
    {\pi\eps\over \sin(\pi\eps)} + \sum_{m=1}^\infty 2\eps^{2m-1} 
    {\Li}_{2m-1}(1-x)\right] \,. 
\nonumber \\
&& 
\label{SpaceOneloopSplitExplicitNeq4}
\end{eqnarray}
In \eqn{SpaceOneloopSplitExplicitNeq4}, the factor $(\mu^2/(-s))^\e$
no longer contains any $i\pi$ terms, as $s$ is now negative.
However, in the expansion of $[-(1-x)]^{-\e}$ we must set 
$\ln[-(1-x)] = \ln(1-x) \pm i\pi$.  The presence of imaginary
parts in the space-like region is a bit surprising, but it can be
traced back to integrand denominators that vanish in the interior
of the integration region for $z>1$ --- for example, 
the denominator of $J^{(b)}(1-z)$ in \eqn{Integral_b_final}, 
as discussed in \sect{OneLoopLightConeIntegralExamples}.
The appearance of an imaginary part can be verified from the explicit 
collinear behavior of one-loop amplitudes.  From this verification, one also
learns that the sign of the $i\pi$ term is ambiguous; it depends on 
whether the leg color-adjacent to leg 2 is incoming or outgoing.   
Fortunately, this sign is the same for all imaginary terms, and it drops
out of the interferences required for the computation of evolution
kernels.  The $\pi^2$ factors from the product of two $i\pi$ terms
do survive and are unambiguous.  (Note that $s(1-x)$ has the
same negative sign in both space-like and time-like regions, so
that the $\pi^2$ terms from the first term 
in~\eqn{SpaceOneloopSplitExplicitNeq4} are the same in both regions.)

At the two-loop level, \eqn{SpaceOneloopSplitExplicitNeq4} and
the squaring relation~(\ref{OneloopTwoloopSplit}) provide a convenient 
way to continue $r^{(2),\, \Neqfour}_S(z,s)$ to $z>1$ (through $\Ord(\e^0)$).
We refrain from giving the explicit continuations of the $\Neqone$
and QCD results, though they are straightforward to carry out.

\section{Comparison to Catani's Formula and Finite Remainders}
\label{CataniComparison}

In this section we compare the pole terms in the
two-loop splitting amplitudes to the expectation based on Catani's 
formula~\cite{CataniTwoloop} for the two-loop infrared divergences
of renormalized amplitudes,
in order to establish their mutual consistency.
More importantly, by examining the order $\e^0$ terms in the splitting
amplitudes, we derive relations that the finite remainders in 
the Catani formalism must obey in the collinear limit.

\subsection{Singular Term Comparison}

Catani's general formula includes color-space operators which have
a fairly intricate structure in the trace-based color decomposition.
In this section, to simplify the analysis we shall restrict
attention to the single-trace coefficients $A_n^{(L)}$ in the $n$-gluon
amplitude, given explicitly in~\eqn{LeadingColorAmplitude},
and to the terms in Catani's formula obtained by making the replacement,
\begin{eqnarray}
{\bom T}_i \cdot {\bom T}_j &\to& -{\Nc \over 2} \ {\bom 1}, 
\hskip1cm \hbox{$i,j$ color-adjacent,} 
\nonumber\\
&\to& 0,  
\hphantom{-{\Nc \over 2} \ } \hskip1cm \hbox{otherwise.}
\label{ColorTrivialReplace}
\end{eqnarray}
These terms include all the leading-color terms, as well as 
certain of the subleading-color terms (including all of the 
color dependence in $H_i^{(2)}$ at order $1/\e$).
We refer to these terms as {\it color trivial}.
In appendix~\ref{FullColorCatani} we perform the full color-space
analysis.

First we need to remove the ultraviolet divergences from
the splitting amplitudes presented in section~\ref{ResultSection}.
The relation between the bare coupling $\alphas^u$ (implicitly used
above) and renormalized coupling $\alphas(\mu) = g^2(\mu)/(4\pi)$, 
through two-loop order, may be written as~\cite{CataniTwoloop}
\begin{equation}
\alphas^u \;\mu_0^{2\e} \;S_{\e} = \alphas(\mu) \;\mu^{2\e} 
\left[ 1 - { \alphas(\mu) \over 2\pi } \; { b_0 \over \e } 
         + \biggl( { \alphas(\mu) \over 2\pi} \biggr)^2 
           \left( { b_0^2 \over \e^2 } - { b_1 \over 2\e } \right) 
+ {\cal O}(\alphas^3(\mu)) \right]\,,
\label{TwoloopCoupling}
\end{equation}
where $\mu$ is the renormalization scale, and
$S_\e = \exp[\e (\ln4\pi + \psi(1))]$.
The first two coefficients appearing in the beta function for QCD,
or more generally ${\rm SU}(\Nc)$ gauge theory with $\Nf$ flavors of
massless fundamental representation quarks, are
\begin{equation}
b_0 = {11 C_A - 4 T_R \Nf \over 6} \,, \hskip 2 cm 
b_1 = {17 C_A^2 - ( 10 C_A + 6 C_F ) T_R \Nf \over 6} \,,
\label{QCDBetaCoeffs}
\end{equation}
where $C_A = \Nc$, $C_F = (\Nc^2-1)/(2\Nc)$, and $T_R = 1/2$.
In pure $\Neqone$ super-Yang-Mills theory, the values are
\begin{equation}
b_0^{\Neqone} = {3 \over 2} C_A \,, \hskip 2 cm 
b_1^{\Neqone} = {3 \over 2} C_A^2 \,.
\label{SUSYQCDBetaCoeffs}
\end{equation}
In $\Neqfour$ super-Yang-Mills theory, a conformal theory, the values are
of course
\begin{equation}
b_0^{\Neqfour} = b_1^{\Neqfour} = 0\,.
\label{Neq4QCDBetaCoeffs}
\end{equation}

One can define a `perturbative expansion' of the $g \to gg$ splitting 
amplitude as
\begin{equation}
\SplitRen(\alphas(\mu)) =
g(\mu) \, \biggl[ \Split^{(0)}
+ { \alphas(\mu) \over 2\pi } \SplitRen^{(1)}
+ \biggl( { \alphas(\mu) \over 2\pi } \biggr)^2 
 \SplitRen^{(2)} + \Ord(\alphas^3(\mu)) \biggr] \,, 
\label{RenExpand}
\end{equation}
where $\SplitRen^{(L)}$ is the $L^{\rm th}$
loop contribution.  
Equation~(\ref{TwoloopCoupling}) is equivalent to the following
$\MSbar$ renormalization prescriptions at one and two loops,
\begin{eqnarray}
 \SplitRen^{(1)} 
 &=&  S_\e^{-1} \, \Split^{(1)}
    - { 1 \over 2 \e } {b_0 \over \Nc} \, \Split^{(0)} ,
\label{OneloopCounterterm} \\
 \SplitRen^{(2)}  
 &=&  S_\e^{-2} \, \Split^{(2)}   
  -  {3 \over 2\e} {b_0 \over \Nc} \, S_\e^{-1} 
                    \, \Split^{(1)}
  + \biggl( { 3 \over 8 \e^2 } {b_0^2 \over \Nc^2}
          - {1 \over 4 \e} {b_1 \over \Nc^2} \biggr)
             \, \Split^{(0)} .~~~
\label{TwoloopCounterterm}
\end{eqnarray}
Here $\Split^{(L)}$ governs the collinear limits of
unrenormalized amplitudes, while $\SplitRen^{(L)}$ controls the limits of
renormalized amplitudes.  All quantities appearing in the remainder
of this section are renormalized.  In particular, $A_n^{(L)}$ refers
to renormalized $L$-loop amplitudes, in contrast to the unrenormalized 
ones used implicitly in previous sections.
The renormalization of the splitting ratios $\rsn^{(L)}$ 
follows simply from \eqns{OneloopCounterterm}{TwoloopCounterterm}.
We will denote the renormalized splitting ratios by $\rsnren^{(L)}$

The full color-space forms of Catani's formula at one and two loops 
are given in \eqns{OneloopCatani}{FullColorTwoLoopCatani}.
The color-trivial terms, defined by the
replacements~(\ref{ColorTrivialReplace}),
are obtained by letting the operator
${\bom I}_n^{(L)} \to \Nc^L \, \hat{I}_n^{(L)} {\bom 1}$, 
in its action on the single-trace term $A_n^{(L)}(1,2,\ldots,n)$
in the amplitude.   The factor of $\Nc^L$ is extracted from 
${\bom I}_n^{(L)}$ for consistency with the normalization of $A_n^{(L)}$; 
we want both $A_n^{(L)}$ and ${\hat I}_n^{(L)}$ to be independent of 
$\Nc$ as $\Nc \to \infty$ in the pure-glue case.
At one loop, we have
\begin{equation}
{\hat I}^{(1)}_n(\e)  = - {1 \over 2} {e^{-\e \psi(1)} \over
\Gamma(1-\e)} \sum_{i=1}^n 
\Biggl[ {1 \over \e^2} + {\gamma_g \over \Nc} \,
{1 \over \e} \Biggr] \biggl(\frac{\mu^2}{- s_{i, i+1}} \biggr)^{\e} \,,
\end{equation}
where $\gamma_g$ is given in~\eqn{QCDgluonValues}.

Similarly, $\hat{I}_n^{(2)}$ is given by 
\begin{eqnarray}
{\hat I}^{(2)}_n(\e)
& =& - \frac{1}{2} {\hat I}^{(1)}_n(\e)
\left( {\hat I}^{(1)}_n(\e) + {2 \over \e} {b_0 \over \Nc} \right)
 + {e^{+\e \psi(1)} \Gamma(1-2\e) \over \Gamma(1-\e)} {1\over \Nc}
\left( {b_0 \over \e} + K_\RS \right) I^{(1)}_n(2\e)
\PlusBreak{ }
  { e^{-\e\psi(1)} \over 4\e \, \Gamma(1-\e) }
 { H_g^{(2)} \over \Nc^2 } 
\sum_{i=1}^n \biggl(\frac{\mu^2}{- s_{i, i+1}} \biggr)^{2\e} \,,
\label{I2hatdef}
\end{eqnarray}
where $K_\RS$ is given in~\eqn{CataniK} and $H_g^{(2)}$ is given
in~\eqn{Hgluon}.

The singularities of the single-trace coefficients are,
\begin{eqnarray}
A_n^{(1)} &=& {\hat I}_n^{(1)} A_n^{(0)} +  A_n^{(1)\rm fin}\,, 
\label{LeadingColorCataniOneloop} \\
A_n^{(2)} &=& {\hat I}_n^{(1)} A_n^{(1)} 
            + {\hat I}_n^{(2)} A_n^{(0)} +  A_n^{(2)\rm fin} \,.
\label{LeadingColorCataniTwoloop}
\end{eqnarray}
Things become a bit simpler if we write the
finite remainders in Catani's formula as multiples of the 
corresponding tree amplitudes:
\begin{equation}
A_n^{(L)\rm fin}(1,2,\ldots,n) 
  = F_n^{(L)}(s_{ij}) \times A_n^\tree(1,2,\ldots,n) \,.
\label{LoopF} 
\end{equation}
\Eqns{LeadingColorCataniOneloop}{LeadingColorCataniTwoloop} become
\begin{eqnarray}
A_n^{(1)} &=& ( \hat{I}_n^{(1)} + F_n^{(1)} ) \, A_n^\tree \,,
\label{Sing1New} \\
A_n^{(2)} &=& \Bigl[ \hat{I}_n^{(2)} 
                    + \hat{I}_n^{(1)} ( \hat{I}_n^{(1)} + F_n^{(1)} )
                    + F_n^{(2)} \Bigr] \, A_n^\tree \,,
\label{Sing2New}
\end{eqnarray}
If the tree amplitude vanishes, we cannot perform this step.
However, in this case the entire analysis is much simpler,
essentially equivalent to the one-loop analysis.

Now consider the collinear limits.  We will provide two independent
forms for the pole terms in $\e$ of $\rsnren^{(2)}$.  The first form
is similar to Catani's formula~(\ref{I2hatdef}) for ${\hat I}_n^{(2)}$.
The second form retains finite terms, so it can be used to predict
the collinear behavior of the Catani finite remainders. 

In the first derivation, we note that the ${\hat I}_n^{(2)}$ term 
by itself does not have particularly enlightening collinear limits, 
because of contributions coming from $\Ord(\e)$ parts of 
${\hat I}^{(1)}_n$.  However, this obscure behavior is balanced by
similar behavior in the ${\hat I}_n^{(1)} A_n^{(1)}$ term 
in \eqn{Sing2New}, suggesting
that it is best to combine the first two terms in this equation:
\begin{eqnarray}
A_n^{(2)} &=& \biggl[ {1\over 2} \Bigl( {\hat I}_n^{(1)}(\e) \Bigr)^2
                    + {\hat I}_n^{(1)}(\e) F_n^{(1)}
       - {1 \over \e} {b_0 \over \Nc} {\hat I}_n^{(1)}(\e)
\PlusBreak{}
   {e^{+\e \psi(1)} \Gamma(1-2\e) \over \Gamma(1-\e)}
   {1\over \Nc} \left( {b_0 \over \e} + K_\RS \right) 
      {\hat I}^{(1)}_n(2\e)
\PlusBreak{} 
 { e^{-\e\psi(1)} \over 4\e \, \Gamma(1-\e) }
 \sum_{i=1}^n  { H_i^{(2)} \over \Nc^2 }
     \biggl( { \mu^2 \over -s_{i,i+1} } \biggr)^{2\e} \biggr]
     A_n^\tree  +\Ord(\e^0) \nonumber\\
   &=& \biggl[ {1\over 2} \Bigl( {\hat I}_n^{(1)}(\e) + F_n^{(1)} \Bigr)^2 
      - {1 \over \e} {b_0 \over \Nc}
           \Bigl( {\hat I}_n^{(1)}(\e) + F_n^{(1)} \Bigr)
\PlusBreak{}
  {e^{+\e \psi(1)} \Gamma(1-2\e) \over \Gamma(1-\e)}
   {1\over \Nc} \left( {b_0 \over \e} + K_\RS \right)
   \Bigl( {\hat I}^{(1)}_n(2\e) + F_n^{(1)} \Bigr)
\PlusBreak{}
   {n \over 4\e } { H_g^{(2)} \over \Nc^2 } \biggr] A_n^\tree 
   + \Ord(\e^0) \,.
 \label{CollinearTwoLoopLeading}
\end{eqnarray}
In the last step we added finite pieces, in particular ones proportional to
$[F_n^{(1)}]^2$ and $K_\RS \, F_n^{(1)}$.  Two singular terms 
proportional to $(1/\e) \times b_0 \, F_n^{(1)}$ cancel against
each other.

Now the combination ${\hat I}_n^{(1)}(\e) + F_n^{(1)}$ is just
$A_n^{(1)}(\e)/A_n^\tree$, so it has simple collinear limits.
Indeed, inserting \eqn{OnelooprS} into \eqn{OneloopSplit},
we find that
\begin{equation}
 {\hat I}_n^{(1)}(\e) + F_n^{(1)}
  \mathop{\longrightarrow}^{a \parallel b}
 {\hat I}_{n-1}^{(1)}(\e) + F_{n-1}^{(1)} + \rsnren^{(1)}(\e) \,.
\label{OneloopNormLimit}
\end{equation}
We use the behavior~(\ref{OneloopNormLimit}) to take the collinear limit
of \eqn{CollinearTwoLoopLeading}, and compare the result 
with \eqn{TwoloopSplit}. The terms quadratic and linear in 
${\hat I}_{n-1}^{(1)}(\e) + F_{n-1}^{(1)}$ belong to
$A_n^{(2)} \times \Split^{(0)}$ and $A_n^{(1)} \times \SplitRen^{(1)}$.
Most of the remaining terms belong to 
$A_n^{(0)} \times \SplitRen^{(2)}
 = A_n^{(0)} \times \Split^{(0)} \times \rsnren^{(2)}(\e)$.
Collecting them, we see that the divergent parts of the two-loop 
splitting amplitude should be given by
\begin{eqnarray}
\rsnren^\twoloop(\e)
 & = &
{1\over 2} 
(\rsnren^\oneloop(\e))^2
- {1\over \e} {b_0 \over \Nc} \rsnren^\oneloop(\e)
 + {e^{+\e \psi(1)} \Gamma(1-2\e) \over \Gamma(1-\e)}
  {1\over \Nc} \left( {b_0 \over \e} + K_\RS \right) \rsnren^\oneloop(2 \e)
\PlusBreak{ }
  {1 \over 4\e }  {H_g^{(2)} \over \Nc^2} \, 
  + \Ord(\e^0) \,.
\label{TwoloopSplitLeading}
\end{eqnarray}

In the second derivation of the pole terms in $\rsnren^\twoloop(\e)$,
we shall retain the finite terms.  Consider first the one-loop case.
\Eqn{OneloopNormLimit} can be rewritten as
\begin{equation}
\rsnren^{(1)} = \Bigl[  \hat{I}_n^{(1)} + F_n^{(1)} \Bigr] \Big|_{a\parallel b}
 - \hat{I}_{n-1}^{(1)} - F_{n-1}^{(1)} \,,
\label{OneLoopConsistency2}
\end{equation}
which allows one to predict the singular terms in
$\rsnren^{(1)}$ in terms of the collinear behavior of $\hat{I}_n^{(1)}$:
\begin{equation}
\rsnren^{(1)} \Big|_{\e\ {\rm pole}} = 
  \hat{I}_n^{(1)} \Big|_{a\parallel b}
 - \hat{I}_{n-1}^{(1)} \,. 
\label{ConsistencyFinalSingOneLoop}
\end{equation}
But we can also solve~\eqn{OneLoopConsistency2} for the collinear
behavior $F_n^{(1)}|_{a\parallel b}$, 
using also \eqn{ConsistencyFinalSingOneLoop}.
For definiteness, we will assume that $a=1$ and $b=2$.  We have then
\begin{equation}
F_n^{(1)}\Big|_{1||2} = F_{n-1}^{(1)} + \xi^{(1)}(z,s_{nP},s_{P3},s),
\label{Fn1Limit0}
\end{equation}
where
\begin{equation}
\xi^{(1)}(z,s_{nP},s_{P3},s) 
\equiv \rsnren^{(1)} - \rsnren^{(1)} \Big|_{\e\ {\rm pole}} \,.
\label{Fn1Limit1}
\end{equation}
Note that the Mandelstam invariants involving the gluons which
are color-adjacent to $a=1$ and $b=2$ appear, namely gluons $n$ and $3$.
In evaluating \eqn{Fn1Limit1}, $\rsnren^{(1)}|_{\e\ {\rm pole}}$ is given by 
\eqn{ConsistencyFinalSingOneLoop}, including all terms at order
$\e^0$.  

At two loops, the collinear limit as $a\parallel b$ of 
\eqn{Sing2New} is, using \eqn{TwoloopSplit},
\begin{eqnarray}
A_n^{(2)}\big|_{a\parallel b}
           &=& \Bigl[ \hat{I}_n^{(2)} 
                    + \hat{I}_n^{(1)} ( \hat{I}_n^{(1)} + F_n^{(1)} )
                    + F_n^{(2)} \Bigr] \Big|_{a\parallel b}
  \times \Split^\tree \times A_{n-1}^\tree
\nonumber \\
&=& \Bigl[ \hat{I}_{n-1}^{(2)} 
         + \hat{I}_{n-1}^{(1)} ( \hat{I}_{n-1}^{(1)} + F_{n-1}^{(1)} )
         + F_{n-1}^{(2)}
\PlusBreak{ \BiglB }
    \rsnren^{(1)} ( \hat{I}_{n-1}^{(1)} + F_{n-1}^{(1)} )
\PlusBreak{ \BiglB }
    \rsnren^{(2)} \Bigr] \times \Split^\tree \times A_{n-1}^\tree \,.
\label{TwoLoopConsistency}
\end{eqnarray}
Solving \eqn{TwoLoopConsistency} for $\rsnren^{(2)}$, we find 
\begin{eqnarray}
\rsnren^{(2)} &=& \Bigl[ \hat{I}_n^{(2)} 
                    + \hat{I}_n^{(1)} ( \hat{I}_n^{(1)} + F_n^{(1)} )
                    + F_n^{(2)} \Bigr] \Big|_{a\parallel b}
\MinusBreak{ }
   \hat{I}_{n-1}^{(2)} 
 - \hat{I}_{n-1}^{(1)} ( \hat{I}_{n-1}^{(1)} + F_{n-1}^{(1)} )
 - F_{n-1}^{(2)}
 - \rsnren^{(1)} ( \hat{I}_{n-1}^{(1)} + F_{n-1}^{(1)} ) \,.
\label{TwoLoopConsistency2}
\end{eqnarray}
\Eqn{TwoLoopConsistency2} contains the one-loop finite remainder $F_n^{(1)}$
multiplied by a singular factor $\hat{I}_n^{(1)}$.  However, the pole 
terms in $\rsnren^{(2)}$ should not depend on any one-loop finite
parts.  Therefore we use \eqn{OneLoopConsistency2} to eliminate the
$\hat{I}_n^{(1)} \, F_n^{(1)}$ term.  After a little algebra, we obtain
\begin{eqnarray}
\rsnren^{(2)} &=& \Bigl[ \hat{I}_n^{(2)} 
            + ( \rsnren^{(1)} + \hat{I}_{n-1}^{(1)} ) \hat{I}_n^{(1)}
            + F_n^{(2)} - F_{n-1}^{(1)} F_n^{(1)} \Bigr] 
      \Big|_{a\parallel b}
\MinusBreak{ }
   \hat{I}_{n-1}^{(2)} 
 - ( \rsnren^{(1)} + \hat{I}_{n-1}^{(1)} ) \hat{I}_{n-1}^{(1)} 
 - F_{n-1}^{(2)} + ( F_{n-1}^{(1)} )^2.
\label{ConsistencyFinal}
\end{eqnarray}
Now it is clear that the $\e$ pole terms in $\rsnren^{(2)}$ have a universal 
form,
\begin{equation}
\rsnren^{(2)} \Big|_{\e\ {\rm pole}} = 
 \Bigl[ \hat{I}_n^{(2)} 
  + ( \rsnren^{(1)} + \hat{I}_{n-1}^{(1)} ) \hat{I}_n^{(1)}\Bigr] 
   \Big|_{a\parallel b}
 - \hat{I}_{n-1}^{(2)} 
 - ( \rsnren^{(1)} + \hat{I}_{n-1}^{(1)} ) \hat{I}_{n-1}^{(1)} \,.
\label{ConsistencyFinalSing}
\end{equation}
We have checked that the pole terms in this expression are
equivalent to those in~\eqn{TwoloopSplitLeading} for $\Neqfour$ and 
$\Neqone$ super-Yang-Mills theory, and for QCD.
We have also verified that they agree with the singular parts of
the $g\to gg$ splitting amplitudes from our explicit results,
eqs.~(\ref{rSNeq4}), (\ref{rSNeq1ppm}), (\ref{rSNeq1mpm}),
(\ref{QCDppmmpm}) and (\ref{QCDmmm}).
Along with the full color-space discussion in 
appendix~\ref{FullColorCatani},   
this shows that Catani's formula for the singular behavior of 
two-loop amplitudes is completely consistent with the collinear limits.

\subsection{Finite Remainders}

The next step is to use the finite parts of the two-loop
splitting amplitudes to deduce the collinear limits of the 
finite remainders $F_n^{(2)}$ in the color-trivial parts of
Catani's formula.  Again letting
$a=1$ and $b=2$, we rearrange~\eqn{TwoLoopConsistency2}, 
with the help of~\eqn{ConsistencyFinalSing}, to get
\begin{eqnarray}
F_n^{(2)}\Big|_{1||2} 
&=& F_{n-1}^{(2)} 
+ \Bigl( F_n^{(1)} \Big|_{1||2} - F_{n-1}^{(1)} \Bigr) F_{n-1}^{(1)}
+ \biggl( \rsnren^{(2)} - \rsnren^{(2)} \Big|_{\e\ {\rm pole}} \biggr) 
\label{Fn2Limit1} \\
&=& F_{n-1}^{(2)} 
+ \biggl( \rsnren^{(1)} - \rsnren^{(1)} \Big|_{\e\ {\rm pole}} \biggr) F_{n-1}^{(1)}
+ \biggl( \rsnren^{(2)} - \rsnren^{(2)} \Big|_{\e\ {\rm pole}} \biggr), 
\label{Fn2Limit2}
\end{eqnarray}
or
\begin{equation}
F_n^{(2)}\Big|_{1||2}  = F_{n-1}^{(2)} 
+ \xi^{(1)}(z,s_{nP},s_{P3},s)  \,  F_{n-1}^{(1)} 
+ \xi^{(2)}(z,s_{nP},s_{P3},s),
\label{Fn2Limit3}
\end{equation}
where
\begin{equation}
\xi^{(2)}(z,s_{nP},s_{P3},s) 
\equiv \rsnren^{(2)} - \rsnren^{(2)} \Big|_{\e\ {\rm pole}} \,.
\label{Fn2Limit4}
\end{equation}
Here $\rsnren^{(2)}|_{\e\ {\rm pole}}$ is given by 
\eqn{ConsistencyFinalSing}, including all terms at order $\e^0$.  
\Eqn{Fn2Limit3} provides a useful check on finite remainders
of two-loop scattering amplitudes, as any two external gluons
become collinear.

Now we present the values of $\xi^{(1)}$ and $\xi^{(2)}$ for the 
various theories we have been considering.  The values of $\xi^{(1)}$
are:
\begin{eqnarray}
\xi^{(1), \, \Neqfour} &=& 
 {1 \over 2} \biggl( \ln(1-z) \ln\Bigl({ -s_{P3} \over -s }\Bigr) 
                   + \ln(z) \ln\Bigl({ -s_{nP} \over -s }\Bigr) 
                   + \ln(z) \ln(1-z) - \zeta_2 \biggr)
\,,~~~~~~
\label{xi1Neq4}\\
\xi^{(1)\, ++, \, \Neqone} 
&=& \xi^{(1)\, -+, \, \Neqone} =
 \xi^{(1), \, \Neqfour}
  - { b_0^{\Neqone} \over 2 \Nc } ( \ln(z) + \ln(1-z) + \ln(-s) ),
\label{xi1Neq1}\\
\xi^{(1)\, ++, \, {\rm QCD}}
&=&
 \xi^{(1)\, -+, \, {\rm QCD}}
  + {1 \over 6} \Bigl( 1 - { \Nf \over \Nc } \Bigr) z (1-z),
\label{xi1ppQCD}\\
\xi^{(1)\, -+, \, {\rm QCD}} &=&
 \xi^{(1), \, \Neqfour}
   - { b_0 \over 2 \Nc } ( \ln(z) + \ln(1-z) + \ln(-s) ).
\label{xi1mpQCD}
\end{eqnarray}
To get the proper analytic behavior of such expressions, 
one should apply the prescription~(\ref{STimeLike}) 
for logarithms of time-like invariants, after expanding 
the logarithmic ratios,
$\ln((-s_1)/(-s_2)) = \ln(-s_1) - \ln(-s_2)$.

Next we the quote the values of $\xi^{(2)}$.
For $\Neqfour$ super-Yang-Mills theory, $\xi^{(2)}$ obeys
an iterative equation in terms of $\xi^{(1)}$, as a consequence of
\eqn{OneloopTwoloopSplit},
\begin{equation}
\xi^{(2), \, \Neqfour} =
 {1 \over 2} \, \Bigl[ \xi^{(1), \, \Neqfour} \Bigr]^2
- \zeta_2 \, \xi^{(1), \, \Neqfour} - { 11 \over 32 } \, \zeta_4
 \,.
\label{xi2Neq4}
\end{equation}

For $\Neqone$ super-Yang-Mills theory, we first define the auxiliary
function
\begin{eqnarray}
\eta(z,s_{nP},s_{P3},s) &=& 
  - 2 ( \Li_3(z) + \Li_3(1-z) ) 
  + { 1 \over 2 } \Bigl( \ln z  \Li_2(z) + \ln(1-z) \Li_2(1-z) \Bigr)
\MinusBreak{ }
    { 1 \over 2 }\Bigl( \ln z \ln(1-z) - {3\over2} \zeta_2 \Bigr)
                       ( \ln z + \ln(1-z) + \ln(-s) ) 
  + { 85 \over 24} \zeta_3
\MinusBreak{ }
    {1 \over 4} \biggl[ 
      \ln z \ln\Bigl({ -s_{nP} \over -s }\Bigr)
             \Bigl( \ln(-s_{nP}) + 2 \ln(-s) \Bigr)
\PlusBreak{ - {1 \over 4} \bigglB }
    \ln(1-z) \ln\Bigl({ -s_{P3} \over -s }\Bigr)
             \Bigl( \ln(-s_{P3}) + 2 \ln(-s) \Bigr)
\PlusBreak{ - {1 \over 4} \bigglB }
    \ln z \ln(1-z) \Bigl( \ln(-s_{nP}) + \ln(-s_{P3}) - \ln(-s) \Bigr)
\PlusBreak{ - {1 \over 4} \bigglB }
    2 \biggl( \ln^2 z \ln\Bigl({ -s_{nP} \over -s }\Bigr)
           + \ln^2(1-z) \ln\Bigl({ -s_{P3} \over -s }\Bigr) \biggr)
   \biggr]
 \,. 
\label{etadef}
\end{eqnarray}
Then the two functions required in \eqn{Fn2Limit3} are given by
\begin{eqnarray}
 \xi^{(2) \, ++, \Neqone} &=&
     \xi^{(2), \, \Neqfour}
   + { b_0^{\Neqone} \over \Nc } \, \eta
   + { 1 \over 8 } \biggl\{ 
       42 \, \xi^{(1), \, \Neqfour}
\PlusBreak{ ~~~~~~~~~~ }
       3 \Bigl( { b_0^{\Neqone} \over \Nc } \Bigr)^2 
         \Bigl[ ( \ln z + \ln(1-z) + \ln(-s) )^2 
              - {8\over3} \ln z \ln(1-z) \Bigr]
\PlusBreak{ ~~~~~~~~~~ }
      {75 \over 4} \zeta_2
    - 6 \, ( \ln z + \ln(1-z) + \ln(-s) ) 
    - { 40 \over 3 } \biggr\}
 \,,
\label{xi2ppNeq1}\\
 \xi^{(2) \, -+, \Neqone} &=&
 \xi^{(2) \, ++, \Neqone} 
+ {3 \over 4} \Biggl\{
    2 \Bigl( \Li_3(1-z) - \zeta_3 \Bigr) 
        - \ln(1-z) \Bigl( \Li_2(1-z) - \zeta_2 \Bigr) 
\MinusBreak{ ~~~~~~~~~~~~~~~~~~~~ }
         { z \over 1-z } \biggl[ 2 \Bigl( \Li_3(z) - \zeta_3 \Bigr) 
                    - \ln(z) \Bigl( \Li_2(z) - \zeta_2 \Bigr) \biggr] 
  \Biggl\}
 \,.~~~~
\label{xi2mpNeq1}
\end{eqnarray}
\Eqn{xi2mpNeq1} follows directly from \eqn{rSNeq1mpm},
because \eqn{ConsistencyFinalSing} for the subtraction term 
$\rsnren^{(2)}|_{\e\ {\rm pole}}$ is the same for $++$ and $-+$.

For QCD, we define one more auxiliary function,
\begin{eqnarray}
 \gamma^{{\rm QCD}} &=&
     \xi^{(2), \, \Neqfour}
   + { b_0 \over \Nc } \, \eta
   + { 1 \over 8 } \biggl\{ 
       2 \Bigl( {83\over3} + {2\over3} \delta_R 
              - {64\over9} {\Nf \over \Nc} 
              + {4\over9} {\Nf^2 \over \Nc^2} \Bigr)  \xi^{(1), \, \Neqfour}
\PlusBreak{ }
    3 \Bigl( { b_0 \over \Nc } \Bigr)^2 
         \Bigl[ ( \ln z + \ln(1-z) + \ln(-s) )^2 
              - {8\over3} \ln z \ln(1-z) \Bigr]
\PlusBreak{ }
      \Bigl( {925\over36} + {5\over12} \delta_R 
           - {15\over2} {\Nf \over \Nc}
           + {5\over9} {\Nf^2 \over \Nc^2 } \Bigr) \zeta_2
\MinusBreak{ }
      \Bigl( {34\over3} - {13\over3} {\Nf \over \Nc}
           + {\Nf \over \Nc^3} \Bigr) \, ( \ln z + \ln(1-z) + \ln(-s) ) 
\MinusBreak{ }
      {1169\over81} + {1\over3} \delta_R 
    + \Bigl( {89\over81} - {8\over27} \delta_R \Bigr) {\Nf \over \Nc}
    \biggr\}
 \,.
\label{gammaQCD}
\end{eqnarray}
In terms of this function, we have
\begin{eqnarray}
 \xi^{(2) \, ++, {\rm QCD}} &=&
 \gamma^{\rm QCD}
 + z (1-z) \biggl\{ 
     {1\over6} \Bigl( 1 - {\Nf \over \Nc}\Bigr) 
       \Bigl( \xi^{(1), \, \Neqfour} 
\MinusBreak{ ~~~~~~~~~~~~~~~~~~~~~~~~~~~~ }
       {b_0 \over 2 \Nc} ( \ln z + \ln(1-z) + 3 \ln(-s) )
     + {445\over36} + {\delta_R\over6} \Bigr)
\MinusBreak{ }
       {1\over8} \biggl[ 1 + {\Nf \over \Nc^3} 
                       + {20\over27} \Bigl( 1 - {\Nf^2 \over \Nc^2} \Bigr) 
                 \biggr] \biggr\}
\PlusBreak{ }
        {1\over12} \Bigl( 1 - {\Nf \over \Nc}\Bigr) 
             \Bigl( z \ln z + (1-z) \ln(1-z) \Bigr)
 \,,
\label{xi2ppQCD}\\
 \xi^{(2) \, -+, {\rm QCD}} &=&
 \gamma^{\rm QCD}
    - {z\over12} \biggl[ \Bigl( 1 - {\Nf \over \Nc}\Bigr) 
           { 12 - 21 z + 11 z^2 \over (1-z)^3 }
       + {9 \over 1-z} {\Nf \over \Nc} \biggr]
\TimesBreak{ \gamma^{\rm QCD} - {z\over12} }
     \biggl[ 2 \Bigl( \Li_3(z) - \zeta_3 \Bigr) 
                    - \ln z \Bigl( \Li_2(z) - \zeta_2 \Bigr) \biggr]
\PlusBreak{ }
      {b_0 \over 2 \Nc } \biggl[ 2 \Bigl( \Li_3(1-z) - \zeta_3 \Bigr) 
            - \ln(1-z) \Bigl( \Li_2(1-z) - \zeta_2 \Bigr) \biggr]
\PlusBreak{ }
      {1\over6} \Bigl( 1 - {\Nf \over \Nc}\Bigr) 
       \biggl[ { z \over (1-z)^2 } \Bigl( \Li_2(1-z) - 2 \zeta_2 \Bigr)
    - {1\over2} \Bigl( { z \ln z \over 1-z } - \ln(1-z) \Bigr) 
       \biggr]
 \,.~~~~~~~~~
\label{xi2mpQCD}
\end{eqnarray}

We can check some limiting properties of these results
as $z\to0$ and $z\to1$, using simple facts about soft
limits of amplitudes.  For example, in the limit
$z \to 0$, leg 1 becomes soft.  The helicity of the hard leg
should be conserved in the soft limit, and the limit should be 
independent of the helicity of the soft leg.   
Thus the cases $P^+ \to 1^+2^+$ ($++$) and 
$P^+ \to 1^-2^+$ ($-+$) should behave identically as $z\to0$.
The tree splitting amplitudes~(\ref{TreeSplitAmplitudepp}) and 
(\ref{TreeSplitAmplitudemp}) are the same in this limit (up to an 
external phase associated with the soft external state).
Hence the $\rsn$ factors, and also the Catani-subtracted
$\xi$ functions, for $++$ and $-+$ should behave identically as $z\to0$.
This property is obvious for $\Neqfour$ super-Yang-Mills theory;
the two $\rsn$ factors are identical for all $z$ due to the $\Neqfour$ 
supersymmetry Ward identity.  For pure $\Neqone$ super-Yang-Mills theory,
one can inspect \eqn{rSNeq1mpm} or \eqn{xi2mpNeq1},
recalling that $\Li_n(1) = \zeta_n$, to see that
the quantity in braces indeed vanishes as $z\to0$.
For QCD, the $\rsn^{(2)}$ factors in \sect{QCDResults} are not written 
in a particularly convenient way for checking the limit.  However, one can 
easily compare $\xi^{(2) \, ++, {\rm QCD}}$ in \eqn{xi2ppQCD} with 
$\xi^{(2) \, -+, {\rm QCD}}$ in \eqn{xi2mpQCD}.
They do approach the same limit as $z\to0$, namely the
limiting behavior of $\gamma^{\rm QCD}$. 

For the $++$ case, the limit $z\to1$ is the same as the limit $z\to0$.
For the $-+$ case, in the limit $z\to1$ the helicity of the hard leg,
now leg 1, flips from $+$ to $-$; hence this splitting amplitude should
be suppressed.  \Eqn{TreeSplitAmplitudemp} shows that it is suppressed 
by two powers of $1-z$ at tree-level.
In the $\Neqfour$ and $\Neqone$
supersymmetric cases, there is no additional $1/(1-z)$ singularity
in the $\rsn$ or $\xi$ factors.  (There is an apparent $1/(1-z)$ 
singularity in the quantity in braces in \eqns{rSNeq1mpm}{xi2mpNeq1},
but again recalling that $\Li_n(1) = \zeta_n$, one sees that it cancels.)
So, up to logs, the soft behavior is the same at the loop-level
as at tree-level.
In the case of QCD, the limiting behavior of $\xi^{(2) \, -+, {\rm QCD}}$
in \eqn{xi2mpQCD} looks quite singular as $z\to1$, since powers
of $1/(1-z)^3$ and $1/(1-z)^2$ appear.  However, these cancel, 
and the actual behavior is
\begin{equation}
 \xi^{(2) \, -+, {\rm QCD}} \to 
    - {1\over 12} \left(1- {\Nf\over \Nc}\right) {1 \over 1-z } 
    + \cdots \,.
\label{xi2mpQCDzto1}
\end{equation}
Taking into account the $(1-z)^2$ behavior of the tree-level 
splitting amplitude, the one power of $1/(1-z)$ in \eqn{xi2mpQCDzto1} 
still means that the soft limit $z\to1$ for $-+$ is suppressed by a 
power of $1-z$, relative to that of $++$, as expected from the helicity 
flip on the hard line.

For the term corresponding to the helicity-flip splitting amplitude,
$P^+ \to 1^- 2^-$, we should not remove the tree-amplitude factors.
Instead we write
\begin{equation}
A_n^{(2){\rm fin}}\Big|_{1||2} 
= \Split_{R,-}^{(1)}(z;1^-,2^-) A_{n-1}^{(1){\rm fin}}
 + \, \sqrt{z(1-z)} { \spa{1}.{2} \over {\spb{1}.{2}}^2 }
   \ \xi^{(2) \, --, {\rm QCD}} \,  A_{n-1}^\tree \,,
\label{Fn2FlipLimit1}
\end{equation}
where
\begin{eqnarray}
\xi^{(2) \, --, {\rm QCD}} &=&
   {1\over6} \Bigl( 1 - {\Nf \over \Nc}\Bigr) 
     \biggl[ \xi^{(1), \, \Neqfour}
           + { \ln z \over 1-z } + { \ln(1-z) \over z }
\MinusBreak{ {1\over6} \Bigl( 1 - {\Nf \over \Nc}\Bigr) }
             {b_0 \over 2 \Nc} ( \ln z + \ln(1-z) + 3 \ln(-s) ) 
           + {199\over18} + {\delta_R \over 6}
            - {5\over9} {\Nf \over \Nc} \biggr]
\MinusBreak{ }
    {1\over8} { \Nc^2 + 1 \over \Nc^3 } \Nf \,.
\label{Fn2FlipLimit2}
\end{eqnarray}
For this helicity configuration, the tree-level splitting amplitude
vanishes, so the one-loop renormalization is trivial:
$\Split_{-R}^{(1)}(z;1^-,2^-) = \Split_{-}^{(1)}(z;1^-,2^-)$.
The expectation as $z\to0$ (or equivalently, $z\to1$) is that
$\xi^{(2) \, --, {\rm QCD}}$ should have no power-law $1/z$ singularity,
since its prefactor $\sqrt{z(1-z)}$ in \eqn{Fn2FlipLimit1}
has only one power of $z$ suppression relative to the $++$ case.  
Indeed, in \eqn{Fn2FlipLimit2} $\ln(1-z)/z$ is finite as $z\to0$.

\Eqns{Fn1Limit0}{Fn2Limit3} govern the collinear behavior of finite 
remainders $F_n^{(L)}$ for the case where only one intermediate 
helicity $\lambda$ can contribute.  If both helicities contribute, 
one needs to sum over the two, according 
to~\eqns{OneloopSplit}{TwoloopSplit}.  Before doing this,
it is best to multiply back by the tree amplitudes 
$A_{n-1}^\tree(\lambda)$, because they are different 
for the two terms.

\section{Dressing with Color}
\label{FullColorSection}

In this section we present the full color structure in the collinear
limits.  We will do this in the context of the trace-based
color decomposition discussed in section~\ref{SplittingSection}
(and reviewed in refs.~\cite{MPReview,LoopReview}).
An alternative, color-space language has been used by Catani to
predict~\cite{CataniTwoloop} and describe the infrared structure of
two-loop amplitudes.  In \app{CataniCollinearSection} we re-express
the collinear behavior in the color-space language, in order
to demonstrate that the divergent parts of our
splitting amplitudes are fully compatible with the structure of
infrared singularities predicted by Catani~\cite{CataniTwoloop}.  
The color dressing here applies equally well to renormalized and
unrenormalized amplitudes, but in the discussion in
\app{CataniCollinearSection}, all quantities are understood to be
renormalized.  

At tree level the trace-based color decomposition for an $n$-gluon 
amplitude is 
\begin{equation}
{\cal A}_n^{\tree \, a_1 a_2  \ldots a_n} = 
 \sum_{\sigma\in S_n/Z_n} 
\Tr(T^{a_{\si(1)}} \ldots T^{a_{\si(n)}})
 A^\tree_n(\si(1),\ldots,\si(n)) \,, 
\label{TreeAmplitude}
\end{equation}
where the $A_n^\tree$ are tree-level color-ordered partial amplitudes
and the notation is defined below \eqn{LeadingColorAmplitude}.  On the
left-hand side we have suppressed the dependence on helicities and momenta
but have left the color dependence explicit, since that is what we focus on
in this section.   

The fully color-dressed
tree amplitude has the following factorization property as the
momenta of legs $1$ and $2$ become collinear,
\begin{eqnarray}
{\cal A}_n^{\tree\, a_1 a_2  \ldots a_n} 
\; \mathop{\longrightarrow}^{1 \parallel 2} &&
i  \, \tilde f^{a_1 a_2 a_P}
    \Split^\tree_{-\lambda}(z; 1^{\lambda_1}, 2^{\lambda_2})\,
{\cal A}_{n-1}^{\tree\, a_P a_3 \ldots a_n} \,,
\label{SplitTreeFullColor}
\end{eqnarray}
where $\Split_{-\lambda}^\tree$ is the tree-level splitting amplitude
(\ref{TreeSplit}) defined for the color-ordered amplitude, and there is
an implicit sum over the helicity $\lambda$ of the intermediate
state.   Here $\tilde f^{a_1 a_2 a_P}$ is the ${\rm SU}(\Nc)$ 
structure constant corresponding to our normalization of the generators, 
$\Tr(T^a T^b) = \delta^{ab}$: 
\begin{equation}
\tilde f^{a_1 a_2 a_P} = -{i}
        \Bigl( \Tr\bigl( T^{a_1} T^{a_2} T^{a_P} \bigr)
                       - \Tr\bigl( T^{a_2} T^{a_1} T^{a_P} \bigr) \Bigr) \,,
\label{FundConvert}
\end{equation}
so that $\tilde f^{a b c}  = \sqrt{2} f^{abc}$, where $f^{abc}$ is the 
structure constant with conventional $T^a$ normalizations.

By inserting \eqn{TreeAmplitude} into both sides of 
\eqn{SplitTreeFullColor}, and using ${\rm SU}(\Nc)$ Fierz identities on the
right-hand side, we can see that this equation is
equivalent to the collinear behavior of color-ordered amplitudes 
given in \eqn{TreeSplit}.
A given term in the amplitude (\ref{TreeAmplitude}) will 
contribute to the collinear limit only when the collinear 
legs $1$ and $2$ are cyclicly adjacent in the associated color trace.
Alternatively, we can derive the color structure of 
\eqn{SplitTreeFullColor} directly 
from the fact that the only diagrams contributing to the collinear limit
are of the type shown in \fig{CollinearTreeFigure}, where
$a=1$ and $b=2$. The factor of $\tilde f^{a_1 a_2 a_P}$ is precisely 
the color factor of the vertex joining legs $1,2$ and $P$.

At one loop the full color decomposition of an $n$-gluon amplitude is,
\begin{equation}
{\cal A}_n^{\oneloop\, a_1 a_2 \ldots a_n} =
\sum_{c=1}^{\lfloor{n/2}\rfloor+1} \hskip -.2 cm
      \sum_{\sigma \in S_n/S_{n;c}} \hskip -.2 cm 
     \Gr_{n;c} ( \si(1), \si(2), \ldots, \si(n) )\, 
      A_{n;c}( \si(1), \si(2), \ldots, \si(n)) \,,
\label{OneLoopGenColor}
\end{equation}
where ${\lfloor{x}\rfloor}$ is the largest integer less than or equal
to $x$. The leading-color structure factor
\begin{equation}
\Gr_{n;1}(1,2, \ldots, n) = \Nc \Tr ( T^{a_1}\ldots T^{a_n} ) \,, 
\end{equation}
is just $\Nc$ times the tree color factor, and the subleading-color
structures are given by
\begin{equation}
\Gr_{n;j}(1,2, \ldots, n) = \Tr ( T^{a_1}\ldots T^{a_{j-1}} )\,
\Tr(T^{a_j}\ldots T^{a_n}) \,, \quad j > 1 \,.
\end{equation}
Again $S_n$ is the set of all permutations of $n$ objects, 
and $S_{n;c}$ is the subset leaving $\Gr_{n;c}$ invariant. 
In the trace-based color decomposition,
fundamental-representation quark loops will contribute only to
$A_{n;1}$, with a relative factor of $\Nf/\Nc$ compared to 
adjoint-representation gluon loops.  

At one loop, the subleading-color amplitudes $A_{n;c}$ are
completely determined in terms of the leading-color ones $A_{n;1}$ (see
eq.~(7.2) of ref.~\cite{Neq4Oneloop}).   The collinear behavior of
the leading-color amplitudes was given in \eqn{OneloopSplit}, where 
the notation `$A_n$' was used instead of `$A_{n;1}$'.   The collinear
behavior of the subleading-color amplitudes can be determined 
from the previously-mentioned relation, or from the observation that the
two-particle collinear limit cannot reduce a pair of traces to a single
trace. Therefore the tree-level $(n-1)$-point amplitude 
$A_{n-1}^\tree$ cannot enter into the limit.
This implies that the one-loop splitting amplitude cannot appear either,
and so the limit is (for $c > 1$),
\begin{eqnarray}
A_{n;c}^{\oneloop}(\ldots,a^{\lambda_a},b^{\lambda_b},\ldots)
 \mathop{\longrightarrow}^{a \parallel b} &&
\sum_{\lambda=\pm} 
  \Split^{\tree}_{-\lambda}(z; a^{\lambda_a},b^{\lambda_b})\,
      A_{n-1;c-1}^{\oneloop}(\ldots,P^\lambda,\ldots) \, , 
      \quad  a,b<c \,, \nonumber \\
A_{n;c}^{\oneloop}(\ldots,a^{\lambda_a},b^{\lambda_b},\ldots)
 \mathop{\longrightarrow}^{a \parallel b} &&
\sum_{\lambda=\pm} 
  \Split^{\tree}_{-\lambda}(z; a^{\lambda_a},b^{\lambda_b})\,
      A_{n-1;c}^{\oneloop}(\ldots,P^\lambda,\ldots) \, , 
   \hskip .8 cm     a,b\ge c \,. \hskip 1 cm 
\label{OneloopSplitsublead}
\end{eqnarray}
Legs $a$ and $b$ must be cyclicly adjacent within the same color
trace, otherwise the collinear limit is finite.  In particular, the
collinear limit is finite if the two legs lie in different traces.

We can reassemble these properties of $A_{n;c}^{\oneloop}$ into a 
description of the collinear behavior of the full color-dressed amplitude, 
\begin{eqnarray}
{\cal A}_n^{\oneloop\, a_1 a_2 a_3 \ldots a_n} 
\mathop{\longrightarrow}^{1 \parallel 2} &&
   i \tilde f^{a_1 a_2 a_P}
\biggl[ 
\Split^\tree_{-\lambda}(z; 1^{\lambda_1}, 2^{\lambda_2})\, 
 {\cal A}_{n-1}^{\oneloop\, a_P a_3 \ldots a_n}  
\PlusBreak{ i \sqrt{2} f^{a_1 a_2 a_P} }
  \Nc \,
\Split^\oneloop_{-\lambda}(z; 1^{\lambda_1}, 2^{\lambda_2}) \, 
 {\cal A}_{n-1}^{\tree\, a_P a_3 \ldots a_n}  
\biggr] \,, \hskip 1 cm 
\label{SplitOneloopFullColor}
\end{eqnarray}
where $\Split^\oneloop_{-\lambda}(z; 1^{\lambda_1}, 2^{\lambda_2})$ are
precisely the color-stripped splitting
amplitudes~(\ref{OneloopSplitExplicitpp})--(\ref{OneloopSplitExplicitNeq4}).
Note that the $c>1$ contributions in \eqn{OneloopSplitsublead}
all go into assembling ${\cal A}_{n-1}^\oneloop$ in the first term 
in brackets.
We can also derive \eqn{SplitOneloopFullColor} directly from
argumentation using the unitarity sewing rules (or by using 
light-cone gauge, ignoring prescription issues).  The diagrams in
\fig{CollinearOneloopFigure} correspond to a factorization of the
process including color indices.  The first term in the brackets
corresponds to the left diagram, and the second term to the right
diagram.  The color factor in the first term is precisely the
tree-level one. The color factor in the second term is that of a
one-loop vertex, $\Nc \, \tilde f^{a_1 a_2 a_P}$ in the pure-glue case.  
(From the diagrammatic point of view, there are three types of color
contributions obtained from color dressing the diagrams in
\fig{LCGaugeOneloopDiagFigure}.  However, all these color factors are
proportional to $\Nc \, \tilde f^{a_1 a_2 a_P}$. Including fermions
in the loop gives the $\Nf/\Nc$ terms in \eqn{OneloopSplitExplicitpp}.)

If we examine the cuts of a color-stripped $1\rightarrow 2$ splitting
amplitude, each contribution has the product of a $(j+2)$-point
scattering amplitude on the left and a $1 \to j$ lower-loop splitting 
amplitude on the right, where $j$ is the number of particles crossing 
the cut. (See, for example, \fig{TwoLoopSplittingCutFigure}.)  
We can exchange the two final-state
particles by reflecting through the horizontal axis.  This produces a
factor of $(-1)^{j+2}$ on the left and $(-1)^{j+1}$ on the right
because of the properties of the amplitudes under
reflection~\cite{MPReview,OneloopColor}.  If the two final-state
particles are gluons we get the same number of Fermi minus signs on
the left and right from interchanging cut internal fermions, if there are any.
Overall we always get a factor of $-1$; that is, the color-stripped
splitting amplitude is anti-symmetric when we exchange the two
arguments, including their helicities and $z \leftrightarrow (1-z)$.
Bose symmetry then implies that the color factor must also be
antisymmetric. The color-dressed splitting amplitude must be
proportional to $f^{a_1 a_2 a_P}$, because there is no other
antisymmetric invariant tensor. This argument holds to all loop orders.

Note from eqs.~(\ref{OneloopHelFlip}), (\ref{OneloopSplitExplicitpp})
and (\ref{OneloopSplitExplicitmp}) that there are no subleading-color
corrections to the pure-glue ($\Nf=0$) contributions 
to the one-loop splitting amplitude $\Split^\oneloop$, 
which appears in the full-color collinear 
limit~(\ref{SplitOneloopFullColor}).  
To understand this fact, and the factor of $\Nc$ in front of the 
splitting amplitude, consider the color factors in a diagrammatic 
calculation. One can use the Jacobi
identity to show that the result of the color algebra in any diagram
is a linear combination of the results for its various possible parent
diagrams.  It is therefore sufficient to consider the parent diagrams.
At one loop, for pure-glue contributions, there is only one kind of
parent diagram, the triangle diagram show in
\fig{LCGaugeOneloopDiagFigure}(a).  Now, as far as the color algebra
is concerned, any triangle subdiagram can replaced by a factor of
$-{1\over2} C_A$ times a three-point color vertex. Similarly, any bubble
subdiagram can be replaced by a factor of $-C_A$.  Thus at one loop
we just obtain an overall coefficient of $C_A = \Nc$, up to
$\Nc$-independent factors.

Now we turn to the two-loop case.  The color decomposition 
generalizes in an obvious way,
\begin{eqnarray}
{\cal A}_n^{(2) \, a_1 a_2 \ldots a_n} &=&
\sum_{c_1 = 1}^{^{\lfloor{n/3}\rfloor+1}}
  \sum_{ c_2 \ge 2 c_1 -1}^{^{\lfloor{ (n + c_1 + 1)/ 2}\rfloor}}
     \sum_{\sigma \in S_n/S_{n;c_1,c_2}}
    \Gr_{n;c_1;c_2} (\si(1), \si(2), \ldots, \si(n) )
\TimesBreak{ ~~~~~~~~~~~~~~~~~~~~~~~~~~}
    A_{n;c_1;c_2}(\si(1), \si(2), \ldots, \si(n)) \,,
\label{TwoLoopGenColor}
\end{eqnarray}
where
\begin{equation}
\Gr_{n;c_1;c_2}(1,2, \ldots, n) = \Tr (T^{a_1}\ldots T^{a_{c_1-1}} ) \,
\Tr( T^{a_{c_1}}\ldots T^{a_{c_2-1}}) \,
\Tr( T^{a_{c_2}}\ldots T^{a_{n}}) \,,
\end{equation}
and $\sigma$ runs over the set of permutations $S_n$, modulo those
in $S_{n;c_1;c_2}$ which leave $\Gr_{n;c_1;c_2}$ invariant.
We identify $\Tr(1) = \Nc$ so, for example, 
\begin{eqnarray}
\Gr_{n;1;1}(1,2,\ldots, n) &=& \Nc^2 \, \Tr( T^{a_1}\ldots T^{a_n} ) \,, 
\\
\Gr_{n;1;j}(1,2, \ldots, n) &=& \Nc \Tr (T^{a_1}\ldots T^{a_{j-1}} )
                              \Tr (T^{a_j}\ldots T^{a_{n}} ) \,,
\qquad j > 1.
\end{eqnarray}
The color-ordered amplitudes contain terms which depend on the number
of quark flavors, for example, with factors $\Nf/\Nc$ and
$(\Nf/\Nc)^2$.  In addition, at two loops $A_{n;1;1}$ also contains
terms of order $1/\Nc^2$ arising from both planar and non-planar
diagrams.  In contrast, while the two-loop gluon splitting amplitude
does contain contributions of $\Ord(\Nf/\Nc)$, $\Ord(\Nf^2/\Nc^2)$,
and $\Ord(\Nf/\Nc^3)$, as we shall see below, it contains no
contributions of $\Ord(1/\Nc^2)$.

The collinear behavior of the leading-color amplitudes $A_{n;1;1}$ was
given in \eqn{TwoloopSplit}, where the notation `$A_n$' was used
instead of `$A_{n;1;1}$'.  At two loops the subleading-color amplitudes
cannot be determined solely from the leading-color ones.  However,
just as at one loop, consideration of the trace structure leads to 
the following collinear limits for the subleading-color amplitudes.
The double-trace partial amplitudes behave as,
\begin{eqnarray}
A_{n;1;c}^{\twoloop}(\ldots,a^{\lambda_a},b^{\lambda_b},\ldots)
 \mathop{\longrightarrow}^{a \parallel b} &&
\sum_{\lambda=\pm} \biggl(
  \Split^{\tree}_{-\lambda}(z; a^{\lambda_a},b^{\lambda_b})\,
      A_{n-1;1;c-1}^{\twoloop}(\ldots,P^\lambda,\ldots) 
\PlusBreak{\sum\biggl()}
 \Split^{\oneloop}_{-\lambda}(z; a^{\lambda_a},b^{\lambda_b})\,
      A_{n-1;c-1}^{\oneloop}(\ldots,P^\lambda,\ldots) \biggr) \, , 
      \quad  a,b<c, \nonumber \\
A_{n;1;c}^{\twoloop}(\ldots,a^{\lambda_a},b^{\lambda_b},\ldots)
 \mathop{\longrightarrow}^{a \parallel b} &&
\sum_{\lambda=\pm} \biggl(
  \Split^{\tree}_{-\lambda}(z; a^{\lambda_a},b^{\lambda_b})\,
      A_{n-1;1;c}^{\twoloop}(\ldots,P^\lambda,\ldots) 
\PlusBreak{\sum\biggl()}
 \Split^{\oneloop}_{-\lambda}(z; a^{\lambda_a},b^{\lambda_b})\,
      A_{n-1;c}^{\oneloop}(\ldots,P^\lambda,\ldots) \biggr)\,, 
   \hskip .8 cm     a,b\ge c.
\nonumber\\
&&\hskip-1cm
\label{TwoloopSplitsubleadA}
\end{eqnarray}
while the  triple-trace partial amplitudes behave as,
\begin{eqnarray}
A_{n;c_1;c_2}^{\twoloop}(\ldots,a^{\lambda_a},b^{\lambda_b},\ldots)
 &&\mathop{\longrightarrow}^{a \parallel b} 
\nonumber\\ &&\hskip -15mm
\sum_{\lambda=\pm} 
  \Split^{\tree}_{-\lambda}(z; a^{\lambda_a},b^{\lambda_b})\,
      A_{n-1;c_1-1;c_2-1}^{\twoloop}(\ldots,P^\lambda,\ldots) \, , 
      \quad  a,b<c_1 \,, \nonumber \\
A_{n;c_1;c_2}^{\twoloop}(\ldots,a^{\lambda_a},b^{\lambda_b},\ldots)
 &&\mathop{\longrightarrow}^{a \parallel b} 
\nonumber\\ &&\hskip -15mm
\sum_{\lambda=\pm} 
  \Split^{\tree}_{-\lambda}(z; a^{\lambda_a},b^{\lambda_b})\,
      A_{n-1;c_1;c_2-1}^{\twoloop}(\ldots,P^\lambda,\ldots) \, , 
   \hskip .8 cm     c_1 \le a,b <  c_2 \,, \hskip 1 cm \nonumber \\
A_{n;c_1;c_2}^{\twoloop}(\ldots,a^{\lambda_a},b^{\lambda_b},\ldots)
 &&\mathop{\longrightarrow}^{a \parallel b}
\nonumber\\ &&\hskip -15mm
\sum_{\lambda=\pm} 
  \Split^{\tree}_{-\lambda}(z; a^{\lambda_a},b^{\lambda_b})\,
      A_{n-1;c_1;c_2}^{\twoloop}(\ldots,P^\lambda,\ldots) \, , 
   \hskip 12mm     a,b\ge c_2 \,. \hskip 2mm 
\end{eqnarray}
As at one loop, legs $a$ and $b$ must be cyclicly adjacent within the
same color trace, otherwise the collinear limit is finite.  

At two loops, the generalization of the splitting amplitude from the
leading-color structure to the fully color-dressed version would seem 
to be more complicated, because of the appearance of non-planar 
contributions, such as the ones depicted in \fig{NonPlanarColor}.  
However, as already mentioned in \sect{SplittingIntegrandsSection}, 
all such non-planar contributions vanish for the $g \rightarrow gg$ case.
The fully color-dressed splitting behavior is given simply in terms of the
color-stripped splitting amplitudes of \sect{ResultSection} as,
\begin{equation}
{\cal A}_n^{\twoloop\, a_1 a_2 a_3 \ldots a_n} 
\, \mathop{\longrightarrow}^{1 \parallel 2} \, 
   i \tilde f^{a_1 a_2 a_P} 
\biggl[ \sum_{l=0}^2
 \Nc^l \, 
\Split^\lloop_{-\lambda}(z; 1^{\lambda_1}, 2^{\lambda_2})\, 
 {\cal A}_{n-1}^{(2-l)\, a_P a_3 \ldots a_n} \biggr] \,. \hskip 1 cm \null 
\label{SplitLloopFullColor}
\end{equation}

The proportionality of the color dressing to $\tilde f^{a_1 a_2 a_P}$
follows from the diagrammatic argument given after 
\eqn{SplitOneloopFullColor}, generalized to two loops.  To
see why the pure-glue contributions to the splitting amplitude just
have an overall factor of $\Nc^2$ in front
(as evidenced by \eqns{QCDppmmpm}{QCDmmm}), consider again the parent
diagrams.  We have already seen in \sect{SplittingIntegrandsSection}
that the non-planar diagrams do not contribute at two loops to the
splitting amplitude to gluons, so we focus on the planar diagrams.  At
two loops there are four kinds of planar parent diagrams.  Each of
these has either a bubble or triangle subdiagram (the two parents with
triangle subdiagrams are shown in \fig{PlanarTriFigure}).  Upon
replacing the subdiagram factor of $-C_A$ or $-{1\over2} C_A$ times a
vertex, respectively, we are left with a triangle diagram, which
generates a second factor of $C_A$, as at one loop.  All pure-glue
contributions are therefore homogeneous in $\Nc$ of degree two.  
This latter argument does not extend to higher loops.  Beginning
at three loops there are pure-glue diagrams, both planar and non-planar, 
which have no surviving triangle or bubble subdiagrams at some stage 
of this reduction.  The simplest example is depicted in \fig{ThreeLoopColor}.
Such diagrams generally give rise to $1/\Nc^2$ suppressed terms.
They correspond to the existence of other ${\rm SU}(\Nc)$ group 
invariants, such as $(d^{abc})^2$, where $d^{abc}$ is the fully 
symmetrized trace of three generator matrices, which are not 
homogeneous in $\Nc$.  Of course, the fermion-loop terms
in the $g\to gg$ splitting amplitudes have $1/\Nc^2$ suppressed 
contributions already at two loops, due to the existence of
both $C_A$ and $C_F$ Casimir invariants.

%
\FIGURE[t]{
{\epsfxsize 4.0 truein \epsfbox{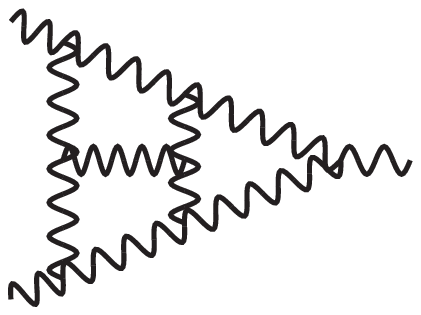}}
\caption{A diagram contributing to the three-loop $g \to gg$ splitting 
amplitude which has subleading-color, $1/\Nc^2$ suppressed terms
in its color factor.}
\label{ThreeLoopColor}}

\section{Conclusions and Outlook}

In this paper, we computed the behavior of a general two-loop
amplitude in massless gauge theory as the momenta of two external
gluons become collinear. (The behavior involving collinear fermions
will be presented elsewhere~\cite{FermionSplit}.) This behavior is
universal, and accordingly governed by a set of $1\rightarrow 2$
splitting amplitudes.  For the $g \rightarrow gg$ splitting there are
three independent helicity configurations (the remainder are related
by symmetry or parity).  One of these configurations vanishes at tree
level.  In an $\Neqfour$ supersymmetric theory, the ratio of a
non-vanishing two-loop splitting amplitude to the corresponding
tree-level amplitude is helicity-independent, and given by
\eqn{rSNeq4}.  In pure $\Neqone$ super-Yang-Mills theory, the helicity
configuration that vanishes at tree level also vanishes to all orders
in perturbation theory.  The ratios for the other two cases are given by
\eqns{rSNeq1ppm}{rSNeq1mpm}.  In QCD, all three helicity amplitudes
are non-vanishing; the ratios to the tree for two of them are given by
\eqn{QCDppmmpm}, and the splitting amplitude for the remaining
configuration by \eqn{QCDmmm}.  The finite remainders of amplitudes,
after subtraction of poles predicted by Catani's formula, also have
simple behavior in collinear limits in the `color-trivial' case.
These limits are given at one and two loops by
\eqns{Fn1Limit0}{Fn2Limit3}, respectively.  The functions
entering these limits are given by
eqs.~(\ref{xi1Neq4})--(\ref{xi1mpQCD}), (\ref{xi2Neq4}),
(\ref{xi2ppNeq1}), (\ref{xi2mpNeq1}),
and~(\ref{xi2ppQCD})--(\ref{Fn2FlipLimit2}).

These splitting amplitudes can be used as a check on 
calculations of higher-point two-loop QCD amplitudes; the collinear
limits of these amplitudes must satisfy \eqns{QCDppmmpm}{QCDmmm}.  
One can also apply the checks to the finite terms after subtraction of
the poles predicted by Catani's formula, using \eqn{Fn2Limit3}.
As mentioned previously, the compatibility of the divergent parts 
of our splitting amplitudes with Catani's formula provides a 
inductive proof of the latter, modulo some assumptions about the
analytic behavior of the $\e$-singular terms in amplitudes.

The two-loop splitting amplitudes are also one of the ingredients
required for computing the NNLO corrections to the Altarelli--Parisi
kernel in an infrared approach.  At NLO, one would add two 
terms~\cite{KosowerUwerSplit}: the one-loop $1\rightarrow 2$
splitting amplitude interfered with its tree counterpart; and the
tree-level $1\rightarrow3$ splitting amplitude squared, integrated
over the phase space of the unobserved partons.  At NNLO, there are
three ingredients:  (a) the $1\rightarrow 2$ splitting amplitude 
computed here, interfered with its tree counterpart; 
(b) the interference of the one-loop~\cite{CDFR}
and tree-level~\cite{CampbellGlover,CataniGrazzini} 
$1\rightarrow3$ splitting amplitudes, integrated
over the unobserved phase space; and (c) the $1\rightarrow4$ 
splitting amplitude squared~\cite{DDFM}, integrated over the 
four-particle unobserved phase space.

In $\Neqfour$ supersymmetric gauge theories, the splitting amplitude
has a remarkable property: it can be written as a polynomial in the
one-loop and tree-level splitting amplitudes.  This led to the
conjecture that a similar property holds for two-loop
amplitudes~\cite{TwoloopN4}.  Explicit calculations showed that this
conjecture is true for the four-point amplitude.  This simplicity suggests
that substantial parts of the theory may be solvable.
We confirm a leading-transcendentality relation between 
the $\Neqfour$ supersymmetric and QCD splitting amplitudes at two loops; 
this posited relation was already used~\cite{Lipatov} to extract the $\Neqfour$
leading-twist anomalous dimensions from those in QCD~\cite{MVVNNLO}.

We performed this computation using the unitarity-based sewing method.
The method has demonstrated many advantages over conventional
diagrammatic techniques over the years.  It has made possible, for
example, the computation of series of amplitudes for arbitrarily
many partons~\cite{Neq4Oneloop,Neq1Oneloop}.  The present computation
demonstrates another advantage: it offers a pathway featuring the
physical-projection advantages of light-cone gauge, while avoiding 
the need for cumbersome prescriptions for dealing with 
the ill-defined integrals of the latter.  Because it effectively combines
many diagrams at an early stage of the calculation, it also
simplifies integrands considerably, and so minimizes the complexity 
of reducing loop integrals to master integrals.


\acknowledgments We thank Babis Anastasiou for key contributions
at an early stage of this work. We also thank Thomas Gehrmann,
Zoltan Kunszt and Marc Schreiber for helpful comments.  
L.D. thanks Cambridge University
for hospitality when this work was begun, and we all
thank the Kavli Institute for Theoretical Physics for its generous
hospitality during completion of this paper.
This research was supported in part by the National Science Foundation 
under Grant No. PHY99-0794.


\appendix

\section{Color-Space Collinear Limit of Catani's Formula}
\label{CataniCollinearSection}

In \sect{CataniComparison} we demonstrated the consistency of the
divergent part of the splitting amplitudes with the Catani formula at
leading order in $\Nc$, or more generally, for the color-trivial
single-trace terms in the formula.  In this appendix we demonstrate that
the consistency also holds for the terms containing non-trivial color 
correlations.  Because of these correlations, the collinear limits 
of the full Catani formula are a bit intricate. In this appendix,
all quantities are understood to be renormalized.

\subsection{Color-Space Notation}

In the color-space language of
refs.~\cite{CataniTwoloop,CataniGrazziniSoft} an amplitude is
expressed in terms of an abstract vector in color space, 
$|\ca_n ^\Lloop\ra$.
To convert to the more standard color notation one may use an orthogonal
basis of unit vectors $| a_1, a_2, \ldots, a_n \ra$ with the property
that
\begin{equation}
\ca_n^{\Lloop\, a_1 a_2 \ldots a_n}(k_1, \lambda_1; k_2, \lambda_2;
  \ldots; k_n, \lambda_n) 
\equiv  \la a_1, a_2, \ldots, a_n | \ca_n^\Lloop \ra \,.
\end{equation}
Color interactions are represented by associating a color charge
${\bom T}_i$ with the emission of a gluon from each parton $i$.  The
color charge ${\bom T}_i = \{T^a_{Ri}\}$ is a vector with respect to
the generator label $a$, and an ${\rm SU}(\Nc)$ color matrix in the
representation $R$ of the outgoing parton $i$.  For external gluons,
$R$ is the adjoint representation $A$, and
$T^a_{Acb} = i f^{cab}$.  (Note that the normalization of fundamental
representation charge matrices is different in
ref.~\cite{CataniTwoloop} from that used elsewhere in the paper.  In
this appendix we normalize the ${\bom T}_i$ according to 
ref.~\cite{CataniTwoloop}.)

In this notation each vector $| \ca_n \ra$ is a color 
singlet, so color conservation is simply 
\begin{equation}
\sum_{i=1}^n {\bom T}_i| \ca_n \ra = 0 \,,
\label{ColorConservation}
\end{equation}
independent of the color representation of each leg.  This identity
incorporates the Jacobi identity and its generalizations very simply.
Typical operators encountered in the discussion involve the combination
\begin{equation}
({\bom T}_i)^{a} ({\bom T}_j)^{a} \equiv {\bom T}_i \cdot {\bom T}_j  \,.
\label{TidotTj}
\end{equation}
For $i=j$, \eqn{TidotTj} reduces to a Casimir operator,
${\bom T}_i^2 = C_i = C_A = \Nc$ if leg $i$ is a gluon, and
${\bom T}_i^2 = C_i = C_F = (\Nc^2 - 1)/(2\Nc)$ if leg $i$ is a quark
or anti-quark (with $T_R = 1/2$).  A useful property is
\begin{equation}
{\bom T}_i \cdot {\bom T}_j = 
{\bom T}_j \cdot {\bom T}_i \,,
\end{equation}
which holds because charge matrices act on different index spaces.

In the collinear limit where $k_1 \to z k_P$, $k_2 \to (1-z) k_P$, 
a tree-level color-space amplitude satisfies
\begin{equation}
| \ca_n^{(0)} \ra \rightarrow
 \sum_{\lambda=\pm} \SplitOp^{\tree}_{-\lambda}
  \; | \ca_{n-1}^{(0)}(\lambda) \ra\,,
\end{equation}
where $\lambda$ denotes the helicity of $P$ and the splitting
functions $\SplitOp^\tree_{-\lambda}$ are now operators acting on the
color space.
The operator $\SplitOp^{\tree}_{-\lambda}$ links the color space
with $n-1$ legs to that of $n$ legs.  That is, we define
\begin{eqnarray}
&&\langle a_1, a_2, a_3, \ldots, a_n| \SplitOp^\tree_{-\lambda}
| \ca_{n-1}^{(1)}(\lambda) \ra
\nonumber\\
&&\hskip0cm
= i \tilde f^{a_1 a_2 a_P} \Split^\tree_{-\lambda} 
      \ca_{n-1}^{\tree \, a_P a_3 \ldots a_n}
      (k_P,\lambda; k_3,\lambda_3; \ldots ; k_n,\lambda_n),
\end{eqnarray}
where $\Split^\tree_{-\lambda}$ is the color-ordered
splitting function defined in \eqn{TreeSplit}. 
We define an operator
\begin{equation}
\SplitOp^\tree_{-\lambda}  = \F \Split^\tree_{-\lambda}\,,
\end{equation}
where for the pure-glue case we have 
\begin{equation}
\langle a_1, a_2, a_3, \ldots, a_n | \F | a_P, a_3, \ldots, a_n  \rangle = 
i \tilde f^{a_1 a_2 a_P} \,.
\end{equation}
The operator $\F$ is distinct from
the operator ${\bom T}_{\! P}$ because it links color spaces of
different dimensions.  The action of the operator $\F$ on an amplitude
is illustrated in \fig{FoperatorFigure}.

Color conservation on the three-vertex implies that 
\begin{equation}
 \F {\bom T}_{\! P} = ({\bom T}_1 + {\bom T}_2) \F \,,
\label{ColorConservationThreeVertex}
\end{equation}
which is a special case of \eqn{ColorConservation} (up to a 
sign due to a swap of incoming and outgoing indices). 
This equation is equivalent to the Jacobi identity.

%
\FIGURE[t]{
\centerline{\epsfxsize 2. truein \epsfbox{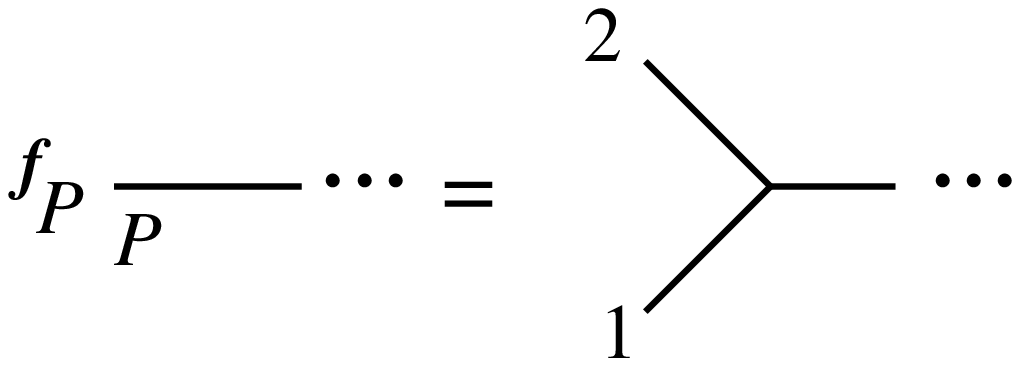}}
\caption[]{
\label{FoperatorFigure}
\small The action of $\F$ on the color index of leg $P$ of an amplitude.  
The vertex represents a factor of $i{\tilde f}^{a_1 a_2 a_P}$.}
}

From \eqn{SplitOneloopFullColor}, it
is apparent that the behavior of the one-loop amplitude in color space 
in the collinear limit is very similar to that of the tree amplitude,
\begin{equation}
| \ca_n^{(1)} \ra \rightarrow
 \sum_{\lambda=\pm} \SplitOp^{\tree}_{-\lambda}
  \; | \ca_{n-1}^\oneloop(\lambda) \ra
+ \sum_{\lambda=\pm} \SplitOp^\oneloop_{-\lambda} 
  | \ca_{n-1}^\tree(\lambda) \ra  \,,
\label{OneloopSplitColorSpace}
\end{equation}
where 
\begin{equation}
\SplitOp^\oneloop_{-\lambda} (1^{\lambda_1}, 2^{\lambda_2}) 
 =  \Nc 
   \Split^\oneloop_{-\lambda} (1^{\lambda_1}, 2^{\lambda_2}) \, \F \,,
\label{OneloopSplitOp}
\end{equation}
with $\Split^\oneloop_{-\lambda}$ the color-stripped splitting
amplitude.
For helicity contributions where the 
tree splitting amplitude does not vanish,
it is convenient to re-express this as 
\begin{equation}
\SplitOp^\oneloop_{-\lambda} (1^{\lambda_1}, 2^{\lambda_2}) 
=  \Nc \,  \rsnren^{\oneloop, \, \lambda_1\lambda_2}(z,s) \, 
  \SplitOp^\tree_{-\lambda} (1^{\lambda_1}, 2^{\lambda_2}) \,,
\label{OneloopSplitOprS}
\end{equation}
where $\rsnren^\oneloop$ is given as $\rsn^\oneloop$ in 
eqs.~(\ref{OneloopSplitExplicitpp})--(\ref{OneloopSplitExplicitNeq4}),
and is renormalized according to \eqn{OneloopCounterterm}.  

The $L$-loop generalization is the obvious one, 
\begin{equation}
| \ca_n^{(L)} \ra \rightarrow
\sum_{l=0}^L \sum_{\lambda = \pm} 
\SplitOp^{\lloop}_{-\lambda}
  \; | \ca_{n-1}^{\Llloop}(\lambda) \ra \,,
\label{LloopGenObvious}
\end{equation}
where
\begin{equation}
\SplitOp^\Lloop_{-\lambda}  = \Nc^L \Split^\Lloop_{-\lambda} \F \,.
\end{equation}
In the two-loop case, $\Split^\twoloop_{-\lambda}$ is the color-trivial
splitting amplitude defined in \eqn{TwoloopSplit}.  

Next we shall demonstrate that the full color-space dependence of 
Catani's formula for the divergences of one- and two-loop
amplitudes is completely compatible with the divergent parts of our
color-dressed splitting amplitudes.

\subsection{One-Loop Warm Up}

At one loop, the infrared divergences of renormalized
$n$-point amplitudes can be written compactly in color-space notation
as~\cite{CataniTwoloop}, 
\begin{equation}
| \ca_n^{(1)} \ra_{\RS} = {\bom I}_n^{(1)}(\e)
  \; | \ca_n^{(0)} \ra_{\RS}
  + |\ca_n^{(1){\rm fin}} \ra_{\RS} \,.
        \label{OneloopCatani}
\end{equation}
The color operator ${\bom I}_n^{(1)}$ is 
\begin{equation}
{\bom I}_n^{(1)}(\e) = \frac{1}{2} {e^{-\e \psi(1)} \over
\Gamma(1-\e)} \sum_{i=1}^n \, \sum_{j \neq i}^n \, {\bom T}_i \cdot
{\bom T}_j \Biggl[ {1 \over \e^2} + {\gamma_i \over {\bom T}_i^2 } \,
{1 \over \e} \Biggr] \Biggl( {\mu^2 \over -s_{ij}} \Biggr)^{\e} \,.
\label{CataniGeneral}
\end{equation}
The sum runs over pairs of external legs. 
For external gluons $i$, we set $\gamma_i$ equal to 
\begin{equation}
\gamma_g = {11 C_A - 4 T_R \Nf \over 6} \,.
\label{QCDgluonValues}
\end{equation}
(For external fermions, the ratio $\gamma_q/{\bom T}_i^2  = 3/2$
is independent of the representation.) 
The subscript $\RS$ indicates that a quantity depends on the regularization
and renormalization scheme.

Now consider the limit of the ${\bom I}_n^{(1)}$ operator as the momenta of
legs 1 and 2 become collinear.
Inserting the decomposition of the amplitude into divergent and finite 
parts, \eqn{OneloopCatani}, into both sides of \eqn{OneloopSplitColorSpace}
and taking the collinear limit, we find that the divergent parts must
satisfy
\begin{eqnarray}
{\bom I}^{(1)}_n(\e) \SplitOp^{\tree}_{-\lambda} 
  | \ca_{n-1}^{(0)}(\lambda) \ra_{\RS} 
 & = & \SplitOp^{\oneloop}_{-\lambda} 
 |\ca_{n-1}^{(0)}(\lambda) \ra_{\RS} 
\PlusBreak{ }
   \SplitOp^{\tree}_{-\lambda} 
{\bom I}^{(1)}_{n-1}(\e) | \ca_{n-1}^{(0)}(\lambda) \ra_{\RS}
  \ +\ \hbox{finite},
\end{eqnarray}
where collinear kinematics are implicit.
On the left-hand side of the equation ${\bom I}_n^{(1)}(\e)$
acts on an $n$-dimensional color space, while on the right-hand side
${\bom I}_{n-1}^{(1)}(\e)$ acts on an $(n-1)$-dimensional 
color space.  A compact way to write the same equation is
\begin{equation}
\SplitOp^{\oneloop}_{-\lambda} = 
{\bom I}^{(1)}_n(\e) \SplitOp^{\tree}_{-\lambda} 
- \SplitOp^{\tree}_{-\lambda} {\bom I}_{n-1}^{(1)}(\e) 
+ \hbox{finite}.
\label{CompactSplitOneloop}
\end{equation}

To make this more explicit, consider an $n$-gluon
amplitude, for simplicity. 
Inserting the explicit forms of the operators, we have that
the infrared-divergent parts of the renormalized one-loop splitting
amplitude must be
\begin{eqnarray}
\SplitOp^{\oneloop}_{-\lambda}  
& = & 
{e^{-\e \psi(1)} \over
\Gamma(1-\e)} \Biggl[ {1 \over \e^2} + {\gamma_g \over C_A} \,
{1 \over \e} \Biggr] 
\TimesBreak{ ~~~~~~~~~~~~~~~ }
\Biggl\{ \Biggl[ \sum_{j=3}^n \,
  {\bom T}_1 \cdot {\bom T}_j 
  \Biggl( \frac{\mu^2}{-z s_{Pj}} \Biggr)^{\e} 
+ \sum_{j=3}^n \,
  {\bom T}_2 \cdot {\bom T}_j 
 \Biggl( \frac{\mu^2}{-(1-z) s_{Pj}} \Biggr)^{\e}
\PlusBreak{ ~~~~~~~~~~~~~~~ } 
   {\bom T}_1 \cdot {\bom T}_2 
 \Biggl( \frac{\mu^2}{- s_{12}} \Biggr)^{\e}
+ \frac{1}{2}  \sum_{i\not=j =3}^n \, {\bom T}_i \cdot
{\bom T}_j \Biggl( \frac{\mu^2}{-s_{ij}} \Biggr)^{\e} 
 \Biggr] \SplitOp^\tree_{-\lambda}  
\MinusBreak{a\null} 
 \SplitOp^\tree_{-\lambda} \Biggl[ \sum_{j=3}^n \, 
  {\bom T}_P \cdot {\bom T}_j 
 \Biggl( \frac{\mu^2}{-s_{Pj}} \Biggr)^{\e}
+ \frac{1}{2}  \sum_{i\not=j=3}^n \, {\bom T}_i \cdot
{\bom T}_j \Biggl( \frac{\mu^2}{-s_{ij}} \Biggr)^{\e} \Biggr]
 \Biggr\} 
\PlusBreak{\SplitOp^\tree_{-\lambda} \BigglBl \sum_{j=3}^n \, 
  {\bom T}_P \cdot {\bom T}_j }
 \hbox{finite},
\end{eqnarray}
where we have separated the terms into those involving the collinear
legs and those not involving them.  The terms that do not involve the
collinear legs simply cancel, since $\SplitOp^\tree_{-\lambda}$
commutes with any ${\bom T}_j$ where $j > 3$.  After separating
terms that involve $s_{Pj}$ from those that do not (allowing 
for shifts in the finite terms), we find,
\begin{eqnarray}
\SplitOp^{\oneloop}_{-\lambda}  
& = & {e^{-\e \psi(1)} \over
\Gamma(1-\e)}
 \Biggl[ {1 \over \e^2} + {\gamma_g \over C_A} \, {1 \over \e} \Biggr]
 \Biggl\{ \Biggl[ 
 - \e \Biggl(
  \ln z \sum_{j=3}^n \, {\bom T}_1 \cdot {\bom T}_j 
  + \ln(1-z) \sum_{j=3}^n \, {\bom T}_2 \cdot {\bom T}_j 
      \Biggr) 
\PlusBreak{{e^{-\e \psi(1)} \over  \Gamma(1-\e)}
 \BigglBl {1 \over \e^2} + {\gamma_g \over C_A} \, {1 \over \e} \BiggrBr
 \BigglBl \BigglBl }
 {\bom T}_1 \cdot {\bom T}_2 
 \Biggl( \frac{\mu^2}{- s_{12}} \Biggr)^{\e} \Biggr] 
   \SplitOp^\tree_{-\lambda} 
\PlusBreak{\null}
 \sum_{j=3}^n \,
  \biggl( ({\bom T}_1 +  {\bom T}_2) \cdot {\bom T}_j 
      \SplitOp^\tree_{-\lambda} - 
    \SplitOp^\tree_{-\lambda}  {\bom T}_P \cdot {\bom T}_j   \biggr)
  \Biggl( \frac{\mu^2}{-s_{Pj}} \Biggr)^{\e}  
\Biggr\}  \nonumber \\
&& \hskip 2 cm 
 +\ \hbox{finite}\,. 
\label{SplitOneloopColorEvaluate}
\end{eqnarray}
In the above, we have been careful in the placement 
of $\SplitOp^\tree_{-\lambda}$
since it links $(n-1)$-point color space to $n$-point color space.
To simplify \eqn{SplitOneloopColorEvaluate} we use
\eqn{ColorConservationThreeVertex}.  
Furthermore, using color conservation, \eqn{ColorConservation}, we have
\begin{eqnarray}
 \sum_{j=3}^n \, 
  {\bom T}_1 \cdot {\bom T}_j 
  \SplitOp^\tree_{-\lambda} 
& = &
- {\bom T}_1 \cdot ({\bom T}_1 + {\bom T}_2) 
  \SplitOp^\tree_{-\lambda} \nonumber \\
& = &
- {1\over2}
 \Bigl( ({\bom T}_1 + {\bom T}_2)^2 + {\bom T}_1^2 - {\bom T}_2^2 \Bigr)
  \SplitOp^\tree_{-\lambda} \nonumber \\
& = &  - {1\over2} C_A \SplitOp^\tree_{-\lambda} \,,
\label{Identity1j}
\end{eqnarray}
with a similar equation with leg 1 replaced by 2.  
Also, 
${\bom T}_1 \cdot {\bom T}_2 \to 
{1\over2} ( ({\bom T}_1 + {\bom T}_2)^2 - {\bom T}_1^2 - {\bom T}_2^2 )
\to -{1\over2} C_A$.
Inserting these relations into the divergent parts of the 
renormalized splitting amplitude gives us
\begin{eqnarray}
\SplitOp^{\oneloop}_{-\lambda}   
& = & 
- C_A \Biggl\{
  \frac{1}{2} {e^{-\e \psi(1)} \over
\Gamma(1-\e)} \ \, 
\Biggl[ {1 \over \e^2} + {\gamma_g \over C_A} \,
{1 \over \e} \Biggr] \Biggl( \frac{\mu^2}{-z (1-z) s_{12}} \Biggr)^{\e} 
\, \Biggl\}\, \SplitOp^{\tree}_{-\lambda} 
\PlusBreak{C_A \BigglBl  \frac{1}{2} {e^{-\e \psi(1)} \over \Gamma(1-\e)} }
 \, \hbox{finite}
 \,. \label{OneloopColorSplit}
\end{eqnarray}
This result is in complete agreement with the divergent parts of the
one-loop color-space splitting amplitude, as given in
\eqns{OnelooprS}{OneloopSplitOprS}, after renormalization according
to \eqn{OneloopCounterterm}.  (Renormalization produces the term
proportional to $\gamma_g = b_0$ in \eqn{OneloopColorSplit}.)  
A similar analysis using mixed
quark-gluon amplitudes yields the same result for the $g \rightarrow
gg$ splitting amplitudes. This demonstrates that the expression
(\ref{OneloopCatani}) for the divergences of one-loop amplitudes is
fully compatible with the splitting amplitudes.

\subsection{Review of Catani's Two-Loop Divergence Formula}

In a beautiful yet mysterious paper, Catani expressed the
infrared-divergent parts of two-loop amplitudes as~\cite{CataniTwoloop}
\begin{equation}
| \ca_n^{(2)} \ra_{\RS} =  
  {\bom I}^{(2)}_{n, \, \RS}(\e) \; | \ca_n^{(0)} \ra_{\RS}
+ {\bom I}_n^{(1)}(\e) \; | \ca_n^{(1)} \ra_{\RS} 
+ |\ca_n^{(2){\rm fin}} \ra_{\RS} \,,
\label{TwoloopCatani} 
\end{equation}
where $|\ca_n^{(2){\rm fin}} \ra_{\RS} $ is the finite
remainder and the operator ${\bom I}^{(2)}_{n, \, \RS}(\e)$ is
\begin{eqnarray}
{\bom I}^{(2)}_{n, \, \RS}(\e)
& =& - \frac{1}{2} {\bom I}_n^{(1)}(\e)
\left( {\bom I}_n^{(1)}(\e) + {2 b_0 \over \e} \right)
  \PlusBreak{}
 {e^{+\e \psi(1)} \Gamma(1-2\e) \over \Gamma(1-\e)}
\left( {b_0 \over \e} + K_\RS \right) {\bom I}_n^{(1)}(2\e,\mu;\{p\})
  + {\bom H}^{(2)}_{n, \, \RS}(\e) \,.
\label{CataniGeneralI2}
\end{eqnarray}
An argument for this general structure has been given in ref.~\cite{StermanIR}.
The quantity $K_\RS$ depends on the variant of dimensional regularization
through the parameter $\delta_R$~\cite{BDDgggg}.
It is given by~\cite{CataniTwoloop,BDDgggg}
\begin{eqnarray}
K_\FDH^{\Neqfour} &=& - \zeta_2 \, C_A 
   \,,  
  \label{CataniKNeq4}\\
K_\FDH^{\Neqone} &=& \left[ 3 - \zeta_2 - {4\over9} \e \right] C_A
   \,,  
  \label{CataniKNeq1}\\
K_\RS^{\rm QCD} &=& \left[ \frac{67}{18} - \zeta_2
    - \biggl( {1\over6} + {4\over9} \e \biggr) (1-\delta_R) \right] C_A
 - \frac{10}{9} T_R \Nf \,.  
  \label{CataniK}
\end{eqnarray}
Note that in passing to the $\Neqfour$ case we have switched away from the 
conventions of ref.~\cite{BDDgggg}, with regard to assigning 
$\e$-dependent terms to $K_\FDH$.  (We did not want to destroy
the uniform transcendental weight which all functions have in the 
$\Neqfour$ case.)

The function ${\bom H}^{(2)}_{n, \, \RS}$ contains only {\em single} poles.
It splits into two types of terms,
\begin{equation}
{\bom H}^{(2)}_{n, \, \RS}(\e) = 
{ e^{-\e\psi(1)} \over 4\e \, \Gamma(1-\e) } 
  \Biggl\{ 
  - \sum_{i=1}^n \sum_{j\neq i}^n \, {\bom T}_i \cdot {\bom T}_j
     { H_i^{(2)} \over {\bom T}_i^2 } \, 
    \Bigl( { \mu^2 \over -s_{ij} } \Bigr)^{2\e}
       + \hat{\bom H}_n^{(2)} \Biggr\} \,,
\label{OurH}
\end{equation}
where $H_i^{(2)}$ is either $H_g^{(2)}$ or $H_q^{(2)}$, depending
on whether particle $i$ is a gluon or a quark. 
The constants $H_g^{(2)}$ and $H_q^{(2)}$ are given 
by~\cite{GOTYqqqq,GOTYqqgg,GOTYgggg,BDDgggg,BDDqqgg}
\begin{eqnarray}
H_q^{(2)} &=& 
\biggl( {13\over2} \zeta_3 - {23\over8} \zeta_2 + {245\over216} \biggr) 
 C_A C_F
+ \biggl( - 6 \zeta_3 + 3 \zeta_2 - {3\over8} \biggr) C_F^2
+ \biggl( {\zeta_2\over2} - {25\over54} \biggr) C_F T_R \Nf 
\PlusBreak{ \zeta_3 }
 \biggl( - {4\over3} C_A C_F + {1\over2} C_F^2 
           + {1\over6} C_F T_R \Nf \biggr) (1-\delta_R)\,,
\label{Hquark}
\end{eqnarray}
and (including also the supersymmetric cases here~\cite{BDDgggg,BDDqqgg}),
\begin{eqnarray}
H_g^{(2), \, \Neqfour} &=& {\zeta_3\over2}  C_A^2 \,,
\label{HgluonNeq4}\\
H_g^{(2), \, \Neqone} &=& 
\biggl( {\zeta_3\over2} + {3\over8} \zeta_2 - {2\over9} \biggr) C_A^2 \,,
\label{HgluonNeq1}\\
H_g^{(2), \, {\rm QCD}} &=& 
\biggl( {\zeta_3\over2} + {11\over24} \zeta_2 + {5\over12} \biggr) C_A^2
- \biggl( {\zeta_2\over6} + {58\over27} \biggr) C_A T_R \Nf
+ C_F T_R \Nf + {20\over27} T_R^2 \Nfsq 
\PlusBreak{\zeta_3}
 \biggl( - {11\over36} C_A^2 + {1\over9} C_A T_R \Nf \biggr) (1-\delta_R).
\label{Hgluon}
\end{eqnarray}
There is no real proof that \eqn{OurH} is the precise form of the 
$n$-point divergences, but previous calculations strongly indicate 
that it is right.  There are several conventions in the literature
for `dressing' the terms proportional to $H_i^{(2)}$ with factors
of $(\mu^2/(-s_{ij}))^\e$ to make them dimensionally consistent.
Different conventions agree at $\Ord(1/\e)$, but generate different finite
remainders, which would have different collinear behavior,
affecting the results of \sect{CataniComparison}.
Here we adopt a convention motivated by Catani's
treatment~\cite{CataniTwoloop} of the $1/\e$ poles proportional to 
$\gamma_i$ at one loop in \eqn{CataniGeneral}.  Note that the $1/\e$
pole part of the terms containing $H_i^{(2)}$ in \eqn{OurH} 
is actually proportional to the color identity matrix ${\bom 1}$, since 
$-\sum_{j\neq i} {\bom T}_i \cdot {\bom T}_j = {\bom T}_i^2$, 
due to~\eqn{ColorConservation}.

Less is known about the function $\hat {\bom H}_n^{(2)}(\e)$, which
contains genuinely non-trivial color structure at $\Ord(1/\e)$.
For the four-point amplitudes it is given by~\cite{BDDgggg,BDDqqgg} 
\begin{equation}
\hat{\bom H}_4^{(2)}(\e) = - 4 \, \ln\biggl( {-s_{12}\over-s_{23}} \biggr)
                            \ln\biggl( {-s_{23}\over-s_{13}} \biggr)
                            \ln\biggl( {-s_{13}\over-s_{12}} \biggr)
      \times \Bigl[ {\bom T}_1 \cdot {\bom T}_2 \,,
                    {\bom T}_2 \cdot {\bom T}_3 \Bigr] \,,
\label{HExtraAlt}
\end{equation}
with $\ln((-s_{12})/(-s_{23})) \to \ln s_{12} - \ln(-s_{23}) - i\pi$ 
in the $s$-channel, {\it etc.}   
A simple ansatz, generalizing \eqn{HExtraAlt} to arbitrary $n$-parton 
amplitudes, is
\begin{equation}
\hat{\bom H}_n^{(2)}(\e) = i \sum_{(i_1,i_2,i_3)}
      f^{a_1a_2a_3} T_{i_1}^{a_1} T_{i_2}^{a_2} T_{i_3}^{a_3} 
      \ln\biggl( {-s_{i_1i_2}\over-s_{i_2i_3}} \biggr)
      \ln\biggl( {-s_{i_2i_3}\over-s_{i_1i_3}} \biggr)
      \ln\biggl( {-s_{i_1i_3}\over-s_{i_1i_2}} \biggr) \,,
\label{HExtraAltN}
\end{equation}
where the sum is over distinct triplets of external legs,
with $i_1\neq i_2 \neq i_3$.
For $n=4$, there are four such triplets (omit any one of the four 
partons).  Using $[T_i^a, T_i^b] = i f^{abc} T_i^c$,
color conservation~(\ref{ColorConservation}), and antisymmetry of 
$f^{abc}$, it is easy to see that each triplet gives an equal contribution,
and \eqn{HExtraAlt} is recovered.

\subsection{Collinear Compatibility with Catani's Two-loop Divergence Formula}
\label{FullColorCatani}

Next let us verify that the $\e$ poles in the collinear
splitting amplitudes match those predicted by Catani for the full
color-space two-loop amplitude. 
In order to analyze these divergent terms,
it is convenient to define the one-loop amplitude
as being a color operator acting on the tree amplitude (assume it
does not vanish),
\begin{eqnarray}
| \ca_n^{(1)} \ra_{\RS} &\equiv& {\bom M}_n^{(1)}(\e) 
  \; | \ca_n^{(0)} \ra_{\RS}  \nonumber\\
& =& ({\bom I}_n^{(1)}(\e) + {\bom M}_n^{(1)\rm fin}(\e) )
  \; | \ca_n^{(0)} \ra_{\RS} \,,
\label{M1Def}
\end{eqnarray}
since we want to combine divergent and finite parts into a single
entity.  We do not actually need the explicit form of the 
${\bom M}_n^{(1)\rm fin}$ operator, although it is straightforward 
to construct this operator in one's favorite color basis.

The full color-space two-loop amplitude is
\begin{eqnarray}
| \ca_n^{(2)} \ra_{\RS} &=& {\bom I}_n^{(1)}(\e) 
  \; | \ca_n^{(1)} \ra_{\RS} 
+ {\bom I}^{(2)}_{n, \, \RS}(\e) \;
         | \ca_n^{(0)} \ra_{\RS}
+ |\ca_n^{(2){\rm fin}} \ra_{\RS}  \hskip .5 cm 
\nonumber \\
&=& {\bom I}_n^{(1)}(\e) 
 \; \Bigl({\bom I}_n^{(1)}(\e) +
          {\bom M}_n^{(1)\rm fin}(\e) \Bigr) 
        | \ca_n^{(0)} \ra_{\RS} 
\PlusBreak{{\bom I}}
\Biggl[  - \frac{1}{2} {\bom I}_n^{(1)}(\e)
\left( {\bom I}_n^{(1)}(\e) + {2 b_0 \over \e} \right)
 + {e^{+\e \psi(1)} \Gamma(1-2\e) \over \Gamma(1-\e)}
\left( {b_0 \over \e} + K_\RS \right) {\bom I}_n^{(1)}(2\e)
\PlusBreak{{\bom I}(\e) \BigglBl}
 {\bom H}^{(2)}_{n, \, \RS}(\e)  \Biggr]
         | \ca_n^{(0)} \ra_{\RS}
 + \hbox{finite} \,. \hskip .5 cm 
\end{eqnarray}
After adding and subtracting finite terms to combine
the $({\bom I}_n^{(1)})^2$ term with finite pieces, 
in much the same way as was done for the color-trivial parts in 
\eqn{CollinearTwoLoopLeading}, we obtain, 
\begin{eqnarray}
| \ca_n^{(2)} \ra_{\RS} &=& 
  \; {1\over 2} \Bigl({\bom M}_n^{(1)}(\e) \Bigr)^2 
        | \ca_n^{(0)} \ra_{\RS} 
\PlusBreak{\null}
   \biggl[ {1\over 2} 
   \Bigl[ {\bom I}_n^{(1)}(\e), {\bom M}_n^{(1)}(\e) \Bigr] 
   + {1\over 4\e} \hat{\bom H}_n^{(2)}(\e)  \biggr]
        | \ca_n^{(0)} \ra_{\RS} 
\PlusBreak{\null}
 \biggl[  -  { b_0 \over \e}  {\bom I}_n^{(1)}(\e)
 +  {e^{+\e \psi(1)} \Gamma(1-2\e) \over \Gamma(1-\e)}
\left( {b_0 \over \e} + K_\RS \right) {\bom I}_n^{(1)}(2\e) \biggr]
        | \ca_n^{(0)} \ra_{\RS} 
\MinusBreak{\null}
    { e^{-\e\psi(1)} \over 4\e \, \Gamma(1-\e) } 
   \sum_{i=1}^n \sum_{j\neq i}^n \, {\bom T}_i \cdot {\bom T}_j
     { H_i^{(2)} \over {\bom T}_i^2 } \, 
    \Bigl( { \mu^2 \over -s_{ij} } \Bigr)^{2\e}
         | \ca_n^{(0)} \ra_{\RS}
 + \hbox{finite} \,. \hskip .5 cm 
\label{FullColorTwoLoopCatani}
\end{eqnarray}

First consider the terms on the penultimate line in
\eqn{FullColorTwoLoopCatani} containing the operator ${\bom I}_n^{(1)}$.
Since the same operator appears as in the one-loop case, the collinear
limit of these terms may be determined in the same way as the one-loop
case, using \eqns{CompactSplitOneloop}{OneloopSplitOprS} in particular.
\Eqn{CompactSplitOneloop} says nothing about the $\Ord(\e^0)$ terms in 
${\bom I}_n^{(1)}$.  Fortunately these terms are identical for 
${\bom I}_n^{(1)}(\e)$ and ${\bom I}_n^{(1)}(2\e)$;
hence their contribution to the singular terms in
\eqn{FullColorTwoLoopCatani}, when they are multiplied by $b_0/\e$,
cancels.  Following the one-loop discussion, we obtain contributions 
to the two-loop splitting amplitude of the form
\begin{equation}
\Nc \biggl( - {b_0\over \e} \rsnren^\oneloop(\e)
 + {e^{+\e \psi(1)} \Gamma(1-2\e) \over \Gamma(1-\e)}
   \left( {b_0 \over \e} + K_\RS \right) \rsnren^\oneloop(2 \e) \biggr)
\SplitOp_{-\lambda}^\tree \, + \;\hbox{finite} \,.
\label{ISplitContributions}
\end{equation}
(Recall that $\rsnren^{(L)}$, introduced in~\sect{CataniComparison}, denotes
the renormalized splitting ratio.)

Now consider the terms containing $H_i^{(2)}$ on the last line of
\eqn{FullColorTwoLoopCatani}.  These terms are written in the same
form as ${\bom I}_n^{(1)}$, but with one less power of $1/\e$,
so their collinear behavior again follows from the one-loop discussion.
Their contribution to the two-loop splitting amplitude is just
\begin{equation}
{ H_g^{(2)} \over 4\e} \, \SplitOp^\tree_{-\lambda}\, + \; \hbox{finite}\,,
\label{H2SplitContributions}
\end{equation}
where $H_g^{(2)}$ is given in \eqn{Hgluon}.

Next consider the 
$\bigl({\bom M}_n^{(1)}(\e)\bigr)^2$ term in~\eqn{FullColorTwoLoopCatani}.  
In the collinear limits for this term we have,
\begin{eqnarray}
{1\over 2} \Bigl({\bom M}_n^{(1)}(\e) \Bigr)^2 
\SplitOp^{\tree}_{-\lambda}  |\ca_{n-1}^{(0)} \ra \,.
\end{eqnarray}
In order to evaluate this we use the collinear relation,
\begin{equation}
{\bom M}^{(1)}_n \SplitOp^{\tree}_{-\lambda} 
    \rightarrow 
\SplitOp^{\tree}_{-\lambda} {\bom M}^{(1)}_{n-1} +
\SplitOp^{\oneloop}_{-\lambda} \,,
\label{M1Limit}
\end{equation}
which is a rewriting of \eqn{OneloopSplitColorSpace} after acting
on $|\ca_{n-1}^{(0)} \ra$ and using \eqn{M1Def}.
Using \eqn{M1Limit}, we find that in the collinear limit,
\begin{equation}
{1\over 2} \Bigl({\bom M}_n^{(1)} \Bigr)^2 
\SplitOp^{\tree}_{-\lambda} 
\rightarrow  {1\over 2} {\bom M}_n^{(1)}
\Bigl(\SplitOp^{\tree}_{-\lambda} {\bom M}_{n-1}^{(1)}
                + \SplitOp^{\oneloop}_{-\lambda} \Bigr) \,.
\label{M1SqrSimpifyA}
\end{equation}
To simplify the $\SplitOp^\oneloop_{-\lambda}$ term we 
may multiply \eqn{M1Limit} by $\Nc \, \rsnren^\oneloop(\e)$ to obtain
\begin{equation}
{\bom M}^{(1)}_n \SplitOp^{\oneloop}_{-\lambda} 
 \rightarrow \SplitOp^{\oneloop}_{-\lambda} {\bom M}^{(1)}_{n-1}
+ \SplitOp^{\oneloop}_{-\lambda} \, \Nc \, \rsnren^\oneloop(\e) \,,
\label{M1LimitrS}
\end{equation}
where we used \eqn{OneloopSplitOprS}. 
Then applying \eqns{M1Limit}{M1LimitrS} to \eqn{M1SqrSimpifyA} yields
\begin{eqnarray}
{1\over 2} \Bigl({\bom M}_n^{(1)} \Bigr)^2 
\SplitOp^{\tree}_{-\lambda} 
&\rightarrow &
 {1\over 2} \Bigl(
     \SplitOp^{\tree}_{-\lambda} {\bom M}_{n-1}^{(1)} {\bom M}_{n-1}^{(1)}
     + 2 \SplitOp^{\oneloop}_{-\lambda} {\bom M}_{n-1}^{(1)} 
       + \SplitOp^{\oneloop}_{-\lambda} \, 
             \Nc \, \rsnren^\oneloop(\e) \Bigr) \nonumber\\
& = & \SplitOp^{\tree}_{-\lambda} \, {1\over 2} \, 
      {\bom M}_{n-1}^{(1)} {\bom M}_{n-1}^{(1)}
   +  \SplitOp^{\oneloop}_{-\lambda} {\bom M}_{n-1}^{(1)}
  + {1\over 2} (\Nc \rsnren^\oneloop(\e))^2 \SplitOp^{\tree}_{-\lambda}
   \,. \nonumber\\
&& \hskip 1.5 cm 
\label{M1SqrSimpifyB}
\end{eqnarray}
The first term on the last line may be identified as a contribution
to $\SplitOp^{\tree}_{-\lambda} |\ca_{n-1}^{(2)}\ra$,
the second as a contribution to 
$\SplitOp^{\oneloop}_{-\lambda} |\ca_{n-1}^{(1)}\ra$,
while the third is the contribution we include in 
$\SplitOp^{\twoloop}_{-\lambda} |\ca_{n-1}^{(0)}\ra$.

Finally we examine the second line of~\eqn{FullColorTwoLoopCatani},
containing the commutator term and $\hat{\bom H}_n^{(2)}$.
After some manipulations involving \eqns{M1Limit}{M1LimitrS}
we may write the part of the commutator term that does not
contribute to $\SplitOp^{\tree}_{-\lambda} |\ca_{n-1}^{(2)}\ra$
as
\begin{eqnarray}
&&
- {1\over2} {\bom M}_n^{(1)}
   \Bigl( {\bom I}_n^{(1)} \SplitOp_{-\lambda}^{(0)}
        - \SplitOp_{-\lambda}^{(0)} {\bom I}_{n-1}^{(1)}
        - \SplitOp_{-\lambda}^{(1)} \Bigr) 
\PlusBreak{\null}
  {1\over2}  \Bigl( {\bom I}_n^{(1)} \SplitOp_{-\lambda}^{(0)}
        - \SplitOp_{-\lambda}^{(0)} {\bom I}_{n-1}^{(1)}
        - \SplitOp_{-\lambda}^{(1)} \Bigr)
  (  {\bom M}_{n-1}^{(1)} + \Nc \rsnren^{(1)} ) \,,
\label{IMH1}
\end{eqnarray}
where the finite combination appearing is,
\begin{eqnarray}
&&{\bom I}_n^{(1)} \SplitOp_{-\lambda}^{(0)}
        - \SplitOp_{-\lambda}^{(0)} {\bom I}_{n-1}^{(1)}
        - \SplitOp_{-\lambda}^{(1)}
\nonumber\\
&=& \Bigl[ \ln z\ {\bom T}_1 + \ln(1-z)\ {\bom T}_2 \Bigr] \cdot
   \sum_{j=3}^n {\bom T}_j \ln(-s_{Pj})  \SplitOp_{-\lambda}^{(0)}
  \ +\ \Ord(\e) \,.
\label{IMH2}
\end{eqnarray}
Only the universal $1/\e$ terms in ${\bom M}_n^{(1)}$ contribute
in \eqn{IMH1} to the order we need, and we may obtain these
terms from ${\bom I}_n^{(1)}$.  Inserting these terms,
and moving them to the right with the help of color commutators, 
the total contribution to 
$\SplitOp^{\twoloop}_{-\lambda} |\ca_{n-1}^{(0)}\ra$
from the commutator term is
\begin{eqnarray}
&& {i\over2\e} \Biggl[
    - \ln\Bigl({z\over1-z}\Bigr)
    \sum_{i=3}^n f^{abc} T_1^a T_2^b T_i^c 
      \ \ln(-s_{Pi}) 
     \Bigl( \ln(-s_{Pi}) + \ln z + \ln(1-z) - \ln(-s_{12}) \Bigr)
\PlusBreak{ {i\over2\e} \Biggl[] }
  \sum_{i=3}^n \sum_{j=3,j\neq i}^n f^{abc} 
    \Bigl( \ln z\ T_1^a + \ln(1-z)\ T_2^a \Bigr) T_i^b T_j^c 
         \ \ln(-s_{Pi}) \ln(-s_{ij}) \Biggr] \SplitOp_{-\lambda}^{(0)}.~~~~
\label{IMH3}
\end{eqnarray}
This result turns out to be precisely the negative of the
corresponding contribution from the $\hat{\bom H}_n^{(2)}$ term.
Thus the second line of \eqn{FullColorTwoLoopCatani} only contributes 
to the tree-level splitting amplitude term,
$\SplitOp^\tree_{-\lambda} | {\cal A}_{n-1}^{(2)}(\lambda) \rangle$,
in \eqn{LloopGenObvious}.

Combining the contributions to $\SplitOp_{-\lambda}^\twoloop$ in
\eqns{ISplitContributions}{H2SplitContributions} and
(\ref{M1SqrSimpifyB}), we obtain,
\begin{eqnarray}
\SplitOp^\twoloop_{-\lambda}(1^{\lambda_1}, 2^{\lambda_2})
 & =&  \Nc^2 \, \Split^\twoloop_{-\lambda}(1^{\lambda_1}, 2^{\lambda_2}) 
           \, \F
\nonumber \\
&=& \Nc^2 \biggl[ {1\over 2} 
(\rsnren^\oneloop(\e))^2
\MinusBreak{~~~~~}
   {b_0\over \Nc\, \e} \rsnren^\oneloop(\e)
 + {e^{+\e \psi(1)} \Gamma(1-2\e) \over \Nc \, \Gamma(1-\e)}
   \left( {b_0 \over \e} + K_\RS \right) \rsnren^\oneloop(2 \e)
\PlusBreak{~~~~~}
  {1 \over 4\e }
  {H_g^{(2)} \over \Nc^2} \; + \; \hbox{finite} \biggr] \,
 \Split^\tree_{-\lambda}(1^{\lambda_1}, 2^{\lambda_2}) \, \F
 \nonumber \\
&=& \Nc^2 \, \rsnren^\twoloop(\e)\, 
\Split^\tree_{-\lambda}(1^{\lambda_1}, 2^{\lambda_2}) \, \F 
\ + \ \hbox{finite}\,,
\end{eqnarray}
where $\rsnren^\twoloop(\e)$ is given in \eqn{TwoloopSplitLeading}.
This in turn agrees with the divergent parts of the
splitting amplitudes for $\Neqfour$ and $\Neqone$ supersymmetric theories,
and for QCD, as given in \sect{ResultSection}, after they are 
renormalized according to~\eqn{TwoloopCounterterm}.
Thus our splitting amplitudes are fully compatible with 
Catani's color-space formula for the infrared divergences of
scattering amplitudes.   

As noted already, this agreement may be turned 
around to prove the validity of Catani's formula inductively, 
up to reasonable assumptions about the analytic structure 
of the $\e$-singular parts, {\it i.e.} such that they do not 
have vanishing collinear limits in all channels.

\section{Relabeling Algorithm}
\label{RelabelingAppendix}

For completeness, in this appendix we review a
simple procedure for mechanically performing relabelings.
For the splitting amplitude problem discussed in this paper, it is not
difficult to identify the diagrammatic structure of a given term and
map the momentum labels used in the cuts to those used in the
integration.  However, for more general problems it may be useful to
perform these steps algorithmically. To do so, first one chooses a set
of parent diagrams and momentum labels used in the integration.  To
sort the terms into diagrams it is useful to collect the terms
on each propagator or light-cone denominator type.  The propagators and
light-cone denominators encode the diagram type, but with labels 
not matching the
ones used in integration.  Assuming the external legs are ordered the
same way in the cut and integration labels (if not, one should permute
the external legs), the problem amounts to finding a mapping between a
given term in the cuts and the diagram to which it belongs.  
For an $n$-point $L$ loop diagram there
are $L$ independent loop momenta and $n-1$ independent external
momenta, under momentum conservation.  If the loop propagators carry
momenta $p_i$ in one set of labelings and $q_i$ in another labeling,
then a change of variables should exist relating the two of the form,
\begin{equation}
q_i = \sum_{j=1}^L a_{ij} p_j + \sum_{j=1}^{n-1} b_{ij} k_j \,, 
\label{ChangeOfVariables}
\end{equation}
where the $a_{ij}$ and $b_{ij}$ are in the set $\{-1,0 ,1\}$.  This
form assumes that for both the original and final momentum labels the
momentum of each internal leg is a sum or difference of the
independent momenta, as naturally arises either in cut or Feynman
diagram momentum routings.  One sweeps over all changes of variables
in \eqn{ChangeOfVariables} until a match is found.  If no change of
variables matches then the subsequent diagrams should be checked until
a match is found.  An efficient way to rule out putative changes of
variables is to first set all external momenta to zero, then re-introduce
them one by one, eliminating candidate diagrams and solving for the $b_{ij}$.

As a simple example, consider the double triangle diagram
with a light-cone denominator occurring in our calculation of the 
two-loop splitting amplitude.
From its origin in the three-particle cuts shown 
in~\fig{TwoLoopLCExampleFigure}(b), where the cut momenta are
$\{ q_1, -q_1-q_2-k_1-k_2, q_2 \}$,   
this diagram is described by the set of propagators and light-cone
denominators
\begin{equation}
\{q_1^2, q_2^2, (q_2 +k_2)^2, (q_2 +k_1+k_2)^2, 
   (q_2+q_1 +k_1 +k_2)^2, (q_1 +k_1+k_2)^2, q_1\cdot n\} \,.
\label{CutLabelsExample}
\end{equation}
The denominator variables used in the integration routines, shown
in~\fig{PlanarTriFigure}(a), are
\begin{equation}
\{p_1^2, (p_1+k_1+k_2)^2, p_2^2, (p_2-k_1-k_2)^2, (p_2-k_1)^2, (p_1+p_2)^2, 
  p_1 \cdot n\} \,
\label{IntegrationLabelsExample}
\end{equation}
where $p_2$ is labeled as $p_3$ in the figure.
After sweeping through the change of variables in
\eqn{ChangeOfVariables} one finds that the two sets of momenta of momenta
are related by
\begin{equation}
q_1 = p_1\,, \hskip 2 cm 
q_2 = p_2 - k_1 - k_2\,.
\end{equation}
With this change of variable the two sets of propagators and
denominators in \eqns{CutLabelsExample}{IntegrationLabelsExample} are
identical, except for the order in which they appear.


\end{document}